\documentclass[12pt,twoside]{article}
\usepackage{amssymb}
\usepackage{amsmath}
\usepackage{latexsym}

\newtheorem{theorem}{Theorem}[section]

\newtheorem{lemma}[theorem]{Lemma}
\newtheorem{proposition}[theorem]{Proposition}

\newtheorem{conjecture}[theorem]{Conjecture}
\def\unit{\mbox{1\hspace{-1mm}l}}

\begin{document}

\begin{titlepage}
\noindent
{\huge Herbert Spohn}\\
\rule[-1.5cm]{7cm}{1mm}\\\vspace{4.5cm}
\begin{center}
{\huge\bf Dynamics of Charged Particles \bigskip\bigskip\\
 and Their Radiation Field}\\
\vspace{1cm}
\end{center}
\newpage
\noindent
Prof. Dr. Herbert Spohn\\
        Zentrum Mathematik und Physik Department\\
        Technische Universit\"{a}t M\"{u}nchen\\
        D-80290 M\"{u}nchen, Germany\\
         spohn@mathematik.tu-muenchen.de\bigskip\\
\end{titlepage}

\noindent
{\bf\Large Preface}\bigskip\\
By intention, my project has two parts. The first one covers the
classical electron theory. It is essentially self--contained and
will be presented in the following chapters. 75 years after the
discovery of quantum mechanics, to discuss only the classical version
of the theory looks somewhat obsolete, in particular since many
phenomena, like the stability of atoms, the existence of spectral
lines and their life time, the binding  of atoms, and many others,
are described only by the quantized theory. Thus
it is a necessity to discuss the quantized version of the
classical models studied here. This is {\it not} quantum
electrodynamics. It is the quantum theory of electrons, stable
nuclei, and photons with no pair production allowed. Well said,
but the quantum part turns out to be a difficult task. There is a
lot of material with the mathematical physics side in flux and
very active at present. Thus it remains to be seen whether the
quantum part will be ever finished. In the meantime I invite the
reader to comments, criticisms, and improvements on the classical
part.

In thank my collaborators, Sasha Komech and Markus Kunze, for
their constant help and insistence. I am very grateful to Joel
Lebowitz. He initiated my interest in tracer particle problems
and I always wanted to apply these ideas to electrons coupled to
the Maxwell field. I am indebted to F. Rohrlich for discussions and
important hints on the literature.
 I acknowledge instructive discussions with
 A.  Arai, V. Bach, D. Bambusi, G. Bauer, J. Bellissard, F.
Bonetto, D. D\"{u}rr, J.-P. Eckmann, L. Erd\"{o}s,
J. Fr\"{o}hlich, L. Galgani, G. Gallavotti,
S. Goldstein, M. Hirokawa, F. Hiroshima, V. Jak$\check{\mathrm s}$i\'{c}, M. Kiessling, E. Lieb,
M. Loss,  D. Noja, C.-A. Pillet, M. Rauscher, L. Rey--Bellet,   H.T. Yau.

I am most thankful
to Mrs. L. V\'{a}squez de Rosswag for her
cheerful and tireless effort in typing.\\

\noindent
Herbert Spohn \hfill M\"{u}nchen, August 1, 1999
\newpage
\hspace{1cm}
\newpage
\tableofcontents
\newpage
\setcounter{section}{-1}
\section{Introduction}\label{sec.in}
``Classical Electron Theory'' is an attempt of building a
dynamical theory of electromagnetic fields coupled to well
concentrated lumps of charges, like electrons and nuclei. As the
name indicates, it is a theory of classical fields in interaction
with classical particles. Two limiting cases have been confirmed
experimentally and are the subject of any course on
electrodynamics: either the currents, i.e. the motion of the
charges, or the electromagnetic fields are given with the task to
predict the behavior of the remaining piece. In stark contrast,
there are only few examples where one is still in the classical
domain and the full power of a coupled theory is required. In
fact, I only know one example, namely the motion of a single
electron in a Penning trap which will be discussed in Section
\ref{sec.g}. Thus the classical electron theory is a mostly
theoretical enterprise, but with this taste it has intrigued
physicists for almost a century.

The motives have varied with time. The founding fathers, like
Abraham and Lorentz, tried to develop a dynamic theory for the
then newly discovered electron. In particular, depending on the
model, they predicted its energy--momentum relation,
cf. Section \ref{sec.c}. This enterprise came to a stand still
with the advent of the
 theory of  special relativity. Based on totally disjoint arguments
it required for any massive particle the relativistically
covariant link between energy and momentum. The classical electron
theory flourished for a second time in the early days of quantum
electrodynamics. The hope was that a refined understanding of the
classical theory would give a hint on how to properly quantize
and how to correctly handle the infinities. The notion of mass
renormalization was taken from the classical theory. But as the
proper quantum theory surfaced, the classical considerations had
little value. In fact, there is no classical analogue of the
renormalization in quantum field theory. There the interaction
with the quantized field reduces the quantum fluctuations of the
electron. The, yet unproven, construction is to counterbalance by
increasing the bare electron fluctuations through letting its
bare mass tend to {\it zero} in such a way that, upon removing the
ultraviolet cut--off, finite
fluctuations remain which are adjusted to yield the experimental
mass of the electron.

At large, the classical electron theory has a poor reputation. One
has to fight with infinities, the bare mass of the electron is
supposedly negative and tends to $-\infty$ in the limit of a point
charge, the theory has instabilities as demonstrated by the runaway
solutions, physical solutions suffer from preacceleration in
contradiction to a causal description, and more. In
contrast, our
approach will be rather conservative with no need for  such spectacular
revisions of the conventional concepts. It is based on two corner stones
\begin{itemize}
 \item a well--defined dynamical theory of extended charges in
interaction with the electromagnetic field,
\item a study of the effective dynamics of the charges under the
condition that the external potentials vary slowly on the scale
given by the size of the charge distribution. This is the {\it
adiabatic limit}.
\end{itemize}
Of course, the familiar objects reappear in disguise with one
important difference. In the adiabatic limit the ratio `mass
induced through the self--interaction with
the electromagnetic field' to `bare mass' remains finite
and does not diverge as it is the case in the traditional point charge limit.

Our approach reflects the great progress which has taken place in the theory of
dynamical systems. After all, a charge coupled to its radiation
field is just one particular case, however with some rather
special features. Perhaps the most unusual one is the appearance
of a central manifold in the effective dynamics if friction
through radiation is included.

A few words on the style and scope of the book are in order.
First of all we systematically develop the theory, no review is
intended. For a subject with a long history, such an attitude
looks questionable. After all, what did the many physicists in that field
contribute? To make up, we included one historical chapter, which
as very often in physics is the history as viewed from our present
understanding. Since there are already excellent historical
studies, we hope to be excused. Further we added at the end of
each chapter {\it Notes and References} which are intended as a guide to
all the material which has been left out.  As prerequisites for
reading, only a basic knowledge of Maxwell's theory of
electromagnetism and of Hamiltonian mechanics is required.

The reader will notice that we state several theorems and give
their proof. In some cases the proof is complete. In other cases
we only present the essential idea. The more technical steps can
be found in the original literature and are therefore not
reproduced. I am convinced that, in particularly in a subject
where there is so little control through experience, one needs
fixed points in terms of mathematical theorems. Over the years a
lot of common knowledge  has been accumulated and we made an effort to
disentangle hard facts from truth by tradition.

The introduction might give the impression as if all problems are
resolved and the classical electron theory is in good shape. This
would be too simplistic an attitude. What I hope is to convince
the reader to view our dynamical problem from a particular
perspective. Once this point has been reached, there are then many loose
ends. To mention only two: our comparison between the true and
approximate particle dynamics could be sharper and a fully
relativistic theory of an extended charge distribution exists only
on a formal level. The greatest reward would be if my notes
encourage further research.
\newpage
\section{A Charge Coupled to its Electromagnetic
Field}\label{sec.a}
\setcounter{equation}{0}
We plan to study the dynamics of a well localized charge,
like an electron or a proton, when coupled to its own electromagnetic field.
The case of several particles is reserved for Chapter \ref{sec.i}.
In a first attempt, one models the particle as a point charge
with some definite mass. If its world line is prescribed, then
the fields are determined through the inhomogeneous Maxwell equations. On
the other hand, if the electromagnetic fields are given, then the
motion of the point charge is governed by
the Lorentz force equation.
While it then seems obvious how to marry the two equations, so to have a
coupled dynamics for the charge and its electromagnetic field,
ambiguities and inconsistencies arise due
to the infinite electrostatic energy of the Coulomb field of the
 point charge. Thus one is forced to introduce a slightly
smeared charge distribution, i.e. an extended charge model.
Mathematically this means that the interaction between particle and field is
cut--off or regularized at short distances, which seems to leave a
lot of freedom. There are strong constraints however. In
particular one has to satisfy local charge conservation, the
theory should be of Lagrangian form, and
it has to reproduce the limiting cases mentioned above.
 In addition, as to be expected for any decent
physical model, the theory should be well--defined and
empirically accurate in its domain of validity. In fact, only two
models have been worked out in some detail: (i) the semi--relativistic Abraham model
of a rigid charge distribution, and (ii) the Lorentz model of a
relativistically covariant extended charge distribution. The aim
of this chapter is to introduce both models at some length. On the
way we recall some properties of the Maxwell equations for later
use.


\subsection{The Maxwell equations}\label{sec.aa}
We prescribe a charge density $\rho(\boldsymbol {x},t)$ and an
associated
current, $\boldsymbol {j}(\boldsymbol {x},t)$, which are linked through the
conservation law
\begin{equation}\label{aa.a}
        \partial_t \,\rho(\boldsymbol {x},t) +
                \nabla \cdot \boldsymbol {j}
        (\boldsymbol {x},t) = 0 \,.
\end{equation}
Of course, $\boldsymbol {x} \in{\mathbb R}^3$, the physical space, and
$t\in{\mathbb
R}$, the time. The Maxwell equations for the electric field $\boldsymbol {E}$ and
the magnetic field $\boldsymbol {B}$ consist of the two evolution equations
\begin{eqnarray}\label{aa.b}
        c^{-1}\partial_t \boldsymbol {B} (\boldsymbol {x},t)
              &=& - \nabla \times
        \boldsymbol {E}(\boldsymbol {x},t) \, ,\nonumber\\[2mm]
        c^{-1}    \partial_t \boldsymbol {E} (\boldsymbol {x},t)
              &=& \nabla \times
        \boldsymbol {B}(\boldsymbol {x},t) - c^{-1}\boldsymbol {j}(\boldsymbol {x},t)
\end{eqnarray}
and the two constraints
\begin{equation}\label{aa.c}
        \nabla \cdot \boldsymbol {E}(\boldsymbol {x},t) =
                \rho(\boldsymbol {x},t)
         \, , \quad \nabla\cdot \boldsymbol {B}(\boldsymbol {x},t)=0  \, .
\end{equation}

We use the Heaviside--Lorentz units. The vacuum susceptibilities
are $\varepsilon_0=1= \mu_0$, which fixes the unit of charge. $c$
is the speed of light. Mostly we will set $c=1$ for convenience,
thereby linking the units of space and time. If needed, one can easily
 reintroduce these natural constants in
the convential way. At some parts we will do this without notice,
so to have the dimensions right and to better keep track of the
order of magnitudes.

We solve  the Maxwell equations as a Cauchy problem, i.e. by
prescribing the fields at time $t=0$. If the constraints (\ref{aa.c})
are satisfied at $t=0$, then by the continuity equation
(\ref{aa.a}) they are satisfied at all times. Thus the initial
data are
\begin{equation}\label{aa.d}
      {\boldsymbol E}(\boldsymbol {x},0) \, ,
          ~ {\boldsymbol B}(\boldsymbol {x},0)
\end{equation}
together with the constraints
\begin{equation}\label{aa.e}
       \nabla \cdot \boldsymbol {E}(\boldsymbol {x},0) =
       \rho(\boldsymbol {x},0) \, ,
        \quad \nabla\cdot
       \boldsymbol {B}(\boldsymbol {x},0) = 0 \, .
\end{equation}
The choice $t=0$ is merely a convention. In some cases it is
preferable to
prescribe the fields either in the distant past or the remote
future. We will only consider physical situations where the fields
decay at infinity and thus have the finite energy
\begin{equation}\label{aa.f}
        {\mathcal E}= \frac{1}{2} \int d^3x\, \big ( \boldsymbol
        {E}(\boldsymbol {x},t)^2 +
         \boldsymbol {B}(\boldsymbol {x},t)^2 \big ) < \infty \,.
\end{equation}
In a thermal state at non--zero temperature, one would be forced
to consider infinite energy solutions. But this is outside the
present scope.

The Maxwell equations (\ref{aa.b}), (\ref{aa.c}) are inhomogeneous wave
equations and thus easy to solve. This will be done in
Fourier space first, where Fourier transform is denoted by\;
$\widehat{}$\;
and defined through
\begin{equation}\label{aa.g}
        \widehat f (\boldsymbol {k}) = (2 \pi)^{-n/2} \int d^n x\,
         e^{-i\boldsymbol {k}\cdot \boldsymbol {x}} f(\boldsymbol {x})\, .
\end{equation}
Then (\ref{aa.b}) becomes
\begin{eqnarray}\label{aa.h}
       \partial_t \boldsymbol {\widehat B}
                (\boldsymbol {k},t) &=& - i
        \boldsymbol {k} \times \boldsymbol {\widehat E}(\boldsymbol
        {k},t)\,, \nonumber\\[2mm]
        \partial_t \boldsymbol {\widehat E}
                (\boldsymbol {k},t) &=& i
         \boldsymbol {k} \times \boldsymbol {\widehat B}(\boldsymbol {k},t)
        - \boldsymbol {\widehat j}(\boldsymbol
        {k},t)
\end{eqnarray}
with the constraints
\begin{equation}\label{aa.i}
i \boldsymbol {k} \cdot \boldsymbol {\widehat E}
(\boldsymbol {k},t) =
\hat \rho(\boldsymbol {k},t)\, , \quad i \boldsymbol {k}\cdot \boldsymbol
{\widehat B}(\boldsymbol {k},t) = 0
\end{equation}
and the conservation law
\begin{equation}
        \partial_t
        \hat\rho(\boldsymbol {k},t) +
        i\boldsymbol {k} \cdot \boldsymbol {\widehat j} (\boldsymbol {k},t)
        = 0\label{aa.j}\,.
\end{equation}
 To solve  the inhomogeneous equations (\ref{aa.h}), as usual, we
rely on the solution of the
homogeneous equations,
\begin{eqnarray}\label{aa.k}
        \left( \begin{array}{{ll}}
        {\boldsymbol{\widehat E}}_{(0)}(\boldsymbol {k},t)\\
        \boldsymbol {\widehat B}_{(0)}(\boldsymbol {k},t)
        \end{array}\right )
        &= & \Big [ \big (\cos|\boldsymbol {k}|t +
       (1- \cos|\boldsymbol {k}| t)\, \frac{1}{\boldsymbol {k}^2}\,
        |\boldsymbol {k}\rangle\langle \boldsymbol {k}|\big )
        \left( \begin{array}{{ll}}
        {\boldsymbol {\widehat E}}(\boldsymbol {k},0)\\
        \boldsymbol {\widehat B}(\boldsymbol {k},0) \end{array}\right )\nonumber\\
        &&+ (\frac{1}{| \boldsymbol {k}|}
        \sin| \boldsymbol {k}| t) i \boldsymbol {k}\times
        \left( \begin{array}{{ccc}}
        {\boldsymbol {\widehat B}}(\boldsymbol {k},0)\\
        -{\boldsymbol{\widehat E}}(\boldsymbol {k},0) \end{array}\right )\Big
        ]\,.
\end{eqnarray}
We insert (\ref{aa.k}) in the time--integrated version of
(\ref{aa.h}).
We impose the
constraints and,  with one partial integration and using the conservation
law, arrive at
\begin{eqnarray}\label{aa.l}
\boldsymbol {\widehat E}(\boldsymbol {k},t)
                &=& (\cos|\boldsymbol {k}|t)
        \boldsymbol {\widehat E}(\boldsymbol {k},0)
                + (|\boldsymbol {k}|^{-1} \sin |\boldsymbol {k}|
        t) i \boldsymbol {k}
                \times \boldsymbol {\widehat B}(\boldsymbol {k},0)\nonumber\\
        &&+ \int\limits_0^t ds \big(-(|\boldsymbol {k}|^{-1}
                \sin |\boldsymbol {k}| (t-s)) i
         \boldsymbol {k}
        \hat\rho(\boldsymbol {k},s) - (\cos|\boldsymbol {k}| (t-s))
                \boldsymbol {\widehat j}
        (\boldsymbol {k},s) \big)\nonumber\\[2mm]
        &=& \boldsymbol {\widehat E}_{(0)} (\boldsymbol {k},t)
                + \boldsymbol {\widehat E}_{
        \mathrm {ret}} (\boldsymbol {k},t)\, ,\\[2mm]
        \boldsymbol {\widehat B}(\boldsymbol {k},t)
                &=& (\cos|\boldsymbol {k}| t)
        \boldsymbol {\widehat B}(\boldsymbol {k},0) -
                (|\boldsymbol {k}|^{-1} \sin
        |\boldsymbol {k}|t) i \boldsymbol {k} \times \boldsymbol {\widehat E}
        (\boldsymbol {k},0)\nonumber\\
         &&+ \int\limits_0^t ds (|\boldsymbol {k}|^{-1}
                 \sin |\boldsymbol {k}| (t-s)) i
         \boldsymbol {k}
         \times \boldsymbol {\widehat j}(\boldsymbol
         {k},s)\nonumber\\[2mm]
        &=& \boldsymbol {\widehat B}_{(0)} (\boldsymbol {k},t) +
                \boldsymbol {\widehat B}_{\mathrm {ret}} (\boldsymbol
                {k},t)\,.\label{aa.m}
\end{eqnarray}
The first terms are
the  initial fields propagated up to time $t$, while the
second terms are the retarded fields. If we would solve the Maxwell equations
into the past, then the retarded fields are to be replaced by the
advanced fields.

Let us introduce the fundamental propagator, $G_t(\boldsymbol {x})$, of
the wave
equation which satisfies
\begin{equation}\label{aa.n}
        \partial_t^2\, G -
        \Delta G = \delta (\boldsymbol{x})\delta (t)
\end{equation}
and is defined as the Fourier transform of $(2 \pi)^{-3/2}\,
|\boldsymbol {k}|^{-1} \sin |\boldsymbol {k}| t$. This means
$G_t (\boldsymbol {x}) = (2 \pi)^{-1}\, \delta (|\boldsymbol {x}|^2
- t^2)$ and for $t \ge 0$
\begin{equation}\label{aa.o}
        G_t(\boldsymbol {x})= \frac{1}{4
                \pi t} \delta (|\boldsymbol{x}|-t)\,.
\end{equation}
 Then in physical space the solution (\ref{aa.l}), (\ref{aa.m})
of the inhomogeneous  Maxwell
equations reads   as
\begin{eqnarray}\label{aa.p}
        \boldsymbol {E}(t) &=& \partial_t G_t * \boldsymbol {E}(0)  +
                \nabla \times G_t
        * \boldsymbol {B}(0)
         -\int\limits_0^t ds \big (\nabla G_{t-s} * \rho(s) + \partial_t G_{t-s} *
        \boldsymbol {j}(s)\big)\nonumber \\
        &=& \boldsymbol {E}_{(0)} (t) + \boldsymbol
                {E}_{\mathrm {ret}} (t)\,,\\
        \boldsymbol {B}(t) &=& \partial_t  G_t * \boldsymbol{B}(0) -
                \nabla \times G_t *
        \boldsymbol {E}(0)
         + \int\limits_0^t ds \nabla \times G_{t-s} *
                 \boldsymbol {j}(s)\nonumber\\
        &=& \boldsymbol {B}_{(0)} (t) +
                \boldsymbol {B}_{\mathrm {ret}} (t) \, .\label{aa.q}
\end{eqnarray}
Here $\ast$
 denotes convolution, i.e. $f_1 \ast f_2(\boldsymbol {x}) =
 \int d^n y  f_1 (\boldsymbol {x}-\boldsymbol {y})  f_2(\boldsymbol {y})
 $.

The expressions (\ref{aa.p}), (\ref{aa.q}) remain meaningful
when  $\rho, \boldsymbol {j}$
are generated by the motion of a single point charge. Let us
denote by $\boldsymbol {q}(t)$ the position and $\boldsymbol {v}(t) =
 \dot{\boldsymbol{q}} (t)$ the velocity
of the particle carrying the charge $e$. Then
\begin{equation}\label{aa.r}
        \rho(\boldsymbol {x},t) = e \delta (\boldsymbol{x}-
        \boldsymbol {q}(t)) ~, ~ \boldsymbol{j}(\boldsymbol {x},t) =
         e \delta(\boldsymbol{x} - \boldsymbol{q}(t)) \boldsymbol {v}(t)\, .
\end{equation}
It is assumed that the particle is relativistic and therefore $|
\boldsymbol{v}(t)| <
1$.
Inserting in (\ref{aa.p}), (\ref{aa.q}) one arrives at the Li\'{e}nard--Wiechert
fields. Since their derivation is handled in any textbook, we do
not repeat the computation here and
discuss only the result. We imagine that the world line, $t \mapsto
\boldsymbol {q}(t)$, of the particle is given for all times. We prescribe the
initial data for the fields at time $t=t_0$ and take the limit $t_0 \to - \infty$
in
(\ref{aa.p}), (\ref{aa.q}). Then, at a fixed space--time
point $(t,\boldsymbol {x})$, the
contribution from the initial fields vanishes and the retarded
fields become the Li\'{e}nard--Wiechert fields. To define them we
introduce the retarded time $t_\mathrm {ret}$, depending on
$t,\boldsymbol {x}$, as the unique solution of
\begin{equation}\label{aa.s}
       t_{\mathrm {ret}} = t - |\boldsymbol {x}-\boldsymbol {q}
       (t_{\mathrm {ret}})|\,.
\end{equation}
 $t_\mathrm {ret}$ is the unique time when the world line crosses the backwards
 light cone with apex at $(t,\boldsymbol {x})$. We also introduce the unit vector
\begin{equation}\label{aa.t}
  \boldsymbol {\widehat n} = \frac{\boldsymbol {x}-
  \boldsymbol {q}(t_{\mathrm {ret}})}{|
        \boldsymbol {x}- \boldsymbol {q}(t_{\mathrm{ret}}) |}\, .
\end{equation}
Then the electric field generated by a moving point charge is
given by
\begin{equation}\label{aa.u}
        \boldsymbol {E}(\boldsymbol {x},t)= (e/4 \pi) \Big [\, \frac{(1-
         \boldsymbol {v}^2)(\boldsymbol
         {\widehat n} - \boldsymbol {v})}
         {(1-\boldsymbol {v} \cdot \boldsymbol {\widehat n})^3
        |\boldsymbol {x}-\boldsymbol {q}|^2}
        + \frac{\boldsymbol {\widehat n} \times[(\boldsymbol {\widehat n}
        -\boldsymbol {v}) \times \dot{\boldsymbol{v}}]}{(1-\boldsymbol {v} \cdot
        \boldsymbol {\widehat n})^3 |\boldsymbol {x}-\boldsymbol {q}|}\,
        \Big ]|_{t=t_{\mathrm {ret}}}
\end{equation}
and the corresponding magnetic field is
\begin{equation}\label{aa.v}
       \boldsymbol {B}(\boldsymbol {x},t) = \boldsymbol {\widehat n}
       \times \boldsymbol {E}(\boldsymbol {x},t)\,.
\end{equation}

(\ref{aa.u}) and (\ref{aa.v}) are less explicit
than the notation suggests, since $t_{\mathrm {ret}}$ depends
through (\ref{aa.s}) on the reference point $(t, \boldsymbol {x})$ and the
particle trajectory.
The first contribution in (\ref{aa.u}) is proportional to
$|\boldsymbol {x}-\boldsymbol {q}|^{-2}$
and independent of the acceleration. This is the near field which
in a certain sense remains attached to the particle. The second
contribution is proportional to $|\boldsymbol {x}-\boldsymbol {q}|^{-1}$
and to the
acceleration. This is the far field, which carries the information
on the radiation field escaping to infinity. If $\boldsymbol{q}(t)$ is smooth
in $t$, then the Li\'{e}nard--Wiechert fields are smooth functions except
at $\boldsymbol{x}=\boldsymbol{q}(t)$, where they diverge as
$|\boldsymbol{x}-\boldsymbol{q}(t)|^{-2}$. The
corresponding potentials have a Coulomb singularity at the world
line of the particle.


\subsection{The Lorentz force equation}\label{sec.ab}
We take now the point of view that the electromagnetic fields
$\boldsymbol {E},\boldsymbol{B}$ are given. The motion of a charged particle,
charge $e$, position $\boldsymbol{q}(t)$, velocity $\boldsymbol{v}(t)$,
is then governed by the Lorentz force equation
\begin{equation}\label{ab.a}
       \frac{d}{dt} (m_0 \gamma \boldsymbol{v}(t)) = e
       \big (\boldsymbol{E}(\boldsymbol{q}(t),t) + c^{-1}\boldsymbol{v}(t) \times
       \boldsymbol{B}(\boldsymbol
       {q}(t),t)\big)\,,
\end{equation}
which as ordinary differential equation has to be supplemented
by the initial conditions $\boldsymbol{q}(0), \boldsymbol
 {v}(0)$. Here $\gamma(\boldsymbol {v})= 1/\sqrt{1-(\boldsymbol {v}/c)^2}$.
 The particle is relativistic with   rest mass $m_0$ as measured
 experimentally.
  Once  the particle is coupled to
the Maxwell field, $m_0$ will attain a new meaning.

The $(\boldsymbol {E},\boldsymbol {B})$ fields in (\ref{ab.a})
 are not completely arbitrary.
They have to be solutions of the Maxwell equations with sources
$(\rho,\boldsymbol {j})$. So to speak, we have separated all
charges into a single
charged particle whose motion is to be determined through
(\ref{ab.a}) and a rest whose motion is taken to be known.

(\ref{ab.a}) is of Hamiltonian form. To see this we introduce the
vector potentials $\phi, \boldsymbol {A}$ such that
\begin{equation}\label{ab.b}
       \boldsymbol {E}(\boldsymbol {x},t) = - \nabla \phi (\boldsymbol {x},t)
       - c^{-1}\partial_t\boldsymbol {A}(\boldsymbol
       {x},t)\, ,
       \quad \boldsymbol {B}(\boldsymbol {x},t) = \nabla \times
       \boldsymbol {A}(\boldsymbol {x},t)\, .
\end{equation}
Then the Lagrangian associated to (\ref {ab.a}) is
\begin{equation}\label{ab.c}
       L(\boldsymbol{q}, \dot{\boldsymbol{q}}) = - m_0 c^2
       (1- c^{-2}\dot{\boldsymbol{q}}^2)^{1/2} - e(\phi (\boldsymbol{q},t)
       - c^{-1} \dot{\boldsymbol{q}} \cdot\boldsymbol {A}(\boldsymbol {q},t))\, .
\end{equation}
Introducing the canonical momentum
\begin{equation}\label{ab.d}
       \boldsymbol {p} = m_0 \gamma (\dot{\boldsymbol{q}})
       \dot{\boldsymbol{q}} + \frac{e}{c} \boldsymbol{A}(\boldsymbol {q},t)
\end{equation}
the Hamiltonian  is given by
\begin{equation}\label{ab.e}
        H(\boldsymbol {q},\boldsymbol {p},t) = (( c \boldsymbol
        {p-}e \boldsymbol {A}(\boldsymbol {q},t))^2
        + m_0^2 c^4 )^{1/2} +e\phi(\boldsymbol {q},t) \,.
\end{equation}
In particular, if the fields are time--independent, then the
energy
\begin{equation}\label{ab.f}
        {\cal E}(\boldsymbol {q},\boldsymbol {v}) = m_0 \gamma (\boldsymbol {v})
        + e \phi(\boldsymbol {q})
\end{equation}
is conserved along the solution trajectories of (\ref{ab.a}).

It should be remarked that in general the solutions to the Lorentz
force equation will have a complicated structure even for
time--independent fields. This has been amply demonstrated for
particular cases. Depending on the external fields the motion ranges from
regular to fully chaotic  with a mixed phase space as a rule.


\subsection{The coupled Maxwell--Lorentz equations}\label{sec.ac}
While, up to a minute fraction, sufficient for all of
electrodynamics,
from a more fundamental point of view it is unsatisfactory that in
the Maxwell and Lorentz force equations so to speak ``one half'' must be prescribed
and one would hope to have a coupled system of equations for the time
evolution of the charged particles together with their electromagnetic
field. If we restrict ourselves to a single particle, it is
obvious how to proceed. From (\ref{aa.b}), (\ref{aa.c}) we have
\begin{eqnarray}\label{ac.a}
       \partial_t \boldsymbol {B}
       (\boldsymbol {x},t) &=& - \nabla \times
       \boldsymbol {E}(\boldsymbol {x},t)\, ,\nonumber\\[3mm]
      \partial_t \boldsymbol {E}
       (\boldsymbol {x},t) &=&
        \nabla \times \boldsymbol {B}(\boldsymbol {x},t) -
       e \delta(\boldsymbol {x}-\boldsymbol {q}(t))
       \boldsymbol {v}(t)
\end{eqnarray}
with the constraints
\begin{equation}\label{ac.b}
       \nabla \cdot \boldsymbol{E}(\boldsymbol{x},t)
        = e \delta(\boldsymbol{x}-\boldsymbol{q}(t))\, , \quad
       \nabla \cdot \boldsymbol{B}(\boldsymbol{x},t) = 0
\end{equation}
and from (\ref{ab.a}) we have
\begin{equation}\label{ac.c}
        \frac{d}{dt} (m_0 \gamma \boldsymbol {v}(t))  =
        e \big (\boldsymbol {E}_{\mathrm {ex}}(\boldsymbol {q}(t)) +
        \boldsymbol {E}(\boldsymbol {q}(t), t) + \boldsymbol {v}(t) \times
        (\boldsymbol {B}_{\mathrm {ex}} (\boldsymbol {q}(t))
         + \boldsymbol {B}(\boldsymbol {q}(t),t))\big ) \, .
\end{equation}
We added in explicitely external electromagnetic fields
$\boldsymbol {E}_{\mathrm{ex}}$, $\boldsymbol {B}_{ex}$, which will
 play a prominent role later on. They are derived from potentials as
\begin{equation}\label{ac.d}
        \boldsymbol{E}_{\mathrm{ex}} = - \nabla\phi_{\mathrm{ex}}\,,\quad
        \boldsymbol {B}_{\mathrm {ex}} = \nabla \times
        \boldsymbol{A}_{\mathrm{ex}}\, .
\end{equation}
 We
assume the potentials  to be time--independent for simplicity, although
 considerable parts
of the theory  to be developed will work also for time--dependent fields. As before,
(\ref{ac.a}), (\ref{ac.c}) are to be solved as initial value
problem. Thus $\boldsymbol {E}(\boldsymbol{x},0), \,
\boldsymbol{B}(\boldsymbol{x},0), \,\boldsymbol{q}(0),$ $
\boldsymbol{v}(0)$ are given.
Note that the continuity equation is satisfied by fiat.

(\ref{ac.a}), (\ref{ac.c}) are the stationary points of a
Lagrangian action, which strengthens our trust in these equations,
since every microscopic classical evolution equation seems to be
of that form. We introduce the electromagnetic potentials as in
(\ref{ab.b}), (\ref{ac.d}). Then the action for (\ref{ac.a}), (\ref{ac.c}) reads
\begin{eqnarray}\label{ac.e}
       S([\boldsymbol {q}, \phi, \boldsymbol {A}]) &=&\int dt
       \big [-m_0
       (1- \dot{\boldsymbol{q}}(t)^2)^{1/2}
       -e \big (\phi_{\mathrm {ex}} (\boldsymbol {q}(t))
        + \phi (\boldsymbol {q}(t),t) \nonumber\\[3mm]
       &&\quad\quad-\dot{\boldsymbol{q}}(t) \cdot (\boldsymbol {A}_{\mathrm{ex}}
       (\boldsymbol {q}(t)) + \boldsymbol {A}(\boldsymbol {q}(t), t))\big )\big ]
          \\[3mm]
       &&  +\frac{1}{2} \int dt \int d^3x \big [
       (\nabla \phi(\boldsymbol {x},t) + \partial_{t}
       \boldsymbol {A}(\boldsymbol {x},t))^2- (\nabla \times \boldsymbol {A}
       (\boldsymbol {x},t))^2 \big ]\, . \nonumber
\end{eqnarray}
The only difficulty is that (\ref{ac.a}), (\ref{ac.c}) make no
sense mathematically. As explained, the solution to the Maxwell
equations is singular at $\boldsymbol{x}=\boldsymbol{q}(t)$ and in
the Lorentz force equation we
are asked to evaluate the fields exactly at that point. One might
be tempted to put the blame on the mathematics which refuses to
handle  equations as singular as (\ref{ac.a}), (\ref{ac.c}).
 However before such a drastic
conclusion is drawn, the physics should be properly understood. The point
charge  carries along with it a potential which at short distances
diverges as the Coulomb potential, cf. (\ref{aa.u}), and which
therefore has the electrostatic energy
\begin{equation}\label{ac.f}
       \frac{1}{2} \int\limits_{\{ |\boldsymbol {x}-
           \boldsymbol{q}(t)| \le R \}}
       d^3 x \boldsymbol {E}(\boldsymbol {x},t)^2
           \simeq \int\limits_0^R dr r^2
       (r^{-2})^2 = \int\limits_0^R dr r^{-2} = \infty \, .
\end{equation}
Taking literally such an object would have an infinite mass and
cannot respond to external forces. It would maintain its velocity
forever, which is not what is observed.

{\it Thus we are forced to regularize the Maxwell--Lorentz equation
(\ref{ac.a}), (\ref{ac.c}) at short distances.}

To carry out such a program there are two in part complementary
points of view. The first one, which we will {\it not} adopt here,
regards the regularization as a mathematical device with the sole
purpose to make sense out of a singular mathematical object through
a suitable limiting procedure. To mention only one prominent
mathematical physics example. The free scalar field, $\phi(\boldsymbol {x})$, in
Euclidean quantum field theory in 1+1 dimensions fluctuates so
wildly at short distances that an interaction as
$\int d^2 x V(\phi(\boldsymbol {x}))$ with $V(\phi) = \phi^2 + \lambda\phi^4$
cannot be properly defined. One way, not necessarily optimal,  to
regularize the theory is to introduce a spatial lattice with
lattice spacing $a$. Such a lattice field theory is well
defined in finite volume. One then carries out the limit
 spacing $a \to 0$
at a simultaneous readjustment of
the interaction potential, $V(\phi) = V_{a}(\phi)$
and obtains a Euclidean invariant,
interacting quantum field theory. Ideally the limit theory should
be independent of the regularization procedure. E.g. we could
start with the free scalar field in the continuum and also
regularize $\phi(\boldsymbol {x}) $ as $\phi \ast g({\boldsymbol {x}})$ with
 $g$ a test function
concentrated at $0$. Then the regularized interaction is
$\int d^2 x V(\phi \ast g(\boldsymbol {x}))$ and in the limit
$g ({\boldsymbol {y}})
 \mapsto \delta(\boldsymbol {y})$
a quantum field theory should be obtained identical to the one of
the lattice regularization.

In the second point of view one argues that there is a physical
cut--off coming from a more refined theory, which is then modelled
in a phenomenological way. While this is a standard procedure, it
is worthwhile to illustrate it in a concrete example. We consider
many $(\cong 10^{23})$ He$^4$ atoms in a container of adjustable size
and we want to compute their free energy according to the rules
of statistical mechanics. The more refined theory is here
non--relativistic quantum mechanics which treats the electrons and
nuclei as point particles carrying a spin $\frac{1}{2}$, resp.
spin  0. As far as we can tell, this model approximately
covers a temperature range $T = 0 ~{^\circ K}$ to $T=10^5~ {^\circ K}$,
way beyond dissociation, and a density range $\rho = 0$ to $\rho =
close~ packing$. Beyond that relativistic effects  must be taken into
account. However there is a more limited range  where we can get away
with a model of classical point particles interacting through an
effective potential of Lennard--Jones type. Once we specify this
pair potential, classical statistical mechanics makes well--defined
predictions at  {\it any} $T, \rho$. There is no limitation in
theory. Only outside a certain range the classical model looses
the correspondence to  the real world. Already from the way we
describe the physical cut--off, there is a fair amount of
vagueness. How much error do we allow in the free energy? What
about more refined properties like density correlations? An
effective potential can be defined quantum mechanically, but it is
temperature--dependent and never strictly pair. Despite all these
imprecisions and shortcomings, the equilibrium theory of fluids
relies heavily on the availability of a classical model.

In the same spirit we modify the Maxwell--Lorentz equations by
introducing an extended charge distribution as a phenomenological
model for the left out quantum electrodynamics. The charge
distribution is stabilized by strong interactions which are
outside the realm of electromagnetic forces. If  the particle is
an electron then on the classical level it looks like an extended
charged object with a size roughly of the order of
its Compton wave length, i.e. $4 \times 10^{-11}$ cm. We impose the
condition that the extended charge distribution has to be adjusted
such that, over the range where classical electrodynamics is
applicable, the  coupled Maxwell--Lorentz equations correctly reproduce the
empirical observations.

Such general clauses seem to leave a lot of freedom. However,
charge conservation and the derivability of the equations of
motion from  an action severely limits the possibilities. In fact,
essentially
only two models of an extended charge distribution have
been investigated so far.\\
(i) {\it The semi--relativistic Abraham model of a rigid charge distribution.} The
$\delta$--function in (\ref{ac.a}), (\ref{ac.b}), and
(\ref{ac.c}) is replaced by a
smooth function $\rho( \boldsymbol{x})$ which is radial and vanishes
 for $|\boldsymbol{x}| \ge
R_\rho$. In Fourier space it means that couplings between  the particle
and
Fourier modes with $|\boldsymbol{k}| \ge R_\rho^{-1}$ are smoothly suppressed.
The Abraham model will be studied in considerable detail
. While defined for all
velocities $|\boldsymbol{v}(t)| < 1$, it becomes empirically incorrect at
velocities close to one. Despite this draw--back we hope that the
Abraham model serves as a blue--print for more realistic cut--off
prescriptions.\\
(ii) {\it The relativistic Lorentz model of  a deformable charge distribution.} The
Abraham model violates relativistic invariance, since
the choice of $\rho$ singles out a specific reference frame. More in
accord with relativity is to require that the charge distribution
is radial in its momentary rest frame.
We will discuss details in Section \ref{sec.ae} and only remark already now
that relativistic invariance forces the equations of motion to be  nonlocal
in time.

We emphasize that for extended charge models the diameter
 $R_\rho$ of the charge distribution
 defines a  length (and upon dividing by $c$ also a
time) scale, relative to which the approximate validity of   effective theories,
like the Lorentz--Dirac equation,
can be
 addressed quantitatively. In fact, apart from the external
 forces, $R_\rho$ is the only length scale available.
\subsection{The Abraham model}\label{sec.ad}
Following Abraham, we model the charge as a spherically
symmetric, rigid body to which the charge is permanently
attached.
To be specific the charge distribution, $\rho$,
with total charge $e$ is assumed to be smooth, radial,
and
 supported in a ball of radius
 $R_\rho$, i.e.\medskip\\
{\it Condition $(C)$}:
\begin{equation}\label{ad.a}
        \rho \in C_0^\infty ({\mathbb R}^3)\,,
                ~\rho(\boldsymbol {x}) =
        \rho_r (|\boldsymbol {x}|)\,, ~\rho(\boldsymbol{x}) = 0
        ~ \mbox{for} ~|\boldsymbol{x}| \ge
        R_\rho\, , \int d^3 x \rho(\boldsymbol {x})= e\, .
\end{equation}
{\it To be definite, we require $e>0$. With the obvious sign change, also $e<0$
is then covered.}\medskip\\
Equivalenty, we could introduce the form factor $f$ such that
\begin{equation}\label{ad.y}
        \rho(\boldsymbol {x}) = e\, f(\boldsymbol {x}^2)~, ~
        \quad \int d^3 x\, f(\boldsymbol {x}^2) = 1\,.
\end{equation}
The equations of motion for the
Abraham model read then
\begin{eqnarray}
       \partial_t\boldsymbol {B}
      (\boldsymbol {x},t)
       & =& - \nabla \times \boldsymbol {E}(\boldsymbol
       {x},t)\,,\nonumber\\[2mm]
       \partial_t
      \boldsymbol {E}(\boldsymbol {x},t) &=&  \nabla \times
      \boldsymbol {B}(\boldsymbol {x},t)- \rho(\boldsymbol {x}-\boldsymbol {q}
      (t)) \boldsymbol {v}(t)\, ,\label{ad.b}\\[3mm]
       \nabla \cdot \boldsymbol {E}(\boldsymbol {x},t) &=&  \rho
        (\boldsymbol {x}-\boldsymbol {q}(t))\, , \quad \nabla \cdot
       \boldsymbol {B}(\boldsymbol {x},t) = 0 \,, \label{ad.c}\\[3mm]
        \frac{d}{d t} (m_{{\mathrm b}} \gamma \boldsymbol {v}(t)) =  \int d^3
        x \!\!\!\!\!\!\!\!\!\!\!\!&& \rho
        (\boldsymbol {x}- \boldsymbol {q}(t))\big [\boldsymbol {E}_{\mathrm {ex}}
         (\boldsymbol {x}) +
        \boldsymbol {v}(t) \times  \boldsymbol {B}_{\mathrm {ex}}
        (\boldsymbol {x})\nonumber\\
        &&+ \boldsymbol {E}(\boldsymbol {x},t) + \boldsymbol {v}(t)
         \times \boldsymbol {B}(\boldsymbol {x},t)\big ]\, .\label{ad.d}
\end{eqnarray}
In contrast to the Lorentz force equation, for the Abraham model we
denote the mechanical mass of the particle by $m_{{\mathrm b}}$ to emphasize
that this bare mass will differ from the observed mass of the
compound object ``particle plus surrounding Coulomb field''.
For the external potentials $\phi_{\mathrm{ex}},
\boldsymbol{A}_{\mathrm{ex}}$ we have
considerable freedom. We require them to be smooth and locally
bounded, including their derivates to avoid too strong local
oscillations. No condition on the increase at infinity is needed,
since $|\boldsymbol {v}(t)| \le 1$. However, it is convenient to have the energy
(\ref{ad.g}) uniformly bounded from below. To keep matters simple
we make the unnecessarily strong assumptions\medskip\\
{\it Condition $(P)$:
\begin{equation}\label{ad.e}
        \phi_{\mathrm {ex}}
       \in C^\infty ({\mathbb R}^3)~, ~ \boldsymbol {A}_{\mathrm
       {ex}}\,
       \in C^\infty ({\mathbb R}^3,{\mathbb R}^3),~
        \phi_{\mathrm{ex}} \ge  \overline \phi> - \infty\,.
\end{equation}
There exists a constant $C$ such that
$|\nabla^k  \phi_{\mathrm{ex}}| \le C, ~ | \nabla^k \, \boldsymbol{A}_{\mathrm{ex}}
       | \le C$ componentwise for $k=1,2,3.$}\medskip

Physically, the Abraham model is not quite consistent. Besides the
center of mass $\boldsymbol {q}(t)$ and its velocity $\boldsymbol {v}(t)$ a rigid body has
also rotational degrees of freedom. As we will see in Chapter
\ref{sec.h}
even if initially non--rotating  the external fields and the
self--interaction necessarily induce a rotation. The translational
 and rotational degrees of freedom are coupled
through the field. This makes the model considerably more
intricate and it is an advisable strategy to understand the
simplified version first.

The Abraham model is derived from the Lagrangian
\begin{eqnarray}
      L &=& - m_{\mathrm{b}} (1- \dot{\boldsymbol{q}}
       ^2)^{1/2}  - ( \phi_{\mathrm{ex}}
       + \phi - \dot{\boldsymbol{q}} \cdot \boldsymbol {A}_{\mathrm{ex}}-
       \dot{\boldsymbol{q}} \cdot \boldsymbol {A}) \ast
       \rho(\boldsymbol{q})\nonumber\\[2mm]
       && + \frac{1}{2} \, \int d^3 x \big[ ( \nabla \phi +
       \partial_t
       \boldsymbol {A})^2 - (\nabla \times \boldsymbol {A})^2 \big
       ]\,.\label{ad.f}
\end{eqnarray}
Correspondingly the energy
\begin{equation}\label{ad.g}
       \mathcal {E}
       (\boldsymbol {E},\boldsymbol {B},
       \boldsymbol {q}, \boldsymbol {v})
       = m_{\mathrm{b}} \gamma (\boldsymbol {v})
       + e \phi_{\mathrm {ex}} \ast\rho\,(\boldsymbol {q})+
       \frac{1}{2}\, \int d^3 x \big(\boldsymbol{E}(\boldsymbol{x})^2
       + \boldsymbol{B}(\boldsymbol{x})^2\big)
\end{equation}
is conserved.

As for any dynamical system, we first have to construct a suitable
phase space. The dynamical variables are $(\boldsymbol{E}(\boldsymbol{x}),
\boldsymbol{B}(\boldsymbol{x}), \boldsymbol{q},\boldsymbol{v}) = Y$
which is called a state of the system. We have $\boldsymbol {q} \in {\mathbb
R}^3, \boldsymbol {v} \in {{\mathbb V}} = \{ \boldsymbol {v}:
|\boldsymbol {v}| < 1\}$. In addition
the field energy (\ref{ad.g}) should be bounded. Thus it is
 natural to introduce the (real) Hilbert space
\begin{equation}\label{ad.h}
       L^2 = L^2 ({\mathbb R}^3, {\mathbb R}^3)
\end{equation}
with norm $||\boldsymbol{E}|| = (\int d^3 x
|\boldsymbol{E}(\boldsymbol{x})|^2)^{1/2}$ and
to define ${\mathcal L}$
as the set of states satisfying
\begin{equation}\label{ad.i}
       \| Y \|_{\mathcal L} = \|\boldsymbol{E} \| + \|\boldsymbol {B}\| +
        |\boldsymbol {q}| + | \gamma(\boldsymbol {v})\boldsymbol {v}| <
       \infty \, .
\end{equation}
In particular the field energy, $\frac{1}{2} (\| \boldsymbol {E}\|^2 +
\|\boldsymbol {B}\|^2)$, is bounded. We equip $\mathcal L$ with the metric
\begin{equation}\label{ad.j}
       d(Y_1, Y_2) = \|\boldsymbol {E}_1 - \boldsymbol{E}_2 \| +
       ||\boldsymbol {B}_1 - \boldsymbol{B}_2|| +
           |\boldsymbol{q}_1 - \boldsymbol{q}_2|
       + |\gamma(\boldsymbol {v}_1)\boldsymbol {v}_1 - \gamma
       (\boldsymbol{v}_2)\boldsymbol{v}_2| \, .
\end{equation}
In addition, one has to satisfy the constraints (\ref{ad.c}). Thus
the phase space, $\mathcal M$, for the Abraham model is the nonlinear
submanifold of $\mathcal L$ defined through
\begin{equation}\label{ad.k}
       \nabla \cdot \boldsymbol {E}(\boldsymbol {x})=
       \rho(\boldsymbol {x}-\boldsymbol {q})\, , \quad \nabla \cdot
       \boldsymbol{B}(\boldsymbol {x})= 0\,.
\end{equation}
$\mathcal M$ inherits its metric from $\mathcal L$.

At several occasions we will need that the system forgets its
initial field data. For this purpose it is helpful to have a little bit of smoothness
and some decay at infinity. Formally we introduce the ``good'' subset
${\mathcal M}^\sigma \subset \mathcal M$, $0 \le \sigma \le 1$,
which consists of fields such that componentwise and outside some
ball of radius $R_0,$ $|\boldsymbol{x}| \ge R_0$, we have
\begin{equation}\label{ad.l}
       |\boldsymbol{E}(\boldsymbol{x})| +
       |\boldsymbol{B}(\boldsymbol{x})| + |\boldsymbol{x}|
       (|\nabla \boldsymbol{E}(\boldsymbol{x})| + |\nabla
       \boldsymbol{B}(\boldsymbol{x})|) \le C\,
       |\boldsymbol{x}|^{-1-\sigma}\,.
\end{equation}
The Li\'{e}nard--Wiechert fields (\ref{aa.u}), (\ref{aa.v})
are in ${\mathcal M}^1$ and ${\mathcal M}^1$ is
dense in $\mathcal M$. However  ${\mathcal M}^\sigma = \emptyset$
for $\sigma > 1$, since $\int d^3 x \rho(\boldsymbol{x}) = e \not=
0$.

The evolution equations (\ref{ad.b}) to (\ref{ad.d}) are of the
general form
\begin{equation}\label{ad.m}
       \frac{d}{dt}\, Y(t)=  F(Y(t))
\end{equation}
with $Y(0) = Y^0 \in \mathcal M $. We have to turn to  the question of the existence and
uniqueness of solutions of the Abraham model (\ref{ad.m}).
\begin{theorem}\label{thm.aa}
 Let conditions $(C)$ and $(P)$ hold and let $Y^0 =
(\boldsymbol {E}^0(\boldsymbol {x}), \boldsymbol {B}^0(\boldsymbol {x}),
 \boldsymbol {q}^0, $ $ \boldsymbol {v}^0)
 \in \mathcal M.$ Then
 the integrated version of Equation (\ref{ad.m}),
 \begin{equation}\label{ad.n}
       Y(t) = Y^0 + \int\limits_0^t ds \, F(Y(s))\,,
 \end{equation}
 has a unique solution $Y(t) = (\boldsymbol {E}(\boldsymbol {x},t),
 \boldsymbol {B}(\boldsymbol {x},t), \boldsymbol {q}(t),
 \boldsymbol{v}(t)) \in \mathcal
M$, which is continuous in $t$ and satisfies $Y(0) = Y^0$. Along
the solution trajectory
\begin{equation}\label{ad.o}
       {\mathcal E} (Y(t)) = {\mathcal E}(Y^0)\, ,
\end{equation}
i.e. the energy is conserved.
\end{theorem}
For short times existence and uniqueness follows through the
contraction mapping principle with constants depending only on
the initial energy. For smooth initial data energy conservation is
verified directly and by continuity it extends to all finite
energy data. Thus we can construct iteratively the solution for
all times.

We first summarize some properties of the inhomogeneous Maxwell
equations. They follow directly from the Fourier and convolution
representations (\ref{aa.l}), (\ref{aa.m}), resp. (\ref{aa.p}),
(\ref{aa.q}).
\begin{lemma}\label{thm.ab}
 In the inhomogeneous Maxwell equations (\ref{aa.b}), (\ref{aa.c})
        let $\rho(\boldsymbol {x},t) =
       \rho(\boldsymbol {x}-\boldsymbol {q}(t)),\; \boldsymbol {j}
       (\boldsymbol {x},t) = \rho(\boldsymbol {x}-
       \boldsymbol {q}(t))\boldsymbol{v}(t)$ with $t \mapsto \boldsymbol {q}(t),
       \boldsymbol {v}(t)$
       continuous. Then (\ref{aa.b}), (\ref{aa.c}) has a unique
       solution in $C({\mathbb R}, L^2 \oplus L^2)$. The solution
       map $(\boldsymbol {E}^0, \boldsymbol {B}^0) \mapsto
        (\boldsymbol {E}(t), \boldsymbol {B}(t))$ depends continuously
       on $\boldsymbol {q}(t),\boldsymbol {v}(t)$.
\end{lemma}
{\it Proof of Theorem 1.1}:
         Let us fix some $b > 0$ and choose initial data such
                 that ${\mathcal E} (Y^0)
         \le b$\,.\\
(i) There exists a unique solution $Y(t) \in C([0, \varepsilon],
                 \mathcal M)$
         for $\varepsilon =\varepsilon (b)$ sufficiently small.

We write (\ref{ad.g}) in the form
\begin{equation}\label{ad.p}
       \frac{d}{dt} (m_{\mathrm {b}} \gamma\, \boldsymbol {v}(t)) =
       \boldsymbol {F}_{\mathrm {ex}} (t) +
       \boldsymbol {F}_{(0)} (t) + \boldsymbol {F}_{\mathrm {self}} (t)
\end{equation}
by inserting $\boldsymbol {E}(\boldsymbol {x},t), \boldsymbol {B}
(\boldsymbol {x},t)$ from the  Maxwell equations according to (\ref{aa.p}),
(\ref{aa.q}). Let
\begin{eqnarray}\label{ad.q}
       W_t(\boldsymbol {x})  &=& \int d^3 k|\hat\rho(\boldsymbol {k})|^2 \,
        e^{i\boldsymbol {k}\cdot \boldsymbol {x}}
       \frac{1}{|\boldsymbol {k}|} \, \sin |\boldsymbol {k}| t\nonumber\\
       &=& (2 \pi)^3 \int d^3 y\, \int d^3 y^\prime \rho(\boldsymbol {y})
       \rho(\boldsymbol {y}^\prime)
       \frac{1}{4 \pi t}\, \delta (|\boldsymbol {y}+ \boldsymbol {x} -
       \boldsymbol {y}^\prime | - t)
       \,.\label{d.n}
\end{eqnarray}
Then
\begin{eqnarray}
\boldsymbol{F}_{{\mathrm ex}}(t) &=& \boldsymbol{E}_{{\mathrm ex}}\ast
\rho (t) + \boldsymbol{v}(t) \times  \boldsymbol{B}_{{\mathrm ex}}\ast
\rho (t)\,,\label{ad.r}\\[3mm]
      \boldsymbol{F}_{(0)}(t) &=&  \int d^3 x\, \rho(\boldsymbol {x}-
           \boldsymbol{q}(t))
       \big [ \partial_t  G_t \ast \boldsymbol {E}^0 (\boldsymbol {x}) +
           (\nabla \times G_t) \ast
       \boldsymbol {B}^0
       (\boldsymbol {x})\nonumber\\
       && \quad + \boldsymbol {v}(t)
       \times \partial_t G_t \ast \boldsymbol {B}^0(\boldsymbol{x})
           - \boldsymbol {v}(t)\times (\nabla \times G_t \ast
       \boldsymbol {E}^0(\boldsymbol {x}))\big ]
           \, ,\label{ad.s}\\[3mm]
        \boldsymbol{F}_{\mathrm {self}}(t) &=&  \int\limits_0^t ds \big [ - \nabla
        W_{t-s} (\boldsymbol {q}(t) - \boldsymbol {q}(s)) - \boldsymbol {v}(s)
         \partial_t  W_{t-s} (\boldsymbol {q}(t)-\boldsymbol
         {q}(s))
        \nonumber\\
        &&\quad + \boldsymbol {v}(t)\times (\nabla\times \boldsymbol {v}(s)
         W_{t-s}(\boldsymbol {q}(t)- \boldsymbol
         {q}(s)))\big ]\,.\label{ad.t}
\end{eqnarray}

We now integrate both sides of (\ref{ad.p})
over the time interval $[0, t]$. The resulting expression is regarded as a map from
the trajectory $t \mapsto \boldsymbol {q}(t), \boldsymbol {v}(t),
 0 \le t \le \delta,$ to the trajectory
$t \mapsto\boldsymbol {\bar  q} (t), \boldsymbol {\bar v} (t)$  and is defined by
\begin{eqnarray}\label{ad.u}
       \boldsymbol {\bar q} (t) &=& \boldsymbol {q}^0 + \int\limits_0^t ds\,
        \boldsymbol {v}(s)\,,\label{d.q}\\
       m_{{\mathrm b}} \gamma (\boldsymbol {\bar v}(t)) \boldsymbol {\bar v}(t)
       &=& m_{{\mathrm b}} \gamma (\boldsymbol {v}^0) \boldsymbol {v}^0 +
       \int\limits_0^t ds \big( \boldsymbol {F}_{\mathrm {ex}} (s) +
           \boldsymbol {F}_{\mathrm{(0)}}
       (s) +\boldsymbol {F}_{\mathrm {self}} (s)\big )\,, \nonumber
\end{eqnarray}
where $\boldsymbol {F}_{\mathrm{ex}} (s), ~\boldsymbol {F}_{(0)}(s),
 ~\boldsymbol {F}_{\mathrm{self}} (s)$
are functionals of $\boldsymbol {q}(\cdot),  \boldsymbol {v}(\cdot)$
according to (\ref{ad.r}) to (\ref{ad.t}).
Since $\rho, W, \phi_{\mathrm {ex}}, \boldsymbol {A}_{\mathrm {ex}}$ are smooth,
this map is a contraction in $C([0,t],\break
{\mathbb R}^3 \times\mathbb V)$,
i.e.
\begin{eqnarray}
&&\sup\limits_{0\le s \le t} \,\big (|\boldsymbol {\bar q}_2 (s) -
 \boldsymbol {\bar q}_1 (s)| +
|\boldsymbol {\bar v}_{2} (s) - \boldsymbol {\bar v}_1 (s)|\big)\nonumber\\[3mm]
 &&\le
c(t,b)\, \sup\limits_{0\le s \le t} \,\big( |\boldsymbol {q}_1 (s) - \boldsymbol {q}_2
(s)| + |\boldsymbol {v}_1(s) - \boldsymbol{v}_2(s)|\big)\,
,\label{ad.v}
\end{eqnarray}
with a constant $c(t,b)$ depending on
$b$ and $c(t,b) < 1$ for sufficiently small $t$. Such a map has a
unique fixed point which is the desired solution $\boldsymbol{q}(t),
\boldsymbol{v}(t)$. By the Maxwell equations also
$\boldsymbol{E}(\boldsymbol{x},t)$,$\boldsymbol{B}(\boldsymbol{x},t)$
is uniquely determined.\\
(ii) The solution map $Y^0 \mapsto Y(t)$ is continuous in $\mathcal
M$.

This follows from Lemma \ref{thm.ab} and the continuity of $\boldsymbol {q}(t),
\boldsymbol {v}(t)$
in dependence on the initial data.\\
(iii) The energy is conserved.

We choose smooth initial fields such that $\boldsymbol {E},
\boldsymbol {B} \in C^\infty ({\mathbb R}^3)$
and
\begin{equation}\label{ad.w}
       |\nabla^\alpha \boldsymbol {E}(\boldsymbol {x})| + |\nabla^\alpha
        \boldsymbol {B}(\boldsymbol {x})| \le
       C (1+|\boldsymbol {x}|)^{-(2+|\alpha|)}\, .
\end{equation}
Here $\alpha = (\alpha_1, \alpha_2, \alpha_3)$ is a multi--index
with $\alpha_i = 0, 1,2, \ldots.$ This set is dense in $\mathcal
M$. By the convolution representation (\ref{aa.p}), (\ref{aa.q}) of  the solution
to  the Maxwell equations we have
$\boldsymbol {E}(\boldsymbol {x},t), \boldsymbol {B}(\boldsymbol {x},t)
\in C^1 ([0,\delta ] \times {\mathbb R}^3)$
and $|\boldsymbol {E}(\boldsymbol {x},t)|+|\boldsymbol {B}
(\boldsymbol {x},t)| \le C (1+|\boldsymbol {x}|)^{-2}$. Also
$\boldsymbol{v}(t) \in C^1 ([0,
\delta])$. Thus we are allowed to differentiate,
\begin{eqnarray}\label{ad.x}
       \frac{d}{dt} \, {\mathcal E} (Y(t)) &=& \gamma^3 \boldsymbol {v} \cdot
       \dot{\boldsymbol{v}} + \boldsymbol {v} \cdot \nabla  \phi_{\mathrm {ex}}
        \ast \rho (\boldsymbol {q})
       + \int d^3 x (\boldsymbol {E} \cdot \partial_t \boldsymbol {E} +
        \boldsymbol {B} \cdot \partial_t \boldsymbol{B})
                \nonumber\\
         &=& \int d^3 x \big(\boldsymbol {E} \cdot (\nabla \times \boldsymbol {B})
        -\boldsymbol {B} \cdot (\nabla
       \times \boldsymbol {E}) \big) = 0 \, ,
      \end{eqnarray}
since the fields decay and hence the surface terms vanish.
Thus ${\mathcal E}(Y(t)) = {\mathcal E}(Y^0)$
for $0 \le t\le \delta$. By continuity this equality extends
to all of $\mathcal M$.\\
(iv) The global solution exists.

>From (iii) we know that ${\mathcal E}(Y(\delta)) = {\mathcal E}(Y^0) \le
b$. Thus we can repeat the previous argument for $\delta \le t \le 2
\delta$,
etc.. Backwards in time we still have the solution (\ref{aa.p}),
(\ref{aa.q}) of the
Maxwell equations, only the retarded fields have to be replaced by the
advanced ones. Thereby we obtain the solution for all times.
$\Box$\medskip

Theorem \ref{thm.aa} ensures the existence and uniqueness of solutions for
 the Abraham model.
For initial data $Y^0 \in \mathcal M$ the solution trajectory
$t \to Y(t)$ lies in the phase space $\mathcal M$,
 is continuous in $t$, and its energy is conserved. We have thus
established the basis for further investigations on the dynamics
of the Abraham model.

\subsection{Appendix: Long time asymptotics}\label{sec.ae}
For dynamical systems one of the first
qualitative question is to understand whether there are general
patterns governing the long time behavior. For the Abraham model
the long time asymptotics is dominated through the loss of energy
radiated to infinity, which is proportional to
$\dot{\boldsymbol{v}}(t)^2$ according to Larmor's formula. Since
the energy is bounded from below, we expect that
\begin{equation}\label{ae.a}
       \lim_{t \to\infty}\dot{\boldsymbol{v}}(t)=0
\end{equation}
under rather general initial conditions. In fact, one would also
expect that
\begin{equation}\label{ae.b}
       \lim_{t \to\infty}\boldsymbol{v}(t)= \boldsymbol{v}_\infty
       \in \mathbb V\,,
\end{equation}
where $\boldsymbol{v}_\infty = 0$ for bounded motion and $\boldsymbol{v}_\infty \not =0$
for a scattering solution.

In this section we will prove (\ref{ae.a}) under the extra
hypothesis\medskip\\
{\it Wiener Condition  $(W)$:}
\begin{equation}\label{ae.c}
        \hat\rho (\boldsymbol{k}) > 0\,.
\end{equation}
 The proof follows rather
closely the physical intuition and leads to an equation of
convolution type which has a definite long time limit only under $(W)$.
$(W)$ means that all modes of the charge distribution couple
to electromagnetic field. According to (\ref{ae.a}) the Abraham model
does not admit then any periodic solution.
Since $\rho$ has compact support, in general, $\hat \rho (\boldsymbol{k})$
may vanish for a discrete set of shell radii  $|\boldsymbol{k}|$. At
present, it remains as an open problem, whether periodic solution
become then possible.

Let us consider a ball of radius $R$ centered at the origin. At time $t$
the field
 energy in this ball and the mechanical energy of the charge  is given
by
\begin{equation}\label{ae.d}
       {\mathcal E}_R (t) = {\mathcal E}(0) -
       \frac{1}{2} \, \int\limits_{\{|\boldsymbol{x}| \ge
       R\}} d^3x \big (\boldsymbol{E}(\boldsymbol{x},t)^2
       + \boldsymbol{B}(\boldsymbol{x}, t)^2\big)
\end{equation}
 for $R$
sufficiently large, using the conservation of total energy.  ${\mathcal E}_R$ changes
in time as
\begin{equation}\label{ae.e}
       \frac{d}{dt}\, {\mathcal E}_R (t) = - R^2
       \, \int d^2 \omega ~ \boldsymbol{\omega} \cdot
       [\boldsymbol{E}(R \boldsymbol{\omega},t) \times
       \boldsymbol{B}(R \boldsymbol{\omega}, t)]\,,
\end{equation}
where $\boldsymbol{\omega}$ is a vector on the unit sphere, $d^2 \omega$ the
surface measure normalized to $4 \pi$, and $\boldsymbol{E}\!\times\!\boldsymbol{B}$
the Poynting vector for the  flux in energy at the surface of the
 ball under consideration. Since the total energy is bounded from below, we
conclude that
\begin{equation}\label{ae.f}
       {\mathcal E}_R (R) -
       {\mathcal E}_R (R+t) = -
       \int\limits_R^{R+t} ds\,  \frac{d}{ds}\,
       {\mathcal{E}}_R (s) \le C
\end{equation}
with the constant $C = {\cal E} (0) - \overline \phi$ independent of $R, t$.

In (\ref{ae.f}) we first take the limit $R \to \infty$, which yields
the energy radiated to infinity during the time interval $[0,t]$
through a large sphere centered at the origin. Subsequently we take
the limit $t \to \infty$ to obtain the total radiated energy. To
state the result we define
\begin{eqnarray}
      \boldsymbol{E}_\infty (\boldsymbol{\omega}, t)  &=&- \frac{1}{4 \pi}\,
       \int d^3 y \, \rho(\boldsymbol{y} -\boldsymbol{q}
       ( t + \boldsymbol{\omega} \cdot
       \boldsymbol{y}))\label{ae.g}\\[2mm]
       && \big [ (1-\boldsymbol{\omega} \cdot \boldsymbol{v})^{-1}
       \dot{\boldsymbol{v}}+ (1- \boldsymbol{\omega} \cdot \boldsymbol{v})^{-2}
       (\boldsymbol{\omega} \cdot \dot{\boldsymbol{v}}) (\boldsymbol{v}-
       \boldsymbol{\omega}) \big ]|_{t+\boldsymbol{\omega}\cdot
       \boldsymbol{y}} \nonumber
\end{eqnarray}
which is a functional of the actual trajectory of the particle.
Whatever its motion we must have
\begin{equation}\label{ae.h}
       \int\limits_0^\infty dt\, \int d^2\omega\, |\boldsymbol{
       E}_\infty ( \boldsymbol{\omega}, t)|^2 \le C < \infty \,.
\end{equation}
Note that the integrand in (\ref{ae.h}) is proportional to $\dot{\boldsymbol{v}}(t)^2$,
 which therefore has to decay to zero for large $t$.

To establish (\ref{ae.h}) is somewhat tedious with pieces of the
argument explained in Sections \ref{sec.dc} and \ref{sec.fd}, \ref{sec.fe}.
 One imagines that
the trajectory $t \mapsto \boldsymbol{q}(t)$ is given and solves
the inhomogeneous Maxwell equations according to (\ref{aa.p}), (\ref{aa.q}). If the
initial fields are in ${\mathcal M}^\sigma, \frac{1}{2} < \sigma \le
1$, then they decay and make no contribution to (\ref{ae.f}) in the
limit $R \to \infty$, cf. our treatment of the initial time slip
in Section \ref{sec.dc}. Next one has to study the asymptotics of the
retarded fields, which is carried out in Sections \ref{sec.fd}, \ref{sec.fe}.
 There $\varepsilon$
is fixed and for our purpose we may set
$\varepsilon =1$. In addition in (\ref{fe.a}) the sphere is centered at
 $\boldsymbol{q}^\varepsilon(t)$, rather than at the origin. This
 means  we can use (\ref{fe.i}, (\ref{fe.d}) with $\boldsymbol{q}^\varepsilon(t)$
 replaced by $0$ for our case.

The real task is to extract from (\ref{ae.h}) that the acceleration
vanishes for long times.
\begin{theorem}\label{thm.ac}
 For the Abraham model satisfying
$(C)$, $(P)$, and the Wiener condition $(W)$ let the initial data
$Y(0) = (\boldsymbol{E}^0, \boldsymbol{B}^0, \boldsymbol{q}^0,
\boldsymbol{v}^0) \in {\mathcal M}^\sigma$ with $\frac{1}{2} < \sigma \le
1$. Then
\begin{equation}\label{ae.i}
       \lim_{t \to\infty} ~ \dot{\boldsymbol{v}}(t) = 0\,.
\end{equation}
\end{theorem}
\noindent
{\it Proof}: By energy conservation $|\boldsymbol{v}(t)| \le \overline
v < 1$. Inserting in (\ref{ad.d}) and  using $(P)$ we conclude that
 $|\dot{\boldsymbol{v}}(t)| \le C$. Differentiating (\ref{ad.d})
 and using again $(P)$ also $|\ddot{\boldsymbol{v}}(t)| \le C $
 uniformly in $t$. Therefore $\boldsymbol{E}_\infty ( \boldsymbol{\omega}, t)$
 is Lipschitz continuous jointly in $ \boldsymbol{\omega}, t$. Since the energy
 dissipation (\ref{ae.h}) is bounded, this implies
\begin{equation}\label{ae.j}
       \lim_{t \to\infty} ~
       \boldsymbol{E}_\infty ( \boldsymbol{\omega},t) = 0
\end{equation}
uniformly in $\boldsymbol{\omega}$.

We analyze the structure of the integrand in (\ref{ae.g}). In the
retarded argument only $y_\parallel = \boldsymbol{\omega} \cdot  \boldsymbol{y}$
appears. Therefore the integration over $\boldsymbol{y}_\perp = \boldsymbol{y}
-y_\parallel  \boldsymbol{\omega}$
can be carried out and we are left with a one--dimensional
integral of convolution type. We set $\rho_a (x_3) = \int dx_1 dx_2 \,
 \rho(\boldsymbol{x})$. Then
 \begin{eqnarray}\label{ae.k}
      \boldsymbol{E}_\infty ( \boldsymbol{\omega},t) & = &\frac{1}{4 \pi}
       \int dy_\parallel\, \rho_a (y_\parallel -
       q_\parallel (t+ y_\parallel))\\
       &&\quad\quad\quad\big [ (1- \boldsymbol{\omega} \cdot \boldsymbol{v})^{-2}
        \boldsymbol{\omega}\times (( \boldsymbol{\omega} - \boldsymbol{v}) \times
       \dot{\boldsymbol{v}}) \big]|_{t+
       y_\parallel}\nonumber\\[2mm]
       &=&\frac{1}{4 \pi} \, \int ds \rho_a
       (t-(s-q_\parallel(s))) \big [(1-  \boldsymbol{\omega} \cdot
       \boldsymbol{v})^{-2}  \boldsymbol{\omega} \times((
       \boldsymbol{\omega}
       -\boldsymbol{v}) \times
       \dot{\boldsymbol{v}})\big]|_s\,.\nonumber
 \end{eqnarray}
 Since $|\dot q_\parallel (s)| < 1$, we can
 substitute $\theta = s- q_\parallel(s)$ and
 obtain the convolution representation
 \begin{equation}\label{ae.l}
        \boldsymbol{E}_\infty ( \boldsymbol{\omega},t)=
        \int d \theta \, \rho_a (t- \theta)\boldsymbol{g}_{\boldsymbol{\omega}}(\theta)
        = \rho_a \ast \boldsymbol{g}_{\boldsymbol{\omega}}(t)\,,
 \end{equation}
 where
\begin{equation}\label{ae.m}
        \boldsymbol{g}_{\boldsymbol{\omega}}(\theta)
        = \frac{1}{4\pi}\, \big [(1-  \boldsymbol{\omega} \cdot
        \boldsymbol{v})^{-2}  \boldsymbol{\omega} \times (( \boldsymbol{\omega}-
        \boldsymbol{v}) \times \dot{\boldsymbol{v}}) \big]|_{s
        (\theta)}\,.
 \end{equation}

>From (\ref{ae.j}) we know that $\lim\limits_{t \to\infty} \, \rho_a \ast
\boldsymbol{g}_{\boldsymbol{\omega}}(t)=0$. If $\hat\rho(\boldsymbol{k}_0) = 0$ for some
 $\boldsymbol{k}_0$, and hence
the Wiener condition would not be satisfied, then the convolution
integral admits a periodic solution and no further progress seems
to be possible. However with $(W)$ and the smoothness of
$\boldsymbol{g}_{
\boldsymbol{\omega}}(\theta)$ already established,
Pitt's extension to the Tauberian
theorem of Wiener ensures us that
\begin{equation}\label{ae.n}
        \lim_{\theta \to\infty} \, \boldsymbol{g}_{\boldsymbol{\omega}} (\theta) = 0\,,
 \end{equation}
 which, since $\theta (t) \to\infty$ as $t \to\infty$, implies
\begin{equation}\label{ae.o}
        \lim_{t \to\infty} \,  \boldsymbol{\omega} \times
        (( \boldsymbol{\omega}-\boldsymbol{v}(t)) \times
        \dot{\boldsymbol{v}}(t)) = 0
 \end{equation}
for every $ \boldsymbol{\omega} $ in the
unit sphere. Replacing $ \boldsymbol{\omega}$
by $- \boldsymbol{\omega}$
and summing both expressions yields $ \boldsymbol{\omega} \times
( \boldsymbol{\omega} \times \dot{\boldsymbol{v}}
(t)) \to 0$ as $t \to\infty$. Since this is true for every
$ \boldsymbol{\omega}$,
the claim follows. $\Box$\medskip

Next we study the comoving electromagnetic fields for large times.
As used already, under our assumptions  the initial fields
decay as $t \to \infty$. Thus we only have to consider the true retarded
fields $\boldsymbol{E}_{\mathrm{ret}} (\boldsymbol{x} + \boldsymbol{q}(t), t)$,
$\boldsymbol{B}_{\mathrm{ret}} (\boldsymbol{x} + \boldsymbol{q}(t),t)$
 centered at the position of the particle. Since
 $\dot{\boldsymbol{v}}
 (t) \to 0$, these fields become almost stationary. In Section
  \ref{sec.ca} we show that there is a unique comoving field with velocity
 $\boldsymbol{v}$.
 These charge soliton fields are denoted by $\boldsymbol{E}_{\boldsymbol{v}},
  \boldsymbol{B}_{\boldsymbol{v}}$ when centered at
 the origin, compare with (\ref{ca.u}),
 (\ref{ca.v}). Thus the true retarded fields $\boldsymbol{E}_{\mathrm{ret}}(\boldsymbol{x}
  + \boldsymbol{q}(t), t),
 \boldsymbol{B}_{\mathrm{ret}} (\boldsymbol{x}+\boldsymbol{q}(t), t)$ have to
  compared with the soliton
 fields $\boldsymbol{E}_{{\boldsymbol{v}}(t)}, \boldsymbol{B}_{{\boldsymbol{v}}(t)}$.
  For this purpose we use the
 representations (\ref{ca.u}), (\ref{ca.v}) for the charge soliton and
(\ref{aa.p}), (\ref{aa.q}) for the retarded fields. We insert the
explicit form (\ref{aa.o}) of the propagator. This yields
\begin{eqnarray}
       \boldsymbol{E}_{\boldsymbol{v}} (\boldsymbol{x}) &=& \int d^3 y\, (4 \pi|\boldsymbol{x}
       -\boldsymbol{y}|)^{-1} \big (|\boldsymbol{x} -
       \boldsymbol{y}|^{-1}  \rho (\boldsymbol{y} -
       \boldsymbol{v}| \boldsymbol{x} -
       \boldsymbol{y}|)\widehat{\boldsymbol{n}}\nonumber\\
        && \quad\quad+  \boldsymbol{v} \cdot
        \nabla\rho(\boldsymbol{y} -\boldsymbol{v}
        |\boldsymbol{x}-\boldsymbol{y}|)(\boldsymbol{v} - \widehat{\boldsymbol{n}})
        \big )\,,\label{ae.p}\\[2mm]
        \boldsymbol{B}_{\boldsymbol{v}} (\boldsymbol{x}) &=& \int d^3 y
        (4\pi|\boldsymbol{x}-\boldsymbol{y}|)^{-1}
        \widehat{\boldsymbol{n}} \times \big (- |\boldsymbol{x} -
        \boldsymbol{y}|^{-1} \rho(\boldsymbol{y} -
       |\boldsymbol{x}-\boldsymbol{y}| \boldsymbol{v})\boldsymbol{v} \nonumber\\
        &&\quad\quad + \boldsymbol{v}\cdot
        \nabla\rho(\boldsymbol{y} -
        |\boldsymbol{x}-\boldsymbol{y}|\boldsymbol{v})\boldsymbol{v}\big
        )\,,\label{ae.q}
\end{eqnarray}
where $\widehat{\boldsymbol{n}} = (\boldsymbol{x} -\boldsymbol{y})/|\boldsymbol{x}-
\boldsymbol{y}|$.
Similarly for the retarded fields
\begin{eqnarray}
        \boldsymbol{E}_{\mathrm{ret}} (\boldsymbol{x} + \boldsymbol{q}(t),
        t)&=&
        \int d^3 y\, \big (4 \pi|\boldsymbol{x}-\boldsymbol{y}|)^{-1}
        \big (|\boldsymbol{x} -
       \boldsymbol{y}|^{-1} \rho (\boldsymbol{y} +
       \boldsymbol{q}(t)- \boldsymbol{q}(\tau)) \widehat{\boldsymbol{n}}\nonumber\\
        && \quad\quad +  \boldsymbol{v}(\tau) \cdot
        \nabla\rho(\boldsymbol{y} +  \boldsymbol{q}(t)-
        \boldsymbol{q}(\tau))(\boldsymbol{v}(\tau) - \widehat{\boldsymbol{n}})\nonumber\\
        &&\quad\quad -\rho(\boldsymbol{y}+
         \boldsymbol{q}(t)-\boldsymbol{q}(\tau))\dot{\boldsymbol{v}}
         (\tau)\big) \,,\label{ae.r}\\[2mm]
        \boldsymbol{B}_{\mathrm{ret}} (\boldsymbol{x} + \boldsymbol{q}(t),t)
        &=&\int d^3 y (4\pi|\boldsymbol{x}-\boldsymbol{y}|)^{-1}
        \widehat{\boldsymbol{n}} \times \big(
        -|\boldsymbol{x}-\boldsymbol{y}|^{-1}
        \rho( \boldsymbol{y}+ \boldsymbol{q}(t)-\boldsymbol{q}(\tau))\nonumber\\
       && \quad\quad
         \boldsymbol{v}(\tau)+
        \boldsymbol{v}(\tau) \cdot\nabla\rho (\boldsymbol{y}+ \boldsymbol{q}(t)
        -\boldsymbol{q}(\tau)) \boldsymbol{v}(\tau)\nonumber\\
        &&\quad\quad -\rho(\boldsymbol{y}+
        \boldsymbol{q}(t)-\boldsymbol{q}(\tau))\dot{\boldsymbol{v}}(\tau)\big)\,,\label{ae.s}
\end{eqnarray}
where $\tau = t - |\boldsymbol{x}-\boldsymbol{y}|$ and $t \ge
t_\rho = 2 R_\rho/(1- \overline v)$.

We compare the fields locally and use that $\lim\limits_{t \to\infty}
\dot{\boldsymbol{v}} (t) = 0$.
Then, for any fixed $R>0$,
\begin{eqnarray}\label{ae.t}
       &&\lim_{t \to\infty} \int\limits_{\{|\boldsymbol{x}|\le R \}} d^3x \Big (
       \big (\boldsymbol{E}
       (\boldsymbol{x} + \boldsymbol{q}(t),t) -
       \boldsymbol{E}_{\boldsymbol{v}(t)}(\boldsymbol{x})\big)^2\nonumber\\
       &&\qquad\qquad\qquad + \big(\boldsymbol{B}
       (\boldsymbol{x} +\boldsymbol{q} (t), t)- B_{v(t)}
       (\boldsymbol{x})\big )^2\Big )  = 0\,.
\end{eqnarray}

Armed with this information we can sketch qualitatively the long
time behavior of the Abraham model. We distinguish bounded and
scattering trajectories. If $\phi_{\mathrm{ex}}$ and $\boldsymbol{A}_{\mathrm{ex}}$
decay sufficiently fast for large $|\boldsymbol{x}|$ and if the
particle escapes into the essentially force free region, then $\lim\limits_{t \to\infty}
\boldsymbol{v}(t) = \boldsymbol{v}_\infty \not= 0$ and the fields are
well approximated by the soliton fields, compare with (\ref{ae.t}).
In particular, if there are no external fields the charge travels
with some definite velocity in the long time limit. This point is
discussed in more detail by a different method in Section
\ref{sec.dc}.

If the motion is bounded,
\begin{equation}\label{ae.u}
       |\boldsymbol{q}(t)| \le \overline q
\end{equation}
for all $t$, then $\dot{\boldsymbol{v}} (t) \to 0$ for $t \to \infty$
implies that
\begin{equation}\label{ae.v}
       \lim_{t \to\infty} \boldsymbol{v}(t)= 0\, .
\end{equation}
Inserting in the Lorentz force equation (\ref{ad.d}) and using that
the fields become soliton--like, one infers that
\begin{equation}\label{ae.w}
       \lim_{t \to\infty} \nabla \phi_{\mathrm{ex}}
       \ast \rho (\boldsymbol{q}(t))= 0\, ,
\end{equation}
i.e.  $\boldsymbol{q}(t)$ approaches the set of
critical points ${\mathcal A}
= \{ \boldsymbol{q}, ~ \nabla  \phi_{\mathrm{ex}}
\ast \rho (\boldsymbol{q}) = 0\}$.

(\ref{ae.v}), (\ref{ae.w}) still leave a lot of freedom. Generically,
one expects that $\boldsymbol{q}(t )$ approaches a definite limit.
This is indeed the case if ${\mathcal A}$ happens
 to be a discrete set. By continuity of solutions in $t, ~\boldsymbol{q}(t )$
has to converge then to some definite $\boldsymbol{q}^\ast \in{\mathcal A}
$. In particular if  $\phi_{\mathrm{ex}} \ast\rho$ is strictly
convex,  the charge will come to rest at the minimum of $\phi_{\mathrm{ex}}
\ast\rho$.
 If ${\mathcal A}$  is not discrete, the long time behavior
 depends on be the specific situation. E.g. let $\boldsymbol{A}_
{\mathrm{ex}}= 0$ and $\phi_{\mathrm{ex}}$ be strictly convex outside
a ball of radius one and let  $\phi_{\mathrm{ex}}=0$ inside this ball.
Each time the particle is reflected by the confining potential it
looses in energy. Thus (\ref{ae.v}) holds, but $\boldsymbol{q}(t )$
has no limit as $t \to \infty$. A more realistic example is a
constant magnetic field and $\phi_{\mathrm{ex}}=0$. Then ${\mathcal A}=
{\mathbb R}^3$, but using the sharper estimates of
Section \ref{sec.dc}
it can be shown that the particle spirals inwards to come to rest
at its center of gyration.

So far (\ref{ae.u}) was an assumption and it would be nice to have
some sufficient criteria. It seems that the only one available is
conservation of energy. Bounded motion requires then sufficiently
deep local minima of $\phi_{\mathrm{ex}}$. Thus we see that
bounded   energy dissipation yields  reasonable results
for electrostatic forces but is rather weak for motion in magnetic
fields, except for the general fact that the acceleration has to
vanish for large $t$.

In general, bounded and unbounded motion coexist. The prime
example is a charge bound by an infinitely heavy nucleus, which is
modelled by an attractive, locally smoothened Coulomb potential.
If the initial field supplies sufficient energy, then the ``atom''
becomes ionized and the charge will travel freely in the long time limit.
Conversely a charge may loose energy through radiation and become
trapped by the external potential.


\subsection{The Lorentz model}\label{sec.af}
To improve on the semi--relativistic Abraham model, with
Lorentz it is natural to assume that the charge distribution is
rigid in its own rest frame. The actual construction of such a
charge distribution requires some effort.  For obvious reasons we will
switch now to a relativistic notation.

We fix a laboratory frame, $\mathcal S$, with coordinates $(t,\boldsymbol{x})
= x^\mu, \mu =
0,1,2,3$,
and invariant length $t^2 - \boldsymbol{x}^2 = t^2 - x_i x_i = x^\mu x_\mu
=  g_{\mu\nu} x^\mu
x^\nu$, where we adopt the standard  summation convention
and use latin indices for three--vectors, greek indices for four--vectors.
 In $\mathcal S$ we
prescribe the world line of a charged particle $t \mapsto z_i
(t), i = 1,2,3$, with velocity $v_i(t) = \frac{d}{dt}z_i(t)$, $
\boldsymbol{v}(t)^2 <
1$. We will also parametrize the world line through its eigentime
 as $s \mapsto z^\mu (s), ~\mu
=0, 1,2,3$.

To construct the extended charge distribution we consider some time $t=t_0$
when the particle has velocity $v_i=v_i(t_0)$. The rest frame for the particle
 at time $t_0$
is denoted by ${\mathcal S}^\prime$ with coordinates
$x^{\prime\mu}$. It is connected to $\mathcal S$ by the Lorentz
transformation
\begin{equation}\label{af.a}
        x^{\prime\mu} = \Lambda(\boldsymbol{v})^\mu_{~\:\nu} x^\nu\, .
\end{equation}
We require the origins of $\mathcal S$ and ${\mathcal S}^\prime$ to
coincide and their spatial axes to be parallel to each other, i.e.
 $\Lambda(\boldsymbol{v})$ is a Lorentz boost and reads explicitly
\begin{eqnarray}
       x_i^\prime &=& x_i + (\gamma^2 / (1+ \gamma)) v_i v_k x_k -
       \gamma v_i t\,,\nonumber\\
       t^\prime &=& \gamma(t-v_k x_k)\,.\label{af.b}
\end{eqnarray}
The Lorentz model assumes that in ${\mathcal S}^\prime$ the four--current
 $j^{\prime\mu}$ is given by
\begin{equation}\label{af.c}
        j^{\prime\mu} = e \big (f((x_j^\prime-
        z_j^\prime(t^\prime))(x_j^\prime- z_j^\prime(t^\prime))), 0 \big )
        |_{t^\prime=t_0^\prime}
\end{equation}
with the relativistic form factor
$f$, where $e f(\boldsymbol{x}^2) =
\rho_{r}(|\boldsymbol{x}|)$ as before.
In particular,  $f$ vanishes for $|\boldsymbol{x}| \ge R_\rho$.

We have to transform $j^{\prime\mu}$ back to the laboratory frame $\mathcal
S$. Using (\ref{af.b}), (\ref{af.c}) yields
\begin{equation}\label{af.d}
       j^\mu = e(\gamma, \gamma \boldsymbol{v}) f((x_j^\prime - z_j
       (t^\prime))(x_j^\prime - z_j(t^\prime)))
       |_{t^\prime=t_0^\prime}\, .
\end{equation}
The condition $t^\prime = t_0^\prime$ means $\gamma (t-v_k x_k) = \gamma(t_0
-v_k z_k(t_0))$, i.e.
\begin{equation}\label{af.e}
       t_0= t - v_k (x_k - z_k(t_0)) \, .
\end{equation}
Therefore
\begin{equation}
       j^\mu (t,\boldsymbol{x}) = e(\gamma, \gamma \boldsymbol{v}) f ((t-t_0)^2 -
       (x_j-z_j(t_0))(x_j-z_j(t_0)))\label{af.f}
\end{equation}
with $t_0= t_0(t,\boldsymbol{x})$ defined through (\ref{af.e}).

An example is shown in Figure 1. $j^\mu(t,\boldsymbol{x})$ vanishes
outside a
tube around the world line. This tube is fibered into cross
sections corresponding to $\{t^\prime = t_0^\prime\}$ in the
momentary rest frame. Along each cross section the charge is
smeared according to the form factor $f$. From the figure we notice
that, when the acceleration becomes too large, the cross sections
overlap, which means that for given $(t,\boldsymbol{x})$ Equation (\ref{af.e}) has
several solutions. (\ref{af.f}) seems to indicate that one should
add the contribution from each solution. This however would
violate charge conservation and the proper prescription is to
reverse the sign of a charge element when it moves backwards in
time. Taking into account multiple solutions to (\ref{af.e}) and
their proper sign leads to the four-current
\begin{equation}\label{af.g}
       j^\mu(x) = \int\limits_{- \infty}^\infty ds \, v^\mu [1- \dot
       v_\nu (x-z)^\nu]  f((x-z)^2)
       \delta(v_\lambda(x-z)^\lambda)
\end{equation}
with $x = (x^{0},\ldots,x^{3})$. Here $z^\mu = z^\mu (s)$ is parametrized
by its eigentime $s$ and
$v^\mu (s) = \dot z^\mu (s) = \frac{d}{ds}  z^\mu (s)$. The
current (\ref{af.g}) satisfies the charge conservation
\begin{equation}\label{af.h}
       \partial_\mu j^\mu  = 0 \,.
\end{equation}

There is an instructive way to rewrite the four--current by using
the Thomas precession, which we recall first. The elementary
observation is that the Lorentz boosts, cf. (\ref{af.b}), do not
form a subgroup of the Lorentz group. Let us consider the inertial
frame ${\mathcal S}^\prime$ with velocity $\boldsymbol{v}^\prime$
relative to $\mathcal S$
and the inertial frame ${\mathcal S}^{\prime\prime}$ with velocity
$\boldsymbol{v}^{\prime\prime}$ relative to  ${\mathcal S}^\prime$
and thus
velocity $\boldsymbol{w}$, the relativistic sum of $\boldsymbol{v}$
and $\boldsymbol{v}^\prime$, relative
to  $\mathcal S$. If we denote by $\Lambda(\boldsymbol{v})$ the
Lorentz boost
with velocity $\boldsymbol{v}$, then
\begin{equation}\label{af.i}
       \Lambda (\boldsymbol{v}^\prime) \Lambda(\boldsymbol{v}) =
           R(\boldsymbol{v},\boldsymbol{v}^\prime) \Lambda(\boldsymbol{w})\,,
\end{equation}
where $R(\boldsymbol{v},\boldsymbol{v}^\prime)$ is a suitable pure
spatial rotation.

To apply this observation to a relativistically rigid charge
distribution we imagine that it has a body fixed frame. We want to
define an inertial frame $\mathcal {K}(s)$ such that relative to this frame
the body axes maintain their orientation throughout time. We agree
that the time axis of $\mathcal {K}(s)$ is parallel to the four--velocity
$v^\mu (s)$
and that the origin of  $\mathcal S$  and $\mathcal {K}(s)$ coincide.
Let $\Lambda (s)$
be the Lorentz transformation from   $ \mathcal S $  to $\mathcal {K}(s)$. To
determine it we subdivide the time axis into little intervals of
length $\Delta s$ and require that $\mathcal {K}((m+1) \Delta s)$ is related
to $\mathcal {K}(m\Delta s)$ by a Lorentz boost with the properly adjusted
velocities, i.e. the space axes of $\mathcal {K}((m+1)\Delta s)$
and $\mathcal {K}(m \Delta s)$
are parallel to each other. Taking the limit $\Delta s \to 0$ one
finds that $\Lambda (s)$ is determined through
\begin{equation}\label{af.j}
       \dot \Lambda^\mu_{~\:\nu} = \eta^\mu_{~\:\lambda}
       \Lambda^\lambda_{~\:\nu} \,, \quad \Lambda^\mu _{~\:\nu} (0) =
       \Lambda^\mu _{~\:\nu} (\boldsymbol{v}(0))\,,
\end{equation}
where
\begin{equation}\label{af.k}
       \eta^\mu _{~\:\nu} = v^\mu \dot v_\nu - v^\nu \dot v_\mu \, .
\end{equation}
Using this definition and
noting that $v_\nu   \xi^\nu \delta (v_\lambda \xi^\lambda) = 0$
the four--current (\ref{af.g}) equals

\begin{equation}\label{af.l}
       j^\mu(x) = \int\limits_{- \infty}^\infty ds [v^\mu -
       \eta^\mu _{~\:\nu} (x - z)^\nu]  f ((x-z)^2) \, \delta
       (v_\lambda (x-z)^\lambda) \, ,
\end{equation}
in which form it has a transparent physical meaning. The current
is  the sum of two contributions. There is a
 translational part  proportional to $v^\mu$ as expected from
the nonrelativistic limit. In addition there is a rotational
component due to the Thomas precession. The minus sign is a
convention. With our definition a vector
time--independent in $\mathcal {K}(s)$ has components
$x^\mu (s)$ in $\mathcal S$ and they change as
\begin{equation}\label{af.m}
       \dot x^\mu = - \eta^\mu _{~\:\nu}  x^\nu \, .
\end{equation}

Before proceeding to the action for the dynamics, we should
understand whether the current (\ref{af.g}) conforms with the
naive physical intuition. An instructive example is a uniformly
accelerated charge, the so--called hyperbolic motion. We assume
that the particle is accelerated along the positive 1--axis
starting from rest at the origin. In the orthogonal direction the
current traces out a tube of diameter $2 R_\rho$ and it suffices
to treat the two--dimensional space--time problem. The center,
$C$, of the charge moves along the orbit
\begin{equation}\label{af.x}
       C= (t, g^{-1}\, (\sqrt{1+g^2t^2}- 1)), ~
       t\ge 0\,,
\end{equation}
where $g>0$ is the acceleration. The curves traced by the right
and left ends, $C_+$ and $C_-$, are determined from (\ref{af.b}) and are
given in parameter form as
\begin{equation}\label{af.y}
       C_{\pm}= \big( (1 \pm R_\rho g)t ,
      g^{-1}((1 + R_\rho g)
       \sqrt{1+g^2t^2}-1)\big), ~  t\ge 0\,.
\end{equation}
The equal--time distance between the center and $C_+$ is $t^{-1} ((R_\rho
g)^2 + 2 R_\rho g)/ ( 2g^2$ $(1+ R_\rho g))$ for large $t$
and thus well localized. However  the left end motion
depends crucially on the magnitude of $R_\rho g$. If
$R_\rho g < 1$, then the distance to the center is
$t^{-1} ((R_\rho g)^2 - 2 R_\rho g)/ (2 g^2 (1- R_\rho g))$
for large $t$. On the other hand for $R_\rho g > 1$, the left end
moves into the past, cf. (\ref{af.y}), and the current density looks
strangely distorted. To have a feeling for the order of magnitudes
involved we insert the classical electron radius. Then
\begin{equation}\label{af.z}
       g > \frac{c^2}{R_\rho}= 10^{31}\, [m/\sec^2]\,,
\end{equation}
which is huge and way beyond the domain of the validity of the
theory. Of course, one would hope that for reasonable initial data
such accelerations can never be reached. But the mere fact that
charge elements may move backwards in time is an extra difficulty.

Armed with the four--current (\ref{af.l}) we can write down the relativistically
covariant action, $ \textsf{S}$, as a functional of the the four--potential $A^\mu$
and the particle coordinates $z^\mu$.
The action of the Lorentz model has four pieces,
\begin{equation}\label{af.n}
       \textsf{S} =  \textsf{S}_0 +  \textsf{S}_{\mathrm {f}} +
           \textsf{S}_{\mathrm {int}} + \textsf{S}_{\mathrm {ex}}\,.
\end{equation}

$ \textsf{S}_0$ is the mechanical action,
\begin{equation}\label{af.o}
       \textsf{S}_0 = m_{{\mathrm b}} \int\limits_{s_1}^{s_2} ds \, .
\end{equation}
$ \textsf{S}_{\mathrm {f}}$ is the field action,

\begin{equation}\label{af.p}
        \textsf{S}_{\mathrm {f}} = - \frac{1}{4}  \int\limits_\Omega d^4x
         F^{\mu\nu}  (x) F_{\mu\nu}(x)\,,
\end{equation}
where the field tensor $F^{\mu\nu}$ is defined through
\begin{equation}\label{af.q}
       F^{\mu\nu} = \partial^\mu A^\nu - \partial^\nu A^\mu \, .
\end{equation}
The interaction is bilinear in  the current and the
field,
\begin{equation}\label{af.r}
        \textsf{S}_{\mathrm {int}} =
                \int\limits_\Omega d^4x   A^\mu (x) j_\mu (x)\,,
\end{equation}
and correspondingly for the interaction with the external
potentials
\begin{equation}\label{af.s}
       \textsf{S}_{\mathrm {ex}} = \int\limits_\Omega d^4 x \, A_{\mathrm
       {ex}}^\mu (x) j_\mu(x)\, .
\end{equation}
The world line $z^\mu(s)$ is specified for $s_1 \le s \le s_2$.
Consequently $\Omega$ is the volume in $ \mathcal{S}$ bounded by the two
hyperplanes  $v^\mu(s_1) (x - z (s_1))_\mu = 0$ and
$v^\mu(s_2) (x- z (s_2))_\mu = 0$.

The actual dynamical trajectory is a stationary point of the
action $ \textsf{S}$ at fixed endpoints. The variation with respect to the
potentials $A^\mu$ leads to the inhomogeneous Maxwell equations
\begin{equation}\label{af.u}
       \partial_\nu F^{\nu\mu} = j^\mu \, .
\end{equation}
The constraints (\ref{aa.c}) are automatically satisfied, since the field tensor is
derived from potentials according to (\ref{af.q}). Next we vary
the world line $z^\mu$. From $ \textsf{S}_0$ we obtain the mechanical
acceleration $m_{{\mathrm b}}  \dot v^\mu$.
The variation of $ \textsf{S}_{\mathrm {int}}$
is somewhat lengthy and deferred to  Section \ref{sec.ag}. The final
evolution equation reads
\begin{eqnarray}\label{af.v}
       m_{{\mathrm b}} \frac{d}{ds}  v^\mu (s)&
           =& e \int d^4 \xi \, f(\xi^2) \delta(v_\lambda
       (s) \xi^\lambda)  [ F^{\mu\nu} (z(s) + \xi)
      + F^{\mu\nu}_{\mathrm{ex}} (z(s)
        + \xi)]\nonumber\\
        && \quad\quad\quad(v_\nu(s) - \eta_{\nu
       \sigma} (s) \xi^\sigma) \, .
\end{eqnarray}

At first glance the equations of motion (\ref{af.v}) look rather
similar to its semi--relativistic sister. It seems natural to specify
then
$z(0), \dot z (0)= v(0)$ and the field tensor on the hyperplane
determined by $(x-z(0))^\lambda v_\lambda (0)=0.$ If we assume
that
$R_\rho|\dot v| < 1$, then the future is decoupled from the past
and one would hope to have a unique solution. For short times one
can presumably copy the proof given for the Abraham model. We are
not aware of mathematical results which ensure the existence
global in time. If the acceleration becomes large and
$R_\rho|\dot v| > 1$ for some time span, it is not even clear how
to properly specify the initial data. Since such
questions remain largely unexplored, the only option is to
proceed as if a solution is well defined.

As seen from the laboratory frame $\mathcal S$ the rigid charge
distribution Thomas precesses. To be physically consistent we have to allow
then also for a rotation of the body fixed frame relative to
$\mathcal{K}(s)$. For sure, such a rotation will be induced
through the back reaction of the electromagnetic field onto the
relativistically rigid charge. Let us denote by
\begin{equation}\label{af.w}
       \omega^\mu _{~\:\nu} (s)
\end{equation}
the angular velocities of the motion of the body fixed frame
relative to $\mathcal{K}(s)$. From the point of view of the laboratory system they  simply
have to be added to $\eta^\mu _{~\:\nu}$, i.e. in the current
(\ref{af.l}) $\eta^\mu _{~\:\nu}$ has to be substituted by $\eta^\mu _{~\:\nu} +
\omega^\mu _{~\:\nu}$.
The variation of the action is now with respect to $A^\mu,
z^\mu,$ and the Euler angles of the body fixed frame relative to
$\mathcal{K}(s)$. This results in the inhomogeneous Maxwell equations, as before, and in a
coupled set of translational and rotational equations of motion which are
nonlocal in time.


\subsection{Appendix: Variation of the action}\label{sec.ag}
We carry out the variation  $\delta {\textsf{S}}_{\mathrm {int}}$
of (\ref{af.r}). Let us first list the necessary identities. Since $(ds)^2 = dz^\mu \, dz_\mu$
we have
\begin{equation}\label{ag.a}
       \delta(ds) = \frac{dz^\sigma}{ds} \delta dz_\sigma =
       v^\sigma \, ds ~\frac{d}{ds}~ \delta z_\sigma
\end{equation}
and
\begin{equation}\label{ag.b}
       \delta v^\mu = \delta \frac{dz^\mu}{ds} = \frac{dz^\mu + \delta dz^\mu}{ds+ \delta ds} -
       \frac{dz^\mu}{ds}\,  = \frac{d}{ds} \, \delta z^\mu -
       v^\mu \, v_\sigma \frac{d}{ds}\, \delta z^\sigma\, .
\end{equation}
Similarly
\begin{equation}\label{ag.c}
       \delta \dot v ^\mu = \frac{d}{ds}\,\delta
       v^\mu - \dot v^\mu v_\sigma\, \frac{d}{ds}\,  \delta
       z^\sigma\, .
\end{equation}
In the variation of $\eta^\mu_{~\:\nu}$ we have to maintain the
frames at $z^\mu(s)$ and $z^\mu(s) + \delta z^\mu(s)$ without
relative spatial rotation. This leads to
\begin{equation}\label{ag.d}
       \delta \eta^\mu_{~\:\nu} =  v^\mu \delta \dot v_\nu -  v^\nu \,
       \delta \dot v_\mu = v^\mu\,  \frac{d}{ds}\,
         \delta v_\nu - v^\nu \, \frac{d}{ds}\, \delta v_\mu
         - \eta^\mu_{~\:\nu}\, v_\sigma \, \frac{d}{ds}\,
         \delta z^\sigma\, .
\end{equation}
If in (\ref{ag.d}) we would also vary $v^\mu$, then the constraint of zero spatial rotation
is no longer satisfied.

In the following we will assume that (\ref{af.e}) has a unique
solution within the tube traced out by the form factor. For this
we need $\dot v^\nu (x - z)_\nu < 1$ inside the tube,
equivalently $R_\rho |\dot v| < 1$, i.e. the acceleration has to
be sufficiently small on the scale of $1/ R_\rho$.
Using this condition, $\textsf{S}_{\mathrm {int}}$
can be written more explicitly as
\begin{equation}\label{af.t}
       \textsf{S}_{\mathrm {int}} = e \int\limits_{s_1}^{s_2} ds \int d^4
       \xi\, f(\xi^2)  \delta(v_\lambda(s)\xi^\lambda)  A^\mu (z(s)+ \xi)
       (v_\mu (s) - \eta_{\mu\nu}(s) \xi^\nu) \,.
\end{equation}
Then
\begin{eqnarray}
       \delta {\textsf{S}}_{\mathrm {int}} &=& e \int\limits_{s_1}^{s_2} ds\,
       (\frac{d}{ds}\, \delta z^\sigma) v_\sigma \int d^4 \xi \,
       f(\xi^2) \delta(v_\lambda \xi^\lambda)\, A^\mu (z+ \xi)
       (v_\mu - \eta_{\mu\nu}\, \xi^\nu)\nonumber\\
       && +  e \int\limits_{s_1}^{s_2} ds \int d^4 \xi \,  f(\xi^2)
       \delta  (v_\lambda \xi^\lambda) \,
        A^\mu (z+ \xi)\,\nonumber\\
        &&\qquad\qquad\qquad[\delta v_\mu  - v_\mu \frac{d}{ds}\, \delta
        v_\nu \xi^\nu + v_\nu \frac{d}{ds}\, \delta v_\mu
        \xi^\nu
        + \eta_{\mu\nu}\, \xi^\nu v_\sigma  \frac{d}{ds}\, \delta
        z^\sigma] \nonumber\\
        &&+ e \int\limits_{s_1}^{s_2} ds \, \int d^4 \xi \,
       f(\xi^2)\, [\delta(v_\lambda \xi^\lambda)\, \delta
       z_\sigma\,
       \partial^\sigma\, A^\mu (z+ \xi) \nonumber\\
       &&\qquad\qquad\qquad+ \delta^\prime (v_\lambda \xi^\lambda) \delta
       v_\sigma\,\xi^\sigma\,
       A^\mu(z+ \xi)] (v_\mu - \eta_{\mu\nu}\, \xi^\nu) \,.\label{ag.e}
\end{eqnarray}
We define
\begin{equation}\label{ag.f}
       Q^\mu = e \int d^4 \xi \, f(\xi^2)\,
       \delta(v_\lambda\xi^\lambda) \, A^\nu (z+ \xi) v_\nu\,
       \xi^\mu \, .
\end{equation}
Then
\begin{eqnarray}
       \delta {\textsf{S}}_{\mathrm {int}} &=& - \int\limits_{s_1}^{s_2} ds\,
       \frac{d}{ds}\, (Q^\mu \delta v_\mu) + \int\limits_{s_1}^{s_2}
       ds \, [\dot Q^\mu -
       v^\mu v_\nu \dot Q^\nu]\frac{d}{ds}\, \delta z_\mu \, \nonumber\\
       && + e \int\limits_{s_1}^{s_2} ds \int d^4 \xi \, f(\xi^2)
       \delta(v_\lambda \xi^\lambda) \, A^\mu (z+ \xi)\,
       \frac{d}{ds}\, \delta z_\mu\nonumber\\
       && + e \int\limits_{s_1}^{s_2} ds \int d^4 \xi\, f(\xi^2)
       \delta^\prime(v_\lambda \xi^\lambda)\, A^\nu (z+ \xi)
       (v_\nu - \eta_{\nu\sigma} \xi^\sigma)
        (\xi^\mu - \xi^\sigma
       v_\sigma v^\mu)\, \frac{d}{ds}\, \delta z_\mu\nonumber\\
        && + e \int\limits_{s_1}^{s_2}ds \, \delta z_\mu \int d^4 \xi
        \, f(\xi^2) \delta(v_\lambda \xi^\lambda) F^{\mu\nu}
        (z+\xi)
        (v_\nu - \eta_{\nu\sigma} \xi^\sigma)\nonumber\\
        && + e \int\limits_{s_1}^{s_2}ds\, \delta z_\mu \int d^4 \xi
        \, f(\xi^2) \delta (v_\lambda \xi^\lambda) \, \partial^\nu
        A^\mu (z+\xi) (v_\nu - \eta_{\nu\sigma} \xi^\sigma)\,
        .\label{ag.g}
\end{eqnarray}
The last term can be rewritten as, using $x \delta (x)= 0$,
\begin{eqnarray}
       && \int d^4 \xi\,f(\xi^2)\, \delta(v_\lambda \xi^\lambda)\,
                \partial^\nu\, A^\mu (z+ \xi)\, (v_\nu - \eta_{\nu\sigma}
                \xi^\sigma)\nonumber\\
       &=& - \int d^4  \xi \, A^\mu (z+ \xi)\, \partial^\nu
        \,[ f(\xi^2)\, \delta(v_\lambda \xi^\lambda) (v_\nu - v_\nu
        \dot v_\sigma \xi^\sigma)]\nonumber\\
       &=& \int d^4 \xi\, A^\mu (z+ \xi)f(\xi^2)\,\delta(v_\lambda \xi^\lambda)
       v_\nu \dot v^\nu\nonumber\\
        && - \int d^4 \xi\, f(\xi^2)\, \delta^\prime(v_\lambda
        \xi^\lambda)\, A^\mu (z+ \xi) v^\nu v_\nu (1- \dot
        v_\sigma\xi^\sigma)\nonumber\\
        &=& - \int d^4 x \, f((x-z)^2)\,
        \delta^\prime(v_\lambda(x-z)^\lambda)\,
        A^\mu (x) (1- \dot v_\sigma (x-z)^\sigma)\nonumber\\
        &=& \frac{d}{ds}\, \int d^4 x\, f((x-z)^2)\,
        \delta(v_\lambda (x-z)^\lambda) A^\mu(x)\nonumber\\
        &=& \frac{d}{ds}\, \int d^4 \xi\, f(\xi^2)\,
        \delta(v_\lambda \xi^\lambda)\, A^\mu (z+\xi)\label{ag.h}
\end{eqnarray}
which cancels against the second term of (\ref{ag.g}).
Therefore
\begin{eqnarray}
       \delta {\textsf{S}}_{\mathrm {int}}  &=& -\int\limits_{s_1}^{s_2} ds\,
       \frac{d}{ds}\, (Q^\mu\, \delta v_\mu) + \int\limits_{s_1}^{s_2}
       ds \, \delta z_\mu \, \Big (- \frac{d}{ds}\, (\dot Q^\mu -
       v^\mu v_\nu \dot Q^\nu)\nonumber\\
       && - e \frac{d}{ds}\, \int d^4 \xi \,
       f(\xi^2)\,\delta^\prime(
       v_\lambda \xi^\lambda) \, A^\nu (z+ \xi)(v_\nu -
       \eta_{\nu\sigma} \xi^\sigma)(\xi^\mu- v^\mu v_\sigma\,
       \xi^\sigma)\nonumber\\
       && + e \int d^4 \xi\, f(\xi^2) \,\delta(v_\lambda
       \xi^\lambda)\, F^{\mu\nu}(\xi+z) (v_\nu -
       \eta_{\nu\sigma}\, \xi^\sigma)\Big )\, .\label{ag.i}
\end{eqnarray}
Under variation of $\delta z_\mu(s)$ the term in the big
brackets has to vanish.

The term containing $F^{\mu\nu}$ is already in its final form.
The remaining summands are the time derivative of  $I$, where
\begin{equation}\label{ag.j}
       I = \dot Q^\mu - v^\mu v_\nu \dot Q^\nu + e \int d^4 \xi \,
       f(\xi^2)\, \delta^\prime(v_\lambda \xi^\lambda)\, A^\nu
       (z+\xi)(v_\nu - \eta_{\nu\sigma} \,\xi^\sigma) (\xi^\mu -
       v^\mu  v_\sigma\, \xi^\sigma)\,.
\end{equation}
We have, using partial integration with respect to $\xi$, $x\delta(x)=0$,
and $ x \delta^\prime(x) = - \delta(x)$,
\begin{eqnarray}
       \dot Q^\mu &=& e \int d^4 \xi\, f(\xi^2) \delta (v_\lambda
       \xi^\lambda) (A^\nu (z+\xi) \dot v_\nu \xi^\mu +
       \partial^\sigma A^\nu(z+\xi) v_\sigma v_\nu
       \xi^\mu)\nonumber\\
       && +  e \int d^4 \xi\, f(\xi^2) \delta^\prime (v_\lambda
       \xi^\lambda) A^\nu (z+\xi) \dot v_\sigma\xi^\sigma
         v_\nu\xi^\mu \nonumber\\
       &=& e \int d^4 \xi\, f(\xi^2) \delta (v_\lambda
       \xi^\lambda) A^\nu (z+\xi) (\dot v_\nu \xi^\mu -
       v^\mu  v_\nu)\nonumber\\
       && - e \int d^4 \xi\, f(\xi^2) \delta^\prime (v_\lambda
       \xi^\lambda) A^\nu (z+\xi)  v_\nu \xi^\mu (1- \dot v_\sigma
         \xi^\sigma)\,,\label{ag.k}\\
         v_\nu \dot Q^\nu&=&  - e \int d^4 \xi\, f(\xi^2)
         \delta(v_\lambda \xi^\lambda)\, A^\nu (z+\xi) v_\nu \dot
         v_\sigma \xi^\sigma
         \,, \label{ag.l}\\
       \dot Q^\mu - v^\mu v_\nu \dot Q^\nu &=& e
       \int d^4 \xi\, f(\xi^2)
         \delta(v_\lambda \xi^\lambda)\, A^\nu (z+\xi) \dot v_\nu
         \xi^\mu\nonumber\\
         && -  e \int d^4 \xi\, f(\xi^2)
         \delta(v_\lambda \xi^\lambda)\, A^\nu (z+\xi) v_\nu v^\mu(1-
         \dot v_\sigma \xi^\sigma)\nonumber\\
         && - e \int d^4 \xi\, f(\xi^2)
         \delta^\prime(v_\lambda \xi^\lambda)\, A^\nu (z+\xi) v_\nu
         \,\xi^\mu (1- \dot v_\sigma \xi^\sigma)\,.\label{ag.m}
\end{eqnarray}
We turn to the second term in (\ref{ag.j}) and use $x \delta^\prime(x)
 = - \delta(x)$, $ x^2 \delta^\prime
(x)=0$,
\begin{eqnarray}
       && e \int d^4 \xi \, f(\xi^2)\, \delta^\prime(v_\lambda
       \xi^\lambda)\, A^\nu (z+\xi) (v_\nu - v_\nu \dot  v_\sigma \xi^\sigma +
       \dot v_\nu v_\sigma \xi^\sigma)(\xi^\mu - v^\mu v_\sigma
       \xi^\sigma)\nonumber\\
       &=& e \int d^4 \xi \, f(\xi^2)\, \delta^\prime(v_\lambda
       \xi^\lambda)\, A^\nu (z+\xi) (v_\nu \xi^\mu - v_\nu \xi^\mu \dot v_\sigma
       \xi^\sigma)\nonumber\\
       && + e \int d^4 \xi \, f(\xi^2)\, \delta(v_\lambda
       \xi^\lambda)\, A^\nu(z+ \xi)(v_\nu v^\mu - v_\nu
       v^\mu \dot v_\sigma \xi^\sigma - \dot v_\nu
       \xi^\mu)\, .\label{ag.n}
\end{eqnarray}
Thus $I = 0$.

We conclude that the variation
\begin{equation}\label{ag.o}
        \delta \big ({\textsf{S}}_0 + {\textsf{S}}_{\mathrm {int}}\big ) = 0
\end{equation}
at fixed endpoints leads to the Lorentz force equation
\begin{equation}\label{ag.p}
       m_{\mathrm b}\, \frac{d}{ds}\, v^\mu = e \int d^4\, \xi\, f(\xi^2)
       \delta(v_\lambda \xi^\lambda)\, F^{\mu\nu}(z+ \xi) (v_\nu -
       \eta_{\nu\sigma} \xi^\sigma)\,.
\end{equation}\bigskip


\subsection*{Notes and References}
{\it ad \ref{sec.aa}, ad \ref{sec.ab}:} The material discussed can be found in any
textbook. Particularly useful I find Jackson (1999) and Scharf
(1994).
\bigskip\\
{\it ad \ref{sec.ac}:} In our history chapter, Section \ref{sec.b}, we
discuss other approaches which cannot be subsumed under short
distance regularization. In the literature the size of a classical electron,
$r_c$, is usually determined through equating the rest mass with
the Coulomb energy, $m_e c^2 = e^2/r_c$, which gives $r_c=3 \times
10^{-13}$cm. This is really a lower bound in the sense that an
even smaller radius would be in contradiction to the experimentally
observed mass of the electron (assuming a positive bare mass, cf.
the discussion in Section \ref{sec.de}). Milonni (1994) argues that
due to quantum fluctuations the electron appears to have a
classical extent, which is given by its Compton wave length $\lambda_c= r_c/\alpha$
with $\alpha$ the fine structure constant. Renormalization in
Euclidean quantum field theory is explained by Glimm,  Jaffe
(1987). Effective potentials for classical fluids are discussed,
e.g., in Huang (1987).
\bigskip\\
{\it ad \ref{sec.ad}:} The Abraham model was very popular at the beginning
of this century and was studied by Lorentz (1892,1915), Abraham (1903,1905),
Sommerfeld (1904,1905), and Schott (1912),
amongst others. Apparently a rotating rigid charge had  been
considered only much later Frenkel (1926), Bhabha,  Corben (1941). The proof
on the existence and uniqueness of the dynamics is taken from
Komech,  Spohn (1999), where a much wider class of external potentials
are allowed. A somewhat different technique is used by Bauer,
D\"{u}rr (1999). They cover also the case of a negative bare mass
and discuss the smoothness of solutions in terms of smoothness of
initial data. \bigskip\\
{\it ad \ref{sec.ae}:} The long time asymptotics is treated in
Komech,
Spohn (1999),  where the details of the proof can be
found.  Pitt's version of the Wiener theorem is proved in Rudin
(1977), Theorem 9.7(b). We remark that Theorem
1.3 gives no rate of convergence.
Thus to investigate the asymptotics of the
velocity and position always requires some extra considerations.
 Komech, Spohn (1998) study the long time asymptotics for zero
external potentials in the case of a scalar wave  field.
In particular, by these methods one can handle the asymptotics of
scattering trajectories.
Presumably these results extend to the Maxwell field. They require
the Wiener condition. In Section \ref{sec.dc} we give  a
different proof without Wiener condition $(W)$ but for a
sufficiently small charge. This indicates that $(W)$ is an
artifact of the proof. One might wonder, how bounded energy
dissipation works in the case of several particles. The result is
somewhat disappointing. One concludes only that the center of mass
acceleration vanishes in the long time limit. To show that also the relative
motions come to rest requires novel techniques.

In the literature, Bohm,  Weinstein (1948), in particular
the review by Pearle (1982),  periodic solutions of the Abraham model have
been reported repeatedly for the case of a charged sphere, i.e.
$\rho(\boldsymbol{x}) = e (4\pi a^2)^{-1} \delta(|\boldsymbol{x}|-a)$, which is not covered
by Theorem \ref{thm.ac} since $(W)$ is violated. These
computations invoke certain approximations and it is not clear
whether the full model, as defined by (\ref{ad.b}) to
(\ref{ad.d}), has periodic solutions. Kunze (1998) excludes
periodic solutions in some small part of phase space for the scalar
field without the Wiener condition $(W)$.\bigskip\\
{\it ad \ref{sec.af}:} This section is based on the monumental work of
Nodvik (1964). Nodvik includes the rotation of the body fixed
frame. A complementary discussion is given by Rohrlich (1990),
Chapter 7--4. The relativistic Thomas precession is discussed in
M{\o}ller (1952).

Of course relativistic theories have been studied much before, e.g. Born (1909).
One difficulty is to write down the proper variant of the Lorentz force equation,
which may be circumvented by considering the Li\'{e}nard--Wiechert fields
generated by a point
charge and to infer through the balance of momentum the actual motion of the charge, e.g.
 Rohrlich (1990), Teitelbom et.al. (1980). One then looses the notion of
 a true trajectory and its approximation through a comparison
 dynamics. In particular, the selection of physical solutions can
  come only through an additional postulate.

The current generated by a point charge can be written as
\begin{equation}\label{ar.a}
       j^\mu (x) = \int\limits_{-\infty}^\infty
       ds v^\mu (s) \delta (x-z(s))\,.
\end{equation}
McManus  (1948) proposed to smear out the $\delta$-function as
\begin{equation}\label{ar.b}
       j^\mu (x) = \int\limits_{-\infty}^\infty ds v^\mu(s) f((x-z
       (s))^2)\,,
\end{equation}
which is to be inserted in the action (\ref{af.r}), (\ref{af.s}).
The resulting equations of motion are nonlocal in space--time.
McManus did not identify the conserved four-momentum. The case of
rectilinear motion only is discussed by Schwinger (1983).
  It seems to us that Nodvik (1964) is the
only worked out example of an extended relativistic charge model.
As to be explained, for slowly varying external potentials it
agrees with experience over the full range of allowed velocities.
 \bigskip\\
{\it ad \ref{sec.ag}:} The variation of Nodvik (1964), which includes the
Euler angles of the body fixed frame, is adapted here to the
restricted variation over the world line of the particle only.
\newpage
\section{Historical Notes}\label{sec.b}
\setcounter{equation}{0}
\subsection{Extended charge models (1897-1912)}\label{sec.ba}
When in 1897 J.J. Thomson identified  the cathode rays as
consisting of particles with charge $-e$, he had not only
discovered the first elementary particle,  but he also
challenged the theoretical physicists to compute the
energy--momentum relation of the electron. To put it slightly
differently, we write the equations of motion in  crossed $\boldsymbol {E}$ and
$\boldsymbol {B}$ fields as
\setcounter{equation}{0}
\begin{eqnarray}\label{ba.a}
       m(\boldsymbol {v}) \dot{\boldsymbol {v}} = e (\boldsymbol {E}
       + c^{-1} \boldsymbol {v} \times \boldsymbol {B})
\end{eqnarray}
with $m(\boldsymbol {v})$ the velocity dependent mass as a $3 \times 3$ matrix.
The challenge was then to determine the ratio $m(\boldsymbol {v})/e$. In a long
series is of experiments starting in 1897 the
velocity dependent mass was measured by W. Kaufmann (1901). With improving technology the
experiments were repeated and extended to a larger range of
velocities. Latest by 1914 the relativistic
dependence $m(\boldsymbol {v}) = m_0 \gamma (\unit +
\gamma^2 c^{-2} |\boldsymbol {v} \rangle\langle \boldsymbol {v}|), \gamma =
(1- \boldsymbol {v}^2/c^2)^{-1/2}$, and $m_0$
the rest mass of the electron, was regarded as
well confirmed.

So which theory could be used to determine $m(\boldsymbol {v})$? In fact, there
was little choice. Since the phenomenon under consideration is
clearly electromagnetic, one had to use Maxwell equations, and
since the trajectory of a single charge was measured, one had to
couple to the Lorentz force equation. Thus the electron was pictured as
a tiny sphere charged with electricity. In the
inhomogeneous Maxwell equations one had to insert  the current
generated by that moving sphere. On the other hand the
electromagnetic fields back react on the charge distribution
through the Lorentz force equation. Thereby one has introduced what is called
an  extended charge
model. Abraham (1905) adopted first a in the absolute laboratory frame rigid
charge distribution. The corresponding energy--momentum relation
is discussed at length in the second volume of his book  on
electromagnetism, compare with Section \ref{sec.ca}. For the Abraham model,
Sommerfeld (1904,1905) obtained an exact equation of motion for the electron,
which as a complicating and unfamiliar feature
 contained memory terms as a result of integrating over the
retarded fields. Lorentz (1904a,b) advertised a charge
distribution which is rigid in its momentary rest frame and therefore,
as seen from the laboratory frame, is contracted parallel to its
momentary velocity. Of course, it was left  completely open by which forces
this charge distribution is kept in place. Poincar\'e
(1906) developed non--electromagnetic models where additional stresses counteracted the
Coulomb repulsion.

In all extended charge models the velocity dependent mass has the
additive structure $m(\boldsymbol {v})= m_{\mathrm{b}}\unit + m_{\mathrm{f}}
(\boldsymbol {v})$, where $m_{\mathrm{b}}$ is the bare
mechanical mass of the particle, in accordance with Newtonian
mechanics taken to be velocity independent, and where $m_{\mathrm{f}}(\boldsymbol{v})$
is the mass due to the coupling to the field, which had to be
computed from the model charge distribution. Clearly, in the
experiment only the sum $m_{\mathrm{b}} + m_{\mathrm{f}}$ can be measured. Lorentz
apparently favored to set $m_{\mathrm{b}} = 0$. His model yields then the
usual
relativistic velocity dependence,
cf. Notes to
Section \ref{sec.ca}. However,  in these times a fully
relativistic model was out of reach. Rather one relied on the
semi--relativistic Abraham model of a rigid sphere and substituted
at the appropriate places a relativistically contracting charge
distribution.

By 1904 the theoretical predictions were worked out with the
experiments not yet precise enough to clearly distinguish between
them. Nevertheless the whole enterprise came to a rather sudden
end, since Einstein (1905a,b) forcefully argued that just like
electromagnetism in vacuum also the mechanical laws had to be
Lorentz invariant. But if Einstein was right, then the
energy--momentum relation of the electron had to be the
relativistic one, as emphasized independently  by Poincar\'e (1906).
Thus the only free parameter was the rest mass of the electron
which anyhow could not be deduced from theory, since the actual
charge distribution was not known. There was simply nothing left to
compute. By latest in 1913 with the atomic model of Bohr, it was
obvious that a theory based on classical electromagnetism could
not account for the observed stability of atoms nor for the sharp
spectral lines. As a tool to explain properties of atoms,
electrons, and nuclei the classical electron theory was abandoned.

The effective equation of motion for the electron as given by
Equation
(\ref{ba.a}) could not possibly have  been the full story.  Through the
work of Larmor it was already understood that a charge looses energy
through radiation roughly proportional to $\boldsymbol {\dot v}^2$. Lorentz
observed that in the approximation of small velocities this loss
could be written as the friction or radiation reaction force
\begin{equation}\label{ba.b}
        \boldsymbol {F}_{rr} = \frac{e^2}{6 \pi c^3}\, \ddot{\boldsymbol
        {v}}\,,
\end{equation}
which had to be added to  the effective Lorentz force equation (\ref{ba.a}).
In 1904 Abraham obtained this friction force for arbitrary
velocities as
\begin{equation}\label{ba.c}
       \boldsymbol {F}_{rr} = \frac{e^2}{6 \pi c^3}\, [\gamma^4
       c^{-2} (\boldsymbol {v} \cdot \ddot{\boldsymbol {v}}) \boldsymbol {v} + 3
       \gamma^6 c^{-4}(\boldsymbol {v} \cdot \dot{\boldsymbol {v}})^2
       \boldsymbol {v} + 3 \gamma^4 c^{-2}(\boldsymbol {v} \cdot \dot{\boldsymbol {v}})
       \dot{\boldsymbol{v}} + \gamma^2 \ddot{\boldsymbol {v}}]\,,
\end{equation}
$\gamma = (1- \boldsymbol {v}^2/c^2)^{-1/2}$. He argued that energy and momentum
is transported to infinity through  the far field. On that scale the
charge distribution is like a point charge and the electromagnetic
fields can be computed from the Li\'enard--Wiechert potentials.
Using conservation of energy and momentum for the total system he
showed that the loss at infinity could be accounted for by the
friction like force (\ref{ba.c}). Von Laue (1909) recognized
that the radiation reaction is relativistically invariant and can
be written as,
\;$\dot{}$\;
denoting now differentiation with respect to the
eigentime,
\begin{equation}\label{ba.d}
       F^\mu_{rr} = \frac{e^2}{6 \pi c^3}\, [\ddot v^\mu - c^{-2}
         \dot v^\lambda  \dot v_\lambda  v^\mu ]\,.
\end{equation}
This is how the radiation reaction appears in the famous 1921
review article of Pauli on relativity. But apparently, there
was no incentive to study properties of the effective Lorentz
force
equation (\ref{ba.a}) including the full radiation reaction
correction (\ref{ba.c}). According to Schott (1912) after studying the motion
in a uniform electric field: {\it
Hence the effect of the reaction due to radiation is quite
inappreciable in this and probably in all practical cases.}
For applications, simpler phenomenological
approaches sufficed.

The first chapter on the dynamics of classical electrons closes
around 1912 as compiled and worked out in
great detail by Schott (1912). In essence there were
 two results: (i) a relativistically invariant
expression for the radiation reaction and (ii) energy--momentum
relations for the charged particle which were depending on the
particular model charge distribution. However all models were
inconsistent with Einstein's theory of special relativity. In
particular, the rest mass came out to be different from the
electrostatic energy of the charge distribution.
\subsection{The point charge }\label{sec.bb}
Our second chapter consists of a single paper: ``Classical
theory of radiating electrons'' submitted by P.A.M. Dirac on March
15, 1938. But before we have to follow up the intermission during
which some research on the classical electron theory continued. We
mention only  the studies by
Fermi (1922) and Frenkel (1925). Fermi argues that Abraham and Lorentz had not used the
relativistically proper definition of energy and momentum which
explained their  disagreement with Einstein's theory.
Frenkel, apparently influenced  by Ehrenfest, proposes to consider electrons
 as undivisable, therefore without any extent, and studies the general
  structure of equations of motion.
 Of course, the
most important event during the intermission was the development
of quantum mechanics, which almost immediately after its discovery
was applied to quantizing the electromagnetic field. Thereby the
line shape and life time for excited states of atoms could be
determined. Quantum mechanics gave a strong push to the classical
theory. One had to quantize in Hamiltonian form. Thus the
Lagrangian and Hamiltonian structure of the  coupled Maxwell--Lorentz
equations had to be explored, the role of the constraints and of
the gauge freedom had to be understood.

It became apparent fairly soon that the newly born quantum
electrodynamics yields infinities when one tries to remove the
ultraviolet cutoff, i.e. in the limit of a point charge
distribution. Thus a problem which had been dropped over 15 years
before reappeared in a different guise. In the '30 and early
'40 it was a fairly widespread believe that one way to overcome
the difficulties of quantum electrodynamics is a better
understanding of the classical theory of point charges coupled to
their radiation field. Of course, this was only a vehicle to the
final goal, namely a consistent quantized theory. We do not
describe the various attempts, since the proper
formulation of quantum electrodynamics eventually went a very different
route. Dirac's paper was equally motivated by quantum
electrodynamics. However, as such it is concerned only with the
classical electron theory.

We have to report the findings of Dirac in fair detail, since most
further activities start from there. The formal argument in the
original paper can be well followed and alternative versions can be
found  in Rohrlich (1990), Teitelbom et. al. (1980), Thirring (1997).
 So there is no need for
repetition and we concentrate on the conclusions. At first reading
it is mandatory to disregard all philosophical claims and to
concentrate on the equations. But before, let us see how Dirac
himself viewed the 1897-1912 period: {\it The Lorentz model of the
electron as a small sphere charged with electricity, possessing mass
account of the energy of the electric field around it, has proved
very valuable in accounting for the motion and radiation of
electrons in a certain domain of problems, in which the
electromagnetic field does not vary too rapidly and the
accelerations of the electrons are not too great.} Dirac wanted to
construct quantum electrodynamics. There the electron is regarded
as an elementary particle with, almost by definition, no internal
structure. Thus Dirac had to dispense with model charges and to
develop a theory of {\it point} electrons.

What did Dirac really accomplish?
Of course, he assumes the validity of the inhomogeneous Maxwell
equations. The current is generated by a point charge whose motion
is yet to be determined. Mechanically this point charge is
relativistic with bare mass $m_{\mathrm{b}}$. There is no explicit back
reaction of the field onto the charge, since at no stage Dirac
would invoke the Lorentz force equation. Rather
conservation of energy and momentum should suffice to fix the true trajectory
 of the point charge.
Note that this is very different
from the extended charge models where the starting point is a
closed systems of equations for the particle and the Maxwell field.
Dirac studies the flow of energy and momentum through a thin tube
of radius $R$ around the world line of the particle. The
computation simplifies by writing  the retarded fields generated
by the motion of the point charge as
\begin{equation}\label{bb.a}
       F_{\mathrm{ret}} = \frac{1}{2}\, (F_{\mathrm{ret}} +
       F_{\mathrm{adv}}) + \frac{1}{2}\, (F_{\mathrm{ret}} - \
       F_{\mathrm{adv}})
\end{equation}
in all of space--time. The difference term turns out to be finite
on the world line of the charge  and yields
in the limit $ R \to 0$,  through a balancing of energy and
momentum, the relativistic radiation reaction (\ref{ba.d}). Thus in
retrospect
one can understand  why in the semi--relativistic Abraham
model the radiation reaction is nevertheless of relativistic form.

The more delicate term in (\ref{bb.a}) is the sum, which is divergent
on the world line of the particle. At the expense of ignoring
other divergent terms, cf. Thirring (1997), Equation (8.4.16), Dirac obtains
the expected result, namely
\begin{equation}\label{bb.b}
       - \frac{e^2}{4 \pi R c^2}\, \dot v^\mu = - m_{\mathrm{f}} \dot v^\mu \, .
\end{equation}
 Adding the radiation reaction (\ref{ba.d}) and equating with the
  mechanical four--momentum, the final result is an equation of motion
   which determines the trajectory
of the particle,
\begin{equation}\label{bb.c}
       (m_{\mathrm{b}} + m_{\mathrm{f}}) \dot v^\mu = m_{\mathrm{exp}}  \dot v^\mu =
       \frac{e}{c} F^{\mu\nu}_{\mathrm{ex}}\, v_\nu +
       \frac{e^2}{6 \pi  c^3}\, \big [ \ddot v^\mu - c^{-2}\,
       \dot v^\lambda \dot v_\lambda v^\mu \big ] + \mathcal{O}(R)
\end{equation}
with an error of the size of the tube, where we  have added in the
prescribed tensor
$ F^{\mu\nu}_{\mathrm{ex}}\, $ of the external fields.

To complete the argument Dirac has to take the limit $R \to 0$.
Since $m_{\mathrm{f}} \to \infty$, this amounts to
\begin{equation}\label{bb.d}
       m_{\mathrm{b}} \to - \infty~, ~  m_{\mathrm{f}} \to \infty~, ~ m_{\mathrm{exp}}  =
       m_{\mathrm{b}}+ m_{\mathrm{f}} \quad \mbox{fixed~,}
\end{equation}
where $m_{\mathrm{exp}}$ is adjusted such that it agrees with the
experimentally determined mass of the charged particle. (\ref{bb.d}) is
the classical charge renormalization.

Dirac admits that {\it such a model is hardly a plausible one
according to current physical ideas but this is not an objection
to the theory provided we have a reasonable mathematical scheme.}

Equation (\ref{bb.c}), dropping the terms $\mathcal{O}(R)$, is the
Lorentz--Dirac equation. Within the framework of Dirac it makes
no sense to ask whether the Lorentz--Dirac equation is ``exact'',
since there is nothing to compare to. The Lorentz--Dirac equation
comes as one package, so to speak. One only could compare with
real experiments, which is difficult since the radiation reaction
is so small, or one could compare with higher level theories as
quantum electrodynamics. But this has never been seriously
attempted, since it would require to have a well defined
relativistic quantum field theory which is a difficult task to
begin with.

The Lorentz--Dirac equation is identical to the effective
equations of motion obtained from extended charge models, if we
ignore for a moment that the kinetic energy might come out
differently depending on which model charge is used. In this sense
Dirac has recovered the classical results through a novel approach. However there
 is an important distinction. For extended charge models one
 has a true solution for the position of the charged particle,
 say $ \tilde{\boldsymbol{q}} (t)$. One can compare then $\tilde{\boldsymbol{q}}  (t)$ with
 a solution of the Lorentz--Dirac equation and hope for agreement
 in asymptotic regimes, like slowly varying potentials. In addition for
 an extended charge model a negative bare mass might have drastic
 consequences which cannot be ignored.

Dirac continues with an observation which shattered the
naive trust in the classical electron theory. He observes that even for zero external
fields (\ref{bb.c}) has solutions where $|\boldsymbol {v}(t)/c| \to 1$ as $t \to \infty$
and $|\dot{\boldsymbol{v}}(t)|$ increases without bound. Such unphysical
solutions he called runaway. If one inserts numbers, then runaways
grow very fast. E.g. for an electron one has $\dot{\boldsymbol{v}}(t) =
\dot{\boldsymbol{v}}(0) e^{t/\tau}$
with $\tau= 10^{-23}$ sec. If the Lorentz--Dirac equation (\ref{bb.c})
is a valid approximation in an extended charge model, which after
all was the main consensus of the 1897-1912 period, then also there
one encounters runaway solutions. It is somewhat  surprising that
apparently runaways went  completely unnoticed before, wich only
proves that no attempt was made to apply the Lorentz--Dirac
equation to a concrete physical problem.

Dirac proposed to eliminate the runaway solutions by requiring the
asymptotic condition
\begin{equation}\label{bb.e}
       \lim_{t \to\infty}\, \dot v^\mu (t) = 0 \, .
\end{equation}
As additional bonus the problem of the missing initial condition
is resolved: Since in (\ref{bb.c}) the third derivative appears, one
has to know $z^\mu (0), \dot z^\mu (0)$, as in any mechanical problem, and in
addition $ \ddot z^\mu (0)$. If one accepts (\ref{bb.e}), the initial
condition $\dot v^\mu(0)$ is replaced by the asymptotic condition
(\ref{bb.e}). Dirac checked that for zero external forces and for a
spatially constant but time--dependent force the asymptotic
condition singles out physically meaningful  solutions.

By the end of 1938 the classical electron theory
was in an awkward shape, in fact in a much worse shape than by the
end of the 1912. Formal, but even by strict standards
careful, derivations yielded an equation with unphysical
solutions. How did they come into existence? While Dirac's
asymptotic condition seemed to be physically sensible, it was very
much ad hoc and imposed post festum to get rid of unwanted guests.
Even those physicists willing to accept the asymptotic condition
as a new principle, like Haag (1955), could not be too
happy. Solutions satisfying the asymptotic condition are acausal
in the sense that the charge starts moving even before any force
is acting. To be sure the causality violation is on the time scale
of $\tau = 10^{-23}$ sec for an electron, and even shorter for a
proton, and thus has no observable consequences. But acausality
remains as a dark spot in a relativistic theory.

The clear recognition of runaway solutions generated a sort of
consensus that the coupled Maxwell--Lorentz equations have
internal difficulties.

 To quote from the preface of the book by
Rohrlich: {\it Most applications treat electrons as
point particles. At the same time, there was the widespread belief
that the theory of point particles is beset with various
difficulties such as infinite electrostatic self--energy, a rather
doubtful equation of motion which admits physically meaningless
solutions, violation of causality, and others. It is not
surprising, therefore, that the very existence of a consistent
classical theory of charge particles is often questioned.}\\
\indent
To quote from Chapter 28 of the Feynman Lectures: {\it Classical
mechanics is a mathematically consistent theory; it just doesn't
agree with experience. It is interesting, though, that the
classical theory of electromagnetism is an unsatisfactory theory
all by itself. The electromagnetic theory predicts the existence
of an electromagnetic mass, but it also falls on its face in doing
so, because it does not produce a consistent theory}.\\
\indent
To quote from the textbook on mathematical physics by Thirring:
 {\it Not all solutions to (\ref{bb.c}) are crazy.  Attempts,
have been made to separate
sense from nonsense by imposing special initial conditions.
It is to be hoped that some day the real  solution of
the problem of the  charge--field interaction will look
differently,
and the equations describing nature will not be so highly unstable
that the balancing act can only succeed by having the system
correctly prepared ahead of time by a convenient coincidence.}\\
\indent
To be sure, these issues were of concern only to theoretical
physicists in search for a secure foundation. Synchroton radiation
sources were built anyhow. The loss in energy of an electron
during one revolution can be accounted for by Larmor's formula.
This is then the amount of energy which has to be supplied in
order to maintain a stationary electron current. The  radiation
emitted from the synchroton source is computed from the
inhomogeneous Maxwell equations with a point charge source, i.e.
from the Li\'enard--Wiechert potentials. No problem.
\subsection{Wheeler--Feynman electrodynamics}\label{sec.bc}
To  avoid the infinities of self--interaction Wheeler and Feynman
(1945,1949) designed a radical solution, at least on the
classical level since the quantized version of their theory was
never accomplished. The Wheeler--Feynman theory departs in two
essential aspects from  standard electrodynamics.\\
(i) The only dynamical variables are the trajectories of the
charges. As such there are no electromagnetic fields, even though
one uses them as a familiar and convenient notational device.\\
(ii) To achieve agreement with observation, the theory requires to
have many particles. For example, in the two--body scattering
problem there is no radiation damping. Such friction forces are
understood as the result of the interaction with the charged
particles in the surrounding matter.

The starting point of the Wheeler--Feynman electrodynamics is an action
first written down by Fokker (1929). Let us consider $N$ particles
with mass $m_i$, charge $e_i$,
 and motion given by the world line $z_{(i)} (\tau_i) ,~ i=1
, \ldots, N$. The world line is parametrized by its eigentime $\tau_i$
and \;$\dot{}$\; denotes differentiation with respect to that eigentime.
The action functional has the form
\begin{eqnarray}\label{bc.a}
        {\sf S} &=& - \sum_{i=1}^N m_i\, c^2 \int \sqrt{(\dot z_{(i)})^2}\, d
        \tau_i\\
        && + \sum_{i,j=1\atop{i} \not = j}^{N} e_ie_j \int\int \delta
        (z_{(i)} - z_{(j)}) \dot z_{(i)} \cdot \dot z_{(j)}
        d \tau_i d \tau_j\,.\nonumber
\end{eqnarray}
A formal variation of $\sf S$ leads to the equations of motion
\begin{equation}\label{bc.b}
       m_i \ddot z_{(i)}^\mu = \frac{e_i}{c} \sum_{j=1\atop{j} \not
       =i}^{N}
       \frac{1}{2} \big ( F_{\mathrm{ret}(j)}^{\mu\nu} (z_{(i)})+
       F_{\mathrm{adv}(j)}^{\mu\nu}  (z_{(i)})\big) \dot
       z_{(i)\nu}\,.
\end{equation}
Here $F_{\mathrm{ret}(j)}^{\mu\nu}, ~ F_{\mathrm{adv}(j)}^{\mu\nu}$
are the retarded and advanced Li\'enard--Wiechert fields generated
by the charge at $z_{(j)}$ and evaluated at $z_{(i)}$. They are derived from the retarded
and advanced   potentials
\begin{eqnarray}\label{bc.c}
       A_{\mathrm{ret}(j)}^\mu (x) &=& e_j
       \dot z_{(j)}^\mu (\tau_{j\mathrm{ret}})/ (x_\sigma -
       z_{(j)\sigma} (\tau_{j\mathrm{ret}})) \dot z_{(j)}^\sigma
        (\tau_{j\mathrm{ret}})\,,\\[2mm]
        A_{\mathrm{adv}(j)}^\mu (x) &=& e_j \dot z_{(j)}^\mu
        (\tau_{j\mathrm{adv}})/
        (x_\sigma - z_{(j)\sigma}
        (\tau_{j\mathrm{adv}})) \dot z_{(j)}^\sigma
        (\tau_{j \mathrm{adv}}) \label{bc.d}
\end{eqnarray}
with $\tau_{j\mathrm{ret}}$, resp. $\tau_{j\mathrm{adv}}$,  the eigentime when the
trajectory $z_{(j)}$ crosses the backward,  resp. the forward, light cone
with apex at $x$.

To transform (\ref{bc.b}) into a familiar form, we
use the decomposition (\ref{bb.a}) and Dirac's
observation that $(F_{\mathrm{ret}} - F_{\mathrm{adv}})/2$ at the
trajectory of the particle yields the radiation reaction. Then
\begin{eqnarray}
       m_i \ddot z_{(i)}^\mu &=& \frac{e_i}{c} \sum_{j=1\atop{j} \not
       =i}^{N}
       F_{\mathrm{ret}(j)}^{\mu\nu} (z_{(i)}) \dot
       z_{(i)\nu}
         +\frac{e_i^2}{6 \pi c^3} (\stackrel{...}{z}_{(i)}^\mu -
        c^{-2}\, \ddot z_{(i)}^\nu\, \ddot z _{(i)\nu} \,\dot z_{(i)}^\mu)\nonumber\\
        && + \frac{e_i}{c} \sum_{j=1}^N \frac{1}{2}
        \big(F_{\mathrm{adv}(j)}^{\mu\nu}  (z_{(i)})-
        F_{\mathrm{ret}(j)}^{\mu\nu}(z_{(i)})\big)
         \dot z_{(i)\nu}\,.\label{bc.e}
\end{eqnarray}
Of course, being symmetric in time, we could have equally
transformed to the advanced fields for the force and a radiation
reaction with reversed sign.

We note that in (\ref{bc.e}) the mass of the particle is not
renormalized. The retarded force is of the usual form. The
radiation reaction has runaways. So one must either impose the
asymptotic condition (\ref{bb.e}) or have the good faith that
(\ref{bc.b}) does not posess such unphysical solutions. The last
term in (\ref{bc.e}) is unwanted and Wheeler and Feynman spend a considerable
effort to argue that for a sum over a large number of charges in
disordered motion this last term vanishes. If it is exactly zero,
the condition of a perfect absorber is satisfied and the standard
equations of motion for  charged particles result.

As with Dirac, one can accept only the whole Wheeler--Feynman
package. Consequently, there has been little further work on the
theory. In particular, it has never been checked how well the
assumption of a perfect absorber is satisfied.

\subsection*{Notes and References}
{\it ad \ref{sec.b}:} A more detailed account on the history of
the classical electron theory can be found in Pais (1972,1982),
Rohrlich (1973), and in the introductory chapters of Rohrlich
(1990). The interconnection with quantum electrodynamics before
the 1947 Shelter  Island  conference is well described in
Schweber (1994).\bigskip\\
{\it ad \ref{sec.bb}}: Kramers (1948) investigations on the mass
renormalization in the classical theory were instrumental for a
correct computation of the Lamb shift. We refer to Dresden (1987 ) and
Schweber (1994).\bigskip\\
{\it ad \ref{sec.bc}}: The two--body problem in Wheeler-Feynman
electrodynamics is discussed by Schild (1963). The existence and
classification of solutions is studied by Bauer (1997). A few
explicit solutions are listed in Stephas (1992).
\newpage
\section{Energy--Momentum
Relation}\label{sec.c}
\setcounter{equation}{0}
For the Abraham model we established already that its energy
 $\cal E$ is conserved, $\mathcal E$ given by Equation  (\ref{ad.g}). If the external fields
vanish, then the dynamics is invariant under spatial translations.
Thus the total momentum, denoted by $\mathcal P$,  must also be  conserved. The minimum of
$\mathcal E$  at fixed $\mathcal P$ defines the energy--momentum
relation.

If the external forces vanish, the simplest solution to the
equations of motion has the  particle travelling at constant velocity
$\boldsymbol{v}$ in company with its  electromagnetic
fields. There seems to be no accepted terminology for this object.
Since it will be used as a basic building block later on, we need
a short descriptive name and we call this particular solution a
{\it charge soliton}, or simply soliton, at velocity $\boldsymbol{v}$,
in analogy to
solitons of nonlinear wave equations. The soliton has an energy
and a momentum which are linked through the energy--momentum relation.

In the following two sections we compute the conserved energy and
momentum, the charge solitons, and the energy--momentum relation for
both the Abraham and the Lorentz model. We will assume
$\phi_{\mathrm{ex}} = 0, ~ \boldsymbol{A}_{\mathrm{ex}} = 0$ throughout.


\subsection{Abraham model}\label{sec.ca}
The mechanical momentum of the particle is given by
\begin{equation}\label{ca.a}
         m_{\mathrm{b}} \gamma \boldsymbol{v}
\end{equation}
and the momentum of the field by
\begin{equation}\label{ca.b}
       \mathcal{P}_f = \int d^3 x  \big(\boldsymbol{E}(\boldsymbol{x})
           \times \boldsymbol{B}
       (\boldsymbol{x})\big)\, .
\end{equation}
Thus we set the total momentum
\begin{equation}\label{ca.c}
       \mathcal{P} = m_{\mathrm b} \gamma \boldsymbol{v} + \mathcal{P}_f
\end{equation}
as a functional on $\mathcal{M}$. One easily checks that $\mathcal{P}$
is conserved by the Maxwell--Lorentz equations (\ref{ad.b}) -
(\ref{ad.d}). The corresponding Lagrangian, compare with
(\ref{ad.f}), is invariant under spatial translations and $\mathcal{P}$
is the conserved quantity which, by N\"{o}ther's theorem,
corresponds to this symmetry.

We want to minimize the energy at fixed total momentum. We
eliminate $\boldsymbol{v}$ between (\ref{ad.g}) and (\ref{ca.c}) and thus have to
minimize
\begin{equation}\label{ca.d}
       \big(m_{\mathrm{b}}^2 + \big (\mathcal{P} - \int d^3 x (
           \boldsymbol{E}   \times \boldsymbol{B})
       \big )^2 \big)^{1/2} +
       \frac{1}{2} \int d^3x  (\boldsymbol{E}^2 + \boldsymbol{B}^2)
\end{equation}
at fixed $\mathcal{P}$ and subject to the constraints $\nabla
\cdot \boldsymbol{E} = \rho\,,  \nabla \cdot \boldsymbol{B} = 0 $. By translation
invariance we can center $\rho$ at an arbitrary $\boldsymbol{q} \in
\mathbb{R}^3$. For $\boldsymbol{q}=0$, say, the  minimizer is unique and given by
\begin{eqnarray}\label{ca.e}
       \boldsymbol{E}_{\boldsymbol{v}} (\boldsymbol{x}) &=& - \nabla \phi_{\boldsymbol{v}}
       (\boldsymbol{x}) + \boldsymbol{v}(\boldsymbol{v}\cdot
       \nabla \phi_{\boldsymbol{v}}(\boldsymbol{x}))\, ,\nonumber\\[2mm]
       \boldsymbol{B}_{\boldsymbol{v}} (\boldsymbol{x}) &=& - \boldsymbol{v}
        \times \nabla \phi_{\boldsymbol{v}} (\boldsymbol{x}) \,,
\end{eqnarray}
where
\begin{equation}\label{ca.f}
       \widehat \phi_{\boldsymbol{v}} (\boldsymbol{k}) = [\boldsymbol{k}^2 -
       (\boldsymbol{v} \cdot \boldsymbol{k})^2]^{-1}
            \hat\rho(\boldsymbol{k})
\end{equation}
and in physical space
\begin{equation}\label{ca.g}
       \phi_{\boldsymbol{v}} (\boldsymbol{x}) =
       \int d^3 y \big ( 4 \pi \sqrt{\gamma^{-2} (\boldsymbol{x}-\boldsymbol{y})^2
       +
       (\boldsymbol{v} \cdot (\boldsymbol{x}- \boldsymbol{y}))^2}
       \,\big )^{-1}  \rho(\boldsymbol{y})\,.
\end{equation}
Here $|\boldsymbol{v}| < 1$, i.e. $\boldsymbol{v} \in \mathbb V$, and
$\boldsymbol{v}$ has to be
adjusted such that $ \mathcal P =  \boldsymbol{P}_{\! s}(\boldsymbol{v})$ with
\begin{eqnarray}
       \boldsymbol{P}_{\! s} (\boldsymbol{v})&=& m_{\mathrm{b}} \gamma
       \boldsymbol{v}
       + \int d^3 k
        | \hat\rho (\boldsymbol{k})|^2
        \big ([\boldsymbol{k}^2 - (\boldsymbol{k} \cdot \boldsymbol{v})^2]^{-1}
       \boldsymbol{v}\nonumber\\
       &&\qquad\qquad\qquad- \gamma^{-2} [\boldsymbol{k}^2 - (\boldsymbol{k} \cdot
       \boldsymbol{v})^2 ]^{-2} (\boldsymbol{k} \cdot
       \boldsymbol{v}) \boldsymbol{k}\big )\nonumber\\[2mm]
        &=& \boldsymbol{v} \big\{m_{\mathrm{b}} \gamma +
       m_{\mathrm{e}}  \,
       |\boldsymbol{v}|^{-2} \, \big [ \frac{1+ \boldsymbol{v}^2}{2|
       \boldsymbol{v}|}\, \log \frac{1+
       |\boldsymbol{v}|}{1-|
       \boldsymbol{v}|} - 1\, \big ] \big \}\, ,
\end{eqnarray}\label{ca.h}
where $m_{\mathrm{e}}$ is the electrostatic energy of the charge distribution
$\rho$,
\begin{equation}\label{ca.i}
       m_{\mathrm{e}}= \frac{1}{2} \int d^3 x\, d^3x^\prime  \rho(\boldsymbol{x})
       \, \rho(\boldsymbol{x}^\prime) (4 \pi
       |\boldsymbol{x}-\boldsymbol{x}^\prime|)^{-1}\, .
\end{equation}
The map
 $\mathbb{V} \ni \boldsymbol{v} \mapsto \boldsymbol{P}_{\! s}(\boldsymbol{v})
 \in \mathbb{R}^3$
is one to one  and therefore $\mathcal{P} = \boldsymbol{P}_{\! s}
(\boldsymbol{v})$
has a unique solution.
The minimizing energy is given by
\begin{eqnarray}\label{ca.j}
        E_s (\boldsymbol{v})&=& m_{\mathrm{b}} \gamma +
        \frac{1}{2}\,\int d^3 k
        | \hat\rho (\boldsymbol{k})|^2\,
        [\boldsymbol{k}^2 - (\boldsymbol{k} \cdot \boldsymbol{v})^2]^{-2}
        \big((1+ \boldsymbol{v}^2 )\boldsymbol{k}^2 - (3-\boldsymbol{v}^2)
        (\boldsymbol{v} \cdot \boldsymbol{k})^2 \big)\nonumber\\
        &=& m_{\mathrm{b}} \gamma + m_{\mathrm{e}} \,
        \big [\frac{1}{|\boldsymbol{v}|}  \log \,
        \frac{1+|\boldsymbol{v}|}{1-|\boldsymbol{v}|} - 1\, \big
        ]\,.
\end{eqnarray}
Eliminating now $\boldsymbol{v}$ between $E_s$ and
 $\boldsymbol{P}_{\! s}$ yields  the {\it
energy--momentum relation}
\begin{equation}\label{ca.k}
       E_\mathrm{{eff}} (\boldsymbol{p}) = E_s (\boldsymbol{v} (\boldsymbol{p}))\,,
\end{equation}
where $\boldsymbol{v}(\boldsymbol{P}_{\! s})$ is the function inverse  to
$\boldsymbol{P}_{\! s}(\boldsymbol{v})$. As to be underlined,  $E_\mathrm{{eff}}$
depends on the charge distribution only through its electrostatic
energy.

We note that
\begin{equation}\label{ca.l}
        \boldsymbol{P}_{\!\! s} (\boldsymbol{v}) = \nabla_{\boldsymbol{v}} T (\boldsymbol{v})
\end{equation}
with
\begin{equation}\label{ca.m}
        T (\boldsymbol{v}) = - m_{\mathrm{b}}\gamma^{-1} +
        \frac{1}{2}\, \gamma^{-2} \int d^3 k
        |\hat\rho(\boldsymbol{k})|^2 \, [\boldsymbol{k}^2 -
        (\boldsymbol{k} \cdot \boldsymbol{v})^2]^{-1}
\end{equation}
and that
\begin{equation}\label{ca.n}
       E_s (\boldsymbol{v}) = \boldsymbol{P}_{\!\!
       s}(\boldsymbol{v})\cdot \boldsymbol{v} - T
      (\boldsymbol{v})\,.
\end{equation}
This suggests that $T$ will play the role of
the inertial term in an effective Lagrangian
 and $E_s$ the role of an effective Hamiltonian as our notation in
(\ref{ca.k}) indicates already. In particular,
\begin{equation}\label{ca.o}
       \boldsymbol{v} = \nabla_{\boldsymbol{p}}
           \, E_{\mathrm{eff}} (\boldsymbol{p})
\end{equation}
 and, equivalently,
\begin{equation}\label{ca.p}
        \boldsymbol{v}\, \frac{{\mathrm{d}} \boldsymbol{P}_{\!
        s}(\boldsymbol{v})}{{\mathrm{d}}  \boldsymbol{v}} = \nabla_{ \boldsymbol{v}}\,
                E_s(\boldsymbol{v})
\end{equation}
 which implies that $\boldsymbol{v}$ is to be
interpreted as velocity.

For a relativistic theory one expects that
\begin{equation}\label{ca.q}
       E_s (\boldsymbol{v}) = (m_{\mathrm{b}} + m_{\mathrm{e}}) \gamma,
        \quad \boldsymbol{P}_{\!\!
       s}(\boldsymbol{v})
       = (m_{\mathrm{b}} + m_{\mathrm{e}}) \gamma \boldsymbol{v}\, .
\end{equation}
Since the Abraham model is semi--relativistic, there is no reason
for  such a property to be satisfied. Still we found that, as in the relativistic
case,  the energy--momentum
relation depends on the charge distribution $\rho$ only through
$m_{\mathrm{e}}$.

For small $\boldsymbol{v}$ we have
\begin{equation}\label{ca.r}
       E_s(\boldsymbol{v}) - E_s(0) \cong \frac{1}{2}\,
       (m_{\mathrm{b}} +  \frac{4}{3}\, m_{\mathrm{e}})  \boldsymbol{v}^2,
       \quad \boldsymbol{P}_{\!\! s}(\boldsymbol{v})
       = (m_{\mathrm{b}} + \frac{4}{3}\, m_{\mathrm{e}})\boldsymbol{v} \, .
\end{equation}
Thus the effective mass in the nonrelativistic approximation is
\begin{equation}\label{ca.s}
       m_\mathrm{{eff}} = m_{\mathrm{b}} + \frac{4}{3}\, m_{\mathrm{e}} \,.
\end{equation}
In Figure 2 we plot $E_s(\boldsymbol{v}), \boldsymbol{P}_{\!\! s}(\boldsymbol{v})
$ for the extreme case  $m_{\mathrm{b}}=0$ and
compare with the relativistic dispersion of mass
$\frac{4}{3}\, m_{\mathrm{e}}$. Clearly at speeds $|\boldsymbol{v}| > 0.3$, the Abraham
model looses its empirical validity. One could partially save the
Abraham model by declaring the Compton wave length as the
characteristic size of the charge distribution. Then
 $m_{\mathrm{e}}/m_{\mathrm{b}} \cong 0.01$ and the relativistic
 dispersion is violated only for speeds close to one.

The energy minimizer has a simple dynamical interpretation. We
look for a solution, $S_{\boldsymbol{q},\boldsymbol{v}}$, of (\ref{ad.b})
, (\ref{ad.d})
travelling at constant
velocity $\boldsymbol{v}$ and find
\begin{equation}\label{ca.t}
       S_{\boldsymbol{q},\boldsymbol{v}} (t) = (\boldsymbol{E}_{\boldsymbol{v}}
       (\boldsymbol{x}-\boldsymbol{q}-\boldsymbol{v}t), \, \boldsymbol{B}_{\boldsymbol{v}}
       (\boldsymbol{x} - \boldsymbol{q}-\boldsymbol{v} t)
       , \boldsymbol{q}+\boldsymbol{v} t, \boldsymbol{v})
\end{equation}
with $\boldsymbol{v} \in \mathbb{V}, ~ \boldsymbol{q}\in
\mathbb{R}^3$,
and $\boldsymbol{E}_{\boldsymbol{v}}, \boldsymbol{B}_{\boldsymbol{v}}$
from (\ref{ca.e}). $S_{\boldsymbol{q},\boldsymbol{v}}$ is the {\it charge
soliton}
labeled by
its center $\boldsymbol{q}$ and velocity $\boldsymbol{v}$. It has the energy
 $\mathcal{E} (S_{\boldsymbol{q},\boldsymbol{v}}) = E_s(\boldsymbol{v})$
and momentum $\mathcal{P}(S_{\boldsymbol{q},\boldsymbol{v}}) =
\boldsymbol{P}_{\! s}(\boldsymbol{v})$.

There is an instructive alternate way to represent the charged
soliton. We consider the inhomogeneous Maxwell equations (\ref{ad.b})
and prescribe the initial data at time $\tau$. We require that the
particle travels along the straight line $\boldsymbol{q}= \boldsymbol{v} t$.
 If we let $\tau \to - \infty$
and consider the solution at time $t=0$, then in (\ref{aa.p}), (\ref{aa.q})
 the initial fields will
have escaped to infinity and only the retarded fields survive.
Using (\ref{aa.p}), (\ref{aa.q}) this leads to
\begin{eqnarray}\label{ca.u}
       \boldsymbol{E}_{\boldsymbol{v}}(\boldsymbol{x}) &=& -
        \int\limits_{-\infty}^0 dt \, \int d^3 y \,  \big (
       \nabla G_{-t}\, (\boldsymbol{x}-\boldsymbol{y}) \, \rho(\boldsymbol{y}
       -\boldsymbol{v} t)\nonumber\\
        && \qquad\qquad\quad\quad+ \partial_t \, G_{-t}\,
         (\boldsymbol{x}-\boldsymbol{y}) \boldsymbol{v}
        \rho(\boldsymbol{y}-\boldsymbol{v} t) \big )\, ,\\
       \boldsymbol{B}_{\boldsymbol{v}} (\boldsymbol{x})
        &=& \int\limits_{- \infty}^0 dt\, \int d^3 y\, \nabla
       \times G_{-t}\, (\boldsymbol{x}-\boldsymbol{y})\, \boldsymbol{v}
        \rho (\boldsymbol{y}- \boldsymbol{v} t)\,,\label{ca.v}
\end{eqnarray}
which can be checked either directly in Fourier space or as being
a solution of the Maxwell equations travelling at constant velocity
$\boldsymbol{v}$.


\subsection{Lorentz model}\label{sec.cb}
We look for a solution travelling at
constant velocity $\boldsymbol{v}$. Since the model is relativistic, we first
determine the four--potential of the charge soliton  in its rest frame which yields
\begin{equation}\label{cb.a}
       A^\mu(x^\prime) = (  g(x^{\prime \,2}), ~ 0)
\end{equation}
with $g(|\boldsymbol{x}|^2) = - (\Delta^{-1} \rho) (\boldsymbol{x})$. Then in
the laboratory frame $\mathcal S$
\begin{equation}\label{cb.b}
       A^\mu(x) = v^\mu g(x^2- (v_\lambda x^\lambda)^2)
\end{equation}
and the electromagnetic field tensor has the form
\begin{eqnarray}\label{cb.c}
       F^{\mu\nu}(x) &=& \partial^\mu A^\nu (x) - \partial^\nu
       A^\mu(x)\\[2mm]
      & =& 2 \,\big[(x^\mu- (v_\lambda x^\lambda) v^\mu)
       v^\nu - (x^\nu - (v_\lambda x^\lambda) v^\nu) v^\mu\big]\, g^\prime (x^2
       - (v_\lambda x^\lambda)^2)\,.\nonumber
\end{eqnarray}
$F^{\mu\nu}$ indeed satisfies the Maxwell equations (\ref{af.u}) with
the current
\begin{equation}\label{cb.d}
       j^\mu (x) = v^\mu \, e f(x^2 - (v_\lambda x^\lambda)^2)\,,
\end{equation}
where $e f(|\boldsymbol{x}|^2) =
\rho(\boldsymbol{x})$,
in accordance with (\ref{ad.y}), (\ref{af.c}).

Expressed in terms of electric and magnetic fields we have
\begin{eqnarray}\label{cb.e}
       \boldsymbol{E}(\boldsymbol{x},t) &=& - \nabla
       \phi_{\boldsymbol{v}}
       (\boldsymbol{x}-\boldsymbol{q}- \boldsymbol{v} t) + \boldsymbol{v}
        (\boldsymbol{v} \cdot \nabla \phi_{\boldsymbol{v}} (\boldsymbol{x}-
        \boldsymbol{q}-\boldsymbol{v} t))\,,\nonumber\\[2mm]
       \boldsymbol{B} (\boldsymbol{x},t) &=& - \boldsymbol{v}
       \times\phi_{\boldsymbol{v}} (\boldsymbol{x}-\boldsymbol{q}-\boldsymbol{v} t)   \,,
\end{eqnarray}
where
\begin{equation}\label{cb.f}
       \widehat\phi_{\boldsymbol{v}} (\boldsymbol{k}) = [\boldsymbol{k}^2 -
       (\boldsymbol{v} \cdot \boldsymbol{k})^2]^{-1} \hat\rho
       (\gamma^{-1} \boldsymbol{k}_{\|} + \boldsymbol{k}_\perp)
\end{equation}
with $\boldsymbol{k}_{\|}$ parallel and $\boldsymbol{k}_\perp$
 orthogonal to $\boldsymbol{v}$. In contrast to the nonrelativistic
 coupling,  the charge distribution is now Lorentz contracted
 as seen from the laboratory frame, compare with (\ref{ca.f}),
 where we note that $ \boldsymbol{k}^2 -
 (\boldsymbol{v} \cdot  \boldsymbol{k})^2=
 \gamma^{-2}  \boldsymbol{k}_{\|}^2 + \boldsymbol{k}_\perp^2$
 and $\hat \rho(\gamma^{-1}  \boldsymbol{k}_{\|} +
  \boldsymbol{k}_\perp) = \hat \rho_r ([ \boldsymbol{k}^2 -
  ( \boldsymbol{v} \cdot  \boldsymbol{k})^2]^{1/2})$.

To determine  energy and momentum of the relativistic soliton
we first have to find out how these quantities are even defined.
We start from the energy--momentum tensor of the electromagnetic
field
\begin{equation}\label{cb.g}
       T^{\mu\nu} = F^{\nu\lambda}\, F_\lambda \:^\mu - g^{\mu\nu}\,
       F^{\alpha\beta}\, F_{\alpha\beta} \, .
\end{equation}
>From the Maxwell equations it satisfies the local balance
\begin{equation}\label{cb.h}
       \partial_\nu \, T^{\mu\nu} = F^{\mu\nu} j_\nu\, .
\end{equation}
We now claim that
\begin{equation}\label{cb.i}
       \frac{d}{ds} \, \big (m_{\mathrm{b}} v^\mu + \int d^4 x\, T^{\mu\nu} (x)
       v_\nu \, \delta(v_\lambda(x-z)^\lambda)\big ) = 0 \, .
\end{equation}
Thus it is natural to regard
\begin{equation}\label{cb.j}
       P^\mu = m_{\mathrm{b}} v^\mu + \int d^4 x \, T^{\mu\nu} (x)
       v_\nu \, \delta(v_\lambda(x-z)^\lambda)
\end{equation}
as the conserved four--momentum.

To derive (\ref{cb.i}) we multiply (\ref{cb.h}) by $[1- \dot v_\sigma (x-z)^\sigma
] \, \delta (v_\lambda(x-z)^\lambda)$ and integrate over all
space--time. Then
\begin{eqnarray}\label{cb.k}
        && \int d^4 x\, F^{\mu\nu} (x) j_\nu(x)\,
        [1- \dot v_\sigma (x-z)^\sigma] \, \delta (v_\lambda(x-z)^\lambda)\\
       &=&\int d^4 x \, \partial_\nu\,
       T^{\mu\nu}(x)\, [1- \dot v_\sigma (x-z)^\sigma]
       \, \delta (v_\lambda(x-z)^\lambda)\nonumber\\
       &=& - \int d^4 x \, T^{\mu\nu}(x) \big[ - \dot v_\nu
        \, \delta (v_\lambda(x-z)^\lambda)
        + (1- \dot v_\sigma (x-z)^\sigma)v_\nu \, \delta^\prime
         (v_\lambda(x-z)^\lambda)\big]\, .\nonumber
\end{eqnarray}
We  have
\begin{eqnarray}
       & &\!\!\!\!\!\!\!\!\!\frac{d}{ds} \int d^4 x\, T^{\mu\nu} (x)
       v_\nu \, \delta(v_\lambda(x-z)^\lambda)\nonumber\\
      &=& \int d^4 x \, \,T^{\mu\nu}(x)\,
       \big ( \dot v_\nu
        \, \delta (v_\lambda(x-z)^\lambda)
        - v_\nu ( 1 - \dot v_\sigma (x-z)^\sigma) \, \delta^\prime
         (v_\lambda(x-z)^\lambda)\big )\nonumber\\
         &=&\int d^4 x\, F^{\mu\nu}(x) j_\nu(x)
        (1- \dot v_\sigma (x-z)^\sigma)
        \, \delta (v_\lambda(x-z)^\lambda)\nonumber\\
         &=& \int d^4 x\, F^{\mu\nu}(x)
       (1- \dot v_\sigma (x-z)^\sigma) \, \delta
         (v_\lambda(x-z)^\lambda)\,
         \nonumber\\
       &&\times \int d s^\prime \,[(v_\nu - v_\nu
       \dot v_\sigma (x-z)^\sigma) f((x-z)^2) \,
       \delta(v_\lambda
       (x-z)^\lambda)](s^\prime)\, .\label{cb.l}
\end{eqnarray}
Under our assumption of not too large an acceleration, compare
with (\ref{af.h}) below, the hyperplane $\{ x:~ v_\lambda(s) (x-z(s))^\lambda = 0 \}$
intersects the hyperball $\{x:~ v_\lambda(s^\prime)~(x-z(s^\prime))^\lambda = 0$,
$f((x-z(s^\prime))^2) > 0 \}$ only if $s=s^\prime$ and
\begin{eqnarray}
       &&\!\!\!\!\!\!\!\!\!\!(1- \dot v_\sigma (s) (x-z(s))^\sigma) \, \delta (v_\lambda(s)
       (x-z(s))^\lambda)\, \delta (v_\lambda (s^\prime)
       (x-z (s^\prime))^\lambda)\nonumber\\
       &=& \delta (v_\lambda(s) (x-z(s))^\lambda)
       \delta(s-s^\prime)\,.\label{cb.m}
\end{eqnarray}
Thus
\begin{eqnarray}
        (\ref{cb.l}) \quad &=& - \int d^4 x\, F^{\mu\nu}(x)
        \, \delta(v_\lambda(x-z)^\lambda)\, (v_\nu - \eta_{\nu \sigma}
        (x-z)^\sigma) \, f((x-z)^2)\nonumber\\
        &=& - m_{\mathrm{b}} \, \frac{d}{ds}\, v^\mu(s)\, ,\label{cb.n}
\end{eqnarray}
where we used the equations of motion (\ref{af.v}). This proves
(\ref{cb.i}).

The expression (\ref{cb.j}) for $P^\mu$ is covariant.
Thus we are allowed to work out the integral in the frame $\mathcal{S}^\prime$
traveling with velocity $\boldsymbol{v}$ relative to $\mathcal S $.
In this frame
\begin{equation}\label{cb.o}
        P^{\prime\, 0} = \gamma \,\big(m_{\mathrm{b}} + \frac{1}{2} \,
       \int d^3 x^\prime\,
       \boldsymbol{E}(\boldsymbol{x}^\prime)^2\big), ~
       P^{\prime\,\mu} = 0 \quad \mbox{for} \quad \mu= 1,2,3
\end{equation}
and thus
\begin{equation}\label{cb.p}
        P^\mu = (m_{\mathrm{b}} + m_\mathrm{e}) \, \gamma \,
       v^\mu\,,
\end{equation}
which shows that the Lorentz model has the physically correct relativistic
four--momentum. Of course, experimentally only the sum, $m_{\mathrm{b}} +
m_\mathrm{e}$, can be observed.
\subsection*{Notes and References}
{\it ad \ref{sec.ca}, \ref{sec.cb}:} Abraham (1905) computed
 the energy--momentum relation in essence
along the same route as outlined here (except for the variational
characterization). Sommerfeld (1904,1905) used the expansion of the
exact self--force, as will be explained in Chapter \ref{sec.e}. Lorentz
(1904a) proposed a model charge which relativistically contracts
along its momentary velocity.
Thus provisionally we replace
the charge distribution $\rho( \boldsymbol{x})$ by its Lorentz contracted version
\begin{eqnarray}\label{cc.a}
        \rho_L( \boldsymbol{x}) &=& \gamma \rho_r  ( [
                 \boldsymbol{x}^2 + \gamma^2 ( \boldsymbol{x} \cdot
         \boldsymbol{v})^2  ]^{1/2} )\,,\\
        \hat\rho_L ( \boldsymbol{k}) &=& \hat\rho_r ( [
                \boldsymbol{k}^2 - ( \boldsymbol{v} \cdot  \boldsymbol{k})^2
         ]^{1/2} )\,.\label{cc.b}
\end{eqnarray}
This expression is substituted in (\ref{ca.f}) and gives the
electromagnetic fields comoving with the charge at velocity $ \boldsymbol{v}$.
Their energy and momentum is computed as before with the result
\begin{eqnarray}\label{cc.c}
        \boldsymbol{P}_L ( \boldsymbol{v}) &=&
                 \boldsymbol{v} \big (m_{\mathrm{b}} \gamma( \boldsymbol{v})
                 + \frac{4}{3} \,
       m_{\mathrm{e}} \gamma ( \boldsymbol{v})\big )\,,\\
       E_L ( \boldsymbol{v}) &=& m_{\mathrm{b}} \gamma( \boldsymbol{v})
           + m_{\mathrm{e}}
       \gamma( \boldsymbol{v}) \big ( 1 + \frac{1}{3} \boldsymbol{v}^2 \big )\label{cc.d}\,.
\end{eqnarray}
The momentum has the anticipated form, except for the factor 4/3
which should be 1. The energy has an unwanted $ \boldsymbol{v}^{2}/3$. In
particular the relation (\ref{ca.p}) does not hold, which implies
that the power equation
$\frac{d}{dt}\, E_L (\boldsymbol{v})$ differs from the force equation $\boldsymbol{v}
\cdot \frac{d}{dt}\, \boldsymbol{P}_L(\boldsymbol{v})$.
We refer to Yaghjian (1992) for a thorough discussion.

Schott (1912) employed as a model charge a deformable elastic
medium. To compute the velocity dependent mass he used in
principle the same method as Sommerfeld, an exact self--force and
an expansion in the charge diameter. Schott considered also
electron models different from those of Abraham and Lorentz.

There have been various attempts to improve on the oversimplistic version (\ref{cc.a})
 of the Lorentz
model.
Fermi (1922) argues that in a relativistic theory energy and
momentum have to be redefined. His argument has been rediscovered
several times  and is explained in Rohrlich
(1990). Poincar\'e (1906) takes the elastic stresses into account. His
theory  is excellently presented in
Yaghjian (1992).
The material of Section \ref{sec.cb} is adapted from of Nodvik (1964).
\newpage
\section{Adiabatic Limit}\label{sec.d}
\setcounter{equation}{0}
If we assume that the mass of an electron is purely
electromagnetic, then by equating its rest energy and  electrostatic
 Coulomb energy  the charge distribution $\rho$
must be concentrated in a ball of radius
\begin{equation}\label{d.a}
       R_\rho = \frac{e^2}{m c^2} = 3 \times 10^{-13}~ \mathrm{cm}
\end{equation}
which is the so called classical electron radius.
Quantum mechanically one argues that through fluctuations the
electron appears to have an effective size of the order of the
Compton wave length
$\lambda_c = \hbar m/c = (e^2/\hbar c)^{-1}\, R_\rho = 137\,
R_\rho$.
Electromagnetic fields which can be manipulated in the laboratory vary
little over that length scale. $R_\rho$ defines a time scale  through the time it
takes light to cross
the diameter of the charge distribution,
\begin{equation}\label{d.b}
       t_\rho = R_\rho/c = 10^{-23} ~ \mathrm{sec}~,\quad \mbox{equivalently as a
       frequency,}~\,
       \omega_\rho = 10^{23} ~ \mbox{Hz} \,.
\end{equation}
Again, manufactured frequencies are much smaller than $\omega_\rho$.
Space--time variations as fast as (\ref{d.a}) and (\ref{d.b}) lead us
deeply into the quantum regime. Thus it is a natural and
physically a  mandatory problem  to study the dynamics of a charged
particle under external potentials which vary {\it slowly} on the
scale of $R_\rho$. This means we have to introduce a scale
 of
potentials an enquire about an approximately autonomous particle
dynamics with an error depending on the scale under consideration. We will introduce
such a scheme formally in the following section. The resulting problem
has many similarities with the derivation of hydrodynamics from
 Newtonian particle dynamics -- with the most welcome addition
that it is simpler mathematically by many order of magnitudes.
Still, the comparison is instructive.


\subsection{Scaling limit}\label{sec.da}
We assumed that in the Lorentz force equation there are in addition to
dynamical fields
$\boldsymbol{E}(\boldsymbol{x},t), ~ \boldsymbol{B}(\boldsymbol{x},t)$
also prescribed external fields acting on the particle, which are
the gradients of the external potentials  $\phi_{\mathrm{ex}} (\boldsymbol{x}),
       \boldsymbol{A}_{\mathrm{ex}} (\boldsymbol{x})$,
       compare with Equation (\ref{ad.d}).
We want to impose that $\phi_{\mathrm{ex}}$ and $\boldsymbol{A}_{\mathrm{ex}}$
are slowly varying on the scale of $R_\rho$. Formally we introduce
a small {\it dimensionless} parameter $\varepsilon$ and consider the
potentials
\begin{equation}\label{da.b}
       \phi_{\mathrm{ex}} (\varepsilon \boldsymbol{x})~,
       \quad \boldsymbol{A}_{\mathrm{ex}}
       (\varepsilon \boldsymbol{x})  \,,
\end{equation}
which are slowly varying in the limit $\varepsilon \to 0$.
Most of our results extend to potentials which vary also  slowly in
time. But for simplicity we restrict ourselves to
time--independent potentials here. Clearly, $\varepsilon$
 appears as a parameter of the potential,
just like $\omega_0$ is a parameter of the harmonic potential
$\frac{1}{2} \, m \omega_0^2 x^2$.
But one really
should think of $\varepsilon$ as a book keeping device which
orders the magnitude of the various terms and the space--time
scales in powers of $\varepsilon$. Such a scheme is familiar in
very diverse contexts and appears whenever one has to deal with
a problem involving scale separation.

So how small is $\varepsilon$ ? From the discussion above one might
infer that if $\phi_{\mathrm{ex}}, \boldsymbol{A}_{\mathrm{ex}}$ vary over a
scale of 1 mm, then $\varepsilon = 10^{-12}$. This is a strictly
meaningless statement, because $e \phi_{\mathrm{ex}},~ \frac{e}{c}
\boldsymbol{A}_{\mathrm{ex}}$
have the dimension of an energy and thus the variation depends on
the adopted energy scale. In (\ref{da.b}) we fix the energy scale and
merely stretch the spatial axes by a factor $\varepsilon^{-1}$.
Since from experience this point is likely to be confusing, let us
consider  the specific example of a charge circling in the uniform
magnetic field $(0,0, B_0)$. Since the corresponding vector
potential is linear in $\boldsymbol{x}$, to introduce $\varepsilon$ as in
(\ref{da.b}) just means that the magnetic field strength equals $\varepsilon B_0$
and the limit $\varepsilon \to 0$ is a limit of small magnetic
field {\it relative} to some reference field $B_0$.
 Thus to obtain $\varepsilon$
we first have to determine the reference field and compare it with
the magnetic field of interest. This shows that in order to fix
$\varepsilon$ we have to specify the physical situation
concretely, in particular the external potentials, the mass and
charge of the particle, $\gamma(\boldsymbol{v})$, and the time span of interest.

The scaling scheme (\ref{da.b}) has the enormous advantage that the
analysis can be carried out in generality. In a second step one
has to figure out $\varepsilon$ for a {\it concrete} situation,
which leads to a quantitative estimate on the error terms. E.g. if
in the case above we consider an electron with velocities such
that  $\gamma \le 10 $, then, by comparing the Hamiltonian term and the
friction term, the reference field turns out to  be $B_0 = 10^{17}$
Gauss. Laboratory magnetic fields are less than $10^5$ Gauss and
thus $\varepsilon < 10^{-12}$.
In practice, $\varepsilon$ is always very small, less than
$10^{-10}$. This means that, firstly, all corrections
{\it beyond} radiation reaction are negligible.  Secondly, we do
not have
to go each time through the scheme indicated above and may as well set
$\varepsilon = 1$ thereby returning to the conventional units. Still on
an theoretical level the use of the scale parameter $\varepsilon$
is very convenient. In
Section \ref{sec.dc} we will work out the example of a constant magnetic field
more  explicitly. If
the reader feels uneasy about the scaling limit, (s)he should
consult this example first.

Adopting (\ref{da.b}), the Lorentz force equation reads now
\begin{eqnarray}
       \frac{d}{dt}\, \big(m_{\mathrm{b}} \gamma \boldsymbol{v}(t)\big) &=&  \int d^3 x\, \rho
       (\boldsymbol{x}-\boldsymbol{q}(t)) \, \big[\varepsilon
       \boldsymbol{E}_{\mathrm{ex}} (\varepsilon \boldsymbol{x}) + \boldsymbol{E}
        (\boldsymbol{x},t)
        \nonumber\\[2mm]
        &&\qquad +
       \boldsymbol{v}(t) \times \big (\varepsilon \boldsymbol{B}_{\mathrm{ex}}
       (\varepsilon \boldsymbol{x}) +
       \boldsymbol{B}(\boldsymbol{x},t)\big)\big]\,, \label{da.c}
\end{eqnarray}
where
\begin{equation}\label{da.d}
       \boldsymbol{E}_{\mathrm{ex}} = - \nabla\phi_{\mathrm{ex}} ~ , ~
       \boldsymbol{B}_{\mathrm{ex}} = \nabla \times \boldsymbol{A}_{\mathrm{ex}} \, .
\end{equation}
It has to be supplemented with the Maxwell equations (\ref{ad.b}),
(\ref{ad.c}). Our goal is to understand the structure of the
solution for small $\varepsilon$ and as a first qualitative step
one should discuss the rough order of magnitudes in powers of
$\varepsilon$. But before we have to specify the initial data.
We give ourselves $\boldsymbol{q}^0,  \boldsymbol{v}^0$ as initial position and velocity of
the charge. The initial fields are assumed to be Coulombic,
centered at $\boldsymbol{q}^0$ with velocity $\boldsymbol{v}^0$,
i.e.\medskip\\
{\it Condition $(I)$}:
\begin{equation}\label{da.e}
     Y(0) = S_{\boldsymbol{q}^0, \boldsymbol{v}^0}
 (0)\,,\medskip
\end{equation}
compare with (\ref{ca.t}). Equivalently, according to (\ref{ca.u}),
(\ref{ca.v}), we may say that the particle
has travelled freely with velocity $\boldsymbol{v}^0$ for the infinite time
span $(- \infty, 0]$. At time $t=0$ the external potentials are
turned on. More geometrically we define the six--dimensional
charge
soliton manifold ${\sf S} = \{ S_{\boldsymbol{q},\boldsymbol{v}}, ~
\boldsymbol{q} \in \mathbb{R}^3, ~ \boldsymbol{v}
 \in \mathbb{V}\}$ as a submanifold of the
phase space $\mathcal M$. Then our initial data are exactly on
$\sf S$.
If there are no external forces, the solution remains on $\sf S$
and moves along a straight line. For slowly varying external
potentials as in (\ref{da.b}) we will show that the solution remains
$\varepsilon$--close to $\sf S$ in the local energy distance.

On general grounds one may wonder whether such specific
initial data are really required. In analogy to hydrodynamics, we
call this the initial slip problem. In times of order $t_\rho$, the
fields close to the charge acquire their Coulombic form. However,
during that period the particle might gain or loose in momentum
and energy and the data at time $t_\rho$ close to the particle are approximately of the
form $S_{\tilde{\boldsymbol{q}}, \tilde{\boldsymbol{v}}}$,
 where $\tilde{\boldsymbol{q}}$
and $\tilde{\boldsymbol{v}}$ are to be computed from the full solution. Of
course, at a distance $ct$ away from the charge, the field still
remembers its $t=0$ data. Thus we see that the initial
slip problem translates into the long time asymptotics of a charge
at {\it zero} external potentials but with general initial field data.
We study this point in more detail in Section \ref{sec.dc}.
At the moment we just circumvent the initial slip by fiat.

Let us discuss the three  relevant time scales, where we recall that
$t_\rho = R_\rho/c$.\medskip\\
(i) {\it Microscopic scale}, $t= \mathcal{O} (t_\rho), ~
\boldsymbol{q} = \mathcal{O}(R_\rho)$. On that scale the particle moves along
an essentially straight line. The electromagnetic fields adjust
themselves to their  comoving Coulombic form. As we will see, they do
this with a precision $\mathcal{O}(\varepsilon)$ in the energy
norm.\medskip\\
(ii) {\it Macroscopic scale}, $t= \mathcal{O} (\varepsilon^{-1}
t_\rho),~
\boldsymbol{q} = \mathcal{O} (\varepsilon^{-1}R_\rho)$. This scale is defined
by the variation of the potentials, i.e. on that scale the
potentials are
$\phi_{\mathrm{ex}} (\boldsymbol{x}), \boldsymbol{A}_{\mathrm{ex}}(\boldsymbol{x})$.
The particle follows the external forces. Since it is in company
with the almost Coulombic fields, the particle responds to the forces
according to the effective energy--momentum relation, which we
determined in the previous section. On the macroscopic scale the
motion is Hamiltonian up to errors of order $\varepsilon$. There
is no dissipation of energy and momentum.\medskip\\
(iii) {\it Friction scale}. Accelerated charges loose energy through radiation, which means
 that there must be friction corrections to the effective
 Hamiltonian motion. According to Larmor's formula the radiation
 losses are proportional to $\dot{\boldsymbol{v}}(t)^2$. Since the
 external forces are of the order $\varepsilon$, these losses are
 proportional to $\varepsilon^2$ when measured in microscopic
 units. Integrated over a time span $\varepsilon^{-1} t_\rho$ the
 friction results in an effect of order $\varepsilon$. Thus we
 expect order $\varepsilon$ dissipative corrections to the
 conservative motion on the macroscopic scale.
 Followed over the even longer time scale $\varepsilon^{-2} t_\rho$,
  the radiation reaction results in ${\cal O}(1)$ deviations from
  the Hamiltonian trajectory.\medskip

On the friction time scale the motion either comes to a stand
still or stays uniform. In addition, as to be shown,
 the dissipative effective
 equation has the same long time behavior as the true solution. Thus we expect no further
qualitatively distinct time scale beyond the friction scale.

>From our description, in a certain sense, the most natural scale
is the macroscopic scale and we transform the Maxwell--Lorentz
equations to this new scale by setting
\begin{equation}\label{da.f}
       t^\prime = \varepsilon t ~, ~ \boldsymbol{x}^\prime = \varepsilon
       \boldsymbol{x} \, .
\end{equation}
We have the freedom of how to scale the amplitudes of the dynamic part of the
 electromagnetic fields. We require that their energy is
 independent of $\varepsilon$. Then
\begin{equation}\label{da.g}
       \boldsymbol{E}^\prime (\boldsymbol{x}^\prime, t^\prime) = \varepsilon^{-3/2}\,
       \boldsymbol{E}(\boldsymbol{x},t)~, ~ \boldsymbol{B}^\prime
       (\boldsymbol{x}^\prime, t^\prime) =
       \varepsilon^{-3/2}\, \boldsymbol{B}(\boldsymbol{x},t)\, .
\end{equation}
Finally the new position and velocity are
\begin{equation}\label{da.h}
       \boldsymbol{q}^\prime ( t^\prime) = \varepsilon \boldsymbol{q}(t)~,
      ~ \boldsymbol{v}^\prime (t^\prime)  =
       \boldsymbol{v}(t)\, ,
\end{equation}
so that  $\frac{d}{dt^\prime} \, \boldsymbol{q}^\prime = \boldsymbol{v}^\prime$.
 There is little risk in omitting the prime.
We denote then
\begin{equation}\label{da.i}
       \boldsymbol{q}^\varepsilon(t)  =
       \varepsilon\boldsymbol{q}(\varepsilon^{-1} t)~, ~
       \boldsymbol{v}^\varepsilon (t) = \boldsymbol{v}(\varepsilon^{-1} t)~,
       ~
       \rho_\varepsilon(\boldsymbol{x}) = \varepsilon^{-3}\, \rho
       (\varepsilon^{-1} \boldsymbol{x}) \, ,
\end{equation}
which means that $\int d^3 x\, \rho_\varepsilon(\boldsymbol{x}) = e$
 independent of $\varepsilon$
 and $\rho_\varepsilon$ is supported in a ball of radius $\varepsilon
 R_\rho$.
In the macroscopic coordinates the Maxwell-Lorentz equations read
\begin{eqnarray}
         \partial_t \boldsymbol{B}(\boldsymbol{x},t) &=& -
         \nabla\times \boldsymbol{E}(\boldsymbol{x},t)\,,\nonumber\\
        \partial_t  \boldsymbol{E}(\boldsymbol{x},t)
        &=& \nabla \times \boldsymbol{B}(\boldsymbol{x},t) -
        \sqrt{\varepsilon} \rho_\varepsilon (\boldsymbol{x}-
        \boldsymbol{q}^\varepsilon(t))
        \boldsymbol{v}^\varepsilon(t)\,, \label{da.j}\\[2mm]
       \frac{d}{dt} \, \big(m_{\mathrm{b}} \gamma \boldsymbol{v}^\varepsilon (t)\big) &=&
        \boldsymbol{E}_{\mathrm{ex}}
        \ast \rho_\varepsilon (\boldsymbol{q}^\varepsilon(t)) +
        \boldsymbol{v}^\varepsilon(t) \times \boldsymbol{B}_{\mathrm{ex}} \ast
        \rho_\varepsilon(\boldsymbol{q}^\varepsilon (t))\nonumber\\
       && + \int d^3 x \, \sqrt{\varepsilon}\, \rho_\varepsilon
       (\boldsymbol{x}-\boldsymbol{q}^\varepsilon
       (t))\, [\boldsymbol{E}(\boldsymbol{x},t) + \boldsymbol{v}^\varepsilon (
       t) \times \boldsymbol{B}(\boldsymbol{x},t)]\nonumber
\end{eqnarray}
together with the constraints
\begin{equation}\label{da.k}
       \nabla\cdot \boldsymbol{E} = \sqrt{\varepsilon}\,  \rho (\cdot -
       \boldsymbol{q}^\varepsilon
       (t)) ~,\quad \nabla \cdot \boldsymbol{B} = 0\, .
\end{equation}
On the macroscopic scale the conserved energy is
\begin{equation}\label{da.l}
       \mathcal E_{\mathrm{mac}} = m_{\mathrm{b}} \gamma (\boldsymbol{v}) +
         \, \phi_{\mathrm{ex}}
      \ast \rho_\varepsilon(\boldsymbol{q})+ \frac{1}{2}
       \int d^3 x \, \big (\boldsymbol{E}(\boldsymbol{x})^2 + \boldsymbol{B}
       (\boldsymbol{x})^2\big) \, .
\end{equation}
Also the initial data have to be transformed and become now\medskip\\
\medskip
{\it Condition $(I_\varepsilon)$:
\begin{equation}\label{da.m}
         Y^\varepsilon(0) = S^\varepsilon_{\boldsymbol{q}^0,
       \boldsymbol{v}^0}= (\boldsymbol{E}_{\boldsymbol{v}^0}
       (\boldsymbol{x}-\boldsymbol{q}^0),\boldsymbol{B}_{\boldsymbol{v}^0}
       (\boldsymbol{x}-\boldsymbol{q}^0),\boldsymbol{q}^0, \boldsymbol{v}^0)
\end{equation}
with
\begin{equation}\label{da.n}
       \boldsymbol{E}_{\boldsymbol{v}} = - \nabla
       \phi_{\boldsymbol{v}}^\varepsilon + \boldsymbol{v}(\boldsymbol{v} \cdot
       \nabla\phi_{\boldsymbol{v}}^\varepsilon)~, ~
       \boldsymbol{B}_{\boldsymbol{v}} = - \boldsymbol{v} \times  \nabla
        \phi_{\boldsymbol{v}}^\varepsilon\,,
\end{equation}
where now
\begin{equation}\label{da.o}
        \widehat\phi_{\boldsymbol{v}}^\varepsilon (\boldsymbol{k}) =  \,
        \sqrt{\varepsilon} \,
        \hat \rho(\varepsilon \boldsymbol{k})/ [\boldsymbol{k}^2-
        (\boldsymbol{v}\cdot \boldsymbol{k})^2] \, .
\end{equation}}

On the macroscopic scale, the scaling parameter $\varepsilon$ can
be absorbed into the ``effective'' charge distribution
 $\sqrt{\varepsilon}  \rho_\varepsilon$.
 Its electrostatic energy,
\begin{equation}\label{da.p}
       m_{\mathrm{e}} = \frac{1}{2}\, \int d^3 k\,  \varepsilon |\hat\rho_\varepsilon
       (\boldsymbol{k})|^2 \frac{1}{\boldsymbol{k}^2} = \frac{1}{2}\,
       \int d^3 k  |\hat\rho (\boldsymbol{k})|^2\,
       \frac{1}{\boldsymbol{k}^2}~,
\end{equation}
is independent of $\varepsilon$, whereas its charge
\begin{equation}\label{da.q}
       \int d^3  x \, \sqrt{\varepsilon} \, \rho_\varepsilon(\boldsymbol{x}) =
       \sqrt{\varepsilon}\, e
\end{equation}
vanishes as $\sqrt{\varepsilon}$. Recall that $\varepsilon$ is a
book keeping device.

We argued that on the macroscopic scale the  response to
the external potentials in the motion of the
charges is of order one. We thus expect that $\boldsymbol{q}^\varepsilon(t)$
tends to a nondegenerate limit as $\varepsilon \to 0$, i.e.
\begin{equation}\label{da.r}
       \lim_{\varepsilon \to 0}   \, \boldsymbol{q}^\varepsilon (t) = \boldsymbol{r} (t)~,
      ~ \lim_{\varepsilon \to 0}\, \boldsymbol{v}^\varepsilon(t) =
       \boldsymbol{u}(t)\,.
\end{equation}
The position $\boldsymbol{r}(t)$ and velocity $\boldsymbol{u}(t)$ on the macroscopic scale
should be  governed by an effective Lagrangian. In Section \ref{sec.ca} we
determined already the effective inertial term. If the potentials
add in as usual, we have
\begin{equation}\label{da.w}
       L_{\mathrm{eff}} (\boldsymbol{q}, \dot{\boldsymbol{q}}) = T(\dot{\boldsymbol{q}}) -
       e\big(\phi_{\mathrm{ex}} (\boldsymbol{q}) - \dot{\boldsymbol{q}} \cdot
       \boldsymbol{A}_{\mathrm{ex}}(\boldsymbol{q})\big)\,,
\end{equation}
which results in the equations of
motion
\begin{equation}\label{da.u}
       \dot{\boldsymbol{r}} = \boldsymbol{u}~, \quad m(\boldsymbol{u})
        \dot{\boldsymbol{u}} =
       e (\boldsymbol{E}_{\mathrm{ex}} (\boldsymbol{r}) +
       \boldsymbol{u} \times \boldsymbol{B}_{\mathrm{ex}}(\boldsymbol{r}))\,.
\end{equation}
The velocity dependent mass $m(\boldsymbol{u})$ has a bare and a field contribution.
>From (\ref{ca.l}) we conclude that
\begin{equation}\label{da.v}
       m (\boldsymbol{v}) = \frac{{\mathrm{d}} P_s (\boldsymbol{v})}
       {{\mathrm{d}} \boldsymbol{v}}
\end{equation}
as a $3 \times 3$ matrix.
 If instead of the
velocity we introduce the canonical momentum, $\boldsymbol{p}$, then the
effective Hamiltonian reads
\begin{equation}\label{da.s}
       H_{\mathrm{eff}} (\boldsymbol{r},\boldsymbol{p})=
        E_{\mathrm{eff}} (\boldsymbol{p}-e \boldsymbol{A}_{\mathrm{ex}}
       (\boldsymbol{r})) + e \phi_{\mathrm{ex}} (\boldsymbol{r})
\end{equation}
with Hamilton's equations of motion
\begin{equation}\label{da.t}
       \dot{\boldsymbol{r}} = \nabla_{\boldsymbol{p}} H_{\mathrm{eff}}
       ~, \quad \dot{\boldsymbol{p}} = - \nabla_{\boldsymbol{r}} H_{\mathrm{eff}} \, .
\end{equation}

Our plan is to establish the limit (\ref{da.r}) and to investigate
the corrections due to radiation losses.


\subsection{Comparison with the hydrodynamic limit}\label{sec.db}
In hydrodynamics one assumes that a small droplet of fluid
with center $\boldsymbol{r}$ has its
intrinsic velocity, $\boldsymbol{u}(\boldsymbol{r})$, and that relative to the moving frame
the particles are distributed according to thermal equilibrium with density
 $\rho(\boldsymbol{r})$
and temperature $T(\boldsymbol{r})$. For such notions to be reasonably well
defined, the hydrodynamic fields $\rho, \boldsymbol{u}, T$ must be slowly
varying on the scale of the typical interparticle distance. This
is how the analogy to the Maxwell--Lorentz equations arises. As
for them we have three characteristic space--time scales.\medskip\\
(i) {\it Microscopic scale}. The microscopic scale
is measured in units of a collision time, resp.
interatomic distance. On that scale the hydrodynamics fields are
frozen. Possible deviations from local equilibrium relax through
collisions. To prove such a behavior one has to establish a
sufficiently fast relaxation to equilibrium. For Newtonian
particles no method is available. For the Maxwell field the
situation is much simpler. Local deviations from the Coulomb field
are transported off to infinity and are no longer  seen.\medskip\\
(ii) {\it Macroscopic scale}. The macroscopic space--time scale is defined by the variation
of the hydrodynamic fields. If, as before, we introduce the
dimensionless scaling parameter $\varepsilon$, then space--time is
$\mathcal{O}(\varepsilon^{-1})$ in microscopic units. On the
macroscopic scale the time between collisions is $\mathcal{O}(\varepsilon)$,
 the interparticle distance $\mathcal{O}(\varepsilon)$, and the
pair potential between the particles at positions $\boldsymbol{q}_i, \boldsymbol{q}_j$
 is $V(\varepsilon^{-1}
(\boldsymbol{q}_i - \boldsymbol{q}_j))$. On the macroscopic scale
the hydrodynamic fields
evolve according to the Euler equations. These are first order
equations, which must be so, since space and time are scaled in
the same way. The Euler equations are formally of Hamiltonian form. There
is no dissipation, no entropy is produced. In fact, there is a slight
complication here. Even for smooth initial data the Euler
equations develop shock discontinuities. There the assumption of
slow variation fails and shocks are a source of entropy.\medskip\\
(iii) {\it Friction scale}. In a real fluid there are frictional forces which are
responsible for  the relaxation to global equilibrium.
One adds to the Euler equations diffusive like, second order in
spatial derivatives, terms and obtains the compressible
Navier--Stokes equations incorporating the shear and volume viscosity
for friction in momentum transport and  thermal conductivity
for friction in energy transport. On the macroscopic scale these
 corrections are of order $\varepsilon$. In the same spirit,
 based on the full
Maxwell--Lorentz equations, there will be dissipative terms of
order $\varepsilon$  emerging which have to be
added to (\ref{da.u}). Of course, in this context
one only has to deal with ordinary differential equations
as effective dynamics.


\subsection{Initial slip}\label{sec.dc}
We adjusted the initial data for the electromagnetic field to be
exactly on the charge soliton manifold, which physically means
that without external forces the particle would travel forever at
constant velocity accompanied by its comoving Coulombic fields. One may wonder
whether such a rigid assumption is really necessary. Let us
consider then times on the microscopic scale of the order $\varepsilon^{-1+\delta}$
with some small $\delta > 0$. The external forces, which are
$\cal{O}(\varepsilon)$, are still negligible, but the microscopic
time span diverges. Thus we are led to investigate the long time limit
of the Abraham model for {\it zero} external fields. Roughly one
has the following picture: initially there is some exchange of
momentum and energy between particle and field, but, since the
total energy is bounded, eventually the particle relaxes to some
definite velocity and the field builds up its comoving Coulombic shape.
 Thus after a very short macroscopic time the state is
already close to the charge soliton manifold from whereon the adiabatic
dynamics
applies.

To prove such a behavior we need little
bit of preparation. Firstly we must have some decay and
smoothness of the initial fields at infinity. We introduced
already such a set of ``good'' initial data, ${\mathcal
M}^\sigma$, in Section \ref{sec.ae} and therefore require $Y(0) \in
{\mathcal M}^\sigma, 0<\sigma\le 1.$ Secondly, we need a notion of
closeness of the fields. At a given time and far away from the
particle the fields are determined by their initial data. Only
close to the particle they are Coulombic. Therefore it is natural
to measure closeness in the {\it local energy norm} defined by
\begin{equation}\label{dc.j}
       \parallel\!\!(\boldsymbol{E},\boldsymbol{B})\!\!\parallel_R^2 =
       \frac{1}{2}\, \int\limits_{\{|\boldsymbol{x}|\le R\}} d^3x\big(\boldsymbol{E}
       (\boldsymbol{x})^2 + \boldsymbol{B}(\boldsymbol{x})^2\big)
\end{equation}
for given radius $R$.

The true solution is $Y(t) = (\boldsymbol{E}
       (\boldsymbol{x}, t), \boldsymbol{B}(\boldsymbol{x},t),\boldsymbol{q}
       (t),  \boldsymbol{v}(t))$ which is to be compared with the
       charge soliton approximation $\big(\boldsymbol{E}_{\boldsymbol{v}(t)}
        (\boldsymbol{x}-\boldsymbol{q}(t)),\boldsymbol{B}_{\boldsymbol{v}(t)}
       (\boldsymbol{x}-\boldsymbol{q}(t)),$ $\boldsymbol{q}(t),
       \boldsymbol{v}(t)\big) $, cf. (\ref{ca.e}). We set
       $ \boldsymbol{Z}_1  (\boldsymbol{x},t) =  \boldsymbol{E}( \boldsymbol{x},t)
       -  \boldsymbol{E}_{\boldsymbol{v}(t)}(\boldsymbol{x} -
        \boldsymbol{q}(t)), ~ \boldsymbol{Z}_2(\boldsymbol{x},t) =
         \boldsymbol{B}(
         \boldsymbol{x},t)-\boldsymbol{B}_{\boldsymbol{v}(t)}
         (\boldsymbol{x}-\boldsymbol{q}(t)),~
         Z=(\boldsymbol{Z}_1,\boldsymbol{Z}_2)$
       and want to establish that, for fixed $R$, $\parallel \!\!Z
       (\cdot + \boldsymbol{q}(t),t)\!\!\parallel_R \to 0$ for large
       times $t \to\infty$.
\begin{proposition}
For the Abraham model with zero external
potentials and satisfying $(C)$ let $|e| \le \overline e$
according to Theorem \ref{thm.ea} and let the
initial data $Y(0) \in {\mathcal{M}}^\sigma$ for some $\sigma \in (0,1]$.
Then for every $R>0$  we have
\begin{equation}\label{dc.a}
       \| Z(\cdot + \boldsymbol{q}(t), t)\|_R \le C_R (1 + |t|)^{-1-\sigma}\,.
\end{equation}
In particular, the acceleration is bounded as
\begin{equation}\label{dc.b}
       |\dot{\boldsymbol{v}}(t)| \le C (1 + |t|)^{-1-\sigma}
\end{equation}
and there exists a $\boldsymbol{v}_\infty \in \mathbb V$ such that
\begin{equation}\label{dc.c}
       \lim_{t \to\infty} \boldsymbol{v}(t) =
       \boldsymbol{v}_\infty\,.
\end{equation}
\end{proposition}
{\it Proof}: From the Lorentz force equation and since $|\boldsymbol{v}(t)| \le
\overline v < 1$
we have
\begin{equation}\label{dc.d}
       |\dot{\boldsymbol{v}}(t)| \le C \,e\, \| Z(\cdot + \boldsymbol{q}(t),
       t)\|_{R_\rho}\,.
\end{equation}
Therefore (\ref{dc.b}) follows from (\ref{dc.a}). Then $\boldsymbol{v}(t) =
\boldsymbol{v}(0)+
\int\limits_0^t ds \, \dot{\boldsymbol{v}}(s)$ and $|\boldsymbol{v}(t) -
 \boldsymbol{v}_\infty | \le
C\, (1+|t|)^{-\sigma}$.

 It remains to establish (\ref{dc.a}), which
uses the method described in Appendix \ref{sec.ec}.
Since $Z(0) \not= 0,$ (\ref{ec.i}) reads now
\begin{equation}\label{dc.e}
       Z(t) = e^{{\sf{A}}t} Z(0)- \int\limits_0^t ds\, e^{{\sf{A}}(t-s)} g(s)\,.
\end{equation}
For the integrand in the second term we have the bound, compare
with (\ref{ec.j}),
\begin{equation}\label{dc.f}
       \| e^{{\sf{A}}(t-s)} g(s)\|_{R_\rho} \le C(\overline v) e^2
        (1 + (t-s)^2)^{-1}
        \|Z(\cdot + \boldsymbol{q}(s), s) \|_{R_\rho}  \,.
\end{equation}
For the first term of (\ref{dc.e}) we note that
$\boldsymbol{Z}_1 (\boldsymbol{x},0) = \boldsymbol{E}^0(\boldsymbol{x}) -
\boldsymbol{E}_{{\boldsymbol{v}
^0}} (\boldsymbol{x})
\in {\mathcal{M}}^\sigma$
by assumption. Using the solution of the inhomogeneous Maxwell
equations in position space and the bound (\ref{ad.l}) we have
\begin{eqnarray}
       |\boldsymbol{Z}_1 (\boldsymbol{x},t)| +| \boldsymbol{Z}_2(\boldsymbol{x},t)|
       &\le& C \, t^{-2} \int d^3y\, \delta(|\boldsymbol{x} -
       \boldsymbol{y}|- t) \big (|\boldsymbol{Z}_1 (\boldsymbol{y},0)|
       + |\boldsymbol{Z}_2 (\boldsymbol{y},0)| \big)\nonumber\\
        &&\!\!\!\!\!\!\!\!\!\!+ C\,t^{-1}\int d^3 y\, \delta(|\boldsymbol{x} -
       \boldsymbol{y}|- t) \big (|\nabla\boldsymbol{Z}_1
       (\boldsymbol{y},0)|+| \nabla \boldsymbol{Z}_2(\boldsymbol{y},0)|\big )\nonumber\\
       &\le& C\, t^{-2} \int d^3 y\, \delta(|\boldsymbol{x} -
       \boldsymbol{y}|- t)(1+
       |\boldsymbol{y}|)^{-1-\sigma}\nonumber\\
       &&\!\!\!\!\!\!\!\!\!\! + C \, t^{-1} \int d^3y\, \delta(|\boldsymbol{x} -
       \boldsymbol{y}|- t) (1+
       |\boldsymbol{y}|)^{-2-\sigma}\,.\nonumber\\[2mm]
       & \le& C\, (1+t)^{-1-\sigma}\,. \label{dc.g}
\end{eqnarray}
We choose $R\ge R_\rho$. Then from (\ref{dc.f}) and (\ref{dc.g})

\begin{equation}\label{dc.h}
        \|Z(\cdot + \boldsymbol{q}(t), t) \|_R
        \le C(1+t)^{-1-\sigma} + C(\overline v) e^2
        \int\limits_0^t ds\, (1+(t-s)^2)^{-1} \|Z(
        \cdot + \boldsymbol{q}(s), s)\|_R\,.
\end{equation}
Let $\kappa= \sup\limits_{t\ge 0}\, (1+t)^{1+\sigma} \|Z(\cdot +\boldsymbol{q}
(t),t)\|_R$. Then
\begin{equation}\label{dc.i}
      \kappa \le C + C(\overline v) e^2\, \big (\int\limits_0^t ds
       \,(1+(t-s)^2)^{-1} (1+s)^{-1-\sigma} \big) \kappa\,,
\end{equation}
which implies $\kappa < \infty$ provided $C(\overline v)\, e^2$ is
sufficiently small. $\Box$\medskip


\subsection{Appendix: How small is $\varepsilon$ ?}\label{sec.dd}
We consider an electron moving in an external magnetic field
oriented along the $z$-axis, $\boldsymbol{B}_{\mathrm{ex}} = (0, 0, B_0)$. The
corresponding vector potential is $\boldsymbol{A}_{\mathrm{ex}} (\boldsymbol{x}) =
\frac{1}{2} \,B_0 (- x_2, x_1, 0) $. According to our convention
the slowly varying vector potential is given by $\boldsymbol{A}_{\mathrm{ex}} (
\varepsilon \boldsymbol{x}) = \frac{1}{2}\,\varepsilon B_0 (- x_2, x_1, 0) $.
Thus $B_0$ is a reference field strength, which we will determine,
and $B = \varepsilon B_0$ is the physical field strength in the laboratory.
The motion of the electron is assumed to be in the 1-2 plane and we set $ \boldsymbol{v}
= (\boldsymbol{u},0)$.
According to Section \ref{sec.gb}, Example (ii), within a good
approximation the motion of the electron is governed by
\begin{equation}\label{dd.a}
       \gamma \dot{\boldsymbol{u}} = \omega_c (\boldsymbol{u}^\perp -
       \beta
        \omega_c \boldsymbol{u})\,.
\end{equation}
Here $\boldsymbol{u}^\perp =( - u_2, u_1)$,  $\omega_c = e B/m_{0} c$ is the cyclotron
frequency, and $\beta = e^2/6 \pi c^3 m_0$. The first term is the
Lorentz force and the second term accounts for the radiation
reaction.

We choose now the reference field $B_0$ such that both
terms balance, i.e.
\begin{equation}\label{dd.b}
       B_0 = (\beta e/m_{0} c)^{-1}\, .
\end{equation}
For electrons
\begin{equation}\label{dd.c}
       B_0 = 1.1 \times 10^{17}  \mbox{Gauss}
\end{equation}
and even larger by a factor $(1836)^2$  for protons. For a laboratory field of $10^5$
Gauss this yields
\begin{equation}\label{dd.d}
       \varepsilon = 10^{-12}\,.
\end{equation}
Written in units of $B_0$ (\ref{dd.a}) becomes
\begin{equation}\label{dd.e}
        \gamma\, \dot{\boldsymbol{u}}= \varepsilon \omega_c^0 (\boldsymbol{u}^\perp -
        \varepsilon \beta \omega_c^0 \boldsymbol{u})
\end{equation}
with $\beta \omega_c^0 = 1, \quad \omega_c^0 = e\, B_0/m_{0}c =
1.6 \times 10^{28} / $sec. Thus friction is of relative order
$\varepsilon$ and higher order corrections are then of relative
order $\varepsilon^2$. As to be demonstrated, the dimensionless
scaling parameter $\varepsilon$ merely serves as a book keeping device to keep
track of the relative order of the various contributions.


\subsection{Appendix: Point charge  limit,
negative bare mass}\label{sec.de}
The convential point charge limit is to let the diameter of the
total
charge distribution $R_\rho \to 0$ such that the charge remains
fixed. Physically, this means that the charge diameter is small in
units of the variation of the external potential, since this is
the only other length scale available. At first sight, one just
seems to say that the potentials vary slowly on the scale set by
the charge diameter and that hence point charge limit and
adiabatic limit coincide.

To see the difference let us regard $R_\rho$ as a small parameter,
relative to some reference scale. As before we require
that the total charge
\begin{equation}\label{de.b}
        \int d^3 x \, \rho (\boldsymbol{x}) = e
\end{equation}
is independent of $R_\rho$. The electrostatic energy diverges then
as
\begin{equation}\label{de.c}
         \frac{1}{2}\, \int d^3 k\, |\hat\rho
        (\boldsymbol{k})|^2 \, \frac{1}{\boldsymbol{k}^2}
        \cong  R_\rho^{\,-1} \overline m_{\mathrm{e}}
\end{equation}
for small $R_\rho$, where $\overline m_{\mathrm{e}}$ is the electrostatic energy
of the charge distribution
at the reference scale which is independent of $R_\rho$. In
particular, the ratio field mass to bare mass grows as $R_\rho^{\,-1}$
in the point charge limit and remains constant in the adiabatic
limit.

To display the order of magnitude of the various dynamical
contributions we
resort again to our standard example of an electron in a uniform
magnetic field $\boldsymbol{B} =B \boldsymbol{\widehat n}, ~ \boldsymbol{\widehat
n} = (0,0,1)$ with $B$ of the
order of 1 Tesla $= 10^4$ Gauss, say. It suffices to consider
small velocities. In the adiabatic limit we set $B= \varepsilon B_0$
where the reference field $B_0 = 1.1 \times 10^{17}$ Gauss, compare
with Section \ref{sec.gb}. Up to higher order corrections, the motion of
the electron is then governed by
\begin{equation}\label{de.d}
       \big ( m_{\mathrm{b}} + \frac{4}{3}\,
        m_{\mathrm{e}}\big )\dot{\boldsymbol{v}} =\frac{e}{c}\, \varepsilon B_0
        (\boldsymbol{v} \times \boldsymbol{\widehat n}) +
        \frac{e^2}{6 \pi c^3} \,
       \ddot{\boldsymbol{v}} + {\mathcal{O}}(\varepsilon^3)
\end{equation}
on the microscopic scale. Going over to the macroscopic time
scale, $t^\prime = \varepsilon^{-1} t,$ (\ref{de.d}) becomes
\begin{equation}\label{de.e}
       \big (m_{\mathrm{b}} + \frac{4}{3}\, m_{\mathrm{e}}\big )\, \dot{\boldsymbol{v}} =
       \frac{e}{c}\, B_0 (\boldsymbol{v} \times \boldsymbol{\widehat n})
        + \frac{e^2}{6 \pi c^3} \,
       \varepsilon \ddot{\boldsymbol{v}} +
       {\mathcal{O}}(\varepsilon^2)\,.
\end{equation}
Setting $m_0 = m_{\mathrm{b}} + \frac{4}{3}\, m_{\mathrm{e}}$, $ \omega_c^0
= e\, B_0/m_{0} c$, $ \beta= e^2/6 \pi c^3 m_0$, and restricting
to the critical manifold, as will be explained in Chapter \ref{sec.g}, Equation
(\ref{de.e}) becomes
\begin{equation}\label{de.f}
       \dot{\boldsymbol{v}} = \omega_c^0  \big(\boldsymbol{v}
        \times \boldsymbol{\widehat n} + \varepsilon\beta
       \omega_c^0\, (\boldsymbol{v} \times \boldsymbol{\widehat n})
       \times \boldsymbol{\widehat n}\big)  + {\mathcal{O}}(\varepsilon^2)\, ,
\end{equation}
equivalently, on the microscopic time scale
\begin{equation}\label{de.g}
       \dot{\boldsymbol{v}} = \omega_c \big(\boldsymbol{v} \times
       \boldsymbol{ \widehat n} + \beta \omega_c (\boldsymbol{v} \times
        \boldsymbol{\widehat
        n}) \times \boldsymbol{\widehat n}\big)  + {\mathcal{O}}(\varepsilon^3)
\end{equation}
with the cyclotron frequency $\omega_c = e\, \varepsilon B_0/m_{0} c =
e B /m_{0} c.$

For the point charge limit we rely on the Taylor expansion of
Section \ref{sec.eb}. Then, for small velocities,
\begin{equation}\label{de.h}
       \big ( m_{\mathrm{b}} + R_\rho^{\,-1}\, \frac{4}{3}\,\overline m_{\mathrm{e}}
       \big )\, \dot{\boldsymbol{v}} =
       \frac{e}{c}  B (\boldsymbol{v} \times \boldsymbol{\widehat n})
       + \frac{e^2}{6 \pi c^3} \,
       \ddot{\boldsymbol{v}}  + {\mathcal{O}}(R_\rho)\, .
\end{equation}
Since based on the same expansion, as long as no limit is taken,
of course, we can switch back and forth between
(\ref{de.h}) and (\ref{de.d}), resp. (\ref{de.e}), provided
 the appropriate units are used. This can be seen more easily if
 we accept momentarily the differential--difference equation
\begin{equation}\label{de.l}
       m_{\mathrm{b}} \dot{\boldsymbol{v}}(t)=
        e \big (\boldsymbol{E}_{\mathrm{ex}} (\boldsymbol{q}(t))
        + c^{-1} \boldsymbol{v}(t) \times \boldsymbol{B}_{\mathrm{ex}}(\boldsymbol{q}
        (t))\big) +\frac{e^2}{12 \pi c R_\rho^{\,2}} \,
       \big (\boldsymbol{v} (t-2 R_\rho/c) -
       \boldsymbol{v} (t)\big)\,,
\end{equation}
cf. (\ref{ea.h}), as an approximate equation for the motion of the
charge. If we expand in the charge diameter $R_\rho$, then
\begin{equation}\label{de.m}
        \big (m_{\mathrm{b}} +
        \frac{e^2}{6 \pi R_\rho c^2}\big) \dot{\boldsymbol{v}} = e
       (\boldsymbol{E}_{\mathrm{ex}} +
       c^{-1} \boldsymbol{v}
       \times\boldsymbol{B}_{\mathrm{ex}}) +
        \frac{e^2}{6 \pi c^3}\, \ddot{\boldsymbol{v}} +
       {\mathcal O} (R_\rho)\,,
\end{equation}
which is the analogue of (\ref{de.h}). On the other hand, if we
assume that the external fields are slowly varying, as discussed
in Section (\ref{sec.da}), then on the macroscopic scale
\begin{eqnarray}\label{de.n}
      \varepsilon m_{\mathrm{b}}  \dot{\boldsymbol{v}} (t)
       & = &\varepsilon \, e
        \big (\boldsymbol{E}_{\mathrm{ex}}
        (\boldsymbol{q}(t))
         +
        c^{-1}
        \boldsymbol{v}(t) \times\boldsymbol{B}_{\mathrm{ex}}
        (\boldsymbol{q}(t))\big) \nonumber\\
        &&+
        \frac{e^2}{12 \pi c R_\rho^{\,2}}\, \big (\boldsymbol{v}(t-2 \varepsilon
        R_\rho /c) - \boldsymbol{v}(t)\big)\,,
\end{eqnarray}
where $R_\rho$ is now regarded as fixed. Taylor expansion in $\varepsilon$
yields
\begin{equation}\label{de.o}
        \big (m_{\mathrm{b}}+ \frac{e^2}{6 \pi R_\rho c^2}\big)
         \dot{\boldsymbol{v}} = e
        (\boldsymbol{E}_{\mathrm{ex}} + c^{-1} \boldsymbol{v}
       \times\boldsymbol{B}_{\mathrm{ex}}) + \varepsilon\,
        \frac{e^2}{6 \pi  c^3}\, \ddot{\boldsymbol{v}} +
        {\mathcal O}(\varepsilon^2)
\end{equation}
which is the analogue of (\ref{de.e}).

  As can be seen from (\ref{de.h}), in
the point charge limit the total mass  becomes so
large that the particle hardly responds to the magnetic field.
 The only way out seems to formally compensate  the diverging $R_\rho^{\,-1}\,
(4/3) \overline m_{\mathrm{e}}$ by setting $m_{\mathrm{b}} = - R_\rho^{\,-1}\,
( 4/3)  \overline m_{\mathrm{e}} +  m_{\mathrm{exp}}$. But this is asking
for trouble, since the energy (\ref{ad.g}) is no longer bounded from
below and potential energy can be transferred to kinetic mechanical
energy without limit. To see this mechanism in detail we consider
the Abraham model with $\boldsymbol{B} _{\mathrm{ex}} = 0$ and
$\phi_{\mathrm{ex}}$
 varying only along the 1-axis. The bare mass of the particle is
 now
 $- m_{\mathrm{b}}$
with $ m_{\mathrm{b}} > 0$, as before. We set
$\boldsymbol{q}(t) = (q_t, 0, 0)$, $
\boldsymbol{v}(t) = (v_t, 0,0)$, $ \boldsymbol{E}_{\mathrm{ex}} =
 (- \phi^\prime (q), 0,0). ~ \phi$
is assumed to be strictly convex with minimum at $q=0$. Initially
the particle is at rest at the minimum of the potential. Thus
$\boldsymbol{E}(\boldsymbol{x},0) = \boldsymbol{E}_0(\boldsymbol{x})$
from (\ref{ca.e}) and $\boldsymbol{B}(\boldsymbol{x},0) = 0$.
We now give the particle a slight
push to the right, which means $q_0 = 0, v_0 > 0$. By conservation
of energy
\begin{eqnarray}\label{de.i}
       && \!\!\!\!\!\!\!\!\!\!-m_{\mathrm{b}}c^2 \, \gamma (v_t) + e\, \phi(q_t) +
       \frac{1}{2}\, \int d^3 x\, \big (\boldsymbol{E}(\boldsymbol{x},t)^2
        + \boldsymbol{B}(\boldsymbol{x},t)^2\big)
      \nonumber\\
       & =&- m_{\mathrm{b}} c^2 \gamma (v_0) + e \phi (q_0) +
        \frac{1}{2} \int d^3 x\, \boldsymbol{E}(\boldsymbol{x},0)^2 \,.
\end{eqnarray}
We split $\boldsymbol{E}$ into longitudinal and transverse components,
$\boldsymbol{E} = \boldsymbol{E}_\| +
        \boldsymbol{E}_\perp, ~\widehat{\boldsymbol{E}}_\| =
        |\boldsymbol{k}|^{-2} \boldsymbol{k} (\boldsymbol{k} \cdot \widehat{\boldsymbol{E}}).$
Clearly $\int d^3 x\, \boldsymbol{E}_\|
\boldsymbol{E}_\perp = 0$ and therefore
\begin{eqnarray}
       \int d^3 x \, \boldsymbol{E}(\boldsymbol{x},t)^2 &\ge&
        \int d^3 x \, \boldsymbol{E}_\|(\boldsymbol{x},t)^{2}\,=
       \int d^3 k\, |\boldsymbol{k}|^{-2}
       (\boldsymbol{k} \cdot \widehat{\boldsymbol{E}}(\boldsymbol{k},t))^2\nonumber\\
         & = &\int d^3 k\, |\boldsymbol{k}|^{-2}\,
         |\hat \rho(\boldsymbol{k})|^2 = \int d^3 x\,\boldsymbol{
         E}(\boldsymbol{x},0)^2 \,,\label{de.j}
\end{eqnarray}
since the initial field has zero transverse component. Inserting
in (\ref{de.i}) yields
\begin{equation}\label{de.k}
       \dot q_t^2 \ge 1 - \big[\gamma (v_0) +
       (e / m_{\mathrm{b}}c^2) (\phi (q_t) - \phi (q_0))\big]^{-2}\,.
\end{equation}
Since $\gamma (v_0) > 1, ~ \dot q_t > 0$ for short times. As the
particle moves to the right $(\phi(q_t) - \phi(q_0))$ is
increasing and therefore $\dot q_t \to 1$ and $q_t \to \infty $ as $t \to \infty$.
Note that $v_0$ and $m_{\mathrm{b}}$ can be
arbitrarily small.
Not so surprisingly, the Abraham model with a negative bare mass behaves rather unphysically.
A tiny  initial kick suffices to generate a
runaway solution.

The point charge limit is honored through a long tradition, which
however seems to have constantly overlooked that physically it is
more appropriate to have the external potentials slowly varying
on the scale of a fixed size charge distribution. Then there is no
need to introduce a negative bare mass and there are no runaway
solutions.
\subsection*{Notes and References}
{\it ad \ref{sec.da}:} The importance of slowly varying external potentials
has been emphasized repeatedly. It is somewhat surprising then,
that this notion was apparently never transcribed to the
equations of motion. In the context of charges and the Maxwell
field the adiabatic limit was first introduced in Komech, Kunze,
Spohn (1999) and in Kunze, Spohn (1999).
\bigskip\\
{\it ad \ref{sec.db}:} A more detailed discussion of the hydrodynamic limit
can be found in Spohn (1991).
\bigskip\\
{\it ad \ref{sec.dc}:} The initial slip as discussed here is a side--remark
in  Komech, Kunze, Spohn (1999), where the adiabatic limit for a
scalar wave field is studied. Komech, Spohn (1998) prove the
long--time asymptotics without the restriction $|e| < \overline e$
but imposing the Wiener condition instead. Orbital stability is
established by Bambusi, Galgani (1993).
\bigskip\\
{\it ad \ref{sec.dd}:} In the early work on the classical electron theory, one simply
expanded in $R_\rho$. $R_\rho$ was considered to be small, but
finite, say, of the order of the classical electron radius. Schott
(1912) pushed the expansion to include the radiation reaction
which he concluded to be ``quite inappreciable in this
and probably in all practical cases''. According to Frenkel (1925)
 the electrodynamics of point, rather than extended, charges
is an idea of P. Ehrenfest. The point charge limit is at the core
of the famous Dirac (1938) paper, cf. Section \ref{sec.bb}. Since
then the limit $m_{\mathrm{b}} \to - \infty$ is a standard piece
of the theory, reproduced in textbooks and survey articles. The
negative bare mass was soon recognized as a source of instability.
We refer to the review by Erber (1961). On a linearized level
stability is studied by Wildermuth (1955) and by Moniz, Sharp
(1977). Bambusi, Noja (1993) discuss the point charge limit in
the dipole approximation and show that in the limit the true
solution is well--approximated by the linear Lorentz--Dirac
equation with the full solution manifold, physical and unphysical,
explored. The bound (\ref{de.k}) is taken from Bauer,  D\"{u}rr (1999), which
is the only quantitative handling of the instability for the full
nonlinear problem.
\newpage
\section{Self -- Force}\label{sec.e}
\setcounter{equation}{0}

The inhomogeneous Maxwell equations have been solved in (\ref{aa.p}),
(\ref{aa.q}).
Thus it is natural to insert them in the Lorentz force equation in order
to obtain a closed, albeit  memory equation for the position of
the particle.

According to (\ref{aa.p}), (\ref{aa.q}) the Maxwell fields are a sum of initial and retarded
terms. We discuss first the  contribution
from the initial fields. By our specific
choice of initial conditions they have the representation, for
$t \ge 0$,
\begin{eqnarray}\label{e.a}
       \boldsymbol{E}_{(0)} (\boldsymbol{x},t) &=& -
       \int\limits_{- \infty}^0 ds \, \int d^3
       y\, \big ( \nabla G_{t-s}\, (\boldsymbol{x}-\boldsymbol{y})\,
        \rho (\boldsymbol{y}-\boldsymbol{q}^0-\boldsymbol{v}^0 s)\\
       && \qquad\qquad\qquad +\partial_t G_{t-s}\, ( \boldsymbol{x} - \boldsymbol{y} )
        \boldsymbol{v}^0 \, \rho(\boldsymbol{y}-\boldsymbol{q}^0
         - \boldsymbol{v}^0 s) \big )\,,\nonumber\\
       \boldsymbol{B}_{(0)}(\boldsymbol{x},t) &=&\int\limits_{- \infty}^0 ds \, \int d^3
       y\,  \nabla \times G_{t-s}\, (\boldsymbol{x}-\boldsymbol{y})\,
        \boldsymbol{v}^0 \rho (\boldsymbol{y}-\boldsymbol{q}^0-\boldsymbol{v}^0
       s)\,,\label{e.b}
\end{eqnarray}
compare with (\ref{ca.u}), (\ref{ca.v}). (\ref{ca.l}) and (\ref{aa.p}), can be
checked by going to Fourier space and using (\ref{aa.q}) as initial
condition in (\ref{e.a}), (\ref{e.b}). Since $G_t$ is concentrated on the light
cone, we conclude from (\ref{e.a}), (\ref{e.b}) that
$\boldsymbol{E}_{(0)} (\boldsymbol{x},t) = 0, ~ \boldsymbol{B}_{(0)}
 (\boldsymbol{x},t) = 0$ for
$|\boldsymbol{q}^0 - \boldsymbol{x}| \le t - R_\rho. $ If we would have allowed for more
general initial data, such a property would hold only
asymptotically for large $t$.

Next we note that by energy conservation the particle cannot travel
too
far. Using the bound on the potential, we can find a $\overline v < 1$
such that
\begin{equation}\label{e.c}
        \sup_{t \in \mathbb{R}}\, |\boldsymbol{v}^\varepsilon(t)| <
        \overline v < 1 \,,
\end{equation}
cf. Equation (\ref{ec.x}).
The charge distribution vanishes for $|\boldsymbol{x} - \boldsymbol{q}(t)| \ge R_\rho$.
Since $| \dot{\boldsymbol{q}} (t)| \le \overline v$, the initial fields and the
charge distribution have zero overlap once
\begin{equation}\label{e.x}
        t \ge \overline t_\rho = 2 R_\rho /(1- \overline
        v)\,.
\end{equation}
Thus for $t> \overline t_\rho$ the initial fields make no contribution to the
self--force and it remains to discuss the effect of the retarded fields.

We insert (\ref{aa.l}), (\ref{aa.m}) into the Lorentz force equation setting the external
potentials equal to zero for a while. Then on the macroscopic
scale, for $t \ge \varepsilon \overline t_\rho$,
\begin{equation}\label{e.d}
       \frac{d}{dt}\, \big (m_{\mathrm{b}} \gamma \,\boldsymbol{v}^\varepsilon (t)\big)
        = \boldsymbol{F}_{\mathrm{self}}^\varepsilon\, (t)
\end{equation}
with the self--force
\begin{eqnarray}
        \boldsymbol{F}_{\mathrm{self}}^\varepsilon\, (t)
        &=&  \int\limits_0^t ds \, \varepsilon \int d^3 k \,
                |\hat\rho (\varepsilon \boldsymbol{k})|^2\,
        e^{-i \boldsymbol{k} \cdot (\boldsymbol{q}^\varepsilon(t)-
        \boldsymbol{q}^\varepsilon (s))}\nonumber\\
        && \big ( (|\boldsymbol{k}|^{-1}\, \sin |\boldsymbol{k}| (t-s))
         i\boldsymbol{k} - (\cos |\boldsymbol{k}| (t-s))
         \boldsymbol{v}^\varepsilon(s)\nonumber\\[2mm]
         && - (|\boldsymbol{k}|^{-1} \sin |\boldsymbol{k}| (t-s)) \,
         \boldsymbol{v}^\varepsilon (t) \times
         (i \boldsymbol{k} \times \boldsymbol{v}^\varepsilon(s))
         \big )\,.\label{e.e}
\end{eqnarray}

(\ref{e.d}) is exact under the stated conditions on the initial
fields. No information has been discarded. The interaction with
the field has been merely transcribed into a memory term. To make
further progress we have to use a suitable approximation which
exploits that the external forces are slowly varying. Since this
corresponds to small $\varepsilon$, we just have to Taylor expand
 $\boldsymbol{F}_{\mathrm{self}}^\varepsilon\, (t)$,
 which is carried out in Section \ref{sec.eb} with the proper
 justification left for Appendix \ref{sec.ec}. But before, also to
 make contact with previous work, we have a closer look at the memory
 term.


\subsection{Memory equation }\label{sec.ea}
Equation (\ref{e.x}) can be further simplified where
 we set $\varepsilon = 1$ in this subsection. By
partial integration
\begin{eqnarray}
        &&\!\!\!\!\!\!\!\!\!   \int\limits_0^t ds  \int d^3
        k \,|\hat\rho (\boldsymbol{k})|^2\,
        e^{- i\boldsymbol{k} \cdot (\boldsymbol{q}(t)-
        \boldsymbol{q}(s))} \boldsymbol{v}(s) \, \frac{d}{ds}\, |\boldsymbol{k}|^{-1}
        \sin |\boldsymbol{k}| (t-s)\nonumber\\
        &=& - \int\limits d^3 k
         \,|\hat\rho (\boldsymbol{k})|^2\, e^{- i \boldsymbol{k} \cdot
        (\boldsymbol{q}(t)- \boldsymbol{q}(0))}
         \boldsymbol{v}(0) |\boldsymbol{k}|^{-1} \sin
         |\boldsymbol{k}| t \nonumber\\
        && - \int\limits_0^t ds \int d^3
        k \,|\hat\rho (\boldsymbol{k})|^2\, e^{- i \boldsymbol{k} \cdot
        (\boldsymbol{q}(t)- \boldsymbol{q}(s))}
         (|\boldsymbol{k}|^{-1}\sin |\boldsymbol{k}|(t-s))
         \big( \dot{\boldsymbol{v}}(s) + i(\boldsymbol{k}
         \cdot \boldsymbol{v}(s)) \boldsymbol{v}(s)\big)\,.\nonumber\\
         \label{ea.a}
\end{eqnarray}
Since $t \ge \overline t_\rho $, the boundary term
vanishes. Inserting (\ref{ea.a}) into (\ref{e.e}), returning to physical space, and
setting $t-s= \tau$, we have for $t \ge \overline t_\rho$
\begin{eqnarray}\label{7}
        \boldsymbol{F}_{\mathrm{self}}\, (t)
        &=& - \int\limits_0^\infty d \tau \, \big [ \dot{\boldsymbol{v}} (t- \tau) +
        (1-\boldsymbol{v}(t) \cdot \boldsymbol{v}(t-\tau)) \nabla_{\boldsymbol{x}}
        \label{ea.b}\\
        &&\qquad\quad + \boldsymbol{v}
        (t- \tau)(\boldsymbol{v}(t)-
        \boldsymbol{v}(t- \tau)) \cdot \nabla_{\boldsymbol{x}} \big]
         W_t(\boldsymbol{x}) |_{\boldsymbol{x}=\boldsymbol{q}(t)
        -\boldsymbol{q}(t- \tau)} \,, \nonumber
\end{eqnarray}
where, as in (\ref{ad.q}),
\begin{equation}\label{ea.c}
         W_t(\boldsymbol{x}) =\int d^3 k\, |\hat\rho(\boldsymbol{k})|^2\,
           e^{- i\boldsymbol{k}
          \cdot \boldsymbol{x}}
          |\boldsymbol{k}|^{-1} \sin |\boldsymbol{k}| t
          \,.
\end{equation}
In (\ref{ea.b}) we have extended the integration to $\infty$, since the
integrand vanishes for $\tau \ge \overline t_\rho$. Carrying out the angle
integrations in (\ref{ea.c}) we obtain
\begin{eqnarray}\label{ea.d}
        && W_t(\boldsymbol{x})= |\boldsymbol{x}|^{-1} \big (h(|\boldsymbol{x}| + t)
         - h(|\boldsymbol{x}|-t)\big )\,,\\
        && h(w) = 2 \pi \, \int\limits_0^\infty d k \, g(k) \cos kw \label{ea.e}
\end{eqnarray}
with $g(|\boldsymbol{k}|)= | \hat\rho(\boldsymbol{k})|^2$.
Since $\rho$ vanishes for $|\boldsymbol{x}| \ge R_\rho, ~ h(w) =0$ for $|w| \ge 2
R_\rho$. Note that $|\boldsymbol{q}(t) - \boldsymbol{q}(t-\tau)| \le \overline v\, \tau$. Thus
for $t\ge \overline t_\rho$ we indeed have $W_t (\boldsymbol{q}(t) -
\boldsymbol{q}(t-\tau)) = 0$,
 as claimed before. $\boldsymbol{F}_{\mathrm{self}}(t)$
has a finite memory extending backwards in time up to $t- \overline t_\rho$.

To go beyond (\ref{ea.d}) one has use a specific $\rho$. Two, at the
time popular, choices are $\rho_s(\boldsymbol{x}) =e (4 \pi R^2_\rho)^{-1}\, \delta
(|\boldsymbol{x}| - R_\rho)$ and $\rho_b(\boldsymbol{x}) = e\,( 4 \pi R^3_\rho/3)^{-1}$
for $|\boldsymbol{x}| \le R_\rho, ~ \rho_b(\boldsymbol{x}) = 0$ for
 $|\boldsymbol{x}| \ge R_\rho$. For the uniformly charged sphere one finds
\begin{equation}
        h(R_\rho w) = \left \{\begin{array}{cccc}
        &e^2 (8 \pi R_\rho)^{-1} ( 1- |w|/2)\,
        \quad  & \mbox{for} \quad |w| \le 2 \,,\\
        & 0
        \quad & \mbox{for}  \quad |w| \ge 2 \,,
        \end{array}\right.  \label{ea.f}
 \end{equation}
 and for the uniformly charged ball
 \begin{equation}
        h(R_\rho w) = \left \{\begin{array}{cccc}
        &e^2 (8 \pi R_\rho)^{-1}\, \frac{9}{8}\, \widetilde h \ast \widetilde h (w)
        \quad  & \mbox{for} \quad |w| \le 2 \,,\\
        & 0
        \quad & \mbox{for}  \quad |w| \ge 2 \,,
        \end{array}\right.  \label{ea.g}
 \end{equation}
with $\widetilde h(w) = (1-w^2)\unit_{\{|w|\le 1\}}$\,.

For the charged sphere $W_t(\boldsymbol{x})$ is piecewise linear
and, by first taking the gradient of $W$, the time
integrations simplify. In the approximation of
 small velocities the motion of the charged particle is then
governed by the differential--difference equation
\begin{equation}\label{ea.h}
        m_{\mathrm{b}} \dot{\boldsymbol{v}} (t)= e \big(\boldsymbol{E}_{\mathrm{ex}}
        (\boldsymbol{q}(t)) +
        \boldsymbol{v} (t) \times \boldsymbol{B}_{\mathrm{ex}}
        (\boldsymbol{q}(t))\big) + \frac{e^2}{12 \pi
        R^2_\rho}\, \big (\boldsymbol{v} (t- 2 R_\rho) - \boldsymbol{v}(t)\big)\,,
\end{equation}
where we have reintroduced the external fields.

The memory equation (\ref{ea.h}) is of suggestive simplicity.
However, to have a well defined dynamics one has to prescribe $\boldsymbol{q}(0)$
and $\boldsymbol{v}(t)$ for $-2R_\rho \le t \le 0$. No instruction for
that choice is provided by the supporters of
differential--difference equations. More importantly, $R_\rho$ is
a small parameter and we might allow on the top of small
velocities a further error of ${\mathcal O} (R_\rho)$ by Taylor
expanding in (\ref{ea.h}) to obtain
\begin{equation}\label{ea.i}
        m_{\mathrm{b}} \dot{\boldsymbol{v}} = e \big(\boldsymbol{E}_{\mathrm{ex}}
         + \boldsymbol{v}  \times \boldsymbol{B}_{\mathrm{ex}}) -
        \frac{e^2}{6 \pi R_\rho}\, \dot{\boldsymbol{v}}+
         \frac{e^2}{6 \pi}\,\ddot{\boldsymbol{v}} + \mathcal{O}
         (R_\rho)\,.
\end{equation}
(\ref{ea.i}) is a differential equation
and only $\boldsymbol{q}, \boldsymbol{v}, \dot{\boldsymbol{v}}$
are needed as initial data. As to be
discussed in Section \ref{sec.f}, in fact $\dot{\boldsymbol{v}}(0)$ is determined by
$\boldsymbol{q}(0), \boldsymbol{v}(0)$, since the physical solution has to lie on the
critical manifold of (\ref{ea.i}).


\subsection{Taylor expansion}\label{sec.eb}
We return to Equation (\ref{e.d}).
As will be  explained in Section \ref{sec.ec} we know that
\begin{equation}\label{eb.a}
       \sup_{t \in \mathbb{R}}\,|\ddot{\boldsymbol{q}}^\varepsilon (t) | \le
       C,~ \sup_{t \in \mathbb{R}}\,
           |\stackrel{...}{\boldsymbol{q}}^{\varepsilon} (t) | \le
        C, ~ \sup_{t \in \mathbb{R}}\,
        |\stackrel{....}{\boldsymbol{q}}^{\,\varepsilon} (t) | \le C
\end{equation}
uniformly in $\varepsilon$, provided the total charge $e$ is sufficiently small.
This smallness condition only reflects that at present we do not know
how to do better mathematically. Physically we expect (\ref{eb.a}) to
hold no matter how large $e$.

Because of (\ref{eb.a}) we are allowed to Taylor expand in (\ref{e.e}).
To simplify notation we set $\boldsymbol{v}^\varepsilon (t) = \boldsymbol{v}$ and $t-s =
\tau$. Then
\begin{eqnarray}\label{eb.b}
       \boldsymbol{v}^\varepsilon (s) &=& \boldsymbol{v}^\varepsilon
        (t-\tau)= \boldsymbol{v}- \dot{\boldsymbol{v}} \tau +
       \frac{1}{2}\, \ddot{\boldsymbol{v}} \tau^2 +
       \mathcal{O}(\tau^3)\,,\\[2mm]
       e^{-i\boldsymbol{k} \cdot (\boldsymbol{q}^\varepsilon(t)-
       \boldsymbol{q}^\varepsilon(s))}&=& e^{-i\boldsymbol{k}
           \cdot(\boldsymbol{q}^\varepsilon
       (t) - \boldsymbol{q}^\varepsilon(t-\tau))} = e^{-i(\boldsymbol{k}
       \cdot \boldsymbol{v})\tau}
       \Big ( 1 + \frac{1}{2}\, \tau^2 i (\boldsymbol{k} \cdot \dot{\boldsymbol{v}})
       - \frac{1}{6}\, \tau^3\, i (\boldsymbol{k}\cdot \ddot{\boldsymbol{v}})\nonumber\\[2mm]
        &&- \frac{1}{2} \, \big (\frac{1}{2}\, \tau^2 (\boldsymbol{k}
       \cdot \dot{\boldsymbol{v}})
       - \frac{1}{6}\, \tau^3 (\boldsymbol{k} \cdot \ddot{\boldsymbol{v}}) \big)^2 +
       \mathcal{O} ((|\boldsymbol{k}| \tau^2)^3)\Big )\, .\label{eb.c}
\end{eqnarray}
Inserting in (\ref{e.e}) and substituting $s^\prime = \varepsilon^{-1} s,~
\boldsymbol{k}^\prime = \varepsilon \boldsymbol{k}$ yields
\begin{eqnarray}
       \boldsymbol{F}_{\mathrm{self}}^\varepsilon (t) &=&
        \int\limits_0^{\varepsilon^{-1}t} d \tau \,
        \varepsilon^{-1} \, \int d^3 k |\hat\rho(\boldsymbol{k}) |^2\,
        e^{-i(\boldsymbol{k}\cdot \boldsymbol{v}) \tau}\nonumber\\
        && \Big \{(|\boldsymbol{k}|^{-1} \sin |\boldsymbol{k}|
        \tau) i \boldsymbol{k} - (\cos |\boldsymbol{k}| \tau)(\boldsymbol{v}-
        \varepsilon \tau \dot{\boldsymbol{v}} +
        \frac{1}{2}\, \varepsilon^2 \tau^2 \ddot{\boldsymbol{v}} )\nonumber\\[2mm]
        && - (|\boldsymbol{k}|^{-1} \sin |\boldsymbol{k}| \tau)
        \big(\boldsymbol{v} \times (i\boldsymbol{k} \times
        \boldsymbol{v})-
        \boldsymbol{v} \times (i\boldsymbol{k} \times
          \varepsilon \tau \dot{\boldsymbol{v}})
        + \frac{1}{2} \, \boldsymbol{v} \times (i \boldsymbol{k} \times \varepsilon^2
        \tau^2 \ddot{\boldsymbol{v}})\big)\nonumber\\[2mm]
        && +\frac{1}{2} \, \varepsilon \tau^2 i (\boldsymbol{k} \cdot
       \dot{\boldsymbol{v}})
        \big (  (|\boldsymbol{k}|^{-1} \sin |\boldsymbol{k}| \tau)
         i\boldsymbol{k} - (\cos |\boldsymbol{k}| \tau)
        (\boldsymbol{v}- \varepsilon \tau \dot{\boldsymbol{v}}) \nonumber\\[2mm]
        && - (|\boldsymbol{k}|^{-1} \sin |\boldsymbol{k}| \tau)(\boldsymbol{v}
         \times (i\boldsymbol{k} \times \boldsymbol{v}) -
         \boldsymbol{v}
        \times (i\boldsymbol{k} \times \varepsilon \tau \dot{\boldsymbol{v}}))\big
        )\nonumber\\[2mm]
        && + \big (- \frac{1}{6}\, \varepsilon^2 \tau^3 i (\boldsymbol{k} \cdot
        \ddot{\boldsymbol{v}}) - \frac{1}{8} \, \varepsilon^2 \tau^4 (\boldsymbol{k}
         \cdot \ddot{\boldsymbol{v}})^2 \big )
         \big ( (|\boldsymbol{k}|^{-1} \sin |\boldsymbol{k}|
         \tau) i\boldsymbol{k} \nonumber\\[2mm]
        && - (\cos |\boldsymbol{k}| \tau) \boldsymbol{v} -
         (|\boldsymbol{k}|^{-1} \sin |\boldsymbol{k}| \tau)
         (\boldsymbol{v}
        \times (i\boldsymbol{k} \times \boldsymbol{v})) \big ) \Big \} +
        \mathcal{O}(\varepsilon^2)\label{eb.d}
       \, .
\end{eqnarray}

The terms proportional to $\varepsilon^{-1}$ cancel by symmetry. We
sort all other terms,
\begin{eqnarray}
        \boldsymbol{F}_{\mathrm{self}}^\varepsilon (t) &=&
        \int d^3 k \, |\hat\rho(\boldsymbol{k})|^2\nonumber\\
        && \Big \{ \big( -
        (\boldsymbol{v} \cdot
        \dot{\boldsymbol{v}}) \nabla_{\boldsymbol{v}} + \dot{\boldsymbol{v}}(\boldsymbol{v}
         \cdot \nabla_{\boldsymbol{v}}) \big )
         \, \int\limits_0^{\varepsilon^{-1}t} d \tau
        e^{-i(\boldsymbol{k} \cdot\boldsymbol{v})\tau}
         (|\boldsymbol{k}|^{-1} \sin |\boldsymbol{k}| \tau)\nonumber\\
        &&+ \big ( \dot{\boldsymbol{v}} + \frac{1}{2}\,\boldsymbol{v} (
        \dot{\boldsymbol{v}}
         \cdot \nabla_{\boldsymbol{v}})\big )
        \int\limits_0^{\varepsilon^{-1}t} d \tau \, \tau
        e^{-i(\boldsymbol{k} \cdot \boldsymbol{v})\tau}(\cos |\boldsymbol{k}| \tau)\nonumber\\[1mm]
        && + \varepsilon \Big ( \frac{1}{2} \,\big [ - (\boldsymbol{v}^2 - 1)
        (\dot{\boldsymbol{v}}
        \cdot\nabla_{\boldsymbol{v}})\nabla_{\boldsymbol{v}} + \boldsymbol{v}(\boldsymbol{v} \cdot
         \nabla_{\boldsymbol{v}})(\dot{\boldsymbol{v}} \cdot\nabla_{\boldsymbol{v}}) +
         (\boldsymbol{v} \cdot \ddot{\boldsymbol{v}})\nabla_{\boldsymbol{v}}\nonumber\\[1mm]
        && - \ddot{\boldsymbol{v}}( \boldsymbol{v} \cdot \nabla_v)\big ]
         + \frac{1}{6}\,\big [ - (1- \boldsymbol{v}^2)
        (\ddot{\boldsymbol{v}} \cdot \nabla_{\boldsymbol{v}})
        \nabla_{\boldsymbol{v}} - \boldsymbol{v}
        (\boldsymbol{v} \cdot \nabla_{\boldsymbol{v}}) (\ddot{\boldsymbol{v}}
        \cdot \nabla_{\boldsymbol{v}})\nonumber\\[1mm]
        && + 3(\boldsymbol{v} \cdot \dot{\boldsymbol{v}})
        (\dot{\boldsymbol{v}} \cdot \nabla_{\boldsymbol{v}})\nabla_{\boldsymbol{v}} - 3
        \dot{\boldsymbol{v}} (\boldsymbol{v} \cdot \nabla_{\boldsymbol{v}})
        (\dot{\boldsymbol{v}} \cdot \nabla_{\boldsymbol{v}}) \big ]
        + \frac{1}{8} \,\big [ (\boldsymbol{v}^2 - 1) (\dot{\boldsymbol{v}} \cdot
        \nabla_{\boldsymbol{v}})^2 \nabla_{\boldsymbol{v}}
        \nonumber\\
         && - \boldsymbol{v} (\boldsymbol{v} \cdot \nabla_{\boldsymbol{v}})
         (\dot{\boldsymbol{v}}
        \cdot \nabla_{\boldsymbol{v}})^2 \big ]\Big ) \int\limits_0^{\varepsilon^{-1}t}
        d \tau \, \tau e^{-i(\boldsymbol{k} \cdot \boldsymbol{v})\tau}
         (|\boldsymbol{k}|^{-1} \sin |\boldsymbol{k}| \tau)\nonumber\\
        && + \varepsilon \Big ( -\ddot{\boldsymbol{v}} - \frac{1}{6}\, \big [
        \boldsymbol{v}
        (\ddot{\boldsymbol{v}} \cdot \nabla_{\boldsymbol{v}}) + 3
        \dot{\boldsymbol{v}}
        (\dot{\boldsymbol{v}} \cdot
        \nabla_{\boldsymbol{v}})\big ]\Big )\nonumber\\
        &&\int\limits_0^{\varepsilon^{-1}t} d \tau \, \tau^2
        e^{-i(\boldsymbol{k} \cdot \boldsymbol{v})\tau}\cos |\boldsymbol{k}|
         \tau \Big \} +
        \mathcal{O}(\varepsilon^2)\, .
        \label{eb.e}
\end{eqnarray}

To take the limit $\varepsilon \to 0$ we go back to position space
and use the fundamental solution of the wave equation. Then, for
$p=0,1$,
\begin{eqnarray}\label{eb.f}
      \lim_{\varepsilon \to 0}&&\!\!\!\!\!\!\!\!\!\!
       \int\limits_0^{\varepsilon^{-1}t}d \tau \int
       d^3 k |\hat\rho(\boldsymbol{k})|^2\,
        e^{-i(\boldsymbol{k}\cdot \boldsymbol{v}) \tau}
        (|\boldsymbol{k}|^{-1} \sin |\boldsymbol{k}| \tau) \, \tau^p
        \\
       && \int\limits_0^\infty dt \,\int d^3 x\, \int d^3 y\, \rho(\boldsymbol{x})
        \rho(\boldsymbol{y})\, \frac{1}{4 \pi t}\,
        \delta (|\boldsymbol{x} + \boldsymbol{v} t -\boldsymbol{y}
        | - t) \, t^p \nonumber\\
        &=& \left \{\begin{array}{llll}
        &\int d^3 k\, |\hat \rho(\boldsymbol{k})|^2 [\boldsymbol{k}^2 -
        (\boldsymbol{k} \cdot \boldsymbol{v})^2]^{-1}
        \quad  & \mbox{for} \quad p=0\,,\nonumber\\[2mm]
        &\int d^3 x \, \rho(\boldsymbol{x}) \int d^3 y \rho
        (\boldsymbol{y})\,(\gamma^2/4 \pi)
        \quad & \mbox{for}  \quad p=1\,.
        \end{array}\right.
\end{eqnarray}
By the same method
\begin{eqnarray}\label{eb.g}
      &&\!\!\!\!\!\!\!\!\!\!\! \lim_{\varepsilon \to 0}
      \int\limits_0^{\varepsilon^{-1}t}d \tau
      \int d^3 k \,|\hat\rho(\boldsymbol{k})|^2\,
        e^{-i(\boldsymbol{k}\cdot \boldsymbol{v}) \tau}\,\tau^{1+p}\,\frac{d}{d \tau}\,
        (|\boldsymbol{k}|^{-1} \sin |\boldsymbol{k}| \tau )
        \\
       &=&- \big(1+ p + (\boldsymbol{v} \cdot \nabla_{\boldsymbol{v}})\big )\,
       \int\limits_0^\infty dt \,
        \int d^3 k\, |\hat\rho(\boldsymbol{k})|^2\,
         e^{-i(\boldsymbol{k} \cdot \boldsymbol{v})t}
         (|\boldsymbol{k}|^{-1} \sin|\boldsymbol{k}|t) t^p\nonumber\\
        &=& \left \{\begin{array}{llll}
       - &\!\!\!\!\int d^3 k\, |\hat \rho(\boldsymbol{k})|^2 (\boldsymbol{k}^2 +
        (\boldsymbol{k} \cdot \boldsymbol{v})^2)
       [\boldsymbol{k}^2- (\boldsymbol{k} \cdot \boldsymbol{v})^2]^{-2}
        \quad  & \mbox{for}\quad p=0\,, \nonumber\\[2mm]
        -&\!\!\!\!\int d^3 x \, \rho(\boldsymbol{x}) \int d^3 y \rho(\boldsymbol{y})\,
         (2 \gamma^4 /4 \pi) \quad &\mbox{for} \quad p=1\,
        .
        \end{array}\right.
\end{eqnarray}

Collecting all terms the final result reads
\begin{eqnarray}
        \boldsymbol{F}_{\mathrm{self}}^\varepsilon (t) &=& - m_{\mathrm{f}} (\boldsymbol{v})
        \dot{\boldsymbol{v}} +
        \varepsilon (e^2/6 \pi)\,
        \big [\gamma^4 (\boldsymbol{v} \cdot \ddot{\boldsymbol{v}})
         \boldsymbol{v}+3 \gamma^6 (\boldsymbol{v} \cdot \dot{\boldsymbol{
        v}})^2 \boldsymbol{v}\nonumber\\[2mm]
        && + 3 \gamma^4 (\boldsymbol{v} \cdot
        \dot{\boldsymbol{v}}) \dot{\boldsymbol{v}}+ \gamma^2
        \ddot{\boldsymbol{v}}\big]
         + \mathcal{O} (\varepsilon^2) \label{eb.h}
\end{eqnarray}
with
\begin{eqnarray}
         m_{\mathrm{f}}(\boldsymbol{v})  &=& m_{\mathrm{e}}
         \Big[
         \big(|\boldsymbol{v}|^{-4} \gamma^2
          (3-\boldsymbol{v}^2) - (2|\boldsymbol{v}|^5)^{-1} (3 + \boldsymbol{v}^2)
          \, \log \, \frac{1+|\boldsymbol{v}|}{1-|\boldsymbol{v}|} \, \big )
          |\boldsymbol{v} \rangle\langle \boldsymbol{v}| \nonumber\\
         && + \big (-|\boldsymbol{v}|^{-2} + (2 |\boldsymbol{v}|^3)^{-1}\, (1+
          \boldsymbol{v}^2)\, \log
        \frac{1+ |\boldsymbol{v}|}{1-|\boldsymbol{v}|} \big )\unit
         \Big ] \,. \label{eb.i}
\end{eqnarray}
Note that $m_{\mathrm{f}}(\boldsymbol{v})=
{\mathrm{d}}(\boldsymbol{P}_{\mathrm{s}}
 - m_{\mathrm{b}} \gamma \boldsymbol{v})
/ {\mathrm{d}} \boldsymbol{v}$ as a $3 \times 3$
matrix.

Up to order $\varepsilon, ~ \boldsymbol{F}_{\mathrm{self}}^\varepsilon (t)$
consists of
two parts with a rather different character. The term
$- m_{\mathrm{f}} (\boldsymbol{v}) \dot{\boldsymbol{v}}$
is the contribution from the electromagnetic field to the change in total momentum. We
computed this term  already in Section \ref{sec.ca} via a completely different route.
 As emphasized there, since the Abraham model is
semi--relativistic, the velocity dependence of $m_{\mathrm{f}}$ has no reason
to be of relativistic form and indeed it is not. The term
proportional to $\varepsilon$ in (\ref{eb.h}) is the {\it radiation reaction}. Again
there is no a priori reason to expect it to be relativistic, but in fact
it is. Using the four--vector notation of Section \ref{sec.af}, the
radiation reaction can be rewritten as
\begin{equation}\label{eb.j}
       \varepsilon (e^2 / 6 \pi) [ \ddot v^\mu - \dot v^\lambda \dot v_\lambda
       v^\mu ]\, .
\end{equation}


\subsection{Appendix: How to bound the
acceleration?}\label{sec.ec}
We return to the microscopic time scale. From the conservation of
energy together with condition $(P)$, we have
\begin{eqnarray}\label{ec.x}
        E_s(\boldsymbol{v}^0) +  (\phi \ast \rho)(\varepsilon \boldsymbol{q}^0) &=&
       \mathcal{E} (\boldsymbol{E}^0,
        \boldsymbol{B}^0,
        \boldsymbol{q}^0, \boldsymbol{v}^0) = \mathcal{E} (\boldsymbol{E}
        (t), \boldsymbol{B}(t),
       \boldsymbol{q}(t), \boldsymbol{v}(t))\nonumber\\[2mm]
       &\ge&\!\!\! m_{\mathrm{b}} \gamma(\boldsymbol{v}(t)) + e
       \,\phi_{\mathrm{min}}
\end{eqnarray}
and therefore
\begin{equation}\label{ec.a}
\sup_t\, |\boldsymbol{v}(t)| \le
        \overline v < 1 \,.
\end{equation}

The external forces are of order $\varepsilon$. Superficially the
self-force is of order one. However for a Coulombic field the
self--force vanishes. Thus if we could show that the deviations
from the appropriate local soliton field are of order
$\varepsilon$, then altogether
\begin{equation}\label{ec.b}
        \sup_t\, |\dot{\boldsymbol{v}}(t)| \le
        C\, \varepsilon
\end{equation}
with $C$ a suitable constant. This is what we want to prove. We will not keep track of the
constants and the value of $C$ changes from equation to equation.
We make sure however that $C$ depends only on $\overline v$ and is
thus determined by the initial conditions. Of course, to
justify the Taylor expansion of Section \ref{sec.eb}, we also need analoguous
estimates on higher derivatives, which can be obtained with more
effort through the same scheme. Here we want to explain how to get
(\ref{ec.b}) and why we need $e$ to be sufficiently small, at least
at present.

>From the equations of motion we have
\begin{eqnarray}\label{ec.c}
       \dot{\boldsymbol{v}} &=&  m_0 (\boldsymbol{v})^{-1}\,
       \Big[ \varepsilon \int d^3 x \,
        \rho (\boldsymbol{x}-\boldsymbol{q}) \big(\boldsymbol{E}_{\mathrm{ex}}
        (\varepsilon \boldsymbol{x}) + \boldsymbol{v} \times \boldsymbol{B}_{\mathrm{ex}}
        \, (\varepsilon \boldsymbol{x})\big)\nonumber\\
        && \qquad\qquad + \int d^3 x \, \rho(\boldsymbol{x}-\boldsymbol{q})
         \big( \boldsymbol{E}(\boldsymbol{x}) + \boldsymbol{v} \times
         \boldsymbol{B}(\boldsymbol{x})\big)\Big ]\,,
\end{eqnarray}
where $m_{\mathrm{0}}^{-1} (\boldsymbol{v})= (m_{\mathrm{b}} \gamma)^{-1} (\unit -
|\boldsymbol{v}|^{-2}|\boldsymbol{v} \rangle\langle\boldsymbol{v}|)$ is the matrix
inverse of $m_0(\boldsymbol{v})$. Clearly by (\ref{ec.a}) we have
 $\|m_0(\boldsymbol{v})^{-1}\| \le C$
and, by condition $(P)$, the first term is
bounded as
\begin{equation}\label{ec.d}
        \varepsilon \big |\int d^3 x \,
        \rho (\boldsymbol{x}-\boldsymbol{q}) \big(\boldsymbol{E}_{\mathrm{ex}}
        ( \varepsilon \boldsymbol{x}) + \boldsymbol{v} \times
        \boldsymbol{B}_{\mathrm{ex}} (\varepsilon \boldsymbol{x})\big)\big|
        \le C\, \varepsilon\, .
\end{equation}
On the other hand the self--force looks like order one. To reduce it
we have to exploit that $\boldsymbol{E},\boldsymbol{B}$ deviate only
little from $\boldsymbol{E_v}, \boldsymbol{B_v}$
close to the charge distribution, i.e. we rewrite the self--force
as
\begin{equation}\label{ec.e}
       \int d^3 x \,
        \rho (\boldsymbol{x}-\boldsymbol{q}) \big [ \boldsymbol{E}
         (\boldsymbol{x}) - \boldsymbol{E_v}(\boldsymbol{x}) + \boldsymbol{v} \times
        ( \boldsymbol{B}(\boldsymbol{x}) - \boldsymbol{B_v} (\boldsymbol{x})) \big]
\end{equation}
and have to show that the term in the square bracket is of order
$\varepsilon$.

Let us define then
\begin{eqnarray}
       Z(\boldsymbol{x},t) =\left ( \begin{array}{{ll}}
       \boldsymbol{E}(\boldsymbol{x},t) - \boldsymbol{E}_{\boldsymbol{v}
       (t)}(\boldsymbol{x}-\boldsymbol{q}(t))&\\
       \boldsymbol{B}(\boldsymbol{x},t)  -
       \boldsymbol{B}_{\boldsymbol{v}(t)} (\boldsymbol{x}-\boldsymbol{q}(t)) \label{ec.f}
       \end{array}\!\!\!\!\!\right) \, .
\end{eqnarray}
Using Maxwell equations and the relations $(\boldsymbol{v} \cdot \nabla)\,
 \boldsymbol{E_v} = -
\nabla \times \boldsymbol{B_v} + \rho \boldsymbol{v}, ~
(\boldsymbol{v} \cdot \nabla) \boldsymbol{B_v} = \nabla \times \boldsymbol{E_v}$
we obtain
\begin{equation}\label{ec.g}
       \dot Z(t)= {\sf A} Z(t) - g(t)\, ,
\end{equation}
where
\begin{equation}
       {\sf A}= \left ( \begin{array}{{cc}}
       0 &\nabla \times \\
       - \nabla \times  & 0
       \end{array}\right ),~ g(\boldsymbol{x},t) = \left (\begin{array}{{ll}}
       (\dot{\boldsymbol{v}}(t) \cdot \nabla_{\boldsymbol{v}})
       \boldsymbol{E_v} (\boldsymbol{x}-\boldsymbol{q}(t))&\\
        (\dot{\boldsymbol{v}}(t) \cdot \nabla_{\boldsymbol{v}})
         \boldsymbol{B_v} (\boldsymbol{x}-\boldsymbol{q}(t))\label{ec.h}
       \end{array}\!\!\!\!\!\right )\,.
\end{equation}
Therefore (\ref{ec.g}) has again the structure of the inhomogeneous
Maxwell equations. Since by our assumption on the initial data $Z(0) =
0$, we have
\begin{equation}\label{ec.i}
       Z(t)= - \int\limits_0^t ds \, e^{{\sf A}(t-s)} g(s) \, .
\end{equation}

We set $W(t,s) = e^{{\sf A}(t-s)} g(s)$. Below we prove that
\begin{equation}\label{ec.j}
       |\boldsymbol{W}_1 (t,s, \boldsymbol{q}(t) + \boldsymbol{x})
       | + |\boldsymbol{W}_2 (t,s,\boldsymbol{q}(t) + \boldsymbol{x})| \le e C|
        \dot{\boldsymbol{v}}(s)|(1+ (t-s)^2)^{-1}
\end{equation}
 for  $|\boldsymbol{x}| \le R_\rho $. Therefore inserting in (\ref{ec.c}) we obtain
\begin{equation}\label{ec.k}
       | \dot{\boldsymbol{v}} (t)| \le e C \big (\varepsilon + e
       \int\limits_0^t ds\, (1 + (t-s)^2)^{-1}\, | \dot{\boldsymbol{v}}(s)|\big
       )\,.
\end{equation}
Let $\kappa= \sup\limits_{t \ge 0} \, |  \dot{\boldsymbol{v}}(t)|$.
 Then (\ref{ec.k}) reads
\begin{eqnarray}
       \kappa &\le& e\, C \big (\varepsilon + e \kappa \int\limits_0^\infty ds\,
        (1 + s^2)^{-1}\big )\,, \nonumber\\
        \kappa &\le& \frac{e\, C}{1- e^2\, C}\,
        \varepsilon\,.\label{ec.l}
\end{eqnarray}
>From the computation below we will see that $C$ depends on $\overline v$
(and on model parameters like the form factor $f $), but not on $e$. Thus taking $e$
sufficiently small we can ensure $e^2\, C < 1$ and $\kappa \le C \varepsilon$
as claimed.

We still have to establish (\ref{ec.j}). $e^{{\sf A}t}$ is given in
Equation (\ref{aa.l}), (\ref{aa.m}).
Since $\nabla \cdot  \boldsymbol{g}_1(s)= 0 = \nabla \cdot  \boldsymbol{g}_2(s)$, the term
proportional to $| \boldsymbol{k} \rangle\langle \boldsymbol{k}|$ drops out.
In real space $| \boldsymbol{k}|^{-1} \sin| \boldsymbol{k}|t$
becomes $G_t$ from (\ref{aa.o}) and $\cos | \boldsymbol{k}|t$
becomes $\partial_t G_t$.
Therefore
\begin{eqnarray}
    &&  \boldsymbol{W}_1 (t,s, \boldsymbol{x}) =\frac{1}{4 \pi(t-s)^2}\, \int d^3 y\, \delta
        ( | \boldsymbol{x}-\boldsymbol{y}| -
        (t-s))\nonumber\\[2mm]
        &&\qquad\qquad\qquad\qquad\quad[ (t-s) \nabla \times  \boldsymbol{g}_2 ( \boldsymbol{y}, s)
        +  \boldsymbol{g}_1( \boldsymbol{y},s)
        - ( \boldsymbol{x} -\boldsymbol{y}) \cdot \nabla
        \boldsymbol{g}_1( \boldsymbol{y},s) ]\,,\nonumber\\
     & &\boldsymbol{W}_2(t,s, \boldsymbol{x})=\frac{1}{4 \pi(t-s)^2}\, \int d^3y \,
        \delta(| \boldsymbol{x}- \boldsymbol{y}|
        -(t-s))\label{ec.m}\\[2mm]
        &&\qquad\qquad\qquad\qquad\quad[- (t-s) \nabla \times
         \boldsymbol{g}_1( \boldsymbol{y},s)
        +  \boldsymbol{g}_2( \boldsymbol{y},s) - ( \boldsymbol{x}-\boldsymbol{y} )
        \cdot \nabla  \boldsymbol{g}_2 ( \boldsymbol{y},s)]\, . \nonumber
\end{eqnarray}
We insert $g$ from (\ref{ec.h}).
$\boldsymbol{E_v}$ and $ \boldsymbol{B_v}$ are first order
derivatives of the function $\phi_{\boldsymbol{v}}$
which according to (\ref{ca.g}) is given by
\begin{equation}\label{ec.n}
       \phi_{\boldsymbol{v}}( \boldsymbol{x}) = \int d^3 y
        \rho( \boldsymbol{x}- \boldsymbol{y}) (4 \pi)^{-1}\,
       \big[ \big((1- \boldsymbol{v}^2) \boldsymbol{y}^2 +
       ( \boldsymbol{v} \cdot  \boldsymbol{y})^2\big) \big ]^{-1/2}\, .
\end{equation}
Using (\ref{ca.e}) we have  component--wise
\begin{eqnarray}
      |\nabla_{\boldsymbol{v}} \boldsymbol{E_v}
        (\boldsymbol{x})| + |\nabla_{\boldsymbol{v}}
        \boldsymbol{B_v}(\boldsymbol{x})| &\le& C\,(\,|\nabla
       \phi_{\boldsymbol{v}}(\boldsymbol{x})| +
       |\nabla\nabla_{\boldsymbol{v}}
       \phi_{\boldsymbol{v}}(\boldsymbol{x})|)\,,\label{ec.o}\\
         |\nabla\nabla_{\boldsymbol{v}} \boldsymbol{E_v}
        (\boldsymbol{x})| + |\nabla\nabla_{\boldsymbol{v}}
        \boldsymbol{B_v}(\boldsymbol{x})| &\le&
        C\, (\,|\nabla\nabla_{\boldsymbol{v}} \phi_{\boldsymbol{v}}(\boldsymbol{x})| +
        |\nabla\nabla\nabla_{\boldsymbol{v}}\phi_{\boldsymbol{v}}
        (\boldsymbol{x})|)\,.\nonumber
\end{eqnarray}
Taking now successive derivatives in (\ref{ec.n}) we obtain the
bounds
\begin{eqnarray}
     |\nabla\phi_{\boldsymbol{v}} (\boldsymbol{x})| +
       |\nabla\nabla_{\boldsymbol{v}} \phi_{\boldsymbol{v}}(\boldsymbol{x})|  &\le & e \,
       C \,(1+|\boldsymbol{x}|)^{-2}\,,\nonumber\\
      |\nabla\nabla\phi_{\boldsymbol{v}}(\boldsymbol{x})|
         + |\nabla\nabla\nabla_{\boldsymbol{v}}
         \phi_{\boldsymbol{v}}
         (\boldsymbol{x})| &\le &
        e\, C\, (1+ |\boldsymbol{x}|)^{-3}\, ,\label{ec.p}
\end{eqnarray}
which implies
\begin{eqnarray}\label{ec.q}
    |\boldsymbol{g}_1 (\boldsymbol{x},s)| + |\boldsymbol{g}_2
     (\boldsymbol{x},s)|  &\le&  e\,
        C |\dot{\boldsymbol{v}}(s)|(1+ |\boldsymbol{x}-
        \boldsymbol{q}(s)|^2)^{-1}\,,\nonumber\\
    |\nabla \boldsymbol{g}_1(\boldsymbol{x},s)| +
     |\nabla \boldsymbol{g}_2(\boldsymbol{x},s)|& \le& e \,
     C |\dot{\boldsymbol{v}}(s)|
     (1+|\boldsymbol{x}-\boldsymbol{q}(s)|^3)^{-1}\, .
\end{eqnarray}
We insert the bound (\ref{ec.q}) in (\ref{ec.m}) which results in a bound on
$W(t,s,\boldsymbol{q}(t)+\boldsymbol{x})$.
 We use that $|\boldsymbol{x}| \le R_\rho$
and $|\boldsymbol{q}(t)- \boldsymbol{q}(s)|\le \overline v |t-s|$, which finally
yields (\ref{ec.j}).

We summarize our findings as
\begin{theorem}\label{thm.ea}
 For the Abraham model satisfying the conditions $(C), (P),$ and $(I)$
there exist constants $\overline e$ and $C$, depending only on the
initial conditions through $\overline v$, such that on the
microscopic time scale we have
\begin{equation}\label{ec.r}
\sup_t\, | \boldsymbol{v}(t)| \le \overline v < 1, ~
        \sup_t\, \big |\big (\frac{d}{dt}\big)^n  \boldsymbol{v}(t)\big| \le
        C\, \varepsilon^n , ~ n=1,2,3 \,,
\end{equation}
provided the charge is sufficiently small, i.e. $e < \overline
        e.$
\end{theorem}

If we would keep track of the constant $C$, we would get a bound
of the admissable charge in Theorem \ref{thm.ea}. Since we believe this
restriction to be an artifact of the method anyhow, there is no
point in the effort.
\subsection*{Notes and References}
{\it ad \ref{sec.ea}:} Sommerfeld (1904,1905) first used systematically memory
equations. In fact he considered the Abraham model with the
kinetic energy $m_{\mathrm{b}} \boldsymbol{v}^2/2$ for the
particle and wanted to understand what happens when $\boldsymbol{v}(0) >
c$. He argued that the particle rapidly looses its energy to
become slower than $c$ by emitting what we call now Cherenkov
radiation. The differential--difference equation was stated first
by Page (1918) with its relativistic generalization Caldirola
(1956). For reviews we refer to Erber (1961) and Pearle (1982).
Moniz,  Sharp (1974,1977) supplied a linear stability analysis and
showed that the solutions to (\ref{ea.h}) are stable provided $R_\rho$
is not too small. For that reason Rohrlich (1997) regards
(\ref{ea.h}) and its relativistic sister as the fundamental
starting point for the classical dynamics of extended charges. We
take the Abraham model as the basic dynamical theory. Memory
equations are a useful tool in analyzing its properties. \bigskip\\
{\it ad \ref{sec.eb}:} Taylor expansion is taken from Kunze, Spohn
(1999). Such an expansion was first used by Sommerfeld (1904,1905) and
then repeated in various disguises. The traditional expansion
parameter is the size of the charge distribution, which in our
context is replaced by the  scaling parameter $\varepsilon$
\bigskip\\
{\it ad \ref{sec.ec}:} The bound on $\dot{\boldsymbol{v}}^\varepsilon (t)$
comes from Kunze, Spohn (1999) where also higher derivatives
are discussed. The contraction argument first appeared in Komech,
Kunze, Spohn (1999).
\newpage
\section{Comparison Dynamics}\label{sec.f}
\setcounter{equation}{0}
If in (\ref{eb.h}) we simply ignore the error of order
$\varepsilon^2$, then we obtain the following approximate equation
for the motion of the charge,
\setcounter{equation}{0}
\begin{eqnarray}\label{f.a}
        \dot{\boldsymbol{q}} = \boldsymbol{v},&&\!\!\!\!\!\! m(\boldsymbol{v})
        \dot{\boldsymbol{v}}= e \big (\boldsymbol{E}_{\mathrm{ex}}
        (\boldsymbol{q}) + \boldsymbol{v}
          \times
          \boldsymbol{B}_{\mathrm{ex}}(\boldsymbol{q})\big)\\[1mm]
          && \!\!\!\!\!+ \varepsilon (e^2/ 6 \pi) \, \big[\gamma^4(\boldsymbol{v} \cdot
           \ddot{\boldsymbol{v}})\boldsymbol{v} + 3 \gamma^6 (\boldsymbol{v}
            \cdot \dot{\boldsymbol{v}})^2
           \boldsymbol{v} + 3 \gamma^4(\boldsymbol{v}
          \cdot \dot{\boldsymbol{v}})\dot{\boldsymbol{v}} +
           \gamma^2 \ddot{\boldsymbol{v}}\big ]\,.\nonumber
\end{eqnarray}
Here $m(\boldsymbol{v})$ is the effective velocity dependent mass.
It is the sum of the bare mass and the mass (\ref{eb.i}) induced
by the field,
\begin{equation}\label{f.x}
       m(\boldsymbol{v})  =
       m_{\mathrm{b}}(\gamma
       \unit + \gamma^3|\boldsymbol{v} \rangle \langle
       \boldsymbol{v}|)
        + m_{\mathrm{f}}( \boldsymbol{v})  \,.
\end{equation}

As anticipated in Section \ref{sec.ca}, via a distinct route, the
leading contribution to (\ref{f.a}) is derived from the effective
Lagrangian
\begin{equation}\label{f.y}
        L_{\mathrm{eff}}(\boldsymbol{q}, \dot{\boldsymbol{q}}) =
        T( \dot{\boldsymbol{q}}) - e \big(\phi_{\mathrm{ex}}
        (\boldsymbol{q})
        - \dot{\boldsymbol{q}} \cdot
        A_{\mathrm{ex}}(\boldsymbol{q})\big)
      \,,
\end{equation}
equivalently from the Hamiltonian
\begin{equation}\label{f.z}
        E_{\mathrm{eff}}\big(\boldsymbol{p} - e A_{\mathrm{ex}}(\boldsymbol{q})\big)
        + e \phi_{\mathrm{ex}} ( \boldsymbol{q}) \,.
\end{equation}
For later purposes it is more convenient to work with the energy
function
\begin{equation}\label{f.b}
       H(\boldsymbol{q},\boldsymbol{v}) = E_s (\boldsymbol{v}) + e \phi_{\mathrm{ex}}
       (\boldsymbol{q})\,,
\end{equation}
which is conserved by the solutions to (\ref{f.a}) with
$\varepsilon = 0$, compare with (\ref{ca.n})\,.

The term of order $\varepsilon$ in (\ref{f.a})
 describes the radiation reaction. Globally its effect
can be deduced from the energy balance. We add to (\ref{f.b}) the
{\it Schott energy},
\begin{equation}\label{f.c}
       G_\varepsilon(\boldsymbol{q},\boldsymbol{v},\dot{\boldsymbol{v}}) =
       H(\boldsymbol{q},\boldsymbol{v}) - \varepsilon \,(e^2/ 6 \pi)
       \,\gamma^4 (\boldsymbol{v} \cdot \dot{\boldsymbol{ v}})\,.
\end{equation}
Then, along the solution trajectories of (\ref{f.a}),
\begin{equation}\label{f.d}
       \frac{d}{dt}\, G_\varepsilon\, (\boldsymbol{q},\boldsymbol{ v},
       \dot{\boldsymbol{v}}) = -
       \varepsilon \, (e^2 / 6 \pi)\, \big[\gamma^4 \dot{\boldsymbol{v}}^2
       + \gamma^6 (\boldsymbol{v} \cdot \dot{\boldsymbol{v}})^2\big]\,.
\end{equation}
Thus $G_\varepsilon$ is decreasing in time. Integrating both sides
of (\ref{f.d}) we have
\begin{eqnarray}\label{f.e}
       &&\!\!\!\!\!\!\!\!\! - G_\varepsilon \, (\boldsymbol{q}(t), \boldsymbol{v}(t),
       \dot {\boldsymbol{v}}(t)) + G_\varepsilon\,
       (\boldsymbol{q}(0), \boldsymbol{v}(0), \dot{\boldsymbol{v}}(0))\nonumber\\
       &=& \varepsilon \, (e^2/ 6 \pi)\, \int\limits_0^t ds\,
       \big[\gamma^4\, \dot{\boldsymbol{v}}(s)^2 + \gamma^6 \, (\boldsymbol{v}(s)
        \cdot \dot{\boldsymbol{v}}(s))^2
       \big]\,.
\end{eqnarray}
The mechanical energy is bounded from below, but the Schott energy
does not have a definite sign. {\it If} (!) the Schott energy remains
bounded in the course of time, then
\begin{equation}\label{f.f}
       \int\limits_0^\infty dt \, \big[\gamma^4 \, \dot{\boldsymbol{v}}(t)^2 +
       \gamma^6\, (\boldsymbol{v}(t) \cdot
       \dot{\boldsymbol{v}}(t))^2 \big] < \infty
\end{equation}
which implies
\begin{equation}\label{f.g}
       \lim_{t\to\infty}\, \dot{\boldsymbol{v}}(t)=0 \,.
\end{equation}

Equation (\ref{f.f}) corresponds to the finite energy dissipation
(\ref{ae.h}) in Section \ref{sec.ae} and we can repeat verbatim
the discussion there. In essence the limit (\ref{f.g}) allows only two scenarios.\\
(i) $\lim\limits_{t\to\infty}\,\boldsymbol{v}(t)= \boldsymbol{v}_\infty \not= 0
 \,.$ This corresponds
to a scattering situation where the particle escapes into a region
with $\boldsymbol{E}_{\mathrm{ex}} = 0 = \boldsymbol{B}_{\mathrm{ex}}$ and then travels with
velocity $\boldsymbol{v}_\infty$ along a straight line.\\
(ii) $\lim\limits_{t\to\infty}\,\boldsymbol{v}(t)=0$. This corresponds to a
bounded motion where the particle eventually comes to rest. At
such a rest point, $(\boldsymbol{q}^\ast, 0)$, we have
\begin{equation}\label{f.h}
       \nabla \phi_{\mathrm{ex}} \, (\boldsymbol{q}^\ast) = 0
\end{equation}
by (\ref{f.a}).

As noted already in Section \ref{sec.ae}, in  general, (\ref{f.g})
and (\ref{f.h}) carry too little information for determining the true
long--time behavior, as can be seen from the case of the motion in a uniform magnetic
field.

Unfortunately the energy balance does not tell the full story.
As noticed apparently first by Dirac (1938), Equation (\ref{f.a}) has solutions
which run away exponentially fast. There is no contradiction to (\ref{f.d}).
Since the Schott energy does not have a definite sign, in (\ref{f.e})
both $G_\varepsilon(t)$ and the time--integral diverge as $t
\to\infty$. The occurence of runaway solutions can be seen most easily in
the approximation of small velocities, setting $\boldsymbol{B}_{\mathrm{ex}} =
0$, and linearizing $\phi_{\mathrm{ex}}$ around a stable minimum,
say at $\boldsymbol{q}=0$. Then (\ref{f.a}) becomes
\begin{equation}\label{f.i}
       m\dot{\boldsymbol{v}} = - m\,\omega_0^2\, \boldsymbol{q} + \varepsilon
       \,k m\, \ddot{\boldsymbol{v}}
\end{equation}
with $km = e^2/6 \pi$. The three components of the linear equation
(\ref{f.i}) decouple and for each component there are three modes
of the form $e^{zt}$. The characteristic equation is
$z^2 = - \omega_0^2 + \varepsilon\, k z^3$ and to leading order
the eigenvalues are
$z_\pm = \pm\, i \omega_0 - \varepsilon \, (k \omega_0^2 /2), ~ z^3=
(1/ \varepsilon k) + \mathcal{O}(1)$. Thus in the 9--dimensional
phase space for (\ref{f.i}) there is a stable
6--dimensional hyperplane, $\mathcal{C}_\varepsilon$. On $\mathcal{C}_\varepsilon$
the motion is weakly damped, friction coefficient
$\varepsilon \, (k \omega_0^2/2)$, and relaxes as $t \to\infty$
to rest at $\boldsymbol{q}=0$. Transverse to $\mathcal{C}_\varepsilon$ the
solution runs away as $e^{(t/\varepsilon k)}$.

Clearly such runaway solutions violate the stability estimates
(\ref{eb.a}). Thus the full Maxwell--Lorentz equations do not have
runaways. They  somehow appear as an artifact of the Taylor
expansion in (\ref{e.e}). Dirac simply postulated that physical
solutions must satisfy the {\it asymptotic condition}
\begin{equation}\label{f.j}
       \lim_{t\to\infty}\, \dot{\boldsymbol{v}}(t)=0\,.
\end{equation}
In the linearized version (\ref{f.i}) this means that the initial
conditions have to lie on $\mathcal{C}_\varepsilon$. In Theorem
\ref{thm.ac} we proved the asymptotic condition to hold for the
Abraham model. Thus only those solutions to (\ref{f.a}) satisfying
the asymptotic condition can serve as a comparison dynamics to the
true solution. We then have to understand how the asymptotic
conditions arises, even better the global structure of the
solution flow to Equation (\ref{f.a}).

We note that in (\ref{f.a}) the highest derivative is multiplied by
a small prefactor. Such equations have been studied in great
detail
under the heading of (geometric) singular perturbation theory. The
main conclusion is that the structure found for the linear
equation (\ref{f.i}) persists for the nonlinear equation (\ref{f.a}).
Of course the hyperplane $\mathcal{C}_\varepsilon$ is now deformed
to some manifold, the critical (or center) manifold. We plan to
explain the standard example in the following section and then
to apply the theory to (\ref{f.a}).


\subsection{An example for singular perturbation
theory}\label{sec.fa}
As purely mathematical example we consider the coupled system
\begin{equation}\label{fa.a}
       \dot x = f(x,y)~ , \quad \varepsilon\, \dot y = y- h(x)\,.
\end{equation}
$h$ and $f$ are bounded, smooth functions. The phase space is
$\mathbb{R}^2$. One wants to understand how the solutions to
(\ref{fa.a}) behave for small $\varepsilon$. If we just set $\varepsilon =
0$, then $y= h(x)$ and we obtain the autonomous equation
\begin{equation}\label{fa.b}
      \dot  x = f(x,h(x))\,.
\end{equation}
Geometrically this means that the two--dimensional phase space has
been squeezed to the line $y= h(x)$ and the base point, $x(t)$, is
governed by (\ref{fa.b}). $\{ y = h(x), x \in \mathbb{R} \} = \mathcal{C}_0$
is the critical manifold to zero--th order in $\varepsilon$.

To see some motion in the phase space ambient to $\mathcal{C}_0$
we change from $t$ to the slow time scale $\tau = \varepsilon^{-1}
t$. Denoting differentiation with respect to $\tau$ by $^\prime$,
(\ref{fa.a}) goes over to
\begin{equation}\label{fa.c}
       x^\prime = \varepsilon\, f(x,y)~ , \quad y^\prime = y -
       h(x)\,.
\end{equation}
In the limit $\varepsilon \to 0$ we now have $x^\prime = 0$, i.e. $x(\tau)
= x_0$ and $y^\prime = y - h(x_0)$ with solution $y(t)= (y_0 - h(x_0))e^t + h(x_0)$. Thus
on that time scale, $\mathcal{C}_0$
consist exclusively of repelling fixed points. This is why
$\mathcal{C}_0$ is called critical. The linearization at $\mathcal{C}_0$
has the eigenvalue one transverse and the eigenvalue  zero
tangential to $\mathcal{C}_0$. In the theory of dynamical systems zero eigenvalues in the
linearization are linked to  center manifolds and $\mathcal{C}_0$
is also called the center manifold (at $\varepsilon = 0)$.

The basic result of singular perturbation theory is that for small
$\varepsilon$ the critical manifold deforms smoothly into
$\mathcal{C}_\varepsilon$. Thus $\mathcal{C}_\varepsilon$
is invariant under the solution flow to (\ref{fa.a}). Its
linearization at $(x,y) \in \mathcal{C}_\varepsilon$ has an
eigenvalue of  $\mathcal{O}(1)$ with eigenvector tangential to
$\mathcal{C}_\varepsilon$ and an eigenvalue $1/\varepsilon$
with eigenvector transverse to $\mathcal{C}_\varepsilon$. Thus for
an initial condition slightly off $\mathcal{C}_\varepsilon$ the
solution very rapidly diverges to infinity. Since $\mathcal{C}_0$
is deformed by order $\varepsilon$, also $\mathcal{C}_\varepsilon$
is of the form $\{ y=h_\varepsilon (x),~  x \in \mathbb{R} \}$.
According to (\ref{fa.a}) the base point evolves as
\begin{equation}\label{fa.d}
       \dot x =  f(x,h_\varepsilon(x))\,.
\end{equation}

Since $h_\varepsilon$ is smooth in $\varepsilon$ it can be Taylor
expanded as
\begin{equation}\label{fa.e}
       h_\varepsilon (x) =  \sum_{j = 0}^m\, \varepsilon^j\, h_j(x)
       + {\mathcal O}(\varepsilon^{m+1})\, .
\end{equation}
By (\ref{fa.a}) and (\ref{fa.d}) we have the identity
\begin{equation}\label{fa.f}
       \varepsilon \, \partial_x h_\varepsilon(x)\,   f(x,
       h_\varepsilon(x)) = h_\varepsilon (x) - h(x)\, .
\end{equation}
Substituting (\ref{fa.e}) and comparing powers of $\varepsilon$ we
can thus determine recursively $h_j(x)$. To lowest order we obtain
\begin{equation}\label{fa.g}
       h_0 (x) = h(x)~ , ~ h_1(x) = h^\prime(x)\,  f(x,
       h(x))
\end{equation}
and to order $\varepsilon$ the base point is governed by
\begin{equation}\label{fa.h}
       \dot x = f(x, h(x)) + \varepsilon \, \partial_y \, f(x, h(x))\,
       h^\prime (x) \, f(x, h(x))\, .
\end{equation}

Given the geometric picture of the center manifold, the stable (not
runaway) solutions to (\ref{fa.a}) can be determined up to a set
precision.


\subsection{The  critical manifold}\label{sec.fb}
Our task is to cast (\ref{f.a}) into the canonical form used in
singular perturbation theory. We set $(\boldsymbol{x}_1,
\boldsymbol{ x}_2)= \boldsymbol{x}= (\boldsymbol{q}, \boldsymbol{v})
 \in \mathbb{R}^3 \times \mathbb{V}, ~ \boldsymbol{y} =
 \dot{\boldsymbol{v}} \in \mathbb{R}^3$,
 \begin{equation}\label{fb.a}
         \boldsymbol{f}(\boldsymbol{x},\boldsymbol{y}) =
          (\boldsymbol{x}_2, \boldsymbol{y}) \in \mathbb{V} \times \mathbb{R}^3
\end{equation}
and
\begin{eqnarray}
       \boldsymbol{g}(\boldsymbol{x},\boldsymbol{y},\varepsilon) &=&
       \gamma^{-2}
       \kappa(\boldsymbol{x}_2)^{-1}\, \big ((6 \pi/ e^2)
       \,[m(\boldsymbol{x}_2)\boldsymbol{y} -
       \boldsymbol{F}_{\mathrm{ex}}(\boldsymbol{x}) ] \nonumber\\
        && -\varepsilon\, [3 \gamma^6 (\boldsymbol{x}_2 \cdot \boldsymbol{y})^2\,
      \boldsymbol{x}_2 + 3\gamma^4\,
       (\boldsymbol{x}_2 \cdot \boldsymbol{y}) \boldsymbol{y}] \big) \,,\label{fb.b}
\end{eqnarray}
where $\gamma = (1- \boldsymbol{x}_2^2)^{-1/2}$ as before, $\boldsymbol{F}_{\mathrm{ex}}
(\boldsymbol{x}) =
e (\boldsymbol{E}_{\mathrm{ex}} (\boldsymbol{x}_1) + \boldsymbol{x}_2 \times
\boldsymbol{B}_{\mathrm{ex}} (\boldsymbol{x}_1))$, and
$\kappa(\boldsymbol{v})$ is the $3 \times 3$ matrix $\kappa(\boldsymbol{v})=
  \unit + \gamma^2 \, |\boldsymbol{v}
 \rangle\langle \boldsymbol{v}|$
with inverse matrix  $\kappa(\boldsymbol{v})^{-1} = \unit- |\boldsymbol{v}
\rangle\langle  \boldsymbol{v}|$.
With this notation Equation (\ref{f.a}) reads
\begin{equation}\label{fb.c}
       \dot{\boldsymbol{x}} = \boldsymbol{f}(\boldsymbol{x},\boldsymbol{y}),
       ~\varepsilon \, \dot{\boldsymbol{y}} =
       \boldsymbol{g}(\boldsymbol{x},\boldsymbol{y},\varepsilon)\,.
\end{equation}

We set $\boldsymbol{h}(\boldsymbol{x}) = m(\boldsymbol{x}_2)^{-1}\,
\boldsymbol{ F}_{\mathrm{ex}}(\boldsymbol{x})$. Then for $\varepsilon = 0$
the critical manifold, $\mathcal{C}_0$, is given by
\begin{equation}\label{fb.d}
       \mathcal{C}_0 = \{ (\boldsymbol{x},\boldsymbol{h}(\boldsymbol{x}))
       , ~ \boldsymbol{x} \in \mathbb{R}^3 \times
       \mathbb{V} \} = \{ (\boldsymbol{q}, \boldsymbol{v}, \dot{\boldsymbol{v}})
       : m(\boldsymbol{v}) \dot{\boldsymbol{v}} =
       \boldsymbol{F}_{\mathrm{ex}} (\boldsymbol{q},\boldsymbol{v})\}\, ,
\end{equation}
which means that the critical manifold for $\varepsilon = 0$ is spanned by the
solutions of the leading Hamiltonian part of Equation (\ref{f.a}).
Linearizing at $\mathcal{C}_0$ the repelling eigenvalue is
dominated by $\gamma^{-2} \kappa(\boldsymbol{x}_2)^{-1}\, m(\boldsymbol{x}_2)$
which tends to zero as
$|\boldsymbol{x}_2 | \to 1$. Therefore $\mathcal{C}_0$ is not uniformly
hyperbolic, which is needed to use the results from Sakamoto (1990).

To overcome this difficulty we modify $\boldsymbol{g}$ to
$\boldsymbol{g}_\delta, ~ \delta$
small, which agrees with $\boldsymbol{g}$ on $\mathbb{R}^3 \times
 \{ \boldsymbol{v}, |\boldsymbol{v}| \le 1- \delta\} \times \mathbb{R}^3$
and which is constantly extended to values $|\boldsymbol{v}| \ge 1- \delta$.
Thus for $|\boldsymbol{x}_2(t) | \le 1- \delta$ the solution to
$\dot{\boldsymbol{x}} = \boldsymbol{f}, ~ \varepsilon\dot{\boldsymbol{y}}
 = \boldsymbol{g}_\delta$ agrees with the
solution to $\dot{\boldsymbol{x}} = \boldsymbol{f}, ~ \varepsilon
\dot{\boldsymbol{y}} = \boldsymbol{g}$. For sufficiently small
 $\varepsilon$ the modified equation has
then a critical manifold $\mathcal{C}_\varepsilon$ with the
properties as discussed in the example of Section \ref{sec.fa}. We only
have to make sure that the modification is never seen. Thus, for
the initial condition  $|\boldsymbol{v}(0)| \le \overline v$, we have to find a
$\delta = \delta (\overline v)$ such that $|\boldsymbol{v}(t)| \le 1- \delta$
for all times. To do so we need the energy balance (\ref{f.d}).

We consider the modified evolution with vector field
$(\boldsymbol{f}, \boldsymbol{g}_\delta)$
and we choose the initial velocity such that $|\boldsymbol{v}(0)| \le \overline
v < 1$. For $\varepsilon$ small enough this dynamics has a critical
 manifold of the form $\dot{\boldsymbol{v}} = \boldsymbol{h}_\varepsilon
 (\boldsymbol{q},\boldsymbol{v})$ and $|\boldsymbol{h}_\varepsilon
 (\boldsymbol{q},\boldsymbol{v})|
 \le c_1 = c_1 (\delta)$. We start the dynamics on
 $\mathcal{C}_\varepsilon$. According to (\ref{f.d}), for all $t \ge 0$,
\begin{eqnarray}
       G_\varepsilon\, (\boldsymbol{q}(t), \boldsymbol{v}(t),
       \boldsymbol{h}_\varepsilon(t)) &\le&
       G_\varepsilon(0) = H(\boldsymbol{q}(0), \boldsymbol{v}(0)) -
       \varepsilon(e^2 /6 \pi)(\boldsymbol{v}(0) \cdot
       \boldsymbol{h}_\varepsilon(0))\nonumber\\
       & \le &
       E_s (\overline v) + e \phi_{\mathrm{ex}}\, (\boldsymbol{q}(0)) +
       \varepsilon c_1 \, .\label{fb.e}
\end{eqnarray}

We choose now $\delta$ such that $\overline v \le 1-2 \delta$.
Since the initial conditions are on $\mathcal{C}_\varepsilon$, the
solution will stay for a while on $\mathcal{C}_\varepsilon$ until
the first time, $\tau$, when $|\boldsymbol{v}(\tau)| = 1- \delta$. After that
time the modification becomes visible. At time $\tau$ we have,
using the lower bound on the energy and (\ref{fb.e}),
\begin{eqnarray}
       E_s (\boldsymbol{v}(\tau)) + e \overline \phi\, &\le&
         H(\boldsymbol{q}(\tau), \boldsymbol{v}(\tau)) = G_\varepsilon(\tau) +
       \varepsilon(e^2 /6 \pi)\, \gamma^4\, (\boldsymbol{v}(\tau) \cdot
       \boldsymbol{h}_\varepsilon (\tau))\nonumber\\
        &\le&E_s (\overline v) + e \phi_{\mathrm{ex}}\, (\boldsymbol{q}(0)) +
       2 \varepsilon c_1\label{fb.f}
 \end{eqnarray}
 and therefore
 \begin{equation}\label{fb.g}
       E_s (1- \delta) \le E_s (1- 2 \delta) + e \,(\phi_{\mathrm{ex}}\,
       (\boldsymbol{q}(0))
       - \overline\phi) +
       2 \varepsilon c_1\,.
 \end{equation}
 $E_s (1- \delta)\cong 1/ \sqrt{\delta}$ for small $\delta$, which
 implies
\begin{equation}\label{fb.h}
       \frac{1}{\sqrt{\delta}} \le c_2 + 4 \, \varepsilon  c_1
 \end{equation}
 with $c_2= 2e \, (\phi(\boldsymbol{q}(0)) - \overline\phi)$. We choose now
 $\delta$ so small that $1/\sqrt{\delta} \ge c_2 + 1$ and then $\varepsilon$
 so small that $4 \varepsilon c_1 < 1$. Then (\ref{fb.h}) is a
 contradiction to the assumption that $|\boldsymbol{v}(\tau)| = 1- \delta$. We
 conclude that $\tau = \infty$ and the solution trajectory stays
on $\mathcal{C}_\varepsilon$ for all times.

Equipped with this information we have for small $\varepsilon$ the
critical manifold
\begin{equation}\label{fb.i}
        \dot{\boldsymbol{v}} =  \boldsymbol{h}_\varepsilon\,
         (\boldsymbol{q},\boldsymbol{v})\,.
 \end{equation}
 On the critical manifold the Schott energy is bounded and from
 the argument leading to (\ref{f.g}) we conclude that Dirac's
 asymptotic condition holds on $\mathcal{C}_\varepsilon$. On the
 other hand, slightly off $\mathcal{C}_\varepsilon$ the solution
 diverges  with a rate of order $1/ \varepsilon$. Therefore the
 asymptotic condition singles out, for given $\boldsymbol{q}
 (0),\boldsymbol{v}(0)$, the
 {\it unique} $\dot{\boldsymbol{v}}(0)$ on $\mathcal{C}_\varepsilon$.

The motion on the critical manifold is governed by an effective
equation which can be determined in approximation as in Section \ref{sec.fa}. We define
\begin{equation}\label{fb.j}
       \boldsymbol{h}(\boldsymbol{q},\boldsymbol{v}) = m(\boldsymbol{v})^{-1}\,
        e\big(\boldsymbol{E}_{\mathrm{ex}}(\boldsymbol{q}) + \boldsymbol{v}
         \times \boldsymbol{B}_{\mathrm{ex}}
       (\boldsymbol{q})\big)\,.
 \end{equation}
 Then, up to errors of order $\varepsilon^2$,
\begin{eqnarray}\label{fb.k}
       &&m(\boldsymbol{v}) \boldsymbol{\dot v} = e\,
       \big(\boldsymbol{E}_{\mathrm{ex}}(\boldsymbol{q}) +
       \boldsymbol{v} \times \boldsymbol{B}_{\mathrm{ex}}
       (\boldsymbol{q})\big) \\[1mm]
       && + \varepsilon \, (e^2 /6 \pi) \, \big[ \gamma^2\kappa(\boldsymbol{v})
        \big( \boldsymbol{v} \cdot
       \nabla_{\boldsymbol{q}} + \boldsymbol{h} \cdot \nabla_{\boldsymbol{v}}
         \boldsymbol{h} +
       (3 \gamma^6 (\boldsymbol{v} \cdot \boldsymbol{h})^2 \boldsymbol{v}
        + 3 \gamma^4\,
       ( \boldsymbol{v} \cdot \boldsymbol{h})\boldsymbol{h} )\big) \big]\,. \nonumber
\end{eqnarray}
The physical solutions of (\ref{f.a}), in the sense of the
asymptotic condition, are governed by Equation (\ref{fb.k}),
which thus should be
regarded as the true comparison dynamics to the microscopic
equation (\ref{de.j}). Note that the error made in going from
(\ref{f.a}) to (\ref{fb.k}) is of the same order as the error made in
the derivation of Equation (\ref{f.a}).

On a formal level (\ref{fb.k}) is easily deduced from (\ref{f.a}). We
regard $m(\boldsymbol{v}) \dot{\boldsymbol{v}} = e\,
(\boldsymbol{E}_{\mathrm{ex}}(\boldsymbol{q}) + \boldsymbol{v}
 \times \boldsymbol{B}_{\mathrm{ex}}(\boldsymbol{q}))$
as the ``unperturbed'' equation and substitute for the terms
inside the square bracket, which means to replace
$\dot{\boldsymbol{v}}$ by $\boldsymbol{h}$
and $\ddot{\boldsymbol{v}}$ by $\dot{\boldsymbol{h}} = (\boldsymbol{v}
 \cdot \nabla_{\boldsymbol{q}}) \boldsymbol{h} + (\boldsymbol{h}
  \cdot \nabla_{\boldsymbol{v}})
\boldsymbol{h}$. While yielding the correct answer, one misses the geometrical
picture of the motion in phase space and of the critical manifold.

For a numerical integration of the comparison dynamics it is
advantageous to use directly (\ref{fb.k}). The only other
practical option would be to solve (\ref{f.a}) backwards in time.
Then the trajectory is pushed rapidly towards the critical
manifold. On ${\mathcal C}_\varepsilon$ one solves however the
time--reversed dynamics which means a final rather than an initial
value problem. Instead of weakly damped the motion is now slowly
accelerating.


\subsection{Tracking of the true solution}\label{sec.fc}
>From (\ref{da.j}) we have the true solution $\boldsymbol{q}^\varepsilon
 (t), \boldsymbol{v}^\varepsilon(t)$
with initial conditions $\boldsymbol{q}^0, \boldsymbol{v}^0$ and correspondingly adapted
field data. We face the problem of how well this solution is
tracked by the comparison dynamics (\ref{f.a}). Let us first
disregard the radiation reaction. From our a priori estimates we
know that
\begin{equation}\label{fc.a}
        \dot{\boldsymbol{q}}^\varepsilon = \boldsymbol{v}^\varepsilon,~
         m(\boldsymbol{v}^\varepsilon)
        \dot{\boldsymbol{v}}^\varepsilon =  e\,\big(\boldsymbol{E}_{\mathrm{ex}}
        (\boldsymbol{q}^\varepsilon) + \boldsymbol{v}^\varepsilon
         \times \boldsymbol{B}_{\mathrm{ex}}(\boldsymbol{q}^\varepsilon)\big)+ \mathcal{O}
         (\varepsilon)
 \end{equation}
 which should be compared to
\begin{equation}\label{fc.b}
        \dot{\boldsymbol{r}}=\boldsymbol{u}~,~ m(\boldsymbol{u})
         \dot{\boldsymbol{u}} = e\, \big(\boldsymbol{E}_{\mathrm{ex}}
         (\boldsymbol{r}) + \boldsymbol{u}
         \times \boldsymbol{B}_{\mathrm{ex}}(\boldsymbol{u})\big)\,.
 \end{equation}
We switched to the variables $\boldsymbol{r}, \boldsymbol{u}$
 instead of $\boldsymbol{q},\boldsymbol{v}$ so to more
 clearly distinguish between the true and comparison dynamics.
 \begin{theorem}\label{thm.fa}
        For the Abraham model satisfying  the conditions $(C), ~ (P)$,
        and $(I)$ let $e \le \overline e$ and $\varepsilon \le \varepsilon_0$
        be sufficiently small. Let $\boldsymbol{r}(t), \boldsymbol{u}(t)$ be the
         solution to the
        comparison dynamics (\ref{fc.b}) with initial conditions
        $\boldsymbol{r}(0) = \boldsymbol{q}^0,
        \boldsymbol{u}(0) = \boldsymbol{v}^0.$ Then for every $\tau > 0$
         there exist constants $c(\tau)$
        such that
\begin{equation}\label{fc.c}
        |\boldsymbol{q}^\varepsilon(t) - \boldsymbol{r}(t)| \le c(\tau) \varepsilon~, ~
        |\boldsymbol{v}^\varepsilon(t) - \boldsymbol{u}(t)| \le  c(\tau) \varepsilon
\end{equation}
for $0 \le t \le \tau$.
 \end{theorem}
{\it Proof}:
One converts (\ref{fc.a}), (\ref{fc.b}) into a first order equation
in its integral form. The difference is then estimated by using
Gronwall's lemma, which yields an error as $\varepsilon e^{Ct}.$
$\Box$\medskip

Theorem  \ref{thm.fa} states that, up to an error of order $\varepsilon$, the
 true solution is well approximated by the Hamiltonian dynamics
 (\ref{fc.b}).  In the next order the comparison dynamics is
\begin{eqnarray}
        &&\dot{\boldsymbol{r}}^\varepsilon =\boldsymbol{u}^\varepsilon,~
         m(\boldsymbol{u}^\varepsilon)  \dot{\boldsymbol{u}}^\varepsilon
         = e\, \big(\boldsymbol{E}_{\mathrm{ex}}(\boldsymbol{r}^\varepsilon)
          + \boldsymbol{u}^\varepsilon
         \times
         \boldsymbol{B}_{\mathrm{ex}}(\boldsymbol{r}^\varepsilon)\big)\label{fc.d}\\[1mm]
         && + \varepsilon(e^2/ 6 \pi)\big[\gamma^4\, (\boldsymbol{u}^\varepsilon
         \cdot \ddot{\boldsymbol{u}}^\varepsilon) \boldsymbol{u}^\varepsilon
         + 3 \gamma^6 (\boldsymbol{u}^\varepsilon \cdot
         \dot{\boldsymbol{u}}^\varepsilon)^2\, \boldsymbol{u}^\varepsilon
         + 3 \gamma^4 (\boldsymbol{u}^\varepsilon \cdot \dot{\boldsymbol{u}}^\varepsilon)
         \dot{\boldsymbol{u}}^\varepsilon + \gamma^2
          \ddot{\boldsymbol{u}}^\varepsilon\big]\nonumber
 \end{eqnarray}
restricted to  its critical manifold $\mathcal{C}_\varepsilon$ and one might
 expect that
\begin{equation}\label{fc.e}
         |\boldsymbol{q}^\varepsilon (t) - \boldsymbol{r}^\varepsilon(t)|
         + |\boldsymbol{v}^\varepsilon(t) -
         \boldsymbol{u}^\varepsilon (t)|= \mathcal{O} (\varepsilon^2)\,.
 \end{equation}
 Because of the improved precision one has the possibility to resolve  the
 radiation reaction correction to (\ref{fc.d}).

An alternative option to keep track of the correction would be to
consider longer times, of the order $\varepsilon^{-1}\, t$ on the macroscopic
time scale. Then the
radiative effects add up to deviations of order one from the
Hamiltonian trajectory. Thus
\begin{equation}\label{fc.f}
        |\boldsymbol{q}^\varepsilon (t) - \boldsymbol{r}^\varepsilon(t)|\cong
\mathcal{O} (\varepsilon)\quad \mbox{for}\quad 0 \le t \le \varepsilon^{-1}\,
 \tau\, .
 \end{equation}

One should be somewhat careful here. In a scattering situation the
charged particle reaches after a finite  macroscopic time the
force free region. According to (\ref{fc.e}) the error
in the velocity is then $\mathcal{O}(\varepsilon^2)$ which builds
up to an error in the position of order $\varepsilon$ over a time
span $\varepsilon^{-1}\, \tau$. Thus we cannot hope to do better
than (\ref{fc.f}). On the other hand when the motion remains
bounded, as e.g. in a uniform external magnetic field,
the charge comes to rest at some point $\boldsymbol{q}^\ast$ in
the long time limit and the rest point $\boldsymbol{q}^\ast$  is
the same for the true and the comparison dynamics. At least for an
external electrostatic potential with a discrete set of critical
points we have already established such a behavior and presumably
it holds in general. Thus for large $\tau$ we have
$\boldsymbol{q}^\varepsilon (\varepsilon^{-1} \tau) \cong \boldsymbol{q}^\ast$
and also  $\boldsymbol{r}^\varepsilon(\varepsilon^{-1} \tau) \cong
\boldsymbol{q}^\ast$. Therefore we conjecture that (\ref{fc.f})
holds for {\it all} times.
\begin{conjecture}\label{thm.fb}
For the Abraham model satisfying $(C), (P),$
and $(I)$ let $\boldsymbol{q}^\varepsilon(t)$ be bounded, i.e.
$|\boldsymbol{q}^\varepsilon(t)|
\le C$\, for all \,$t \ge 0, ~  \varepsilon\le  \varepsilon_0$ \,. Then
there exists $(\boldsymbol{r}^\varepsilon(0)\,, \boldsymbol{u}^\varepsilon(0),\\
\dot{\boldsymbol{u}}^\varepsilon(0))
 \in \mathcal{C}_\varepsilon$ such that
\begin{equation}\label{fc.g}
        \sup_{t \ge 0}\,|\boldsymbol{q}^\varepsilon (t) - \boldsymbol{r}^\varepsilon(t)| =
\mathcal{O} (\varepsilon)\,,
 \end{equation}
 where $\boldsymbol{r}^\varepsilon(t)$ is the solution to (\ref{fc.d}) with said
 initial conditions.
\end{conjecture}

At present we are far from such strong results. The problem is
that an error of order $\varepsilon^2$ in (\ref{fc.d}) is
generically amplified as $\varepsilon^2 e^{t/\varepsilon}$.
Although such an increase violates the a priori bounds, it renders
a proof of (\ref{fc.g}) difficult. We seem to be back to (\ref{fc.c})
which carries no information on the radiation reaction.
Fortunately the radiation correction in (\ref{fc.d}) can be seen in
the energy balance.
\begin{theorem}
 Under the assumptions of Theorem \ref{thm.fa} we have
\begin{equation}\label{fd.h}
        \big |[E_s(\boldsymbol{v}^\varepsilon (t))+ e\, \phi_{\mathrm{ex}}
        (\boldsymbol{q}^\varepsilon (t))]
        - [E_s(\boldsymbol{u}^\varepsilon(t)) + e
        \phi_{\mathrm{ex}}(\boldsymbol{r}^\varepsilon(t))] \big| \le C c (\tau)
        \varepsilon^2
 \end{equation}
for $0 \le t \le \tau$, where $\boldsymbol{r}^\varepsilon(t),
\boldsymbol{u}^\varepsilon(t)$
is the solution to (\ref{fc.d}) with initial data $\boldsymbol{r}^\varepsilon(0)=
\boldsymbol{q}^0,$ $ \boldsymbol{u}^\varepsilon(0) = \boldsymbol{v}^0,$ $
\dot{\boldsymbol{u}}^\varepsilon(0) =
\boldsymbol{h}_\varepsilon(\boldsymbol{q}^0, \boldsymbol{v}^0)$.
\end{theorem}
{\it Proof}: We use the estimate (\ref{eb.h})
on the self--force,
where $|\boldsymbol{f}^\varepsilon(t)| \le C \varepsilon^2$ for $\varepsilon
\overline t_\rho \le t$
and $|\boldsymbol{f}^\varepsilon(t)| \le C \varepsilon$ for $0 \le t\le \varepsilon
\overline t_\rho$. Then, as in (\ref{f.d}),
\begin{equation}\label{fd.i}
       \frac{d}{dt}\, G_\varepsilon\,
       (\boldsymbol{q}^\varepsilon,\boldsymbol{v}^\varepsilon,
        \dot{\boldsymbol{v}}^\varepsilon) =\boldsymbol{f}^\varepsilon(t)
       \cdot \boldsymbol{v}^\varepsilon
        - \varepsilon\, (e^2/6 \pi) [\gamma^4(\dot{\boldsymbol{
       v}}^\varepsilon)^2 + \gamma^6 (\boldsymbol{v}^\varepsilon \cdot
        \dot{\boldsymbol{v}}^\varepsilon)^2]
\end{equation}
and therefore
\begin{eqnarray}
      && |H(\boldsymbol{q}^\varepsilon,\boldsymbol{v}^\varepsilon) -
       H(\boldsymbol{r}^\varepsilon,\boldsymbol{u}^\varepsilon)|\nonumber\\[2mm]
      &&\le \varepsilon\, (e^2/6 \pi) |\gamma(
       \boldsymbol{v}^\varepsilon)^4  (\boldsymbol{v}^\varepsilon \cdot
        \dot{\boldsymbol{v}}^\varepsilon)- \gamma
        (\boldsymbol{u}^\varepsilon)^4 (\boldsymbol{u}^\varepsilon
        \cdot \dot{\boldsymbol{u}}^\varepsilon)| \nonumber\\
       && + \int\limits_0^t ds\, \big(
       |\boldsymbol{f}^\varepsilon \cdot  \boldsymbol{v}^\varepsilon| + \varepsilon
       \, (e^2/6 \pi) | \gamma(\boldsymbol{v}^\varepsilon)^4\, (\dot{\boldsymbol{
       v}}^\varepsilon)^2 +\gamma(\boldsymbol{v}^\varepsilon)^6 \,
         (\boldsymbol{v}^\varepsilon
       \cdot \dot{\boldsymbol{v}}^\varepsilon )^2\nonumber\\
       && \qquad\quad\quad- \gamma(\boldsymbol{u}^\varepsilon)^4\,
        (\dot{\boldsymbol{u}}^\varepsilon)^2 -
       \gamma(\boldsymbol{u}^\varepsilon)^6\, (\boldsymbol{u}^\varepsilon \cdot
        \dot{\boldsymbol{u}}^\varepsilon)^2|\big)\,.\label{fd.j}
\end{eqnarray}
Since $|\boldsymbol{v}^\varepsilon |, |\boldsymbol{u}^\varepsilon|$ remain bounded away from
1, we can use the bound
$|\boldsymbol{v}^\varepsilon (t) - \boldsymbol{u}(t)| \le c (\tau)
\varepsilon$ from
Theorem \ref{thm.fa}.  Reinserting (\ref{fc.c})
 into (\ref{fc.a}) and (\ref{fc.b}) we obtain
 $|\dot{\boldsymbol{v}}^\varepsilon (t) - \dot{\boldsymbol{u}}^\varepsilon (t)|
 \le c (\tau)\varepsilon.$
 Furthermore $\int\limits_0^t ds\,
|\boldsymbol{f}^\varepsilon (s)|
\le C t \varepsilon^2   $. We conclude that
\begin{equation}\label{fd.k}
     |H(\boldsymbol{q}^\varepsilon(t),\boldsymbol{v}^\varepsilon(t)) -
     H(\boldsymbol{r}^\varepsilon
      (t),\boldsymbol{u}^\varepsilon(t))| \le  C (t+ c (t))\varepsilon^2\,.
\end{equation}
$\Box$\medskip

\subsection{Electromagnetic fields in the adiabatic
limit}\label{sec.fd}
So far we have concentrated on the Lorentz equation with retarded
fields and have obtained approximate evolution equations for the
charged particle. Such an approximate solution can be reinserted
into the inhomogeneous Maxwell equations in order to obtain the
electromagnetic fields in the adiabatic limit.

As before, let $\boldsymbol{q}^\varepsilon(t),\boldsymbol{v}^\varepsilon(t)),
 t \ge 0 $, be the true
solution. We extend it to $\boldsymbol{q}^\varepsilon(t)=\boldsymbol{q}^0 +
 \boldsymbol{v}^0 t,~ \boldsymbol{v}^\varepsilon(t)
= \boldsymbol{v}^0$ for  $t \le 0$. According to (\ref{ca.u}),
(\ref{ca.v}) and using the scaled
fields as in (\ref{da.g}), we have
\begin{equation}\label{fd.a}
     \frac{1}{\sqrt{\varepsilon}}\, \boldsymbol{E}(t) = - \int\limits_{-\infty}^t ds\,
     \int d^3 y\, \big(\nabla
     G_{t-s} \ast \rho_\varepsilon(s) +  \partial_t  G_{t-s}\ast
      \boldsymbol{j}_\varepsilon(s)\big)
\end{equation}
with $\rho_\varepsilon(\boldsymbol{x},t) = \rho_\varepsilon
(\boldsymbol{x}- \boldsymbol{q}^\varepsilon(t)),
\boldsymbol{j}_\varepsilon(\boldsymbol{x},t)
= \rho_\varepsilon(\boldsymbol{x}-\boldsymbol{q}^\varepsilon(t))
\boldsymbol{v}^\varepsilon(t)$.
Inserting from (\ref{aa.o}) and by partial integration
\begin{eqnarray*}
   &&\frac{1}{\sqrt{\varepsilon}}\, \boldsymbol{E}(\boldsymbol{x},t)
    =  - \int\limits_{-\infty}^t ds\,\int d^3 y\,
     \frac{1}{4 \pi(t-s)}\, \delta(|\boldsymbol{x}-\boldsymbol{y}| - (t-s))\,\nabla
     \rho_\varepsilon(\boldsymbol{y},s)\nonumber\\
      && -  \int\limits_{-\infty}^t ds\,\int d^3 y \, \frac{1}
      {4 \pi(t-s)^2}\, \delta(|\boldsymbol{x}-\boldsymbol{y}| - (t-s))\,
      [(\boldsymbol{y}-\boldsymbol{x}) \cdot
      \nabla \boldsymbol{j}_\varepsilon(\boldsymbol{y},s)
        + \boldsymbol{j}_\varepsilon
        (\boldsymbol{y},s)]
\end{eqnarray*}
\begin{eqnarray}\label{fd.b}
      &=& -  \int d^3 y \, \Big( \frac{1}{4 \pi |\boldsymbol{x}-\boldsymbol{y}|}\, \nabla
      \rho_\varepsilon(\boldsymbol{y}-\boldsymbol{q}^\varepsilon
      (t-|\boldsymbol{x}-\boldsymbol{y}|))
      \boldsymbol{v}^\varepsilon(t-|\boldsymbol{x}-\boldsymbol{y}|)\\
      && + \frac{1}{4 \pi|\boldsymbol{x}-\boldsymbol{y}|^2}\,
      \boldsymbol{v}^\varepsilon(t-|\boldsymbol{x}-\boldsymbol{y}|) (1+
      (\boldsymbol{y}
      - \boldsymbol{x}) \cdot \nabla)\,
      \rho_\varepsilon(\boldsymbol{y}-\boldsymbol{q}^\varepsilon(t-|
      \boldsymbol{x}-\boldsymbol{y}|)) \Big)\,.\nonumber
\end{eqnarray}
In the same fashion
\begin{equation}\label{fd.c}
     \frac{1}{\sqrt{\varepsilon}}\, \boldsymbol{B}(\boldsymbol{x},t) = -
     \int d^3 y \, \frac{1}{4 \pi|\boldsymbol{x}-\boldsymbol{y}|}\,
      \boldsymbol{v}^\varepsilon(t-|\boldsymbol{x}-\boldsymbol{y}|)
     \times \nabla \rho_\varepsilon(\boldsymbol{y}-
     \boldsymbol{q}^\varepsilon(t-|\boldsymbol{x}-\boldsymbol{y}|))\,.
\end{equation}
In the limit $\varepsilon \to 0$ we have $\rho_\varepsilon(\boldsymbol{x}) \to
\delta(\boldsymbol{x})$
and, by Theorem \ref{thm.fa}, $\boldsymbol{q}^\varepsilon(t) \to
\boldsymbol{r}(t),$ $
\boldsymbol{v}^\varepsilon (t) \to
\boldsymbol{u}(t)$, where $\boldsymbol{r}(t) = \boldsymbol{q}^0
+ \boldsymbol{v}^0 t,~ \boldsymbol{u}(t) =\boldsymbol{v}^0$ for $t \le 0$. We
substitute $\boldsymbol{y}^\prime = \boldsymbol{y} -
\boldsymbol{q}^\varepsilon(t-|\boldsymbol{x}-\boldsymbol{y}|)$ with volume
element $\det ({\mathrm{d}} \boldsymbol{y}/{\mathrm{d}} \boldsymbol{y}^\prime) =
 [1-\boldsymbol{v}^\varepsilon (t-|\boldsymbol{x}
-\boldsymbol{y}|) \cdot (\boldsymbol{x}-\boldsymbol{y})
/ |\boldsymbol{x}-\boldsymbol{y}|]^{-1}$. Then $\delta
(\boldsymbol{y}^\prime)$ leads to the constraint
$0=\boldsymbol{y}- \boldsymbol{r}(t-|\boldsymbol{x}-\boldsymbol{y}|)$
which has the unique solution $\boldsymbol{y }=
\boldsymbol{r}(t_{\mathrm{ret}})$,
 compare with (\ref{aa.s}). In particular the volume element
  $\det({\mathrm{d}}\boldsymbol{y}/{\mathrm{d}}\boldsymbol{y}^\prime)$
 becomes in the limit
 $[1- \boldsymbol{\widehat n} \cdot \boldsymbol{u}(t_{\mathrm{ret}})]^{-1}~ \mbox{with}~
  \boldsymbol{\widehat n} = \boldsymbol{\widehat n}(\boldsymbol{x},t) =
  (\boldsymbol{x}-\boldsymbol{r}(t_{\mathrm{ret}})) /
  |\boldsymbol{x}-\boldsymbol{r}(t_{\mathrm{ret}})|.$

We conclude that
\begin{eqnarray}\label{fd.e}
   \lim_{\varepsilon \to 0}\, \frac{1}{\sqrt{\varepsilon}}\,
   \boldsymbol{E}(\boldsymbol{x},t)
   &=&\boldsymbol{\overline E}(\boldsymbol{x},t) \,,\\
   \lim_{\varepsilon \to 0}\, \frac{1}{\sqrt{\varepsilon}}\,
   \boldsymbol{B}(\boldsymbol{x},t)
   &=&\boldsymbol{\overline B}(\boldsymbol{x},t) \,,\label{fd.f}
\end{eqnarray}
where $\boldsymbol{\overline E}\,, \boldsymbol{\overline B}$ are the Li\'{e}nard--Wiechert
fields (\ref{aa.u}), (\ref{aa.v}) generated by a point charge moving
along the trajectory $t \mapsto \boldsymbol{r}(t)$. The convergence in
(\ref{fd.e}), (\ref{fd.f}) is pointwise if one excludes the Coulomb
singularity at $\boldsymbol{x} = \boldsymbol{r}(t)$.


\subsection{Larmor's formula}\label{sec.fe}
We want to determine the energy per unit time radiated to infinity
and consider, for this purpose, a ball of radius $R$ centered at $\boldsymbol{q}^\varepsilon
(t)$. At time $t+ R$ the energy in this ball is
\begin{equation}\label{fe.a}
     \mathcal{E}_{R, \boldsymbol{q}^\varepsilon(t)}\, (t+R)=
     \mathcal{E}(0) - \frac{1}{2}\, \int\limits_{
     \{|\boldsymbol{x}-\boldsymbol{q}^\varepsilon
     (t)|\ge R\}} d^3x\, \big (\boldsymbol{E}(\boldsymbol{x},t + R)^2 +
      \boldsymbol{B}(\boldsymbol{x},t+ R)^2\big)
\end{equation}
using conservation of total energy. The radiation emitted from
the charge at time $t$ reaches the surface of the ball at time
$t+R$ and the energy loss per unit time is given by
\begin{eqnarray}\label{fe.b}s
       I_{R,\varepsilon}(t) &=& \frac{d}{dt}\,
       \mathcal{E}_{R, \boldsymbol{q}^\varepsilon(t)}\nonumber\\
       &=& \int d^3 x \, \delta(|\boldsymbol{x}-
       \boldsymbol{q}^\varepsilon(t)|-R)\, \Big (
       \frac{1}{2}\,\big(\boldsymbol{E}(\boldsymbol{x},t+ R)^2 + \boldsymbol{B}
       (\boldsymbol{x},t+R)^2\big)\,\nonumber\\[1mm]
       && (\boldsymbol{n}(\boldsymbol{x}) \cdot
       \boldsymbol{v}^\varepsilon(t))
        + \boldsymbol{E}(\boldsymbol{x},t+R) \cdot [\boldsymbol{n}(\boldsymbol{x})
       \times \boldsymbol{B}(\boldsymbol{x},t+R)]
       \Big)\nonumber\\
       &=& \frac{1}{2}\,R^2 \int d^2 \omega\, \Big( \big
       (\boldsymbol{E}(\boldsymbol{q}^\varepsilon(t) + R \boldsymbol{\omega}, t + R)^2 +
       \boldsymbol{B}(\boldsymbol{q}^\varepsilon(t) + R \boldsymbol{\omega}, t+ R)^2\big )
      \nonumber\\[2mm]
      &&(\boldsymbol{\omega} \cdot \boldsymbol{v}^\varepsilon(t))
      + 2
       \boldsymbol{E}(\boldsymbol{q}^\varepsilon(t) + R\boldsymbol{\omega}, t + R)\cdot
       [\boldsymbol{\omega}
       \times \boldsymbol{B}(\boldsymbol{q}^\varepsilon(t) + R \boldsymbol{\omega}, t+ R)]
       \Big)\,,\nonumber\\
\end{eqnarray}
where $\boldsymbol{n}(\boldsymbol{x})$ is the outer normal of the ball
and $|\boldsymbol{\omega}|=1$ with $d^2 \omega$
the integration over the unit sphere. (\ref{fe.b}) holds for
sufficiently large $R$, since we used that $\{ \boldsymbol{x}|~
|\boldsymbol{x}-\boldsymbol{q}^\varepsilon(t)| \ge R\}
\cap \{ \boldsymbol{x} |~ |\boldsymbol{x}-\boldsymbol{q}^\varepsilon(t+R)|
 \le \varepsilon R_\rho\} =
\emptyset$, which is the case for $(1 - \overline v)R \ge \varepsilon
R_\rho$.

(\ref{fe.b}) still contains the reversible energy transport between the
considered ball and its complement. To isolate that part of the
energy which is irreversibly  lost we have to take the limit $R\to
\infty$. For this purpose we first partially integrate in
 (\ref{fd.b}), (\ref{fd.c}) by using the identity
\begin{equation}\label{fe.c}
       \nabla\rho = \nabla_{\boldsymbol{y}}\,\rho - \frac{\boldsymbol{y}-
       \boldsymbol{x}}{|\boldsymbol{y}-\boldsymbol{x}|}\, \big(1 +
       \frac{(\boldsymbol{y}-\boldsymbol{x}) \cdot \boldsymbol{v}^\varepsilon}
       {|\boldsymbol{y}-\boldsymbol{x}|}\big)^{-1}\,
       (\boldsymbol{v}^\varepsilon \cdot \nabla_{\boldsymbol{y}}) \rho
\end{equation}
at the argument $\boldsymbol{y}-\boldsymbol{q}^\varepsilon
(t- |\boldsymbol{y}-\boldsymbol{x}|)$. For large $R$ the
fields in (\ref{fe.b}) become then
\begin{eqnarray}
      && R \boldsymbol{E}(\boldsymbol{q}^\varepsilon(t)+R \boldsymbol{\omega}, t +  R) \cong
       \sqrt{\varepsilon} \int d^3y\, \frac{1}{4 \pi}\, \rho_\varepsilon
       (\boldsymbol{y}-\boldsymbol{q}^\varepsilon)\nonumber\\[2mm]
      &&\big [- (1-\boldsymbol{\omega}\cdot \boldsymbol{v}^\varepsilon)^{-1}\,
       \dot{\boldsymbol{v}}^\varepsilon - (1- \boldsymbol{\omega}\cdot \boldsymbol{v}^\varepsilon)^{-2}\,
       (\boldsymbol{\omega} \cdot \dot{\boldsymbol{v}}^\varepsilon)(\boldsymbol{v}^\varepsilon
       -\boldsymbol{\omega})\big]|_{t+ \boldsymbol{\omega}\cdot(\boldsymbol{y}-
       \boldsymbol{q}^\varepsilon(t))}\,,\label{fe.i}\\[2mm]
       &&R \boldsymbol{B} (\boldsymbol{q}^\varepsilon(t)+R \boldsymbol{\omega}, t +  R)
       \cong
       \sqrt{\varepsilon} \int d^3y\, \frac{1}{4 \pi}\, \rho_\varepsilon
       (\boldsymbol{y}-\boldsymbol{q}^\varepsilon)\nonumber\\[2mm]
       && \big[- (1-\boldsymbol{\omega} \cdot \boldsymbol{v}^\varepsilon)^{-1}\,
       (\boldsymbol{\omega} \times \dot{\boldsymbol{v}}^\varepsilon) -
       (1- \boldsymbol{\omega} \cdot \boldsymbol{v}^\varepsilon)^{-2}\,
       (\boldsymbol{\omega} \cdot \dot{\boldsymbol{v}}^\varepsilon)
       (\boldsymbol{\omega}\times  \boldsymbol{v}^\varepsilon)\big]|_{t+\boldsymbol{\omega}
       \cdot(\boldsymbol{y}-
       \boldsymbol{q}^\varepsilon(t))} \nonumber\\[2mm]
       &&= \boldsymbol{\omega}\times R \boldsymbol{E}
       (\boldsymbol{q}^\varepsilon(t) + R \boldsymbol{\omega}, t+R)\,,\label{fe.d}
\end{eqnarray}
where we used that  $t+ R- |\boldsymbol{q}^\varepsilon(t) + R \boldsymbol{\omega}
 -\boldsymbol{y}|
= t+\boldsymbol{\omega}\cdot ( \boldsymbol{y}-\boldsymbol{q}^\varepsilon(t))
 + \mathcal{O}(1/R)$ for large $R$. Inserting
in (\ref{fe.b}) yields
\begin{eqnarray}
        &&\!\!\lim_{R\to \infty} I_{R, \varepsilon}(t)=
        I_\varepsilon(t)\nonumber\\
      &&= - \varepsilon \int d^2 \boldsymbol{\omega}\,
       (1- \boldsymbol{\omega} \cdot \boldsymbol{v}^\varepsilon(t))
      \big(R\boldsymbol{E} (\boldsymbol{q}^\varepsilon(t)+ R \boldsymbol{\omega}, t+ R)\big)^2\label{fe.e}\\
     && = - \varepsilon  \int d^2 \boldsymbol{\omega}\, ( 1- \boldsymbol{\omega} \cdot
       \boldsymbol{v}^\varepsilon(t)) \Big( \big[ \, \frac{1}{4 \pi}\,\int d^3y\,
       \rho_\varepsilon (\boldsymbol{y} -\boldsymbol{q}^\varepsilon)(1 -\boldsymbol{\omega}\cdot
       \boldsymbol{v}^\varepsilon)^{-2}( \boldsymbol{\omega}\cdot
       \dot{\boldsymbol{v}}^\varepsilon)\big]^2\nonumber\\
       && - \big [ \frac{1}{4 \pi}\,
       \int d^3y \rho_\varepsilon (\boldsymbol{y}-\boldsymbol{q}^\varepsilon)
       (1- \boldsymbol{\omega} \cdot \boldsymbol{v}^\varepsilon)^{-1}
        \dot{\boldsymbol{v}}_\varepsilon \nonumber\\
        &&\qquad\qquad+ (1 - \boldsymbol{\omega} \cdot
        \boldsymbol{v}^\varepsilon)^{-2}
        (\boldsymbol{\omega} \cdot \dot{\boldsymbol{v}}^\varepsilon)
        \boldsymbol{v}^\varepsilon\big
        ]^2 \Big)|_{t + \boldsymbol{\omega} \cdot (\boldsymbol{y}
        -\boldsymbol{q}^\varepsilon (t))}.\label{fe.f}
\end{eqnarray}
$I_\varepsilon(t)$ is the energy radiated per unit time at $\varepsilon$
fixed. As argued before it is of order $\varepsilon$. The
expression (\ref{fe.e}) shows that $I_\varepsilon(t) \le 0$.

(\ref{fe.f}) is not yet Larmor's formula. For this we have to go to
the adiabatic limit $\varepsilon\to 0$. Then $\boldsymbol{q}^\varepsilon(t) \to
\boldsymbol{r}(t)$. Since $\rho_\varepsilon(\boldsymbol{x}) \to e \delta(\boldsymbol{x})$
 we have $\boldsymbol{y} \cong \boldsymbol{q}^\varepsilon(t) \cong
\boldsymbol{r}(t)$ in (\ref{fe.f}). From the $d^3y$ volume element we
get an
additional factor of $(1-\boldsymbol{\omega} \cdot \boldsymbol{v}^\varepsilon)^{-1}$. Thus
\begin{eqnarray}
        \lim_{\varepsilon\to 0} I_\varepsilon(t) =I(t)&=&
        - e^2 \int d^2 \boldsymbol{\omega}\, (1- \boldsymbol{\omega} \cdot \boldsymbol{u}(t))
        \big(4 \pi( 1- \boldsymbol{\omega}\cdot \boldsymbol{u}(t))^{-3}\big)^2
         \nonumber\\[2mm]
      &&\big((\boldsymbol{\omega}\cdot \dot{\boldsymbol{u}}(t))^2 - [(1 -\boldsymbol{\omega} \cdot
      \boldsymbol{u}(t)) \dot{\boldsymbol{u}}(t) + (\boldsymbol{\omega}\cdot
      \dot{\boldsymbol{u}}(t))\boldsymbol{u}(t)]^2\big)\nonumber\\[2mm]
      &=& - (e^2/6 \pi)\big[\gamma^4 \dot{\boldsymbol{u}}(t)^2 +
      \gamma^6(\boldsymbol{u}(t) \cdot
      \dot{\boldsymbol{u}}(t))^2\big]\nonumber\\[2mm]
      &=& - (e^2/ 6 \pi) \gamma^4 \big[\dot{\boldsymbol{u}}(t)^2 - (\boldsymbol{u}(t) \times
      \dot{\boldsymbol{u}}(t))^2 \big]\,,\label{fe.g}
\end{eqnarray}
which is the standard textbook formula of Larmor. Note that the
same energy loss per unit time was obtained already in (\ref{f.d})
using only the energy balance for the comparison dynamics.

Starting from (\ref{fe.b}) we could  alternatively first take the
limit $\varepsilon^{-1} I_{R, \varepsilon}
(t) \to I_{R,0}(t)$, which is the change of energy in a ball of
radius $R$ centered at the particle's position $\boldsymbol{r}(t)$ in the
adiabatic limit. As before we have to isolate the irreversible
energy loss through
\begin{equation}\label{fe.h}
       \lim_{R \to\infty} I_{R,0}(t) = I(t)\,.
\end{equation}
The energy loss does not depend on the order of limits, as it
should be.
\subsection*{Notes and References}
{\it ad  \ref{sec.f}:} The radiation damped harmonic oscillator is discussed in
Jackson (1999) with a variety of physical applications. The asymptotic condition
is first stated in Dirac (1938). It has been reemphasized by Haag (1955)
in analogy to a similar condition in quantum field theory. \bigskip\\
{\it ad \ref{sec.fa}:} Singular, or geometric, perturbation theory
is a standard tool in the theory of dynamical systems. We refer to
Jones (1995) for a review with many applications. In the context
of synergetics, Haken (1983), one talks of slow and fast variables
and the slaving principle, which means that fast variables are
slaved by the slow ones. Within our context this would correspond to an
attractive critical manifold. The renormalization group flows in
critical phenomena have a  structure similar to the one discovered here.
 The critical surface
corresponds to critical couplings which flow then to some fixed
point governing the universal critical behavior. The critical
surface is repelling and slightly off that surface the trajectory
moves towards either the high temperature or low temperature fixed
points.\bigskip\\
{\it ad \ref{sec.fb}:} Particular cases have been studied before,
most extensively the one--dimensional potential of finite width and with
linear interpolation, Haag (1955), Carati,  Galgani (1993),
Carati et al (1995), Blanco (1995),  in addition head on
collision in the two--body problem, Huschilt and Baylis (1976), and
motion in a uniform magnetic field, Endres (1993).  These
authors emphasize that there can be several solutions to the
asymptotic condition. From the point of view of singular
perturbation theory such a behavior is generic. If $\varepsilon$
is increased, then the critical manifold is strongly deformed and
no longer given as a graph of a function. For specified $\boldsymbol{q}(0),
\boldsymbol{v}(0)$ there are then several $\dot{\boldsymbol{v}}(0)$
on $\mathcal{C}_\varepsilon$ which means that the solution to
the asymptotic condition is not unique. However these authors miss to underline
 that the nonuniquess in the examples occurs only at such
high field strengths where a classical theory has long lost its
empirical validity. At moderate field strengths the worked out examples
confirm our findings. The applicability of singular perturbation
theory for a general class of potentials is first recognized in
Spohn (1998).\bigskip\\
{\it ad \ref{sec.fc}, \ref{sec.fd}, \ref{sec.fe}:} These results
are adapted from Kunze,  Spohn (1999).
\newpage
\section{The Lorentz--Dirac Equation}\label{sec.g}
\setcounter{equation}{0}
In relativistic notation the Lorentz--Dirac equation reads
\setcounter{equation}{0}
\begin{equation}\label{g.a}
        m_0\, \dot v^\mu =   (e/c) F^{\mu\nu} (z)
        v_\nu + (e^2/ 6 \pi c^3)\,
         [\ddot v^\mu - c^{-2} \dot v^\lambda
          \dot v_\lambda \,
          v^\mu]\,,
\end{equation}
where we reintroduced the speed of light, $c.~ m_0$ is the
experimental rest mass of the charged particle.
$F^{\mu\nu}$ is the electromagnetic field tensor of the {\it external}
fields. In this section we omit the index ``ex'' for better
readability. Formally, Equation (\ref{g.a}) can be derived from the
Lorentz model in the adiabatic limit. To conform with the usual
notation we have set the adiabatic scale parameter $\varepsilon =
1$.
But it should be kept in mind that the radiation reaction in
(\ref{g.a}) is a small correction to the Hamiltonian part.

If we fix a frame of reference and go over to three--vectors, then
the Lorentz--Dirac equation becomes
\begin{eqnarray}
        m_0 \, \gamma\, \kappa(\boldsymbol{v}) \dot{\boldsymbol{v}}
         &=& e (\boldsymbol{E} (\boldsymbol{q}) + c^{-1}\boldsymbol{v} \times
         \boldsymbol{B}
        (\boldsymbol{q}))\nonumber\\
        && + (e^2 / 6 \pi c^3)\, \gamma^2 \kappa (\boldsymbol{v})\, [ \ddot{\boldsymbol{v}} + 3
        \gamma^2 \,c^{-2}\, (\boldsymbol{v} \cdot \dot{\boldsymbol{v}})\,
         \dot{\boldsymbol{v}}]\,, \label{g.b}
\end{eqnarray}
with the $3 \times 3$ matrix $\kappa(\boldsymbol{v})
 = \unit + c^{-2}\, \gamma^2\, |\boldsymbol{v} \rangle\langle \boldsymbol{v}|$
and its inverse $\kappa (\boldsymbol{v})^{-1} = \unit - c^{-2} |\boldsymbol{v}
\rangle\langle \boldsymbol{v}|$. The Lorentz--Dirac
equation (\ref{g.a}) differs from (\ref{f.a}) only through a proper
relativistic kinetic energy. Clearly, qualitative properties of
the solution flow should  not depend on such a detail. In the analysis
of (\ref{f.a}) we only used  the critical manifold to be a uniform
repeller, except for $|\boldsymbol{v}|/c$ close to one, and the energy balance
\begin{eqnarray}
        &&\frac{d}{dt}\, \big ( m_0 c^2 \, \gamma (\boldsymbol{v}) +
        e \phi (\boldsymbol{q}) - (e^2 / 6 \pi c^3)\, \gamma^4 (\boldsymbol{v}
         \cdot\dot{\boldsymbol{v}})\big )\nonumber\\
       && \qquad \qquad= - (e^2 / 6 \pi c^3)\, \gamma^4 (\dot{\boldsymbol{v}},
        \kappa (\boldsymbol{v}) \dot{\boldsymbol{v}})\,,\label{g.c}
\end{eqnarray}
which ensured that the solution stays for all times on the
critical manifold, provided the radiation reaction term is
sufficiently small. Thus we can follow the blueprint of Section \ref{sec.fa}
to
obtain an effective second order equation for the motion on the
critical manifold.

 In Section \ref{sec.gb} we
 work out some examples of experimental interest. While
 at present an actual test is rather indirect at best,
  the examples should convince the reader that the effective second
 order equation can be handled with ease. As an extra bonus we
 will make some predictions on the motion of the charge which could not have been guessed
 on the basis of Larmor's formula.


\subsection{Critical manifold, the Landau--Lifshitz
equation}\label{sec.ga}
We write (\ref{g.b}) in the standard form of singular perturbation
theory, compare with Section \ref{sec.fb}. Then
\begin{equation}\label{ga.a}
        \dot{\boldsymbol{x}} = \boldsymbol{f}(\boldsymbol{x}, \boldsymbol{y}),
         \quad \varepsilon \dot{\boldsymbol{y}} = \boldsymbol{g}(\boldsymbol{x},
          \boldsymbol{y}, \varepsilon)
\end{equation}
with
\begin{eqnarray}\label{ga.b}
       \boldsymbol{f}(\boldsymbol{x}, \boldsymbol{y}) &=&
       (\boldsymbol{x}_2, \boldsymbol{y})\,,\\
       \boldsymbol{g}(\boldsymbol{x},\boldsymbol{y},\varepsilon)&=& (6 \pi
        c^3/e^2)\big (m_0 \, \gamma^{-1}\boldsymbol{y}
       - e\, \gamma^{-2} \kappa(\boldsymbol{x}_2)^{-1} (\boldsymbol{E}
       (\boldsymbol{x}_1) + c^{-1}\boldsymbol{x}_2 \times
       \boldsymbol{B}(\boldsymbol{x}_1))\big )\nonumber\\
       && - 3 \varepsilon \gamma^2 c^{-2} (\boldsymbol{x}_2
       \cdot{\boldsymbol{y}}) \boldsymbol{y} \,. \label{ga.c}
\end{eqnarray}
To conform with (\ref{f.a}) we reintroduced the small parameter
$\varepsilon$. At zeroth order the critical manifold is $\{\boldsymbol{y} =
\boldsymbol{h}(\boldsymbol{x})\}$
with $\boldsymbol{h}(\boldsymbol{q},\boldsymbol{v})= (e/m_0) \gamma^{-1}\, \kappa
(\boldsymbol{v})^{-1}\, \big(\boldsymbol{E}(\boldsymbol{q})
+c^{-1}
\boldsymbol{v} \times \boldsymbol{B}(\boldsymbol{q})\big).$
Linearizing (\ref{ga.b}), (\ref{ga.c}) at
$\boldsymbol{y}=\boldsymbol{h}(\boldsymbol{x})$ the  repelling eigenvalue is\linebreak
$(6 \pi c^3/e^2)\, m_0 \gamma^{-1}
 + \mathcal{O}(\varepsilon)$,
which vanishes as $|\boldsymbol{v}|/c \to 1$.
Thus we have to rely on the same construction as in Section \ref{sec.fb}.

To order $\varepsilon$ the effective second order equation is
given by (\ref{fb.k}), except that now $m(\boldsymbol{v}) = m_0\, \gamma
\kappa(\boldsymbol{v})$. We work out the various terms and
set $\varepsilon = 1$. Then the motion on the critical manifold of
the Lorentz--Dirac equation is governed by
\begin{eqnarray}
        \dot{\boldsymbol{q}} &=&
        \dot{\boldsymbol{v}}\,,\nonumber\\
       m_0\, \gamma\,\kappa(\boldsymbol{v}) \dot{\boldsymbol{v}}
        &=& e (\boldsymbol{E} + c^{-1}\boldsymbol{v} \times \boldsymbol{B})
        \nonumber\\
       && + \frac{e^2}{6 \pi c^3}\, \Big[ \frac{e}{m_0}\, \gamma\,
       (\boldsymbol{v} \cdot \nabla_{\boldsymbol{q}}) (\boldsymbol{E}
        + c^{-1}\boldsymbol{v} \times
        \boldsymbol{B})+ \big(\frac{e}{m_0}\big)^2\, c^{-1}\Big((\boldsymbol{E}
         \times
        \boldsymbol{B}) \nonumber\\[1mm]
       &&  + c^{-1} (\boldsymbol{v} \cdot \boldsymbol{E})
       \boldsymbol{E} + c^{-1}(\boldsymbol{v}
        \cdot \boldsymbol{B})\boldsymbol{B}+ \, \big (- \boldsymbol{E}^2 -
       \boldsymbol{B}^2\nonumber\\[1mm]
       &&+ c^{-2} (\boldsymbol{v} \cdot \boldsymbol{E})^2
       + c^{-2}(\boldsymbol{v} \cdot \boldsymbol{B})^2
       + 2 c^{-1} \boldsymbol{v }\cdot (\boldsymbol{E}
        \times \boldsymbol{B})\big) \gamma^2 c^{-1}
         \boldsymbol{v}\Big )\Big ]\,.\label{ga.d}
\end{eqnarray}

While singular perturbation theory provides a systematic method,
Equation (\ref{ga.d}) can also be derived formally. In (\ref{ga.b}) we regard
$m_0 \gamma\, \kappa(\boldsymbol{v}) \dot{\boldsymbol{v}} = e\,
 (\boldsymbol{E}+  c^{-1}\boldsymbol{v} \times \boldsymbol{B})$ as unperturbed
equation, differentiate it once, and substitute $\ddot{\boldsymbol{v}}$ inside the square
brackets of (\ref{g.b}). Resubstituting $\dot{\boldsymbol{v}}$
from the unperturbed equation results in Equation (\ref{ga.d}). This argument
is carried out more easily in the covariant form of the
Lorentz--Dirac equation. The unperturbed part is
\begin{equation}\label{ga.e}
       m_0  \dot v^\mu = (e/c) F^{\mu\nu} (z) v_\nu
\end{equation}
and differentiating with respect to the eigentime,
\begin{equation}\label{ga.f}
      (m_0 c/e)  \ddot v^\mu = v_\lambda \partial^\lambda F^{\mu\nu}
       (z)  v_\nu + F^{\mu\nu} (z) \dot v_\nu \,.
\end{equation}
Substituting (\ref{ga.e}) and (\ref{ga.f})  in (\ref{g.a}) yields
\begin{eqnarray}
       m_0 \dot{v}^\mu &=& \frac{e}{c}  F^{\mu\nu} v_\nu +
       \frac{e^2}{6 \pi c^3}
       \,\Big [ \frac{e}{m_0 c}v_\lambda \partial^\lambda F^{\mu\,\nu}
        v_\nu\nonumber\\[2mm]
        && + \big (\frac{e}{m_0 c}\big)^2 \big(F^{\mu\nu}
        F_{\nu}^{\,\,\lambda} v_\lambda
        +c^{-2} F^{\alpha\lambda}
        F_\lambda{\,^\beta} v_\alpha v_\beta v^\mu\big)\Big ]\,.
        \label{ga.g}
\end{eqnarray}
Written in three--vectors Equation (\ref{ga.g}) coincides with (\ref{ga.d}) together with the
equation for the energy balance.

Of course, the justification of Equation (\ref{ga.d}) comes only from the
structure of the solution flow to (\ref{g.b}). Higher order
corrections, although rather unimportant in our context, would have to be
computed by the method explained in Section \ref{sec.fa}.

Equation (\ref{ga.g}) appears for the first time in the second
volume of the Landau--Lifshitz Course in Theoretical Physics. It
seems to be appropriate to call then Equation (\ref{ga.g}) the
Landau--Lifshitz equation. The error in going from (\ref{g.a}) to
(\ref{ga.g}) is of the same order as the one in the derivation of the
Lorentz--Dirac equation itself. {\it Thus we regard the Landau--Lifshitz
equation as the effective equation governing  the motion of a charged particle in the
adiabatic limit.}


\subsection{Some applications}\label{sec.gb}
(i) {\it Zero magnetic field.} For zero magnetic field the
Landau--Lifshitz equation simplifies to
\begin{eqnarray}
       m_0 \, \gamma\, \kappa(\boldsymbol{v}) \dot{\boldsymbol{v}} &=& e\,
       \boldsymbol{E} +
        \frac{e^2}{6 \pi c^3}\, \big [ \frac{e}{m_0} \, \gamma\,
       (\boldsymbol{v} \cdot \nabla_{\boldsymbol{q}})
       \boldsymbol{E}\label{gb.a}\\[1mm]
       &&  + \big ( \frac{e}{m_0 c}\big )^2
         \big ( (\boldsymbol{v} \cdot \boldsymbol{E}) \boldsymbol{E}
         - \gamma^2 \, \boldsymbol{E}^2 \boldsymbol{v} + \gamma^2 c^{-2}
        (\boldsymbol{v} \cdot \boldsymbol{E})^2 \boldsymbol{v} \big ) \big
        ]\,.\nonumber
\end{eqnarray}

Of interest is a central potential. We set $\boldsymbol{q}
=\boldsymbol{r},~ |\boldsymbol{r}| = r,~
\boldsymbol{\hat r} =
\boldsymbol{r}/|\boldsymbol{r}|, \phi_{\mathrm{ex}} (\boldsymbol{q}) =
\phi(r) $ which implies $\boldsymbol{E}= - \phi^\prime
\boldsymbol{\hat r}$. Then (\ref{gb.a}) becomes
\begin{eqnarray}
       m_0 \, \gamma\, \kappa(\boldsymbol{v}) \dot{\boldsymbol{v}} &=& -
        e\, \phi^\prime \boldsymbol{\hat r} +
        \frac{e^2}{6 \pi c^3}\, \big [ \frac{e}{m_0} \, \gamma\,
        (-  (\boldsymbol{v} \cdot \boldsymbol{\hat r})
        \phi^{\prime\prime}\boldsymbol{\hat r} \nonumber\\[1mm]
        && - \frac{1}{r}(\boldsymbol{v} -(\boldsymbol{v}
         \cdot \boldsymbol{\hat r}) \boldsymbol{\hat r})
         \phi^\prime   + \big (\frac{e}{m_0 c} \big )^2
        \, \phi ^{\prime\, 2} \big ( (\boldsymbol{v}
        \cdot \boldsymbol{\hat r})  \boldsymbol{\hat r} -
        \gamma^2 \, \boldsymbol{v}\nonumber\\[1mm]
        && + \gamma^2 \, c^{-2} ( \boldsymbol{v} \cdot
        \boldsymbol{\hat r})^2\, \boldsymbol{v} \big ) \big ]\,.\label{gb.b}
\end{eqnarray}
The angular momentum $\boldsymbol{L} = \boldsymbol{r} \times m_0
 \gamma \boldsymbol{v}$
satisfies
\begin{equation}\label{gb.c}
       \dot{\boldsymbol{L}} = \frac{e^2}{6 \pi c^3}\, \big [ -\frac{e}{m_0}\,
        \frac{1}{r}
       \, \phi^\prime - \big ( \frac{e}{m_0 c}\big )^2\,
       \gamma^2\,  (1- c^{-2} (\boldsymbol{v} \cdot \boldsymbol{\hat
       r})^2 ) \phi^{\prime \,2}\,\big ] \boldsymbol{L}\,.
\end{equation}
Thus the orientation of $\boldsymbol{L}$ is conserved and the motion lies in
the plane orthogonal to $\boldsymbol{L}$. No further reduction seems to be
possible and one would have to rely on a numerical integration.
Only for the harmonic oscillator, $\phi(r)=
 \frac{1}{2}\, m_0 \omega^2_0 r^2$,
a closed form solution can be achieved.

A somewhat more tractable case is to assume that $\phi_{\mathrm{ex}}$
varies only along the 1--axis. Setting $\boldsymbol{v}=
 (v, 0,0), ~ \boldsymbol{q}=
(x,0,0)$, and $\boldsymbol{E}= (-\phi^\prime , 0,0)$, Equation (\ref{gb.a}) becomes
\begin{equation}\label{gb.d}
       m_0\,\gamma^3\, \dot v = - e
       \phi^\prime (x) - \frac{e^2}{6\pi c^3}\,
        \frac{e}{m_0}
       \, \gamma \, \phi^{\prime\prime}(x) v \,.
\end{equation}
The radiation reaction  is proportional to $- \phi^{\prime\prime} (x)
v$, which we can be regarded as a spatially varying friction
coefficient proportional to $\phi^{\prime\prime}(x)$. For a convex
potential, $\phi^{\prime\prime}> 0$, like an
oscillator potential, this friction coefficient is strictly
positive and the resulting motion is damped until the minimum of $\phi$
is reached. In general however, $\phi^{\prime\prime}$ will not
have a definite sign, like the double well potential $\phi(x)
\simeq
(x^2 - 1)^2$, or the washboard potential $\phi (x)\simeq - \cos x$. At
locations where $\phi^{\prime\prime}(x) < 0$ one has antifriction
and the mechanical energy increases. This gain is always
dominated by losses as can be seen from the energy balance
\begin{eqnarray}
        &&\frac{d}{dt}\, \big [m_0\,\gamma + e \phi + \frac{e^2}{6\pi c^3}\,
        \, \frac{e}{m_0}\, \gamma \,\phi^\prime v \big
        ]\nonumber\\
        &&= - \frac{e^2}{6 \pi c^3}\, \big ( \frac{e}{m_0}\big)^2 \,
        \phi ^{\prime\, 2}- \frac{1}{m_0}\, \big ( \frac{e^2}{6 \pi c^3}\,
        \frac{e}{m_0}\big )^2\, \gamma \,\phi^\prime \phi^{\prime\prime}
        v\, .\label{gb.e}
\end{eqnarray}
The last term in (\ref{gb.e}) does not have a definite sign. But its
prefactor is down by one order in $\varepsilon$ and therefore it
is outweighed by $- \phi^{\prime \,2}$.

Equation (\ref{gb.d}) has one peculiar feature. If $\phi(x) = - a_0 x, ~ a_0 >
0$, over some  some interval $[a_-, a_+]$, then $\phi^{\prime\prime} = 0$
over that interval and the friction term vanishes. The particle
entering at $a_-$ is uniformly accelerated to the right until it
reaches $a_+$. From Larmor's formula we know that the energy
radiated per unit time equals $(e^2 /6 \pi c^3) (e/m_0)^2 a_0^2$. This
energy must come entirely from the near field without a mechanical
contribution. The same behavior is found for the Lorentz--Dirac
equation. If, locally, $\boldsymbol{E} = const $ and $\boldsymbol{B}=0$, then the
Hamiltonian part is solved by the hyperbolic motion, i.e. a
constantly accelerated relativistic particle. For this solution
the radiation reaction vanishes which means that locally the critical manifold happens to
be independent of $\varepsilon$. The radiated energy originates
from the near field only.\medskip\\
(ii) {\it Zero electrostatic field and constant magnetic field.}
We set $\boldsymbol{B} = (0,0,B)$ with constant $B$. Then (\ref{ga.d}) simplifies
to
\begin{eqnarray}\label{gb.f}
        m_0\,\gamma \,\kappa(\boldsymbol{v}) \dot{\boldsymbol{v}}
         &=&  \frac{e}{c} (\boldsymbol{v} \times \boldsymbol{B}) + \frac{e^2}{6\pi c^3}
        \, \big (\frac{e}{m_0 c}\big )^2\,  \big[(\boldsymbol{v} \cdot
        \boldsymbol{B}) \boldsymbol{B} -
        \gamma^2 \boldsymbol{B}^2 \boldsymbol{v}\\[1mm]
        && + \gamma^2\, c^{-2}
        (\boldsymbol{v} \cdot \boldsymbol{B})^2 \boldsymbol{v}\big]\,.\nonumber
\end{eqnarray}
We multiply by $\kappa(\boldsymbol{v})^{-1}$ and obtain
\begin{equation}\label{gb.g}
        m_0\,\gamma \,  \dot{\boldsymbol{v}} =  \frac{e}{c}(\boldsymbol{v}
         \times \boldsymbol{B}) + \frac{e^2}{6\pi c^3}\,
        \big (\frac{e}{m_0 c}\big )^2\,  [(\boldsymbol{v} \cdot \boldsymbol{B})
         \boldsymbol{B} - \boldsymbol{B}^2 \boldsymbol{v}]\,.
\end{equation}
The motion parallel to $\boldsymbol{B}$
decouples with $\dot v_3 = 0$. We set $v_3 = 0$ and $\boldsymbol{v}=
(\boldsymbol{u},0), ~ \boldsymbol{u}^\perp
= (-u_2, u_1)$.
Then the motion in the plane orthogonal to $\boldsymbol{B}$ is governed by
\begin{equation}\label{gb.h}
        \gamma \, \dot{\boldsymbol{u}} =  \omega_c (\boldsymbol{u}^\perp -
        \beta \omega_c \boldsymbol{u})
       \,,
\end{equation}
with cyclotron frequency $\omega_c = e B/m_0 c$ and $\beta = e^2/6\pi c^3
m_0$. (\ref{gb.h}) holds over the entire velocity range. For an
electron $\beta \omega_c = 8.8 \times 10^{-18} B$ [Gauss]. Thus
even for very strong fields the friction is small compared to the
inertial terms.

(\ref{gb.h}) can be integrated as

\begin{equation}\label{gb.i}
         \frac{d}{dt}\,\gamma = - \beta \omega_c^2\, (\gamma^2-1)
\end{equation}
with solution
\begin{equation}\label{gb.j}
         \gamma_t = [\gamma_0 + 1 + (\gamma_0-1)
         e^{-2\beta\omega_c^2 t}][\gamma_0+1 - (\gamma_0 - 1)
         e^{-2\beta \omega_c^2 t}]^{-1}\,,
\end{equation}
which tells us how $\boldsymbol{u}(t)^2$ shrinks to zero. To determine the angular
dependence we introduce polar coordinates as $\boldsymbol{u} = u(\cos \varphi, \sin
\varphi)$. Then
\begin{equation}\label{gb.aa}
         \frac{du}{d\varphi}  =  - \beta \omega_c u~,~
          \frac{d \varphi}{dt}= \gamma^{-1} \omega\,.
\end{equation}
Thus  $u(\varphi)$ shrinks
exponentially,
\begin{equation}\label{gb.l}
           u(\varphi)  = u(0)\, e^{- \beta \omega_c \varphi}\,.
\end{equation}
Since  $\beta\omega_c = 8.8 \times 10^{-18}
B$ [Gauss] for an electron, even for strong fields the change of $u$ in one
revolution is tiny.

To obtain the evolution of the position $\boldsymbol{q}=(\boldsymbol{r}
,0), |\boldsymbol{r}|=r$, we use
that for zero radiation reaction, $\beta = 0$,
\begin{equation}\label{gb.m}
         r =\frac{u}{\omega_c}\,\gamma\,.
\end{equation}
By (\ref{gb.l}) this relation remains approximately valid for
non--zero $\beta$. Inserting $u(t)$ from (\ref{gb.j}) we obtain
\begin{equation}\label{gb.n}
           r(t) = r_0\, e^{- \beta \omega_c^2 t}
           [1 + ((\gamma_0-1)/2)(1- e^{-2 \beta \omega_c^2 t}) ]^{-1}
\end{equation}
with  $r_0$ the initial radius and $u(0)/c = (\gamma_0 - 1)^{1/2} /\gamma_0$
the initial speed which are related through (\ref{gb.m}). In the
ultra--relativistic regime, $\gamma_0 \gg 1$, and for times such
that $\beta \omega_c^2 t \ll 1$, (\ref{gb.n}) simplifies to
\begin{equation}\label{gb.o}
           r(t) = r_0\, \frac{1}{1+ \gamma_0 \beta \omega_c^2 t}
\end{equation}
and the initial decay is according to the power law  $t^{-1}$
rather than exponential.

For an electron $ \beta \omega_c^2 = 1.6 \times
10^{-6}(B ~[\mbox{Gauss}])^2/\mbox{sec}$.
Therefore if we choose a field strength  $B=10^3$ Gauss and an initial
radius of $r_0 = 10$ cm, which corresponds to the ultra--relativistic
case of $\gamma = 6 \times 10^4$, then the radius shrinks within
0.9 sec to $r(t) = 1~ \mu {\mathrm m}$ by which time the electron has made
$2 \times 10^{14}$ revolutions.\medskip\\
(iii) {\it The Penning trap}. An electron can be trapped for a very long
time in the combination of a homogeneous magnetic field and an
electrostatic quadrupole potential, which has come to be known as a Penning
trap. Its design has been optimized towards high precision
measurements of the gyromagnetic $g$-factor of the electron.
Our interest here is that the motion in the plane
orthogonal to the magnetic field consists of two coupled modes,
which means that the damping cannot be guessed  by pure energy
considerations using Larmor's formula. One really needs the full
power of the Landau--Lifshitz equation.

An ideal Penning trap has the electrostatic quadrupole potential
\begin{equation}\label{gb.p}
           e \phi (\boldsymbol{x}) = \frac{1}{2}\,
           m \omega_z^2 (- \frac{1}{2} x_1^2 -\frac{1}{2}\,
           x_2^2 + x_3^2)\,,
\end{equation}
which satisfies $\triangle \phi = 0$, superimposed with the uniform
magnetic field
\begin{equation}
\boldsymbol{B}= (0,0,B)\,.\label{gb.pp}
\end{equation}
The quadrupole field provides an axial restoring force whereas the
magnetic field is responsible for the radial restoring force,
which however could be outweighed by the inverted part of the
harmonic electrostatic potential.

We insert $\boldsymbol{E}=- \nabla\phi$ and $\boldsymbol{B}$ in the Landau--Lifshitz
equation. The terms proportional to $(\boldsymbol{v} \cdot \nabla_{\boldsymbol{q}})
\boldsymbol{E},~ \boldsymbol{E}\times \boldsymbol{B}, ~
(\boldsymbol{v} \cdot \boldsymbol{B})\boldsymbol{B}$, and
 $\boldsymbol{B}^2 \boldsymbol{v}$ are linear in $\boldsymbol{v}$,
 resp. $\boldsymbol{q}$. The remaining
terms are either cubic or quintic and will be neglected. This is
justified provided
\begin{equation}\label{gb.q}
       \frac{|\boldsymbol{v}|}{c} \ll 1
\end{equation}
and
\begin{equation}\label{gb.r}
       (m_0 \omega_z^2/e)\, r_{\mathrm{max}} \ll
        B, ~ \mbox{i.e.} \quad r_{\mathrm{max}} \ll
        c(\omega_c/\omega_z^2)\,,
\end{equation}
if $r_{\mathrm{max}}$ denotes the maximal distance from the
trap center. With these assumptions the Landau--Lifshitz equation decouples
into an in--plane motion and an axial motion governed by
\begin{eqnarray}\label{gb.s}
      \dot{\boldsymbol{u}} &=& \frac{1}{2}\,  \omega_z^2 \boldsymbol{r}
       + \omega_c \boldsymbol{u}^\perp -
       \beta\big [(\omega_c^2 -\frac{1}{2}\,
       \omega_z^2) \boldsymbol{u} + \frac{1}{2}\, \omega_c\omega_z^2 \boldsymbol{r}^\perp
       \big ]\,,\\
       \ddot z &=& - \omega_z^2 z- \beta \omega_z^2 \dot
       z\,.\label{gb.t}
\end{eqnarray}
Here $\boldsymbol{q}= (\boldsymbol{r},z), ~\boldsymbol{v}
= (\boldsymbol{u}, \dot z ), ~ (x_1, x_2)^\perp = (-x_2,
x_1)$.

The axial motion is just a damped harmonic oscillator with
frequency $\omega_z$ and friction coefficient
\begin{equation}\label{gb.u}
       \gamma_z = \beta \omega_z^2 \,.
\end{equation}

The in--plane motion can be written in matrix form  as
\begin{equation}\label{gb.v}
       \frac{d}{dt}\, \psi =  (A +  \beta V)\psi
\end{equation}
with $\psi =(\boldsymbol{r},\boldsymbol{u})$ and $A_{11}=0, ~ A_{12} = \unit, ~ A_{21} =
\omega_z^2 \unit, ~ A_{22} = i \omega_z \sigma_y, ~
V_{11} = 0, ~ V_{12}= 0,~ V_{21} = i \omega_c\omega_z^2
\sigma_y, ~ V_{22}= (\omega_z^2 - \omega_c^2)\unit $, where $\sigma_y$
is the Pauli spin matrix with eigenvectors $\chi_\pm,~ \sigma_y \chi_\pm =
\pm \chi_\pm.$ The unperturbed motion is governed by the
$ 4 \times 4$ matrix $A$. It has the
eigenvectors
 $\psi_{+,\pm} = (\pm i (1/\omega_+) \chi_\mp,  \chi_\mp)$
with eigenvalues $\pm i \omega_+$ and
$\psi_{-,\pm} = (\pm i(1/\omega_-) \chi_\mp, \chi_\mp)$ with
eigenvalues $\pm i \omega_-$, where
\begin{equation}\label{gb.w}
       \omega_\pm = \frac{1}{2}\, \big ( \omega_c \pm
       \sqrt{\omega_c^2- 2 \omega_z^2}\,\,\big )\,.
\end{equation}
The mode with frequency $\omega_+$ is called cyclotron mode and
the one with $\omega_-$ magnetron  mode. Experimentally
$\omega_c \gg \omega_z$ and therefore $\omega_+ \ll \omega_-$.
 The orbit is then an epicycle with rapid
cyclotron and slow magnetron motion.

The adjoint matrix $A^\ast$ has eigenvectors orthogonal to the $\psi^\prime
s$. They are given by $\varphi_{+,\pm} = (\mp i (\omega_z^2/\omega_+)
\chi_\mp, \chi_\mp) $ with eigenvalues $\pm i\omega_+$ and
$\varphi_{-, \pm} = (- (\omega_z^2/\omega_-) \chi_\mp,$ $
\chi_\mp)$ with eigenvalue $\mp i \omega_-$.

Since $\beta$ is small, the eigenfrequencies of $A + \beta V$ can
be computed in first order perturbation. The cyclotron modes
attains a negative real part corresponding to the friction
coefficient
\begin{equation}\label{gb.x}
       \gamma_+ = \frac{e^2}{6 \pi c^3 m_0}\,
       \frac{\omega_+^3}{\omega_+ - \omega_-}
\end{equation}
and the magnetron mode attains a positive real part corresponding
to the antifriction coefficient
\begin{equation}\label{gb.y}
       \gamma_- = \frac{e^2}{6 \pi c^3 m_0}\,
       \frac{\omega_-^3}{\omega_- - \omega_+}\,.
\end{equation}
As the electron radiates it lowers its potential energy by
increasing the magnetron radius.

Experimentally $B= 6 \times 10^4$ Gauss and the voltage drop across the trap
is $10 $V. This corresponds to $\omega_z = 4 \times 10^8~
\mbox {Hz}, ~ \omega_+ = 1.1 \times 10^{12} ~\mbox{Hz}, ~ \omega_- = 7.4 \times 10^4 ~
\mbox{Hz}$. The conditions (\ref{gb.q}), (\ref{gb.r}) are easily
satisfied. For the life--times one obtains $(1/\gamma_z) = 5 \times
10^8~ \mbox{sec}$, $ (1/\gamma_+) = 8 \times 10^{-2} ~\mbox{sec,}~\mbox{and}~
 - (1/\gamma_-)=
2 \times 10^{23}$ sec. Thus the magnetron motion is stable, as
observed by keeping a single electron trapped over weeks. The
cyclotron motion decays within fractions of a second. The axial
motion is in fact damped by coupling to the external circuit  and
decays also within a second.

The variation with the magnetic field is more clearly discussed in
terms of the dimensionless ratio $(\omega_c /\omega_z) = \lambda$.
Then
\begin{eqnarray}
       \omega_\pm &=& \omega_z\, \frac{1}{2}\, (\lambda \pm
       \sqrt{\lambda^2 - 2})\,,\nonumber\\[1mm]
       \gamma_\pm &=& \pm \beta \omega_z^2 (\lambda \pm
       \sqrt{\lambda^2 - 2})^3 /8 \sqrt{\lambda^2 - 2}
       \,.\label{gb.z}
\end{eqnarray}
For large $\lambda, ~\omega_+ \cong \lambda,~ \omega_- \cong
\lambda^{-1}$, whereas $\gamma_+ \cong \lambda^2, \gamma_-\cong
\lambda^{-4}$. As $\lambda \to \sqrt{2}$, we have $\omega_+ = \omega_- =
\omega_z/\sqrt{2}$.
However the friction coefficients diverge as $(\lambda -
\sqrt{2})^{-1/2}$.
Let us call $ B_c$ the critical field at which the mechanical motion
becomes
unstable. For $B > B_c$, one has still periodic motion with
frequency $\omega_z/\sqrt{2}$, but the coming instability is
disclosed through the vanishing lifetime. In the mentioned
experiment $\lambda = 2.7 \times 10^3$ and  for
fixed $\omega_z$ the critical field  strength would be $B_c = 30$
Gauss.
\subsection*{Notes and References}
{\it ad \ref{sec.g}:} The name Lorentz--Dirac is standard but historically
inaccurate. Some authors, e.g. Rohrlich (1997), therefore propose Abraham--Lorentz--Dirac
instead. The radiation reaction term was first derived by Abraham
(1905), compare with Sections \ref{sec.e} and \ref{sec.f} von Laue (1909)
realized its covariant form. In the Pauli Handbuch article on relativity the
equation is stated as in (\ref{g.a}). The contribution of Dirac is explained in
Section \ref{sec.bb}. \bigskip\\
{\it ad \ref{sec.ga}:} The literature on the critical manifold of
the Lorentz--Dirac equation is listed in {\it ad \ref{sec.fc}}. The
 Landau--Lifshitz equation appears
in all editions of their Course in Theoretical Physics.  They provide no
hint on the geometrical picture of the solution flow nor on the
errors involved in their approximation. It is rather surprising
that the contribution of Landau and Lifshitz is ignored in
essentially all discussions of radiation reaction, one notable
exception being Teitelbom et al (1980). For that reason
the Landau--Lifshitz equation was rederived independently in
Spohn (1998). There
have been other attempts to replace the Lorentz--Dirac equation
by a second order equation, Mo,  Papas (1971), Bonnor (1974),
Parrot (1987), Ford, O'Connell (1991,1993).
Based on Ford, O'Connell (1991), Jackson (1999)
uses  the substitution in the case of a radiation
damped harmonic oscillator and discusses  applications. In the general
case only Landau and Lifshitz obtain the correct center manifold
equation.\bigskip\\
{\it ad \ref{sec.gb}:} Uniform acceleration is discussed in
Rohrlich (1990). Constant magnetic field is important for
synchroton sources. Since the electron is kept on its circular
orbit, Larmor's formula is precise enough. Landau and Lifshitz
(1959) give a  brief discussion. The power law for the
ultra--relativistic case is noted in Spohn (1999a). Shen (1972a,1978)
discusses at which field strengths quantum corrections will
become important. His results are only partially reliable, since
he does not start from the Landau--Lifshitz equation. The Penning
trap is reviewed by Brown and Gabrielse (1986), which includes a
discussion of the classical orbits and their life--times. They
state the results (\ref{gb.x}), (\ref{gb.y}) as based on a
quantum resonance computation. Since the result is classical, it
must follow from the Landau--Lifshitz equation, Spohn (1999a). In
the classical framework, more general trap potentials can be
handled through numerical integration routines for ordinary differential equations.
\newpage
\section{Spinning Charges}\label{sec.h}
If an electron is modelled as a classical lump of highly
concentrated charge, then merely  by the interaction with its own
radiation field the charge distribution will start to rotate. A
proper mechanical description must include then the angular
velocity of the internal rotation and an equation for the torque.
The argument seems to leave little choice and in this chapter we
will progress a few steps in the direction of including the
classical spin, which leads to unexplored and interesting
territory.

If we take the quantum mechanical description as starting point,
however, as we should do, then the situation is more ambivalent.
To be a little bit more specific we consider a quantum particle
subject to slowly varying external forces, which is the standard
semiclassical limit. The center of the wave packet evolves then
according to an effective classical evolution equation of the form
(\ref{fc.b}). Of course, the energy--momentum relation
$E_s(\boldsymbol{P})$ has to computed now from the quantum
hamiltonian. Only if the model is fully relativistic, we can be
sure a priori that $E_s(\boldsymbol{P}) = (\boldsymbol{P}^2 +
m^2)^{1/2}$.
Let us assume that in addition the particle carries a spin
$\frac{1}{2}$. The corresponding spinor, $\psi_t \in
\mathbb{C}^2$, is governed by
\begin{equation}\label{h.a}
        i \hbar\, \frac{d}{dt}\, \psi_t = H_S(t) \psi_t
\end{equation}
with $H_S(t)$ the time--dependent spin hamiltonian. In a
relativistic theory we have
\begin{eqnarray}\label{h.b}
        H_S(t)&=& \frac{e}{mc}\, \frac{\hbar}{2}
        \boldsymbol{\sigma} \cdot \big[ \big( \frac{g}{2}
        - 1 + \frac{1}{\gamma}\big) \boldsymbol{B}_{\mathrm{ex}} (\boldsymbol{r}
        (t)) - \big( \frac{g}{2} - 1\big) \, \frac{\gamma}{1+ \gamma}
        c^{-2}\\
        && (\boldsymbol{u}(t) \cdot \boldsymbol{B}_{\mathrm{ex}}
        (\boldsymbol{r} (t))) \boldsymbol{u}(t) - \big (\frac{g}{2}
        - \frac{\gamma}{1+\gamma}\big) c^{-1} \boldsymbol{u}(t)
        \times E_{\mathrm{ex}}
        (\boldsymbol{r}(t))\big]\,.\nonumber
\end{eqnarray}
Here $\boldsymbol{\sigma} = (\sigma_1, \sigma_2, \sigma_3)$ are
the Pauli spin matrices with $\sigma_i^2 = 1$ and commutation
relations $[\sigma_i, \sigma_j] = 2 i \varepsilon_{ijk} \sigma_k,
i, j, k = 1,2,3. $
$g$  is the gyromagnetic ratio of the quantum particle, $g \cong 2$
for an electron. The spin passively adjusts itself to the fields
along the semiclassical orbit $t \mapsto (\boldsymbol{r}(t), \boldsymbol{u}
(t))$ traced out by the particle. To leading order there is no back
reaction onto the translational degrees of freedom. As we will see
this is not the case for a classical spin and both degrees of
freedom are coupled, a property which is shared by
relativistically covariant Lagrangians for a particle with spin.
We conclude that, in contrast to the translational degrees of
freedom, a model including the classical spin serves only within
limits as a phenomenological description for a quantum spin.

There is another, physically more basic objection. $H_S(t)$ is
linear in $\boldsymbol{\sigma}$. Thus defining the average spin
$\boldsymbol{s}_t = < \psi_t, \boldsymbol{\sigma}\psi_t >$ we see
that in (\ref{h.b}) $\hbar$ drops out and $\boldsymbol{s}_t$
satisfies a classical spin equation. In fact, (\ref{h.b}) becomes
the BMT equation for $\boldsymbol{s}_t$, an equation originally
obtained on purely classical grounds. However, such an
approximation by a classical angular momentum is valid only in the
large spin number limit. In the semiclassical the spin degree if
freedom remains fully quantum.
Of course, in the standard polarization experiments, as for
example the high precision measurements of the gyromagnetic ratio,
interference is not probed and the classical picture serves well.


\subsection{Abraham model with spin}\label{sec.ha}
Abraham models the charge as a nonrelativistic rigid body.
Clearly, a complete mechanical description must specify both its
center of mass and its angular velocity, which we denote by $\boldsymbol{\omega}(t) \in
\mathbb{R}^3$. The spinning charge  generates the current
\begin{equation}\label{ha.a}
        \boldsymbol{j}(\boldsymbol{x},t) = \big(\boldsymbol{v}
        (t) + \boldsymbol{\omega}(t) \times (\boldsymbol{x}-\boldsymbol{q}(t))\big)
        \rho(\boldsymbol{x}-\boldsymbol{q}(t))\,.
\end{equation}
Therefore the source term in Maxwell equations is modified as
\begin{eqnarray}
         \partial_t \boldsymbol{B}(\boldsymbol{x},t) &=& - \nabla \times
         \boldsymbol{E}(\boldsymbol{x},t),\nonumber\\
         \partial_t  \boldsymbol{E}(\boldsymbol{x},t) &=&  \nabla \times
         \boldsymbol{B}(\boldsymbol{x},t)- \big(\boldsymbol{v}(t) +
         \boldsymbol{\omega}(t) \times (\boldsymbol{x}-\boldsymbol{q}(t))\big) \rho
         (\boldsymbol{x}-\boldsymbol{q}(t))\,,\nonumber\\
         \nabla \cdot \boldsymbol{E}(\boldsymbol{x},t) &=& \rho(\boldsymbol{x}
         -\boldsymbol{q}(t), ~ \nabla \cdot
         \boldsymbol{B}(\boldsymbol{x},t) = 0 \,,\label{ha.b}
\end{eqnarray}
which satisfies charge conservation, since $\rho $ is radial.

The mass distribution of the rigid body is assumed to have the
same  form factor as the charge distribution,  $\rho_m (\boldsymbol{x})= m_{\mathrm{b}}
f(\boldsymbol{x}^2)$.
The bare moment of inertia is then
\begin{equation}\label{ha.c}
         I_{\mathrm{b}} = \frac{2}{3}\, \int d^3 x \rho_m
         (\boldsymbol{x}) \boldsymbol{x}^2
\end{equation}
with corresponding angular momentum $\boldsymbol{S}=I_{\mathrm{b}}\,\boldsymbol{\omega}$.
The electric
dipole moment of the charge distribution $\rho$ vanishes
by symmetry.  The
magnetic dipole moment of the current (\ref{ha.a}) is given by
\begin{eqnarray}
       \mu &=& \frac{1}{2}\, \int d^3 x \boldsymbol{x} \times \big(\boldsymbol{v}
       + (\boldsymbol{\omega}
       \times \boldsymbol{x})\big) \rho(\boldsymbol{x})\nonumber\\
       &=&\frac{1}{2}\frac{e}{m_{\mathrm{b}}} I_{\mathrm{b}}\, \boldsymbol{\omega}
       =
         g_{\mathrm{b}}
        \frac{e}{2 m_{\mathrm{b}}}\, \boldsymbol{S}\label{ha.d}
\end{eqnarray}
with bare gyromagnetic ratio $g_{\mathrm{b}} = 1$. As in the case
of the bare mass, $I_{\mathrm{b}}$ and $g_{\mathrm{b}}$
will be renormalized through the self--interaction.

The Lorentz force equation comes now in two parts, one for the
linear and one for the angular momentum. To be consistent we
stick to the nonrelativistic form and have
\begin{eqnarray}\label{ha.e}
         \frac{d}{dt}
       m_{\mathrm{b}} \boldsymbol{v}(t) &=& \int d^3 x \rho
       (\boldsymbol{x}-\boldsymbol{q}(t)) \big[
       \boldsymbol{E}(\boldsymbol{x},t) + \big(\boldsymbol{v}(t) +
       \boldsymbol{\omega}(t)\times (\boldsymbol{x}
       -\boldsymbol{q}(t))\big) \times
       \boldsymbol{B}(\boldsymbol{x},t) \big]\,,\nonumber\\
        \frac{d}{dt}
       I_{\mathrm{b}}\, \boldsymbol{\omega}(t) &=& \int d^3 x \rho
       (\boldsymbol{x}-\boldsymbol{q}(t)) ( \boldsymbol{x}-\boldsymbol{q}
       (t)) \times
       \big[\boldsymbol{E}(\boldsymbol{x},t)\\
       &&\qquad\qquad\quad+
       \big(\boldsymbol{v}(t)
        +
       \boldsymbol{\omega}(t)\times (\boldsymbol{x}-\boldsymbol{q}(t))\big)\times
       \boldsymbol{B}(\boldsymbol{x},t) \big]\,.
       \label{ha.f}
\end{eqnarray}
If in addition there are external forces acting on the  charge,
then $\boldsymbol{E}$ and $\boldsymbol{B}$ in (\ref{ha.e}), (\ref{ha.f}) would have to be
replaced by $\boldsymbol{E} + \boldsymbol{E}_{\mathrm{ex}}$ and
 $\boldsymbol{B}+ \boldsymbol{B}_{\mathrm{ex}}$,
respectively.

The Abraham model of Section \ref{sec.ad} is obtained by formally
setting $\boldsymbol{\omega}(t)=0$. Note that this is not consistent with the
Lorentz
torque equation (\ref{ha.f}), since $\dot{\boldsymbol{\omega}} \not= 0$, in
general, even for $\boldsymbol{\omega}=0$.

The Abraham model with spin has the conserved energy
\begin{equation}\label{ha.g}
        \mathcal{E} = \frac{1}{2}\, m_{\mathrm{b}} \boldsymbol{v}^2 +
        \frac{1}{2}\, I_{\mathrm{b}} \boldsymbol{\omega}^2 +
          \frac{1}{2} \int d^3 x (\boldsymbol{E}^2 + \boldsymbol{B}^2)
\end{equation}
and the conserved linear momentum
\begin{equation}\label{ha.h}
       \mathcal{P} = m_{\mathrm{b}}\boldsymbol{v} + \int d^3 x
       \boldsymbol{E} \times \boldsymbol{B}\,.
\end{equation}
In addition the total angular momentum
\begin{equation}\label{ha.i}
       \mathcal{J} = \boldsymbol{q} \times m_{\mathrm{b}} \boldsymbol{v} + I_{\mathrm{b}}
       \boldsymbol{\omega} + \int d^3 x \boldsymbol{x} \times (\boldsymbol{E}
       \times \boldsymbol{B})
\end{equation}
is conserved. Of course, also the Abraham model without spin is
invariant under rotations and must therefore have a
correspondingly conserved quantity. Only it does not have the standard
form of the total angular momentum, which from a somewhat
different perspective indicates that inner rotations must be
included.

As good tradition already, we assume that the external forces are
slowly varying and want to derive in this adiabatic limit an
effective equation of motion for the particle including its spin.
As a first step of this program we have to determine the charge
solitons. We set
\begin{eqnarray}
        \boldsymbol{q}(t) &=& \boldsymbol{v} t~ ,~
        \boldsymbol{\omega}(t) = \boldsymbol{\omega}~,
        \label{ha.j}\\
        \boldsymbol{E}(\boldsymbol{x}, t) &=& \boldsymbol{E}
        (\boldsymbol{x} - \boldsymbol{v} t)~ , ~
        \boldsymbol{B}(\boldsymbol{x}, t) =
        \boldsymbol{B}(\boldsymbol{x} - \boldsymbol{v}
        t)\label{ha.k}
\end{eqnarray}
and have to determine the solutions of
\begin{eqnarray}
      && - \boldsymbol{v} \cdot \boldsymbol{\nabla} \boldsymbol{B}
       = - \boldsymbol{\nabla} \times
       \boldsymbol{E}~,\nonumber\\
        &&-\boldsymbol{v} \cdot \boldsymbol{\nabla}\boldsymbol{E}=
       \boldsymbol{\nabla} \times \boldsymbol{B} - \rho
       (\boldsymbol{v} + \boldsymbol{\omega} \times
       \boldsymbol{x})\,,\label{ha.l} \\
      && \boldsymbol{\nabla} \boldsymbol{E} =\rho~ ,~
        \boldsymbol{\nabla} \cdot \boldsymbol{B}=0\,,\nonumber\\
       &  &0=\int d^3 x \rho (\boldsymbol{x}) [\boldsymbol{E}
        (\boldsymbol{x}) + (\boldsymbol{v} +\boldsymbol{\omega} \times
        \boldsymbol{x}) \times
        \boldsymbol{B}(\boldsymbol{x})]\,,\label{ha.m}\\
        &&0= \int d^3 x \rho(\boldsymbol{x}) \boldsymbol{x} \times [\boldsymbol{E}
        (\boldsymbol{x}) + (\boldsymbol{v} + \boldsymbol{\omega} \times
        \boldsymbol{x}) \times \boldsymbol{B}(\boldsymbol{x})]\,.\label{ha.n}
\end{eqnarray}
We will solve (\ref{ha.l}) to (\ref{ha.n}) in Fourier space.

The inhomogeneous Maxwell equations (\ref{ha.l}) are solved by
\begin{equation}
       \widehat{\boldsymbol{E}} =\widehat{\boldsymbol{E}}_1 +
       \widehat{\boldsymbol{E}}_2 ~,~ \widehat{\boldsymbol{B}} =
       \widehat{\boldsymbol{B}}_1
       + \widehat{\boldsymbol{B}}_2\label{ha.o}
\end{equation}
with
\begin{eqnarray}
        &&\widehat{\boldsymbol{E}}_1 (\boldsymbol{k}) = - i[\boldsymbol{k}^2 -
         (\boldsymbol{k} \cdot  \boldsymbol{v})^2]^{-1} (\boldsymbol{k}-
         (\boldsymbol{k} \cdot \boldsymbol{v})\boldsymbol{v})
         \hat\rho (\boldsymbol{k})\,,\label{ha.p}\\
        &&\widehat{\boldsymbol{E}}_2(\boldsymbol{k}) = -
        [\boldsymbol{k}^2 - (\boldsymbol{k} \cdot
        \boldsymbol{v})^2]^{-1} ( \boldsymbol{\omega} \times
        \boldsymbol{k}) (\boldsymbol{v} \cdot
        \nabla_{\boldsymbol{k}}) \hat\rho
        (\boldsymbol{k})\,,\label{ha.q}
\end{eqnarray}
and
\begin{eqnarray}
        &&\widehat{\boldsymbol{B}}_1 (\boldsymbol{k}) =  i[\boldsymbol{k}^2 -
         (\boldsymbol{k} \cdot  \boldsymbol{v})^2]^{-1}
         (\boldsymbol{k}\times
         \boldsymbol{v}) \hat\rho(\boldsymbol{k})\,,\label{ha.r}\\
        &&\widehat{\boldsymbol{B}}_2(\boldsymbol{k}) = -
        [\boldsymbol{k}^2 - (\boldsymbol{k} \cdot
        \boldsymbol{v})^2]^{-1} ( \boldsymbol{k} \times
        (\boldsymbol{\omega} \times \nabla_{\boldsymbol{k}}))
       \hat\rho(\boldsymbol{k})\,.\label{ha.s}
\end{eqnarray}
Note that $\widehat{\boldsymbol{E}}_1 ,~\widehat{\boldsymbol{B}}_1$
are odd and $\widehat{\boldsymbol{E}}_2,~\widehat{\boldsymbol{B}}_2$
are even in $\boldsymbol{k}$.

Using that odd terms vanish, in the Lorentz force equation we have
\begin{eqnarray}\label{ha.t}
       &&-\int d^3 k \hat\rho^\ast [ \boldsymbol{k}^2 - (\boldsymbol{v}
        \cdot \boldsymbol{k})^2]^{-1} (\boldsymbol{\omega}
        \times\boldsymbol{k}) (\boldsymbol{v} \cdot
        \nabla_{\boldsymbol{k}})\hat\rho\nonumber\\
        &&-\int d^3 k \hat\rho^\ast [ \boldsymbol{k}^2 - (\boldsymbol{v}
        \cdot \boldsymbol{k})^2]^{-1} \boldsymbol{v} \times (\boldsymbol{k}
        \times (\boldsymbol{\omega} \times
        \nabla_{\boldsymbol{k}}))\hat\rho\nonumber\\
        &&+ \int d^3 k ((\boldsymbol{\omega}\times
        \nabla_{\boldsymbol{k}})\hat\rho^\ast) [\boldsymbol{k}^2-
         (\boldsymbol{v} \cdot \boldsymbol{k})^2]^{-1}\times (\boldsymbol{k}
         \times \boldsymbol{v})
        \hat\rho \nonumber\\
        &=&-\int d^3 k \hat\rho^\ast [ \boldsymbol{k}^2 - (\boldsymbol{v}
        \cdot \boldsymbol{k})^2]^{-1} |\boldsymbol{k}|^{-1}
        \hat\rho_r^\prime\nonumber\\
        &&\qquad\qquad\big( (\boldsymbol{\omega}\times
        \boldsymbol{k}) (\boldsymbol{v}\cdot
        \boldsymbol{k}) + \boldsymbol{v}\times
        (\boldsymbol{k} \times (\boldsymbol{\omega}\times
        \boldsymbol{k})) - ((\boldsymbol{\omega}\times
        \boldsymbol{k}) \cdot \boldsymbol{v})
         \boldsymbol{k}\big)= 0\nonumber\\
\end{eqnarray}
for every $\boldsymbol{v}$ and $\boldsymbol{\omega}$, where we
took into account that $\hat\rho$ is radial.

The Lorentz torque equation requires more work. Using again that
odd terms vanish, we have
\begin{eqnarray}
      && i \int d^3 k \hat\rho^\ast \big (\nabla_{\boldsymbol{k}}
       \times \widehat{\boldsymbol{E}}_1 + \nabla_{\boldsymbol{k}}
       \times (\boldsymbol{v} \times\widehat{\boldsymbol{B}}_1) +
       \nabla_{\boldsymbol{k}} \times ((
       \boldsymbol{\omega}\times
       i \nabla_{\boldsymbol{k}})
       \times\widehat{\boldsymbol{B}}_2\big )\nonumber\\
       &=& - \int d^3
       k|\boldsymbol{k}|^{-1}\hat\rho^{\ast\prime}_r
       [\boldsymbol{k}^2 -
       (\boldsymbol{k}\cdot\boldsymbol{v})^2]^{-1}
       \hat\rho\nonumber\\
       && \qquad\qquad\big (  \boldsymbol{k}\times (\boldsymbol{k}- (\boldsymbol{k}
       \cdot \boldsymbol{v})\boldsymbol{v}) - \boldsymbol{k}\times
       (\boldsymbol{v} \times (\boldsymbol{k} \times
       \boldsymbol{v}))\big)\nonumber\\
       &&+  \int d^3 k |\boldsymbol{k}|^{-1}
       \hat\rho^{\ast\prime}_r
       \boldsymbol{k}\times ((\boldsymbol{\omega}
       \times \nabla_{\boldsymbol{k}}) \times
       \widehat{\boldsymbol{B}}_2)\nonumber\\
       &=&\int d^3
       k|\boldsymbol{k}|^{-1}\hat\rho^{\ast\prime}_r
       \boldsymbol{k} \times (\nabla_{\boldsymbol{k}}( \boldsymbol{\omega}
       \cdot \widehat{\boldsymbol{B}}_2) -  \boldsymbol{\omega}
        \nabla_{\boldsymbol{k}} \cdot
        \widehat{\boldsymbol{B}}_2)\nonumber\\
        &=& -\int d^3
       k|\boldsymbol{k}|^{-1}\hat\rho^{\ast\prime}_r
       (\boldsymbol{k} \times \boldsymbol{\omega})
       \nabla_{\boldsymbol{k}} \cdot
        \widehat{\boldsymbol{B}}_2\,.\label{ha.u}
\end{eqnarray}
For the divergence of $ \widehat{\boldsymbol{B}}_2$ we find
\begin{equation}
        \nabla_{\boldsymbol{k}} \cdot
        \widehat{\boldsymbol{B}}_2= 2 [\boldsymbol{k}^2 -
        (\boldsymbol{k} \cdot \boldsymbol{v})^2]^{-2}
        \boldsymbol{k}^2 (\boldsymbol{\omega}
        \cdot\nabla_{\boldsymbol{k}} - (\boldsymbol{v} \cdot
        \boldsymbol{\omega}) (\boldsymbol{v} \cdot
        \nabla_{\boldsymbol{k}})) \hat\rho\label{ha.v}
\end{equation}
and therefore zero Lorentz torque results in the equation
\begin{equation}
       \int d^3 k |\nabla_{\boldsymbol{k}} \hat\rho|^2 2
       [\boldsymbol{k}^2 - (\boldsymbol{k} \cdot
       \boldsymbol{v})^2]^{-2} (\boldsymbol{k} \times
       \boldsymbol{\omega}) (\boldsymbol{\omega}
       \cdot\boldsymbol{k} - (\boldsymbol{v} \cdot
       \boldsymbol{\omega}) (\boldsymbol{v}\cdot \boldsymbol{k}))
       = 0\,.\label{ha.w}
\end{equation}
Taking into account that $\hat\rho$ is radial, the torque vanishes
only if either $\boldsymbol{\omega}\parallel \boldsymbol{v}$ or
$\boldsymbol{\omega}
\bot \boldsymbol{v}.$ If $\boldsymbol{v} = 0$, the torque
vanishes always. For $\boldsymbol{\omega}$ oblique to $\boldsymbol{v}$
Equations (\ref{ha.l}) to (\ref{ha.n}) have no solution.

Physically the charge distribution is rigid, but the
electromagnetic fields are Lorentz contracted along
$\boldsymbol{v}$.
This mismatch yields a nonvanishing torque unless $\boldsymbol{\omega}
\parallel \boldsymbol{v}$, resp. $\boldsymbol{\omega}
\bot\boldsymbol{v}$. Clearly, this mismatch is an artifact of the
semi--relativistic Abraham model. For a relativistic extended charge
distribution there is a charged soliton for every $\boldsymbol{v}$
and $\boldsymbol{\omega}$.

In the adiabatic limit, there must be two disjoint effective
equations of motion. If $\boldsymbol{\omega} \parallel \boldsymbol{v}$
initially, then it will remain so approximately and there is a
closed equation for $\boldsymbol{q}, \boldsymbol{v}$ and $\boldsymbol{\omega} \cdot
\boldsymbol{v}$. Similarly, if $\boldsymbol{\omega} \bot \boldsymbol{v}$
initially, this property is almost preserved in time. $\boldsymbol{v}$
and $\boldsymbol{\omega}- (\boldsymbol{v} \cdot\boldsymbol{\omega})\boldsymbol{\omega}$
vary slowly on the same time scale. We expect that through the emission of radiation
 an $\boldsymbol{\omega}$ initially
oblique to $\boldsymbol{v}$
will rapidly relax to either being parallel or orthogonal to
$\boldsymbol{v}$,
depending on the initial conditions. This admittedly rather
sketchy picture raises some interesting dynamical questions, in
particular how precisely the spinning particle succeeds  in slowly
turning its axis of rotation. We have not persued this issue,
since it leads away from the quantum spin.


\subsection{Relativistic dynamics of charged particle with
spin}\label{sec.hb}
To be supplied.

\subsection*{Notes and References}
{\it ad \ref{sec.h}}: BMT is an acronym for Bargmann, Michel,
Telegdi (1955). The BMT equation is explained in Jackson (1999).
Bailey, Picasso (1970) is an informative article on how the BMT equation is used in
the analysis of the high precision measurements of the electron
and muon  $g$-factor. The BMT equation with $g=2$ is the
semiclassical limit of the Dirac equation,  Rubinow, Keller (1963),
Bolte, Keppeler (1999), Spohn (1999c).\bigskip\\
{\it ad \ref{sec.ha}}: Kiessling (1999) observes that the usual
form of the total angular momentum is conserved only if the inner
rotation of the charged particles is included. The charge soliton
solutions with spin are determined in Spohn (1999b).\bigskip\\
{\it ad \ref{sec.hb}}: Just as for translational degrees of
freedom, one way to guess the correct effective spin dynamics is
to impose Lorentz invariance. In addition, one could require that
the
equations of motion come from a Lagrangian action. In full
generality, including an electric dipole moment, this program was
carried out by Bhabha, Corben (1941). Alternative approaches are
compared in Corben (1961), Nyborg (1962). A concise and useful summary is given by Barut
(1964) who discusses also how the BMT equation fits into the
general scheme. In the relativistic extended charge model of
Nodvik (1964) spin is included and the effective equation of
motion is derived formally. Recent work on this model is Appel,
Kiessling (1999).
\newpage
\section{Many Charges}\label{sec.i}
\setcounter{equation}{0}
There is little effort in extending the Abraham model to several
particles. We label their positions and velocities as
$\boldsymbol{q}_j(t), \boldsymbol{v}_j (t), j=1, \ldots, N$. The
$j$-th particle has the bare mass $m_{\mathrm{b}j}$ and the
charge $e_j$, where for simplicity all particles have the same
form factor $f$, i.e. $\rho_j(\boldsymbol{x}) = e_j f(\boldsymbol{x}^2)$. The
motion of each particle is governed by the Lorentz force equation,
as before, and the current in the Maxwell equations becomes now
the sum over the single particle currents. Therefore the equations
of motion read
\setcounter{equation}{0}
\begin{eqnarray}
        \partial_t \boldsymbol{B}(\boldsymbol{x}, t) &=& - \nabla
        \times \boldsymbol{E}(\boldsymbol{x}, t)\,,\nonumber\\
        \partial_t \boldsymbol{E}(\boldsymbol{x}, t) &=&  \nabla
        \times \boldsymbol{B}(\boldsymbol{x}, t)- \sum_{j=1}^N\,
         \rho_j (\boldsymbol{x}-
         \boldsymbol{q}_j(t)) \boldsymbol{v}_j (t)\,,\label{i.a}\\
         \nabla \cdot  \boldsymbol{E}(\boldsymbol{x}, t) &=&  \sum_{j=1}^N\,
         \rho_j (\boldsymbol{x}- \boldsymbol{q}_j(t))~, ~ \nabla \cdot \boldsymbol{B}
         (\boldsymbol{x},t) = 0\,,\nonumber\\
         \frac{d}{dt} \big( m_{\mathrm{b}i}\,
         \gamma_j\boldsymbol{v}_j (t)\big) &=& \int d^3 x
         \rho(\boldsymbol{x}-\boldsymbol{q}_i(t)) \big(
         \boldsymbol{E}(\boldsymbol{x}, t) + \boldsymbol{v}_i(t)
         \times
         \boldsymbol{B}(\boldsymbol{x},t)\big)\,,\label{i.b}
\end{eqnarray}
$i=1, \ldots, N,$ with $\gamma_i = (1- \boldsymbol{v}_i^2)^{-1/2}$.

There are no external forces. Thus the force acting on a given
particle is due to the other particles as mediated through the
Maxwell field. In addition, there is the self--force which we have
discussed already at length. Physically we trust our model only if
particles are very far apart on the scale set by $R_\rho$. If two
particles are at a distance of a few $R_\rho$, then there are
strong forces which depend on the details of the phenomenological
and unknown charge distribution. Thus we assume that initially
\begin{equation}\label{i.c}
       |\boldsymbol{q}_i^0 - \boldsymbol{q}_j^0| =
       \mathcal{O} (\varepsilon^{-1} R_\rho)~, ~
       i \not=j\,.
\end{equation}
We emphasize that the scale parameter $\varepsilon$ enters only
through the initial conditions. $\varepsilon^{-1}$ is the typical
distance of particles measured in units of $R_\rho$. For the
initial fields it is natural to again impose the condition of no
slip. Then they are a linear superposition of charge soliton
fields corresponding to
$\boldsymbol{q}_i^0,~ \boldsymbol{v}_i^0,~ i=1, \ldots, N$.

We expect (\ref{i.c}) to remain valid at least over a certain
macroscopic time span and we want to understand whether in this
limiting regime there is a closed dynamics for the particles
by themselves. Since the particles are far apart, it takes a time
of order $\varepsilon^{-1} R_\rho/c = \varepsilon^{-1} t_\rho$ for
light to travel inbetween  and the force on a given particle
depends on the other particles at a macroscopically retarded time.
This means that the effective equations of motion are closed, but
nonlocal with a structure to be explained in the following
section. In many circumstances the velocities can be regarded as
small,
$|\boldsymbol{v}_j|/c\ll 1$, and retardation effects are neglible.
To lowest order this yields then the static Coulomb interaction.
Somewhat unexpected even the first order correction has still the
form of an effective Lagrangian.
\subsection{Retarded interaction}\label{sec.ia}
We insert the solution of the inhomogeneous Maxwell equation
(\ref{i.a}) into the Lorentz force equation (\ref{i.b}). The
forces are additive and the force on particle $i$ naturally splits
into self--force $(j=i)$ and a mutual force $(j\not=i)$. For the
self--force we use the Taylor expansion of Section \ref{sec.e}.
Thereby the mass is renormalized and the next order is the
radiation reaction. For the mutual force we recall that in Section
\ref{sec.fd} we showed already that to leading order  the field
generated by charge $j$ is the Li\'{e}nard--Wiechert field. Thus,
ignoring radiation reaction, we obtain as retarded equations of
motion
\begin{equation}\label{ia.a}
        m_i( \boldsymbol{v}_i)
          \dot{\boldsymbol{v}_i}= \sum_{j=1\atop{j} \not
       =i}^{N}\, e_j \big( \boldsymbol{E}_{\mathrm{ret}j}
       (\boldsymbol{q}_i, t) +\boldsymbol{v}_i \times
       \boldsymbol{B}_{\mathrm{ret}j} (\boldsymbol{q}_i,
       t)\big)~,~ t \ge 0\,.
\end{equation}
Here $m_i$ is the effective mass of particle $i$ as defined in
(\ref{f.a}). $\boldsymbol{E}_{\mathrm{ret}j}(\boldsymbol{x},t)$
equals (\ref{aa.u}) with $\boldsymbol{q}$ replaced by $\boldsymbol{q}_j$
and $t_{\mathrm{ret}}$ replaced by  $t_{\mathrm{ret}j}$ which is
defined by
\begin{equation}\label{ia.b}
        t_{\mathrm{ret}j}= t - |\boldsymbol{x}- \boldsymbol{q}_j
        (t_{\mathrm{ret}j})|\,.
\end{equation}
For $\boldsymbol{x}=\boldsymbol{q}_i$ the retarded time is $\mathcal{O}(\varepsilon^{-1}
t_\rho)$.
Similarly $\boldsymbol{B}_{\mathrm{ret}j} (\boldsymbol{x},t)$
equals (\ref{aa.v}) with $\boldsymbol{q}$ replaced by $\boldsymbol{q}_j$
and $t_{\mathrm{ret}}$ replaced by $t_{\mathrm{ret}j}$. Note that
the equations of motion in (\ref{bc.e}) have the same structure.

To solve (\ref{ia.a}) one needs the trajectories for the whole
past. Our assumption of no initial slip is equivalent to
\begin{equation}\label{ia.c}
        \boldsymbol{q}_i(t) +\boldsymbol{q}_i^0 + t\boldsymbol{v}_i^0
        ~,~ i=1, \ldots, N~, ~ t \le 0
\end{equation}
which must be added to (\ref{ia.a}).

Using (\ref{ia.a}) we can estimate the size of the various
contributions. The far field contributions to
$\boldsymbol{E}_{\mathrm{ret}j}$ and
$\boldsymbol{B}_{\mathrm{ret}j}$ are of
$\mathcal{O}(\varepsilon^2)$ and the near field contributions are
$\mathcal{O}(\varepsilon)\dot{\boldsymbol{v}}_j$. Thus $\dot{\boldsymbol{v}}_i=
\mathcal{O}(\varepsilon^2)$
and the first order correction from the
near field is $\mathcal{O}(\varepsilon^3)$. In the next order we
see the radiation reaction which is proportional to
$\dot{\boldsymbol{v}}_i^2$ and thus $\mathcal{O}(\varepsilon^4)$.
In (\ref{ia.a}) we would have to add the Lorentz--Dirac term of
(\ref{f.a}) for particle $i$. As a consequence (\ref{ia.a}) will
pick up runaway solutions and we have to restrict to the critical
manifold. Thereby the friction force becomes retarded. This is not
surprising. The friction comes from acceleration which is due to
the retarded motion of all other charges.

The issues raised here remain largely unexplored, at present. One
would like to know how well (\ref{ia.a}) approximates the true
dynamics and over what time scale.


\subsection{Limit of small velocities} \label{sec.ib}
We impose that the initial velocities are small. The natural scale
turns out to be
\begin{equation}\label{ib.a}
       |\dot{\boldsymbol{v}}_j| =\mathcal{O}(\sqrt{\varepsilon} c)
\end{equation}
and, of course, we have to show that this order is maintained, at
least for a certain time span. To preserve the relation
$\dot{\boldsymbol{q}}=\boldsymbol{v}$ we
have to adjust the time scale as $\varepsilon^{-1/2}$
relative to the macroscopic time scale. The accumulated force is then of order
$\sqrt{\varepsilon}$ which just balances the velocity in
(\ref{ib.a}). Therefore we arrive at the following scale
transformation
\begin{eqnarray}
       &&t= \varepsilon^{-3/2} t^\prime ,~\boldsymbol{q}_j =
       \varepsilon^{-1}\boldsymbol{q}_j^\prime~
       , ~ \boldsymbol{v}_j = \sqrt{\varepsilon}
       \boldsymbol{v}_j^\prime\,,\label{ib.b}\\
       &&\boldsymbol{x}= \varepsilon^{-1} \boldsymbol{x}^\prime ,~
       \boldsymbol{E}=
       \varepsilon^{3/2} \boldsymbol{E}^\prime ~,~ \boldsymbol{B}
       =\varepsilon^{3/2} \boldsymbol{B}^\prime\,,\nonumber
\end{eqnarray}
where the primed quantities are considered to be of ${\mathcal
O}(1)$. The field amplitudes are scaled by $\varepsilon^{3/2}$ so
to preserve the field energy.

There is little risk of confusion in omitting the primes. We set
\begin{equation}\label{ib.c}
         \boldsymbol{q}_j^\varepsilon (t)   = \varepsilon
         \boldsymbol{q}_j \big(\varepsilon^{-3/2} t\big)~,\quad
       \boldsymbol{v}_j^\varepsilon(t) = \varepsilon^{-1/2}
         \boldsymbol{v}_j (\varepsilon^{-3/2} t)\,.
\end{equation}
Then the rescaled Maxwell--Lorentz equations are
\begin{eqnarray}
      \sqrt{\varepsilon}\, \partial_t
      \boldsymbol{B}(\boldsymbol{x},t)&=&
      - \nabla \times \boldsymbol{E}(\boldsymbol{x},t)\,,\nonumber\\
      \sqrt{\varepsilon}\, \partial_t
      \boldsymbol{E}(\boldsymbol{x},t)&=& \nabla \times
      \boldsymbol{B}(\boldsymbol{x},t)
      - \sum_{j=1}^N \sqrt{\varepsilon} \boldsymbol{v}_j^\varepsilon (t)
      \sqrt{\varepsilon} \rho_{j,\varepsilon} (\boldsymbol{x}-
      \boldsymbol{q}_j^\varepsilon(t))\,,\label{ib.d}\\
      \nabla \cdot\boldsymbol{E}(\boldsymbol{x},t)&=&\sum_{j=1}^N
      \sqrt{\varepsilon}\rho_{j,\varepsilon} (\boldsymbol{x}-
      \boldsymbol{q}_j^\varepsilon(t)) ~ , ~ \nabla \cdot
      \boldsymbol{B}(\boldsymbol{x},t)=0\,,\nonumber
\end{eqnarray}

\begin{eqnarray}
      \lefteqn {\varepsilon \frac{d}{dt}\big ( m_{\mathrm{b}i} (1-
      \varepsilon \boldsymbol{v}_i^\varepsilon (t)^2)^{-1/2}
      \boldsymbol{v}_i^\varepsilon(t)\big) } \label{ib.e}\\
     &=& \int d^3 x \sqrt{\varepsilon} \rho_{\varepsilon,i}
      (\boldsymbol{x}- \boldsymbol{q}_i^\varepsilon(t))
      \big (\boldsymbol{E}(\boldsymbol{x},t)
      + \sqrt{\varepsilon} \boldsymbol{v}_i^\varepsilon(t) \times
      \boldsymbol{B}(\boldsymbol{x},t)\big)\,.\nonumber
\end{eqnarray}
On the new scale the velocity of light tends to infinity as $c/\sqrt{\varepsilon}$
and the charge distribution has total charge $\sqrt{\varepsilon}$,
finite electrostatic energy $m_{\mathrm{e}}$, and shrinks to a
$\delta$-function as $\sqrt{\varepsilon} \rho_{\varepsilon,j}
(\boldsymbol{x})
= \sqrt{\varepsilon} \,\varepsilon^{-3} \rho_j (\varepsilon
\boldsymbol{x})$. Recall that the scale parameter $\varepsilon$ is
just a convenient way to order the magnitudes of the various
contributions.

Before entering into more specific computations, it is useful to
first sort out what should be expected. We follow our practice
from before and denote positions and velocities of the comparison
dynamics by $\boldsymbol{r}_j, \boldsymbol{u}_j, j=1, \ldots, N,$ i.e. $
\boldsymbol{q}_j^\varepsilon (t) \cong \boldsymbol{r}_j (t),
\boldsymbol{v}_j^\varepsilon (t) \cong \boldsymbol{u}_j(t)$. Since the velocities are small,
the kinetic energy takes its nonrelativistic limit
\begin{equation}\label{ib.f}
       T_0 (\boldsymbol{u}_j) = \frac{1}{2} \big (
       m_{\mathrm{b}j}+
       \frac{4}{3}\, m_{\mathrm{e}j}  \big ) \boldsymbol{u}_j^2\,,
\end{equation}
up to a constant, compare with (\ref{ca.s}). Note that the mass  of the particle is renormalized
through the interaction with the field. For small velocities
magnetic fields  are small and retardation effects can be
neglected. Thus the potential energy of the effective dynamics
should be purely Coulombic and be given by
\begin{equation}\label{ib.g}
       U_0(\boldsymbol{r}_1, \ldots, \boldsymbol{r}_N) = \frac{1}{2} \sum_{i\not=
       j=1}^N \frac{e_i e_j}{4 \pi |\boldsymbol{r}_i-\boldsymbol{r}_j|}~.
\end{equation}

So what is the next order? For the kinetic energy we merely expand
in (\ref{ca.s}) with the result
\begin{equation}\label{ib.h}
       T_1 (\boldsymbol{u}_j) =   \varepsilon
       \big( \frac{1}{8} m_{\mathrm{b}j} + \frac{2}{15}
       m_{\mathrm{e}j}\big)\boldsymbol{u}_j^4 \,.
\end{equation}
The next order correction to the Coulomb forces requires more
explicit considerations, which will be explained in the following
section. There are corrections due to retardation and to the
magnetic field, which combine into a velocity dependent potential
as
\begin{equation}\label{ib.i}
       U_1(\boldsymbol{r}_1, \boldsymbol{u}_1,\ldots,
        \boldsymbol{r}_N,\boldsymbol{u}_N) = -
       \varepsilon \frac{1}{4} \sum_{i\not=
       j=1}^N \frac{e_i e_j}{4 \pi |\boldsymbol{r}_i-\boldsymbol{r}_j|}
       \big ( \boldsymbol{v}_i \cdot\boldsymbol{v}_j +
       (\boldsymbol{v}_i\cdot
        \widehat{\boldsymbol{r}}_{ij})
        (\widehat{\boldsymbol{r}}_{ij} \cdot \boldsymbol{v}_j)\big)
\end{equation}
with $\widehat{\boldsymbol{r}}_{ij} = (\boldsymbol{r}_i - \boldsymbol{r}_j)
 /|\boldsymbol{r}_i- \boldsymbol{r}_j|\,.$

In principle we could continue the expansion. It is of interest to
see  at what scale radiation effects will be important. They are
proportional to $\dot{\boldsymbol{v}}^2$. On the microscopic scale
$\dot{\boldsymbol{v}} \cong \varepsilon^2$, as argued before. In
rescaled velocities and accumulated over the time span $\varepsilon^{-3/2}$
this results in a loss of energy of the order $\varepsilon^{3/2}$.
Thus the next order correction to the comparison dynamics is
dissipative and of order $\varepsilon^{3/2}$.

We recall that $|\boldsymbol{v}_j
|/c = {\mathcal O} (\sqrt{\varepsilon})$. Thus we may set $\varepsilon = 1$
at the expense of reintroducing the velocity of light, $c$. Then
up to an error of order  $(|\boldsymbol{v}_j|/c)^{3/2}$ the
effective dynamics of the $N$ charges is conservative and is
governed by the Lagrangian
\begin{eqnarray}\label{ib.j}
        && L_{\mathrm{Darwin}} = \sum_{j=1}^N
        \big( ( m_{\mathrm{b}j} + \frac{4}{3} m_{\mathrm{e}j})
        \frac{1}{2}\boldsymbol{u}_j^2 +
        (\frac{1}{8}  m_{\mathrm{b}j} + \frac{2}{15}
        m_{\mathrm{e}j}) c^{-2}\boldsymbol{u}_j^4 \big)
        \nonumber\\
        &&- \frac{1}{2}\sum_{i\not=
       j=1}^N
        \frac{e_i e_j}{4 \pi |\boldsymbol{r}_i-\boldsymbol{r}_j|}
       \big [ 1 - \frac{1}{2 c^2}\,
       \big(\boldsymbol{u}_i \cdot \boldsymbol{u}_j  +(\boldsymbol{u}_i \cdot
        \widehat{\boldsymbol{r}}_{ij})
        (\boldsymbol{u}_j \cdot
        \widehat{\boldsymbol{r}}_{ij})\big)\big ]\,.\nonumber
\end{eqnarray}
$ L_{\mathrm{Darwin}}$ is known as the Darwin Lagrangian and  widely used in
plasma physics.


\subsection{The Darwin Lagrangian}\label{sec.ic}
As can be seen from ({\ref{i.a}), (\ref{i.b}) the
 forces
are additive. Thus it suffices to consider two particles only.
As initial conditions we choose the linear superposition of the two charge
 solitons corresponding to the initial data $\boldsymbol{q}_i^0,
 \boldsymbol{v}_i^0~,~
 i=1,2$. We
solve Maxwell equations and insert in the Lorentz force. As already
explained, in the self--interaction the contribution from the
initial fields vanishes for $t \ge \overline t_\rho$. In the
mutual interaction the initial fields take a time of order $\sqrt{\varepsilon}$
to reach the other particle and their contribution vanishes for
$t\ge \sqrt{\varepsilon}|\boldsymbol{q}_1^0 -
\boldsymbol{q}_2^0|$.
Thus for larger times we are allowed to insert in (\ref{i.b}) the
retarded fields only, which yields
\begin{eqnarray}\label{ic.a}
        \varepsilon\frac{d}{dt} \big (
       m_{\mathrm{b}1}  \, \gamma_1
       \boldsymbol{v}_1^\varepsilon(t)\big)
           &=& \boldsymbol{F}_{\mathrm{ret},11} (t) + \boldsymbol{F}_{\mathrm{ret},12}
           (t)\,,\\
         \varepsilon\frac{d}{dt} \big (
       m_{\mathrm{b}2}  \, \gamma_2 \boldsymbol{v}_2^\varepsilon(t)\big)
           &=& \boldsymbol{F}_{\mathrm{ret},21} (t) + \boldsymbol{F}_{\mathrm{ret},22}
           (t)\,,\label{ic.b}
\end{eqnarray}
where
\begin{eqnarray}
          \boldsymbol{F}_{\mathrm{ret},ij}(t) &=& \int\limits_0^t ds \int d^3 k
          \hat\rho_i^{\,\ast} (\varepsilon  \boldsymbol{k}) \hat\rho_j (\varepsilon
           \boldsymbol{k})
          e^{i \boldsymbol{k} \cdot (\boldsymbol{q}_i^\varepsilon (t) -
          \boldsymbol{q}_j^\varepsilon(s))}\nonumber\\
          && \Big( - \varepsilon^{1/2}(|\boldsymbol{k}|^{-1} \sin (|\boldsymbol{k}|
          (t-s)/\sqrt{\varepsilon}) ) i\boldsymbol{k} - \varepsilon
          (\cos(|\boldsymbol{k}|(t-s)/\sqrt{\varepsilon}))
          \boldsymbol{v}_j^\varepsilon(s)\nonumber\\[2mm]
          &&+ \varepsilon^{3/2}   (|\boldsymbol{k}|^{-1} \sin (|\boldsymbol{k}| (t-s)/
          \sqrt{\varepsilon}) \boldsymbol{v}_i^\varepsilon (t)
          \times \big (i\boldsymbol{k} \times
          \boldsymbol{v}_j^\varepsilon (s)\big)\Big )\label{ic.c}\,,
\end{eqnarray}
$i,j=1,2$.

For the self--interaction we set $\varepsilon \boldsymbol{k}
=\boldsymbol{k}^\prime \,, \varepsilon^{-3/2} t =
t^\prime$. Then
\begin{eqnarray}
        \boldsymbol{F}_{\mathrm{ret},11}(t)&=& \varepsilon^{3/2}
        \int\limits_0^\infty d \tau \int d^3 k |\hat \rho_1(\boldsymbol{k})
        |^2 e^{i\boldsymbol{k}\cdot(\boldsymbol{q}_1^\varepsilon (t)-
        \boldsymbol{q}_1^\varepsilon (t - \varepsilon^{3/2}
        \tau))/\varepsilon}\nonumber\\
        && \Big( - \sqrt{\varepsilon} ( |\boldsymbol{k}|^{-1} \sin |\boldsymbol{k}|
        \tau ) i\boldsymbol{k} - \varepsilon (\cos|\boldsymbol{k}|\tau)
        \boldsymbol{v}_1^\varepsilon
       (t- \varepsilon^{3/2} \tau)\nonumber\\[2mm]
        &&+ \varepsilon^{3/2} ( |\boldsymbol{k}|^{-1} \sin |\boldsymbol{k}| \tau)
        \boldsymbol{v}_1^\varepsilon (t) \times \big(i
        \boldsymbol{k} \times \boldsymbol{v}_1^\varepsilon (t-\varepsilon^{3/2} \tau)
        \big)\Big)\,.\label{ic.d}
\end{eqnarray}
We Taylor expand as
\begin{eqnarray}
        \varepsilon^{-1} (\boldsymbol{q}_1^\varepsilon(t) -
        \boldsymbol{q}_1^\varepsilon (t- \varepsilon^{3/2}\tau ) )
        &=& \varepsilon^{1/2}
        \tau \boldsymbol{v}
        - \frac{1}{2} \varepsilon^2 \tau^2
        \dot{\boldsymbol{v}}\,,\nonumber\\
        \boldsymbol{v}_1^\varepsilon (t-\varepsilon^{3/2} \tau)
        &=& \boldsymbol{v}
        - \varepsilon^{3/2}\tau
        \dot{\boldsymbol{v}}\,.\label{ic.e}
\end{eqnarray}
Then, up to errors of order $\varepsilon^{5/2}$,
\begin{eqnarray}
        \boldsymbol{F}_{\mathrm{ret},11}(t)&=&
        \int\limits_0^\infty d \tau \int d^3 k |\hat \rho_1(\boldsymbol{k})
        |^2 \Big \{ \varepsilon \big[ - (|\boldsymbol{k}|^{-1}\sin |\boldsymbol{k}|
        \tau )\, \frac{1}{2}
        \tau^2 (\boldsymbol{k}\cdot \dot{\boldsymbol{v}})
        \boldsymbol{k}\nonumber\\
        &&+ (\cos|\boldsymbol{k}| \tau)\tau \dot{\boldsymbol{v}} \big ] +
        \varepsilon^2 \big[ \big( - (|\boldsymbol{k}|^{-1} \sin |\boldsymbol{k}|
        \tau ) \frac{1}{2} \tau^2 (\boldsymbol{k}
        \cdot\dot{\boldsymbol{v}} )\boldsymbol{k}\nonumber\\
        &&+ (\cos|\boldsymbol{k}|\tau) \tau
        \dot{\boldsymbol{v}}\big ) \big(- \frac{1}{2} \tau^2
        (\boldsymbol{k} \cdot \boldsymbol{v})^2\big )\nonumber\\
        &&- (\cos |\boldsymbol{k}|\tau) \frac{1}{2} \tau^3 (\boldsymbol{k}
        \cdot \dot{\boldsymbol{v}})(\boldsymbol{k}
        \cdot\boldsymbol{v})\boldsymbol{v} + (
        |\boldsymbol{k}|^{-1} \sin |\boldsymbol{k}|\tau
         )\nonumber\\
        &&\big(\tau^2 (\boldsymbol{k}\cdot\boldsymbol{v}) \boldsymbol{v}
        \times (\boldsymbol{k} \times \dot{\boldsymbol{v}})
        + \frac{1}{2} \tau^2
        (\boldsymbol{k}\cdot\dot{\boldsymbol{v}}) (\boldsymbol{v}
        \times (\boldsymbol{k}\times
        \boldsymbol{v}))\big)\big]\Big\}\,,\label{ic.f}
\end{eqnarray}
which, upon integration, agrees with (\ref{ib.f}), (\ref{ib.h}).

For the mutual interaction we leave the $k$-integration and set
$\varepsilon^{1/2} t=t^\prime$.  Then
\begin{eqnarray}
        \boldsymbol{F}_{\mathrm{ret},12}(t) &=& \sqrt{\varepsilon}
        \int\limits_0^\infty d\tau \int d^3 k
          \hat\rho_1^\ast (\varepsilon \boldsymbol{k}) \hat\rho_2 (\varepsilon
          \boldsymbol{k})
          e^{i\boldsymbol{k} \cdot (\boldsymbol{q}_1^\varepsilon (t) -
          \boldsymbol{q}_2^\varepsilon(t-\sqrt{\varepsilon}\tau))}\nonumber\\
          &&\Big( - \varepsilon^{1/2}  (|\boldsymbol{k}|^{-1}
          \sin |\boldsymbol{k}| \tau ) i \boldsymbol{k} -
          \varepsilon (\cos |\boldsymbol{k}|\tau)   \boldsymbol{v}_2^\varepsilon
          (t- \sqrt{\varepsilon} \tau)\nonumber\\
          && + \varepsilon^{3/2}  ( |\boldsymbol{k}|^{-1} \sin
          |\boldsymbol{k}| \tau)
          \boldsymbol{v}_1^\varepsilon(t) \times \big(i \boldsymbol{k}
          \times \boldsymbol{v}_2^\varepsilon (t- \sqrt{\varepsilon}
          \tau)\big)\Big )\,. \label{ic.g}
\end{eqnarray}
We Taylor expand as
\begin{eqnarray}
        &&\boldsymbol{q}_1^\varepsilon (t) -
          \boldsymbol{q}_2^\varepsilon(t-\sqrt{\varepsilon}\tau) =
         \boldsymbol{r}+ \sqrt{\varepsilon} \tau \boldsymbol{v}_2
          - \frac{1}{2}\, \varepsilon \tau^2 \dot{\boldsymbol{v}}_2\,,\nonumber\\
         && \boldsymbol{v}_1^\varepsilon(t) = \boldsymbol{v}_1~, ~
          \boldsymbol{v}_2^\varepsilon (t- \sqrt{\varepsilon}
          \tau) = \boldsymbol{v}_2-  \sqrt{\varepsilon}
          \tau\dot{\boldsymbol{v}}_2 \label{ic.h}
\end{eqnarray}
with $\boldsymbol{r} = \boldsymbol{q}_1^\varepsilon(t) -
\boldsymbol{q}_2^\varepsilon (t)$.
Then, up to errors of order $\varepsilon^{5/2}$,
\begin{eqnarray}
        \boldsymbol{F}_{\mathrm{ret,12}}&=&
        \int\limits_0^\infty d \tau \int d^3 k \hat\rho_1^\ast(\varepsilon\boldsymbol{k})
        \hat\rho_2(\varepsilon\boldsymbol{k})
        e^{i\boldsymbol{k}\cdot \boldsymbol{r}} \Big\{ - \varepsilon (|
        \boldsymbol{k}|^{-1}
        \sin |\boldsymbol{k}|
        \tau )\, i\boldsymbol{k}\nonumber\\
        &&+ \varepsilon^2 \big[ (|\boldsymbol{k}|^{-1}
        \sin|\boldsymbol{k}|\tau) \big(-\frac{1}{2} \, \tau^2
        (\boldsymbol{k} \cdot \dot{\boldsymbol{v}}_2) \boldsymbol{k}
        + \frac{1}{2} \, \tau^2
        (\boldsymbol{k} \cdot \boldsymbol{v}_2)^2
        i \boldsymbol{k}\nonumber\\
        &&+\boldsymbol{v}_1 \times (i \boldsymbol{k} \times
        \boldsymbol{v}_2)\big) + (\cos |\boldsymbol{k}|\tau) (\tau
        \dot{\boldsymbol{v}}_2 - i\tau (\boldsymbol{k} \cdot
        \boldsymbol{v}_2) \boldsymbol{v}_2) \big]\Big
        \}\nonumber\\[2mm]
       &=& (e_1 e_2 /4 \pi) \Big (-\varepsilon \nabla_{\boldsymbol{r}}
       |\boldsymbol{r}|^{-1} + \varepsilon^2
       \big[ \big(\frac{1}{2}\, \nabla_{\boldsymbol{r}}
       (\dot{\boldsymbol{v}}_2 \cdot  \nabla_{\boldsymbol{r}})-
       \frac{1}{2} \nabla_{\boldsymbol{r}}(\boldsymbol{v}_2 \cdot
       \nabla_{\boldsymbol{r}})^2\big) |\boldsymbol{r}|\nonumber\\
       &&-  (\dot{\boldsymbol{v}}_2 - \boldsymbol{v}_2
       (\boldsymbol{v}_2 \cdot \nabla_{\boldsymbol{r}}))
       |\boldsymbol{r}|^{-1} + (\boldsymbol{v}_1 \times
       (\nabla_{\boldsymbol{r}} \times \boldsymbol{v}_2))
       |\boldsymbol{r}|^{-1}\big ]\Big)\,.\label{ic.i}
\end{eqnarray}

We define the potential part of the Darwin--Lagrangian for two particles as
\begin{equation}\label{ic.j}
         L_P = (e_1 e_2/4 \pi)\Big ( - \frac{1}{|\boldsymbol{r}|
         }\,
         + \frac{\varepsilon}{2}\, \frac{\boldsymbol{v}_1
         \cdot \boldsymbol{v}_2}{|\boldsymbol{r}|}
         \, + \frac{\varepsilon}{2} \frac{(\boldsymbol{v}_1
         \cdot \boldsymbol{r})(\boldsymbol{v}_2 \cdot
         \boldsymbol{r})}{|\boldsymbol{r}|^3}\Big)\,.
\end{equation}
Then
\begin{equation}\label{ic.k}
       \boldsymbol{F}_{\mathrm{ret,12}}(t)=- \varepsilon
       \Big( \frac{d}{dt}\, (\nabla_{\!\!\boldsymbol{v}_1} L_P)
       - \nabla_{\boldsymbol{r}} L_P \Big) +
       \mathcal{O}(\varepsilon^{5/2})\,.
\end{equation}
Inserting (\ref{ic.f}) and (\ref{ic.k}) into the Lorentz force
equation (\ref{ic.a}), (\ref{ic.b}), we conclude that, upon
neglecting contributions of order $\varepsilon^{5/2}$, the
dynamics of the charges is governed by the Darwin Lagrangian
(\ref{ib.j}).
To control our Taylor expansion one has to resort to the
contraction argument of Section \ref{sec.ec}. It becomes now
considerably more involved. We note that in the effective
equations of motion there is no mechanism which would preclude
head on collisions. Thus the dynamics governed by $L_{\mathrm{Darwin}}$
can hold only until the first collision. For the subsequent motion
one has to go back to the full microscopic evolution.
\subsection*{Notes and References}
{\it ad \ref{sec.ib} and \ref{sec.ic}:} The Darwin Lagrangian is
discussed in Jackson (1999). In Kunze, Spohn (1999b) the errors
relative to the motion governed by the Darwin Lagrangian are
estimated.
\newpage
\section*{References}\label{ref}
\begin{list}{}
{\setlength{\leftmargin}{0.5cm} \setlength{\itemindent}{-0.5cm}}

\item Abraham M. (1903):
        Prinzipien der Dynamik des Elektrons, Ann. Physik
        {\bf 10}, 105-179.

\item Abraham M. (1904):
        Die Grundhypothesen der Elektronentheorie, Physikalische
        Zeitschrift {\bf 5}, 576-579.

\item Abraham M. (1905):
        Theorie der Elektrizit\"{a}t, Vol II: Elektromagnetische
        Theorie der Strahlung. Teubner, Leipzig, 2nd edition (1908).

\item Appel W. and Kiessling M.K.-H. (1999):
        In preparation.

\item Bailey J. and Picasso E. (1970):
        The anomalous magnetic moment of the muon and related
        topics, Progress in Nuclear Physics {\bf 12}, 43--75.

\item Bambusi D. (1994):
        A Nekhoroshev--type theorem for the Pauli--Fierz model of
        classical electrodynamics, Ann. Inst. H. Poincar\'e, Phys.
        Th\'eor. {\bf 60}, 339-371.

\item Bambusi D. (1996):
        A proof of the Lorentz--Dirac equation for charged point
        particles, preprint, Univ. Milano.

\item Bambusi D. and Galgani L. (1993):
        Some rigorous results on the Pauli--Fierz model of
        classical electrodynamics, Ann. Inst. H. Poincar\'{e}, Phys.
        Th\'eor. {\bf 58}, 155-171.

\item Bambusi D. and Noja D. (1996):
        On classical electrodynamics of point particles and mass
        renormalization, some preliminary results, Lett. Math. Phys.
        {\bf 37}, 449-460.

\item Bargmann V., Michel L., and Telegdi V.L. (1959):
        Precession of the polarization of particles moving in a
        homogeneous electromagnetic field, Phys. Rev. Lett. {\bf
        2}, 435-436.

\item Barut A.O. (1964):
                Electrodynamics and Classical Theory of Fields and
                Particles.
                Dover,  New York.
\item Barut A.O. (1980):
                 Foundations of Radiation Theory and Quantum
                Electrodynamics, ed.. Plenum, New York.

\item Bauer G. (1997):
        Ein Existenzsatz f\"{u}r die Wheeler--Feynman Elektrodynamik.
        Dissertation, LMU M\"{u}nchen, unpublished.

\item Bauer G. and  D\"{u}rr D. (1999):
                The Maxwell-Lorentz system of a rigid charge
                distribution,
        preprint, LMU M\"{u}nchen.

\item Baylis W.E. and Huschilt J. (1976):
        Nonuniqueness of physical solutions to the Lorentz--Dirac
        equation, Phys. Rev. D {\bf 13}, 3237-3239.

\item Bhabha H.J. (1939):
        Classical theory of electrons, Proc. Indian Acad. Sci.
         A {\bf 10}, 324-332.

\item Bhabha H.J. and Corben H.C. (1941):
        General classical theory of spinning particles in a
        Maxwell field.
        Proc. Roy. Soc. (London) {\bf A 178}, 273-314.

\item Blanco R. (1995):
        Nonuniqueness of the Lorentz--Dirac equation with the
        free--particle asymptotic condition, Phys. Rev. E {\bf
        51}, 680-689.

\item Bohm D. and Weinstein M. (1948):
        The self--oscillations of a charged particle, Phys.
        Rev. {\bf 74}, 1789-1798.

\item Bolte J. and Keppeler S. (1999):
        A semiclassical approach to the Dirac equation, Annals
        Physics {\bf 274}, 125-162.

\item Bonnor W.B. (1974):
        A new equation of motion for a radiating charged particle,
        Proc. Roy. Soc. Lond. {\bf A 337}, 591-598.

\item Born M. (1909):
        Die Theorie des starren Elektrons in der Kinematik des
        Relativit\"{a}tsprinzips, Ann. Physik {\bf 30}, 1-56.

\item Brown L.S. and Gabrielse G. (1986):
        Geonium theory: physics of a single electron or ion in a
        Penning trap, Rev. Mod. Phys. {\bf 58}, 233-278.

\item Bucherer A.H. (1909):
        Die experimentelle Best\"{a}tigung des Relativit\"{a}tsprinzips,
        Ann. Physik {\bf 28}, 513-536.

\item Caldirola P. (1956):
        A new model of classical electron, Nuov. Cim. {\bf 3},
        Supplemento 2, 297-343.

\item Carati A., Delzanno P., Galgani L., and Sassarini J. (1995):
        Nonuniqueness properties of the physical solutions of the
        Lorentz--Dirac equation, Nonlinearity {\bf 8}, 65-79.

\item Carati A. and Galgani L. (1993):
        Asymptotic character of the series of classical
        electrodynamics and an application to bremsstrahlung,
        Nonlinearity {\bf 6}, 905-914.

\item Coleman S.  (1982):
        Classical electron theory from a modern standpoint, Chapter 6
        in: Electromagnetism: Paths to Research, D. Teplitz, ed..
        Plenum, New York.

\item Corben H.C. (1961):
        Spin in classical and quantum theory, Phys. Rev. {\bf
        121}, 1833-1839.

\item Cushing J.T. (1981):
        Electromagnetic mass, relativity, and the Kaufmann
        experiments, Am. J. Phys. {\bf 49}, 1133-1149.

\item Dirac P.A.M. (1938):
                Classical theory of radiating electrons,
        Proc. Roy. Soc. A {\bf 167}, 148-169.

\item Dresden M. (1987):
        H. A. Kramers. Between Tradition and Revolution. Springer,
        New York.

\item Einstein A. (1905a): Ist die Tr\"{a}gheit eines K\"{o}rpers von
seinem Energiegehalt abh\"{a}ngig?,
        Ann. Physik {\bf 17}, 639-641,
         translation in: The Principle of
        Relativity. Dover, New York, 1952.

\item Einstein A. (1905b):
        Zur Elektrodynamik bewegter K\"{o}rper,
         Ann.
        Physik {\bf 17}, 891-921, translation in: The Principle of
        Relativity. Dover, New York, 1952.

\item Eliezer C.J. (1950):
        A note on electron theory, Proc. Camb. Phil. Soc. {\bf
        46}, 199-201.

\item Endres D.J. (1993):
        The physical solution to the Lorentz--Dirac equation for
        planar motion in a constant magnetic field, Nonlinearity
        {\bf 6}, 953-971.

\item Erber T. (1961):
        The classical theories of radiation reaction, Fortschritte
        der Physik {\bf 9}, 343-392.

\item Fermi E. (1922):
        \"{U}ber einen Widerspruch zwischen der elektrodynamischen und
        der relativistischen Theorie der elektromagnetischen
        Masse, Physikalische Zeit-\linebreak
        schrift {\bf 23}, 340-344.

\item Feynman R.P., Leighton R.B., and Sands M.  (1963):
        The Feynman Lectures in Physics. Addison--Wesley, Reading,
        Mass.

\item Fokker A.D. (1929):
        Ein invarianter Variationssatz f\"{u}r die
        Bewegung mehrerer elektrischer Massenteilchen, Z. Physik
        {\bf 58}, 386-393.

\item Ford G.W. and O'Connell R.F. (1991):
        Radiation reaction in electrodynamics and the elimination
        of runaway solutions, Phys. Letters A {\bf 157}, 217-220.

\item Ford G.W. and O'Connell R.F. (1993):
        Relativistic form of radiation reaction, Phys. Letters A
        {\bf 174}, 182-184.

\item Frenkel J. (1925):
        Zur Elektrodynamik punktf\"{o}rmiger Elektronen, Z. Physik
        {\bf 32}, 518-534.

\item Frenkel J. (1926):
        Die Elektrodynamik des rotierenden Elektrons, Z. Physik
        {\bf 37}, 243-262.

\item Galgani L., Angaroni C., Forti L., Giorgilli A., and
        Guerra F. (1989):
        Classical electrodynamics as a nonlinear dynamical system,
        Phys. Letters A {\bf 139}, 221-230.

\item Glimm J. and Jaffe A. (1987):
        Quantum Physics, A Functional Integral Point of View.
        Springer, Berlin, 2nd edition.

\item Grotch H., Kazes E., Rohrlich F., and Sharp D.H. (1982):
        Internal retardation, Acta Phys. Austr. {\bf 54},
        31-38.

\item Haag R. (1955):
        Die Selbstwechselwirkung des Elektrons, Z. Naturforsch,
        {\bf 10a}, 752-761.

\item Haken H. (1983):
        Advanced Synergetics. Instability Hierarchies of
        Self--Organiz\-ing Systems and Devices. Springer, Berlin.

\item Herglotz G. (1903):
        Zur Elektronentheorie, Nachr. K. Ges. Wiss. G\"{o}ttingen,
        (6), 357-382.

\item Huang K. (1987):
        Statistical Mechanics. Wiley, New York, 2nd edition.

\item Huschilt J. and Baylis W.E. (1976):
        Numerical solutions to two--body problems in classical
        electrodynamics: head on collisions with retarded fields
        and radiation reaction, Phys. Rev. D {\bf 13}, 3256-3261
        and D {\bf 13}, 3262-3268.

\item Jackson J.D. (1999):
        Classical Electrodynamics.  Wiley, New York, 3rd edition.

\item Jones C. (1995):
        Geometric singular perturbation theory, in Dynamical
        Systems, Proceedings, Montecatini Terme 1994,  Johnson,
        ed.,  Lect. Notes Math.
         1609, 44-118.  Springer, New York.
\item Kaufmann W. (1901):
        Series of papers in Nachr. K. Ges. Wiss. G\"{o}ttingen,
        (2), 143-155 (1901);
        (5), 291-296 (1902);
        (3), 90-103 (1903). Physikalische Zeitschrift
         {\bf 4}, 54-57 (1902). Sitzungsber. K.
        Preuss. Akad. Wiss. {\bf 2}, 449-956
        (1905). Ann.  Physik {\bf 19},
        487-553 (1906).

\item Kiessling M. (1999):
        Classical electron theory and conservation laws, Phys.
        Letters A, to appear.

\item Komech A., Spohn H., and Kunze M. (1997):
        Long--time asymptotics for a classical particle
        interacting with a scalar wave field, Commun.  PDE {\bf
        22}, 307-335.

\item Komech A., Kunze M., and Spohn H.   (1999):
        Effective dynamics of a mechanical particle coupled
        to a wave field,  Commun. Math. Phys. {\bf 203}, 1-19.

\item Komech A. and Spohn H. (1998):
        Soliton--like asymptotics for a classical particle
        interacting with a scalar wave field, Nonlin. Analysis
        {\bf 33}, 13-24.
\item Komech A. and Spohn H. (1999):
        Long--time asymptotics for coupled Maxwell--Lorentz
        equations,    J. Diff. Eq., to appear.

\item Kramers H.A. (1948):
        Nonrelativistic quantum electrodynamics and correspondence
        principle. Collected Scientific Papers,  845-869,
        North--Holland, Amsterdam 1956.

\item Kunze M. (1998):
        Instability of the periodic motion of a particle
        interacting with a scalar wave field,
        Comm. Math. Phys. {\bf 195}, 509-523.

\item Kunze M. and Spohn H. (1998):
        Radiation reaction and center manifolds,   SIAM
        J. Math. Anal., submitted.

\item Kunze M. and Spohn H. (1999a):
        Adiabatic limit of the Maxwell--Lorentz equations,
        preprint, TU M\"{u}nchen.

\item Kunze M. and Spohn H. (1999b):
        In preparation.

\item Landau L.D. and Lifshitz E.M. (1959):
        The Classical Theory of Fields. Addison--Wesley, Reading
        MA, and Pergamon Press, London.

\item Laue M.  von (1909):
        Die Wellenstrahlung einer bewegten Punktladung nach dem
        Relativit\"{a}tsprinzip, Ann. Physik {\bf 28},
        436-442.

\item Levine H., Moniz E.J., and Sharp D.H. (1977):
        Motion of extended charges in classical electrodynamics,
        Am. J. Phys. {\bf 45}, 75-78.

\item Lorentz H.A. (1892):
        La th\'eorie \'electromagnetique de Maxwell et son
        application aux corps mouvants, Arch. N\'eerl. Sci.
        Exactes Nat. {\bf 25}, 363-552.

\item Lorentz H.A. (1904a):
        Electromagnetic phenomena in system moving with any
        velocity less than that of light, Proceedings of the
        Academy of Sciences of Amsterdam, {\bf 6}, 809-831;
        contained in: The Principle of Relativity. Dover, New York,
        (1952).

\item Lorentz H.A. (1904b):
        Weiterbildung der Maxwell'schen Theorie:
        Elektronentheorie, Enzyklop\"{a}die der Mathematischen
        Wissenschaften   V2, 145-280.

\item Lorentz H.A. (1909):
        Versuch einer Theorie der elektrischen und optischen
        Erscheinungen in  bewegten K\"{o}rpern, Teubner, Leipzig,
        orig. Leyden (1895).

\item Lorentz H.A. (1915):
       The Theory of Electrons and its Applications to the
       Phenomena of Light and Radiant Heat, 2nd edition.
       Reprinted by Dover, New York (1952).

\item McManus H. (1948):
        Classical electrodynamics without singularities,Proceed.
        Royal Soc. London {\bf 195},
        323-336.

\item Minkowski H. (1908):
        Die Grundgleichungen f\"{u}r elektromagnetische Vorg\"{a}nge in
        bewegten K\"{o}rpern, G\"{o}ttinger Nachr. 53.

\item Milonni  P.W. (1994):
        The Quantum Vacuum, an Introduction to Quantum
        Electrodynamics. Academic Press, San Diego.

\item M{\o}ller C. (1952):
        The Theory of Relativity. Oxford University Press.

\item Mo T.C. and Papas C.H. (1971):
        New equation of motion for classical charged particles,
        Phys. Rev. D {\bf 4}, 3566-3571.

\item Moniz E.J. and Sharp D.H. (1974):
        Absence of runaways and divergent self--mass in
        nonrelativistic quantum electrodynamics, Phys. Rev. D {\bf
        10}, 113-1136.

\item Moniz E.J. and Sharp D.H. (1977):
        Radiation reaction in nonrelativistic quantum
        electrodynamics, Phys. Rev. D {\bf 15}, 2850-2865.

\item Neumann G. (1914):
        Die tr\"{a}ge Masse schnell bewegter Elektronen, Ann.
        Physik, {\bf 45}, 529-579.

\item Nodvik J.S. (1964):
        A covariant formulation of classical electrodynamics for
        charges of finite extension, Ann. Phys. (N.Y.) {\bf 28},
        225-319.

\item Noja D. and Posilicano A. (1998):
        The wave equation with one point interaction and the
        (linearized) classical electrodynamics of a point
        particle, Ann. Inst. H. Poincar\'e, Phys. Th\'eor. {\bf
        68}, 351-377.

\item Nyborg P. (1962):
        On classical theories of spinning particles, Nuov. Cim.
        {\bf 23}, 47-62.

\item Page L. (1918):
        Is a moving mass retarded by the reaction of its own
        radiation?, Phys. Rev. {\bf 11}, 377-400.

\item Pais A. (1972):
        The early history  of the theory of the
        electron: 1897-1947. In: Aspects of Quantum Theory, A.
        Salam and E.P. Wigner, eds.. Cambridge University Press.

\item Pais  A. (1982):
        `Subtle is the Lord'. The Science and Life of Albert
        Einstein. Oxford University Press.

\item Panofsky W.K.H. and Phillips M. (1962):
        Classical Electricity and Magnetism. Reading,
        MA, Addison--Wesley, 2nd edition.

\item Parrot S. (1987):
        Relativistic Electrodynamics and Differential Geometry.
        Sprin\-ger, Berlin.

\item Pauli W. (1921):
        Relativit\"{a}tstheorie, Enzyklop\"{a}die der Mathematischen
        Wissenschaften ${\mathrm{\bf V}} 19$, 543-775. Translated as: Theory
        of Relativity. Pergamon, New York, 1958.

\item Pearle P. (1977):
        Absence of radiationless motions of relativistically rigid
        classical electron, Found. Phys. {\bf 7}, 931-945.

\item Pearle P. (1982):
        Classical electron models, Chapter 7 in:  Electromagnetism:
        Paths to Research, D. Teplitz, ed.. Plenum, New York.

\item Plass G.N. (1961):
        Classical electrodynamic equations of motion with
        radiative reaction, Rev. Mod. Phys. {\bf 33},
        37-62.

\item Poincar\'e H. (1906):
        Sur la dynamique de l'\'electron, Rendiconti del Circolo
        Matematico di Palermo {\bf 21}, 129-176. Translated by
        H.M. Schwartz, Am J. Phys. {\bf 39}, 1287-1294, {\bf 40}, 862-872, and
        {\bf 40}, 1282-1287.

\item Richardson O.W. (1916):
        The Electron Theory of Matter. Cambridge
        University Press, 2nd edition.

\item Rohrlich F. (1960):
        Self--energy and stability of the classical electron, Am.
        J. Phys. {\bf 28}, 639-643.

\item Rohrlich F. (1973):
        The electron: development of the first elementary particle
        theory. In: The Physicist's Conception of Nature, J.
        Mehra, ed.. D. Reidel, Dordrecht.

\item Rohrlich F. (1990):
        Classical Charged Particles. Addison Wesley, Redwood City,
        CA, 2nd edition.

\item Rohrlich F. (1997):
        The dynamics of a charged sphere and the electron, Am.
        J. Phys. {\bf 65}, 1051-1056.

\item Rubinow S.I. and Keller J.B. (1963):
        Asymptotic solution of the Dirac equation,
        Phys. Rev. {\bf 131}, 2789-2796.

\item Rudin W. (1977):
        Functional Analysis. McGraw Hill, New York.

\item Sakamoto K. (1990):
        Invariant manifolds in singular perturbation problems for
        ordinary differential equations, Proc. Roy. Soc.
        Edinburgh,
        Sect. A {\bf 116}, 45-78.

\item Scharf G. (1994):
        From Electrostatics to Optics. Springer, Heidelberg.

\item Schild A. (1963):
        Electromagnetic two--body problem, Phys. Rev. {\bf 131}, 2762.

\item  Schott G.A. (1912):
        Electromagnetic Radiation. Cambridge University Press.

\item  Schott G.A. (1915):
        On the motion of the Lorentz electron, Phil. Mag. {\bf 29},
        49-62.

\item Schweber S. (1994):
        QED and the Men Who Made It: Dyson, Feynman, Schwin\-ger,
        and Tomonaga. Princeton University Press.

\item Schwinger J. (1949):
        On the classical radiation of accelerated electrons, Phys.
        Rev. {\bf 75}, 1912-1925.

\item Schwinger J. (1983):
        Electromagnetic mass revisited, Found. Phys.
          {\bf 13}, 373-383.

\item Shen C.S. (1972a):
        Magnetic bremsstrahlung in an intense magnetic field,
        Phys. Review D {\bf 6}, 2736-2754.

\item Shen C.S. (1972b):
        Comment on the new equation of motion for classical
        charged particles, Phys. Rev. D {\bf 6}, 3039-3040.

\item Shen C.S. (1978):
        Radiation and acceleration of a relativistic charged particle in an
        electromagnetic field,
        Phys. Rev. D {\bf 17}, 434-445.

\item Sommerfeld A. (1904):
        Zur Elektronentheorie, 1. Allgemeine Untersuchung des
        Feldes einer beliebig bewegten Ladung, 2. Grundlagen f\"{u}r
        eine allgemeine Dynamik  des Elektrons, 3. \"{U}ber
        Lichtgeschiwindigkeits- und
        \"{U}berlichtgeschwin\-digkeitselektronen, Nachr. der Kgl.
        Ges. der  Wiss. G\"{o}ttingen, Math.-phys. Klasse, S. 99-130,
        S. 363-469, (1905) S. 201-235.

\item Sommerfeld A. (1905):
        Simplified deduction of the field and the forces of an
        electron moving in any given way, Akad. van Wetensch. te
         Amsterdam, {\bf 7}, 346-367.

\item Spohn H. (1991):
        Large Scale Dynamics of Interacting Particles. Springer, Berlin.

\item Spohn H. (1998):
                Runaway charged particles and center manifolds,
                unpublished manuscript.

\item Spohn H. (1999a):
        The critical manifold of the Lorentz--Dirac equation,
        preprint.

\item Spohn H. (1999b): privates notes.

\item Spohn H. (1999c)
        Semiclassics of the Dirac equation, in preparation.
\item Stephas P. (1992):
        Analytic solutions for Wheeler--Feynman interaction,
        J. Math. Phys. {\bf 33}, 612.

\item Teitelbom C., Villarroel D., and van Weert Ch. G. (1980):
        Classical electrodynamics of retarded fields and point
        particles. Rev. Nuov. Cim. {\bf 3}, 1-64.

\item Thirring W. (1997):
        Classical Mathematical Physics, Dynamical Systems and Field
         Theory. Springer, New York,
        3rd edition.

\item Thomas L.H. (1926):
        The motion of the spinning electron, Nature {\bf 117},
        514.
\item Thomas L.H. (1927):
        On the kinematics of an electron with an axis, Phil. Mag. {\bf 3},
        1-22.

\item Thomson J.J. (1897):
        Cathode rays, Phil. Mag. {\bf 44}, 294-316.

\item Valentini A. (1988):
        Resolution of causality violation in the classical
        radiation reaction, Phys. Rev. Lett. {\bf 61}, 1903-1905.

\item Wheeler J.A. and Feynman R.P. (1945):
        Interaction with the absorber as the mechanism of
        radiation, Rev. Mod. Phys. {\bf 17}, 157-181.

\item Wheeler J.A. and Feynman R.P. (1949):
        Classical electrodynamics in terms of direct interparticle
        action, Rev. Mod. Phys. {\bf 21}, 425-433.

\item Wildermuth K. (1955):
        Zur physikalischen Interpretation der
        Elektronenselbstbeschleunigung, Z. Naturf. {\bf 10 a}, 450-459.

\item Yaghjian A.D. (1992):
        Relativistic Dynamics of a Charged Sphere. Lect. Notes
        in
        Physics  {\bf m 11},  Springer, Berlin.
\end{list}
\end{document}